RÉPUBLIQUE ALGERIENNE DÉMOCRATIQUE ET POPULAIRE
MINISTÈRE DE L'ENSEIGNEMENT SUPÉRIEUR ET DE LA RECHERCHE SCIENTIFIQUE
UNIVERSITÉ DES SCIENCES ET DE LA TECHNOLOGIE
« HOUARI BOUMEDIENNE »

**FACULTÉ DE PHYSIQUE**

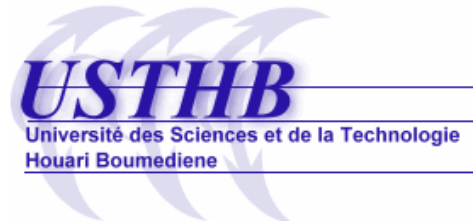

**MÉMOIRE**
Présenté pour l'obtention du diplôme de :
**MAGISTER**

EN: PHYSIQUE

Spécialité : Physique Théorique Basses et Moyennes Energies

Par
**HADJARA MASSINISSA**

**SUJET**

# ÉTUDE ET MISE EN ŒUVRE DES TECHNIQUES TEMPS-DISTANCE EN HÉLIOSISMOLOGIE

Soutenu publiquement le 26/03/2006, devant le jury composé de:

| | | |
|---|---|---|
| M$^r$ H. DJELLOUAH | Professeur (USTHB) | Président |
| M$^r$ N. SEGHOUANI | Chargé de Recherche (CRAAG) | Rapporteur |
| M$^{me}$ N. ALLAL | Professeur (USTHB) | Examinatrice |
| M$^{me}$ Z. IGHEZOU | Professeur (USTHB) | Examinatrice |
| M$^{me}$ F. CHAFA | Maître de Conférences (USTHB) | Examinatrice |
| M$^r$ T. ABEDLATIF | Directeur de Recherche (CRAAG) | Examinateur |
| M$^r$ A. HARFOUCHE | Chargé de Cours (USTHB) | Invité |

# Table des matières

# Dédicace

A celui à qui je dois toute ma force et ma détermination

                A mon cher père : Vava A3ziz.

A celle à qui je dois toute ma générosité et ma tendresse

                A ma chère mère : Yemma Ta3zizth.

A ceux à qui je dois toute ma gaieté et ma joie de vivre

         A mes chers sœur et frère (Kahina & Nourdine) : Oultsma dhi Eyma I3zizine.

" L'Imagination est plus importante que la connaissance. La connaissance est limitée.
L'imagination englobe le monde entier, stimule le progrès, suscite l'évolution.
Sans imagination, pas de progrès."

<div style="text-align: right;">Albert Einstein.</div>

<div style="text-align: center;">"Ce qui ne me tue pas me rend plus fort."</div>

<div style="text-align: right;">Friedrich Nietzsche ; Crépuscule des idoles (1888)</div>

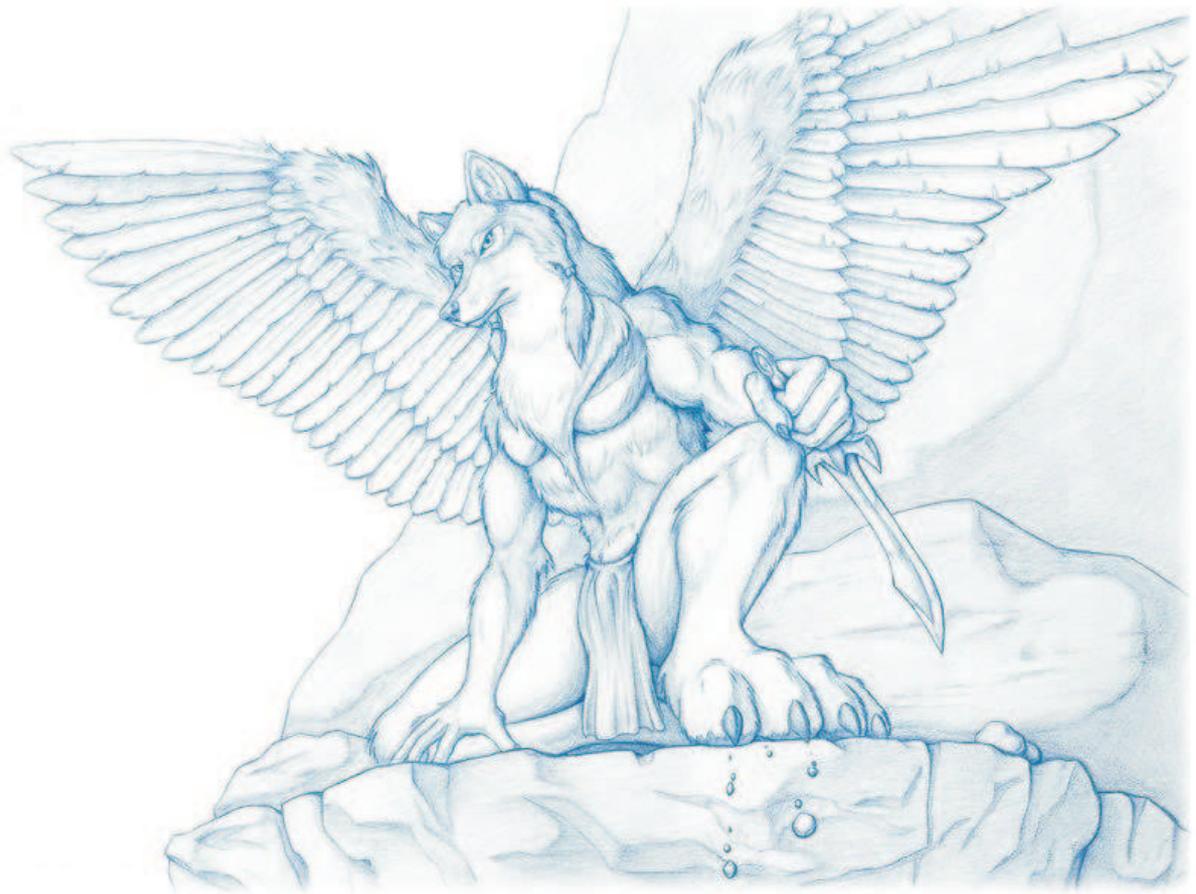

# Chapitre 1

# Introduction générale

Les étoiles vibrent comme les cordes d'un violon. Certaines étoiles particulières baptisées : céphéides, "RR Lyrae", sont connues depuis longtemps pour être des étoiles variables, chères aux astronomes amateurs. Leur luminosité peut varier au cours du temps de façon spectaculaire : une étoile brillante peut ainsi devenir invisible pour reprendre son éclat initial après une période allant de quelques heures à une dizaine de jours. Pendant longtemps, on s'est interrogé sur l'origine de ce comportement singulier. Aujourd'hui, nous savons que ces variations de lumière sont le résultat de variations à grande échelle de ces étoiles : elles se dilatent et se contractent périodiquement.

Au cours des toutes dernières années, les astronomes ont découvert que le phénomène de vibration des étoiles n'est pas rare. En revanche, il est rare que ces vibrations atteignent une amplitude suffisante pour être visibles directement depuis la terre. Les études théoriques montrent, en effet, qu'une majorité d'étoiles présente ce phénomène, mais avec de très faibles amplitudes. Ainsi, notre Soleil lui-même vibre, et cette découverte a donné naissance il y a une dizaine d'années à une nouvelle et fructueuse discipline : "l'Héliosismologie". Les vibrations du Soleil ont des amplitudes plus faibles que celles des céphéides ou des $\delta$ Scuti, mais elles sont également beaucoup plus rapides et plus complexes et elles fournissent des informations inestimables sur l'intérieur solaire. Sans oublier que les étoiles animées de vibrations similaires à celles du Soleil sont beaucoup plus difficiles à observer ; un observateur placé sur une planète autour de l'étoile la plus proche de nous, proxima $\alpha$ du Centaure, aurait ainsi bien du mal à détecter les vibrations solaires. Les astrophysiciens se sont mis en tête de découvrir et mesurer les vibrations rapides des étoiles ; c'est qu'ainsi est née une discipline encore plus récente que l'héliosismologie : l'astérosismologie ou sismologie des étoiles. [1]

Depuis la découverte de certaines des oscillations solaires, on n'a pas cessé d'essayer d'exploiter ce nouveau filon afin de diagnostiquer les intérieurs stellaires. Afin de comprendre le potentiel diagnostique des oscillations solaires, une perspicacité de base dans l'étude des propriétés des oscillations stellaires est exigée. Les oscillations observées ont des petites amplitudes comparées aux magnitudes absolues des étoiles et par conséquent peuvent être décrites en tant que perturbations linéaires, autour des modèles solaires résultant des calculs d'évolution. Les fréquences fournissent un diagnostic direct des propriétés internes du soleil : donnant ainsi un modèle solaire. Les aspects appropriés des fréquences peuvent être calculés avec une très grande précision et les différences entre les fréquences observées et calculées peuvent être liées aux erreurs dans le modèle. L'héliosismologie contribue donc à l'amélioration des modèles, et par conséquent, au renforcement de nos connaissances sur le fonctionnement du Soleil.

---

[1] Ne confondons pas ici 'astéro' qui signifie 'étoile' et 'astro' qui signifie 'astre' dans un sens plus large.

L'avènement de l'héliosismologie a ouvert une nouvelle ère dans la recherche scientifique sur les étoiles en général et le Soleil en particulier. En effet, avant l'héliosismologie, nos connaissances sur l'intérieur du Soleil étaient purement théoriques, se basant essentiellement sur des modèles d'évolution des étoiles. L'héliosismologie a permis, grâce à la mesure précise des ondes acoustiques piégées à l'intérieur du Soleil, d'avoir une connaissance très "fine" des différents paramètres internes du Soleil tels que la rotation, la température et la densité. On peut distinguer, à ce niveau, deux types d'approches : la première considère le Soleil dans sa globalité ; il s'agit de l'héliosismologie globale. La seconde s'intéresse aux "détails" ou aux phénomènes locaux ; il s'agit de l'héliosismologie locale.

Dans ce mémoire, nous nous intéressons plus particulièrement à cette deuxième approche par le biais d'une technique inspirée de la sismologie terrestre et qui relie le temps de parcours de l'onde, pour aller d'un point à un autre à la surface du Soleil, au trajet parcouru par celle-ci. Il s'agit de la technique temps-distance.

## 1.1 Problématique

De la même manière que les sismologues étudient l'intérieur de la terre à partir des ondes générées par les séismes, l'héliosismologie est la science qui s'intéresse à l'étude des ondes oscillatoires à l'intérieur du Soleil. La température, la composition chimique, les vitesses de rotation à différentes profondeurs sont autant de facteurs qui influencent les fréquences d'oscillations des ondes piégées à l'intérieur du Soleil. L'observation et la mesure des fréquences (ou vitesses) d'oscillations des principaux modes, à la surface du Soleil, nous permettent d'étudier les propriétés internes de l'astre.

Dans notre étude, nous utilisons une méthode héliosismique locale, nommée "Time-Distance" ; cette dernière, après un traitement approprié (remapping, tracking, et filtrage) des données, issues du réseau terrestre d'observation GONG (Global Oscillation Network Group), nous permet de déduire, par corrélation des signaux observés, la relation entre le temps de trajet de l'onde et sa distance de parcours [2] entre différents points à la surface en fonction des différents paramètres sub-surfaciques solaires, via la relation de dispersion. Une fois les temps de parcours établis, par approximation de la fonction de corrélation par un paquet d'onde gaussien, et en se basant sur le principe de Fermat, traitant des ondes acoustiques, on trouve la relation entre ces temps et les paramètres internes du milieu traversé. Deux modèles sont considérés dans cette étude. Le premier n'inclut pas les effets du champ magnétique et prend en compte la vitesse d'écoulement et la vitesse du son. Le second, quant à lui, inclut les effets magnétiques, par le biais de la vitesse d'Alfven [3], en plus des paramètres précédemment cités.

La dernière étape consiste à inverser les temps obtenus, via la relation émise par le principe de Fermat, ce qui nous permet ainsi de remonter aux paramètres internes solaires souhaités.

---
[2]Le nom de la méthode vient du fait que la méthode repose essentiellement sur la relation (la fonction de corrélation) qui relie le temps de trajet de l'onde à sa distance de parcours.
[3]La vitesse d'Alfven est la vitesse de perturbation du champ magnétique.

## 1.2 Plan de la thèse

Notre thèse suit le plan suivant :

– Chapitre 1

  Introduction générale, posant des définitions et fixant l'objet du travail.

– Chapitre 2

  Dans le deuxième chapitre, nous introduisons des notions générales sur le Soleil et sur l'héliosismologie.

– Chapitre 3

  Dans le troisième chapitre, nous explicitons les équations de base en héliosismologie.

– Chapitre 4

  Le quatrième chapitre est réservé à la méthode héliosismologique locale "Temps-Distance" et à tous ses aspects théoriques.

– Chapitre 5

  Le cinquième chapitre est consacré aux données expérimentales, à leur méthodes d'analyse et à la détermination des temps de parcours des ondes traitées.

– Chapitre 6

  Le sixième chapitre porte sur l'aspect théorique de l'inversion des temps de parcours via le principe de Fermat ; ce qui nous permet ainsi de remonter aux paramètres internes solaires tels que la vitesse des flux (ce qui fait l'objet de notre étude) et la vitesse du son.

– Chapitre 7

  Nous terminerons enfin par la discussion des résultats obtenus dans le cinquième chapitre ainsi que par les perspectives.

Nous joignons en annexe un complément sur les outils de base utilisés dans cette thèse.

# Chapitre 2

# Généralités

Nous entamons ce chapitre par des généralités sur le soleil, la science qui le régit en général et la science qui étudie ses oscillations en particulier. Ce chapitre est composé de trois sections majeures : la première décrit le Soleil, les modèles mis en vigueur pour son étude et les équations de base auxquelles obéit son évolution. La seconde introduit la notion et l'historique de l'Héliosismologie globale (la science qui étudie les oscillations de Soleil dans sa globalité). La troisième et dernière section, quant à elle, décrit l'Héliosismologie locale (science qui étudie localement les oscillations solaires) et énumère ses différentes techniques.

## 2.1 Le Soleil

Le soleil est une étoile des plus communes parmi les 300 milliards autres qui peuplent la Voie lactée. Avec une masse de $1.989 \times 10^{30} Kg$, il se trouve à mi-chemin entre les étoiles de grande masse (10 à 30 fois la masse solaire, relativement peu nombreuses et très lumineuses) et celles de faible masse (d'un 1/10 à 1/20 de la masse solaire, très nombreuses et de très faible luminosité). Malgré tout, il est aujourd'hui la seule étoile de l'univers dont on puisse extraire, en permanence, des informations, provenant de toutes ses différentes couches et ce malgré les 150 millions de kilomètres qui nous en séparent. Il est situé 250 000 fois plus près de la terre que la seconde plus proche étoile : α du Centaure. Formidable laboratoire, il nous permet d'accroître nos connaissances sur le fonctionnement, l'histoire, l'évolution des étoiles et de l'univers. Agée de 4.5 milliards d'années, notre étoile n'est encore qu'à la moitié de sa vie. D'un rayon qui équivaut à 109 fois celui de la terre ($6.9610^5$ Km), notre astre est composé de 71% d'Hydrogène (noté $X$), de 27% d'Héluim ($Y$) et de 2% de plus de 70 autres éléments plus lourds ($Z$), avec $X + Y + Z = 1$. Le soleil est formé de trois zones principales (voir Fig (2.1)) :

1. **Le cœur :** d'une densité supérieure à 150 fois celle de l'eau et d'une température avoisinant les 150 millions de degrés (température et pression nécessaires à la fusion de l'Hydrogène en Hélium), il occupe 15% du volume solaire et est le siège de réactions thermonucléaires. Une énergie équivalente à celle libérée par l'explosion de 100 milliards de bombes à hydrogène d'une mégatonne y est générée à chaque seconde.

2. **La zone radiative :** qui s'étend sur un demi rayon et dont la densité vaut 15 fois celle de l'eau ; un photon mettra jusqu'à un million d'années pour traverser cette zone,

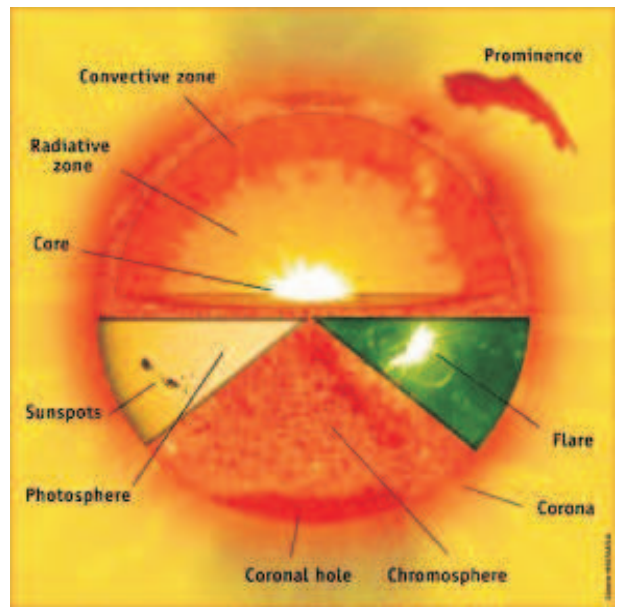

FIG. 2.1 – Coupe schématique du Soleil [SOHO/NASA].

3. **La zone convective :** elle est animée de mouvements tourbillonnaires qui évacuent la chaleur vers l'atmosphère solaire (la photosphère), la partie visible du soleil, là où la température atteint les $6000°K$.

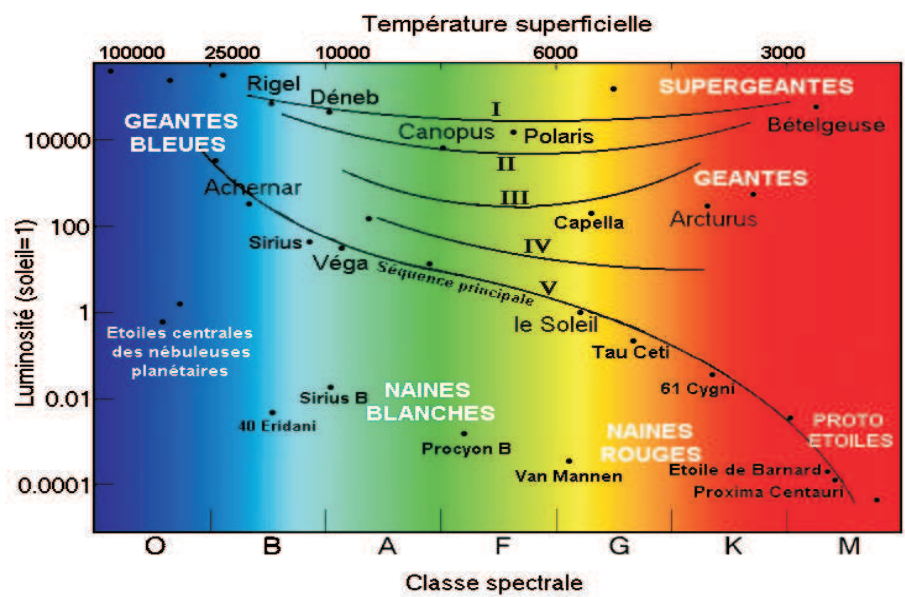

FIG. 2.2 – Classification des étoiles par leurs magnitudes et leur types spectraux [Astrosurf-Magazine].

La figure (2.2) ; communément appelée : "diagramme Hetzsprung-Russell", nous montre la place de notre soleil par rapport aux autres étoiles de l'univers dans la classification de Hetzsprung-Russell [1].

---

[1]Le diagramme de Hetzsprung-Russell résulte d'une combinaison des travaux du danois Ejnar Hetzsprung (1837-1967) et de l'américain Henry Norris Russell (1877-1957) qui travaillaient chacun de son coté à l'élaboration de celui-ci.

La détermination précise de la première étape d'évolution stellaire, qui consiste à transformer l'Hydrogène en Hélium (le cycle p-p, CNO), a été faite par Hans Bethe dans les années 40. A la même époque Chandrasekher et Schwarzschilld [72] mettent au point les bases de l'évolution stellaire, à partir desquelles des modèles, dits 'standards', permettant de suivre et de prédire l'évolution stellaire, ont été élaborés.

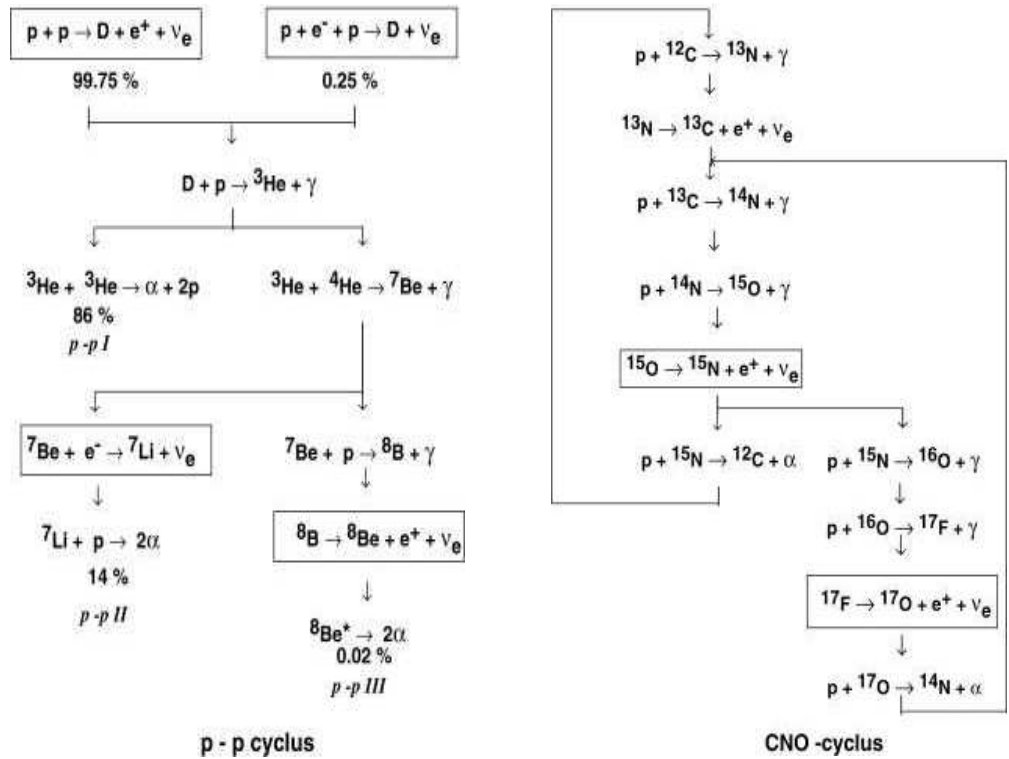

FIG. 2.3 – Les cycles p-p et CNO [Utrecht University (Netherlands)].

En 1769 Alexandre Wilson s'aperçoit que la période de révolution des taches solaires change sensiblement avec leur latitude : elle varie de presque 25 jours à l'équateur solaire à 27,5 jours à une latitude de 40° ; les soupçons de Scheiner d'un soleil fluide sont fondés.

Au début du $18^e$ siècle, les taches solaire apparurent nombreuses. Le phénomène devient cyclique, alternant des périodes de maximum et de minimum d'environ 11 ans. Ce cycle (de 22 ans) qui est dû aux champs magnétiques, fut démontré par S. Schwabe, à partir des observations collectées durant vingt-cinq ans, entre 1826 et 1851.

Par ailleurs, il y a plus de vingt ans, l'américain R. Davis installait, dans une mine du Dakota du sud, un immense réservoir de trichloréthyléne afin de détecter les neutrinos de basse énergie, émis par une réaction nucléaire secondaire au cœur du Soleil. Ces neutrinos transforment, chaque jour, quelques atomes de chlore en argon-37 radioactif. Ils sont patiemment comptabilisés dans les détecteurs. Au fil des années, les résultats de cette expérience unique, ont révélé que le flux de neutrinos détecté est en moyenne trois à quatre fois plus faible que la prédiction théorique du modèle de référence. Or ces neutrinos proviennent bien des régions les plus centrales du Soleil (moins de 10% du rayon) et leur taux de production varie extrêmement vite avec la température. On est donc conduit à penser que la température centrale calculée ($15,5 \times 10^6 K$) pourrait être en fait surestimée de 15 à 20%. L'origine de cette différence préoccupante est très débattue : faut-il remettre en cause ce que nous savons de la structure du Soleil, et par conséquent, celle de toutes les étoiles, i.e. directement ou indirectement toute l'astrophysique ? ou bien, comme celà a été suggéré, la solution réside-t-elle dans des propriétés inédites des neutrinos ? C'est à ce genre de questions que l'Héliosismoligie tente de répondre.

De plus, et afin de bien étudier le soleil dans sa globalité, les scientifiques ont mis au point des modèles dits standard :

## 2.2 Modèles standards

Un nombre de modèles solaires ont été construits, chacun basé sur une modification simple de la physique d'un modéle solaire de référence. En outre, un modèle combinant plusieurs améliorations a été élaboré pour fournir un meilleur modèle solaire. Des améliorations ont été apportées aux taux de réaction nucléaire, à l'équation d'état, aux opacités, et au traitement de l'atmosphère. L'impact de ces améliorations sur la structure et sur les fréquences des bas $\ell$ des modes $p$ (voir la sous-section 2.6.1) du modèle sont en étude actuellement. Nous constatons que le modèle solaire combiné, qui est basé sur la meilleure physique qui nous est disponible, reproduit le spectre observé d'oscillation (pour les bas $\ell$) dans les limites des erreurs acceptables liées aux incertitudes dans le modèle physique (principalement les opacités).

Les modèles sont basés principalement sur les équations d'évolution stellaire et leur hypothèse (que nous retrouverons ci-dessous), ainsi que sur les équations d'états et leur considération qui sont propres à chaque modèle. Il faut savoir que $Y_0$, le taux d'hélium, et $\alpha$ la longueur de mélange, sont considérés comme des paramètres libres, qu'il faut ajuster, afin de retrouver les mêmes résultats observés sur notre Soleil aujourd'hui, tels que la masse, la luminosié...etc.

## 2.3  Équations de base de l'évolution solaire

Le Soleil étant une étoile parmi tant d'autres, il obéit donc aux équations de base de l'évolution stellaire. En supposant l'étoile de symétrie sphérique, sans champ magnétique et sans rotation, les équations s'écrivent, en coordonnées sphériques, comme suit [25] :

Conservation de la masse :
$$\frac{d\mathtt{m}}{dr} = 4\pi r^2 \rho. \tag{2.1}$$

Équilibre hydrostatique :
$$\frac{dp}{dr} = -\frac{G\mathtt{m}\rho}{r^2}. \tag{2.2}$$

Transport de l'énergie :
$$\frac{dT}{dr} = \nabla_T \frac{T}{p} \frac{dp}{dr}. \tag{2.3}$$

Conservation de l'énergie :
$$\frac{d\mathtt{L}}{dr} = 4\pi r^2 \left[ \rho\varepsilon - \rho \frac{d}{dt}\left(\frac{u}{\rho}\right) + \frac{p}{\rho}\frac{dp}{dt} \right]. \tag{2.4}$$

Où $r$ est la distance au centre, $p$ est la pression, $\mathtt{m}$ est la masse à l'intérieur de la sphère de rayon $r$, $\rho$ est la densité, $T$ est la température, $\mathtt{L}$ est la luminosité (le taux d'énergie émergeant de la sphère $r$ par unité de temps), $\varepsilon$ est le taux d'énergie produit par les réactions nucléaires (incluant la perte d'énergie due à l'émission des neutrinos) par unité de temps et de masse, et $u$ le taux d'énergie interne par unité de volume. En outre, le gradient de température est caractérisé par $\nabla_T = d\ln T/d\ln p$ et est déterminé par le mode du transport d'énergie. Quand l'énergie est transportée par rayonnement, par exemple, alors $\nabla_T = \nabla_{rad}$, où le gradient radiatif est donné par :

$$\nabla_{rad} = \frac{3}{16\pi caG} \frac{\kappa p}{T^4} \frac{L(r)}{m(r)} \tag{2.5}$$

Ici $c$ est la vitesse de la lumière, $a$ est la constante de densité du rayonnement et $\kappa$ l'opacité, définie de sorte que $l = 1/(k\rho)$ soit le libre parcours moyen d'un photon ; l'opacité nous permet de calculer le transport d'énergie par transfert radiatif, ainsi que l'énergie déposée dans chaque couche. Dans les régions où $\nabla_T$ excède le gradient adiabatique $\nabla_{ad} = (\partial \ln T/\partial \ln P)_s$, la dérivée étant prise à une entropie spécifique constante $s$, la couche devient instable. Dans ce cas ($\nabla_T \geq \nabla_{ad}$, appelé aussi le critère de Schwarzschild), le transport d'énergie s'effectue principalement par un mouvement convectif. il faut noter que la description détaillée de la convection est très difficile, mais il faut savoir que lorsque l'on est en présence d'une zone convective nous utilisons la théorie de la longueur de mélange (généralement noté MLT Mixing Length Theory) qui remonte à Prandtl en 1925, et qui fait l'analogie entre la convection et le transfert de chaleur moléculaire, en remplaçant les molécules par des éléments de masse macroscopiques dont le libre parcours moyen est la longueur de mélange $l$.

La génération d'énergie au soleil résulte de la transformation de l'hydrogène en hélium, et la réaction s'écrit comme suit :

$$4\,^1H \rightarrow\,^1He + 2e^+ + 2\nu_e \tag{2.6}$$

En satisfaisant les contraintes de la conservation de la charge et du nombre de leptons. Ici les positrons sont immédiatement annihilés, alors que les neutrinos électroniques s'échappent du soleil essentiellement sans réaction avec la matière et représentent donc une perte d'énergie immédiate. Cette réaction implique différents ordres de réactions, selon la température. Ces réactions diffèrent sensiblement dans la perte d'énergie des neutrinos et par conséquent dans l'énergie réellement disponible dans l'étoile.

Le changement de la composition résultant de l'équa.(2.6), gouverne en grande partie l'évolution solaire. Jusqu'à assez récemment, les calculs des modèles solaires "standards" n'ont pas inclus tout autre effet qui change la composition. Cependant, Noerdlinger en 1977 [65] a précisé l'importance de la diffusion de l'hélium dans le soleil. Et depuis lors, l'importance de la diffusion et de l'arrangement de l'hélium a fait ses preuves en Héliosismologie et ces processus sont maintenant généralement inclus dans les calculs. Spécifiquement, le taux de changement de l'abondance d'hydrogène s'écrit ainsi :

$$\frac{\partial X}{\partial t} = R_H + \frac{1}{r^2 \rho} \frac{\partial}{\partial r}\left[r^2 \rho \left(D_H \frac{\partial X}{\partial r} + V_H X\right)\right] \tag{2.7}$$

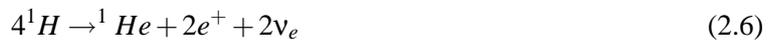

ici $R_H$ est le taux de changement de l'abondance d'hydrogène des réactions nucléaires, $D_H$ est le coefficient de diffusion et $V_H$ est la vitesse d' arrangement. Des équations semblables sont naturellement satisfaisantes pour les abondances d'autres éléments ; l'hydrogène s'élève vers la surface, et les éléments plus lourds comprenant l'hélium descendent vers l'intérieur.

Les équations de base de l'évolution stellaire étant posées, on remarque bien que dans les équa.(2.1)-(2.4), il y a beaucoup plus de variables que d'équations ; le système est ouvert. Une fermeture donc s'impose, elle est faite par les équations d'états et leur hypothèses.

## 2.4 Équation d'état

L'équation d'état permet, à partir de deux des trois paramètres physiques $(p, T, \rho)$ de déterminer, pour un mélange donné, le troisième paramètre physique.

Il existe plusieurs équations d'état pour les modèles stellaires. Deux équations d'état sont les plus utilisées ces dernières années dans les modèles solaires ; l'équation d'état MHD qui a été développée par Mihalas, Hummer et Däppen en 1988 [28], utilisant le schéma chimique, prend en couple les atomes et les ions, considérant l'ionisation comme un processus chimique. Et l'équation d'état OPAL (opacity laboratory), récemment développée par Rogers en 1996 [70], elle suit le schéma physique, en utilisant l'ensemble de grand canonique (mesures), et fait intervenir les électrons et les noyaux comme constituants fondamentaux du plasma. Elles se présentent toutes les deux sous formes de tables pour les différentes valeurs de la fraction de masse d'hydrogène, d'hélium, de la température, de la densité et de la pression. La difficulté principale dans le développement de ces équations d'état vient du fait que l'on est en présence d'un plasma partiellement ionisé.

Récemment, Elliott & Kosovichev en 1998 [37] ont montré que, avec la précision actuelle de l'héliosismologie, les effets relativistes au coeur du soleil ne pouvaient plus être négligés. Actuellement, aucune des équations les plus utilisées dans la construction des modèles solaires ne prend en compte ces effets.

## 2.5 Oscillations solaires

Les oscillations solaires se manifestent dans l'atmosphère de différentes manières : le déplacement fait bouger l'atmosphère, changeant le transport d'énergie dans les couches externes du soleil causant ainsi des oscillations dans le rendement d'énergie ; il en résulte ainsi des oscillations de la température atmosphérique qui sont mises en évidence dans les raies spectrales solaires. Chacun de ces effets peut être employé pour observer les oscillations ; puisqu'elles reflètent toutes les mêmes modes, elles devraient, évidemment, rapporter les mêmes fréquences d'oscillations. Le choix de la technique d'observation est alors déterminé par une combinaison de considérations et de propriétés techniques du bruit de fond, des effets de l'atmosphère terrestre pour des observations au sol et des effets d'autres variations de l'atmosphère solaire.

### 2.5.1 Observation des oscillations

Pour réaliser la suppression de bruit et la résolution exigée de fréquence, les observations sont typiquement analysées avec cohérence sur plusieurs mois. En outre, les lacunes temporelles dans les données présentent des bandes latérales de fréquence dans le spectre de puissance et compliquent la détermination des fréquences. Par conséquent, les données avec des interruptions minimales sont fortement souhaitables. Ceci implique immédiatement le besoin de la combinaison des données de plusieurs emplacements autour de la terre afin de compenser le cycle jour/nuit. Ceci est réalisé grâce à des réseaux d'observations terrestres à l'instar de BISON (BIrmingham Solar Oscillation Network), GONG (Global Oscillation Network Groupe), IRIS (Internationnal Research in the Interior of the Sun),...etc. Par ailleurs, l'observation depuis l'espace via des satellites, permet aussi d'aboutir au même souhait. Nous citerons le plus célébre utilisé à cette fin : SOHO (SOlar and Heliospheric Observatory), qui a à son bord 4 experiences : GOLF (Global Oscillation at Low Frequencies), VIRGO (Variability of solar Iradience Gravity Oscillation), MDI (Michelson-Doppler Imager) et SOI (Solar Oscillation Investigation).

Les observations les plus détaillées des oscillations solaires, ont été effectuées et mesurées à partir de l'effet Doppler des raies dans le spectre solaire. Comme illustré dans la Fig.(2.4), ceci peut être fait en mesurant l'intensité dans deux bandes de chaque côté d'une raie spectrale appropriée. Si les intensités sont enregistrées au moyen d'un détecteur, le résultat est une image de vitesse, mesurant le mouvement de la surface solaire avec une résolution spatiale potentiellement élevée. Alternativement, avec la lumière du soleil passant du filtre au détecteur, on obtient une vitesse moyenne du disque, ce qui revient à observer le soleil comme une étoile.

La technique peut être plus intensivement développée pour l'observation dans l'espace grâce au tachymètre dit de Fourier. Ici la ligne de décalage est obtenue à partir de quatre mesures dans les bandes étroites à travers une ligne spectrale donnée. Ceci permet la définition d'une mesure qui est essentiellement linéaire dans les lignes-de-vue de la vitesse, sur la gamme considérable des vitesses produites au-dessus de la surface solaire. Dans les réalisations réelles, les bandes spectrales sont définies par des interféromètres de Michelson. Des exemples d'images Doppler obtenues en utilisant cette technique sont montrés dans la Fig.(2.5).

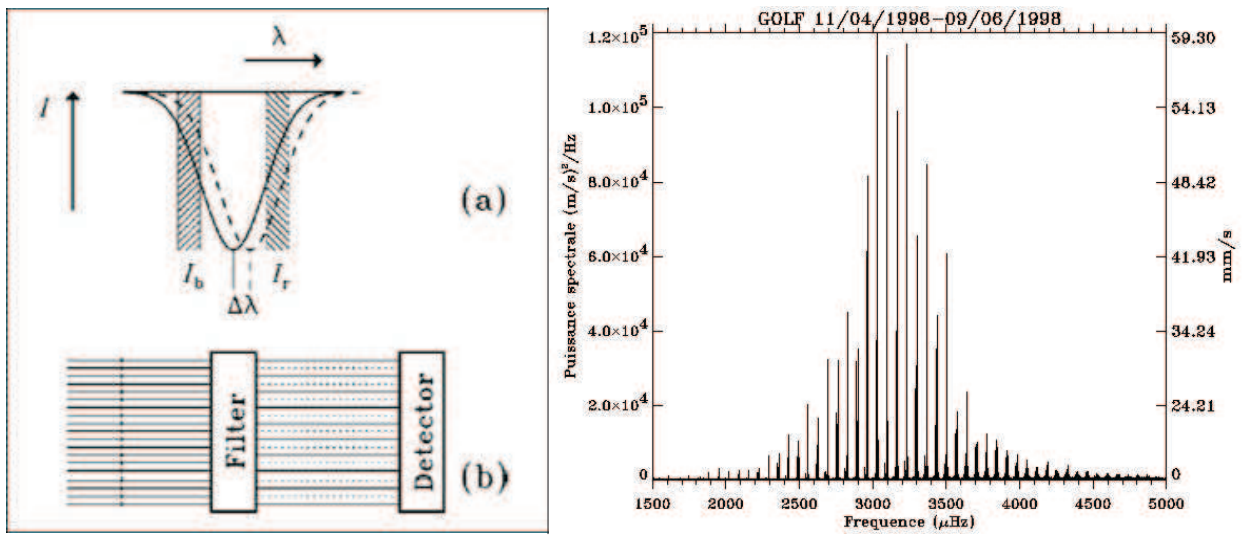

FIG. 2.4 – A gauche : illustration schématique des observations de vitesse Doppler. En a) la vitesse de ligne de vue décalée dans l'axe des longueurs d'onde par $\Delta\lambda$, de la position représentée en ligne continue à celle en discontinue. Ceci change les intensités $I_b$ et $I_r$ mesurées dans l'intervalle étroit des longueurs d'ondes (voir les deux rectangles hachurés), aussi bien que le rapport $(I_b - I_r)/(I_b + I_r)$ fournit une mesure du décalage et par conséquent de la vitesse. b) illustre l'installation expérimentale du filtre qui alterne de laisser passer la lumière à travers les bandes entre $I_b$ et $I_r$. Si un détecteur de formation d'image est employé, les images résultantes en $I_b$ et $I_r$ peuvent être combinées pour former une image Doppler [25]. A droite : le spectre de puissance des fréquences observé sur le soleil par l'instrument GOLF [SOHO].

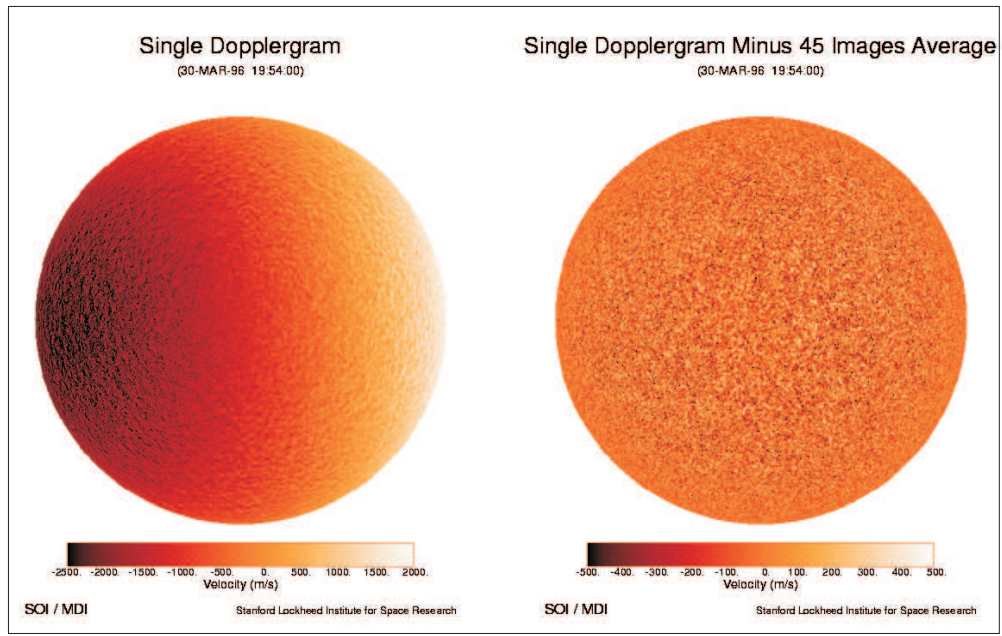

FIG. 2.5 – Image Doppler obtenue avec l'instrument MDI [25].

## 2.6 Héliosismologie globale

Milieu fluide à trois dimensions, la sphère solaire vibre suivant des millions de notes (qui sont graves, du fait de l'immensité du soleil, 17 octaves plus basses que celles qu'on trouve sur un piano) qui varient par leur fréquence et leur structure géométrique. L'Héliosismolgie (combinaison de trois mots : helios : du grec classique signifiant le soleil ou la lumière, seismos : également du grec qui signifie tremblement et logos : raisonnement ou science) est justement la science qui étudie ces vibrations. L'Héliosismologie est un outil très puissant qui nous fournit beaucoup d'informations sur la structure solaire, sa rotation, ses températures et sa densité... etc.

Les premières indications, sur les oscillations solaires, furent données par Plaskett en 1916 [66], qui observa des fluctuations des vitesses Doppler dans les raies solaires, mais il ne pouvait être sûr qu'elles étaient bien propres au soleil ou bien dues aux perturbations atmosphériques terrestres. C'est Hart (1954-1956) [50] [51] qui établit que l'origine était bien solaire, mais la véritable première détection des oscillations solaires est attribuée à Robert Leighton, qui, en 1962 [61], détecta des oscillations périodiques locales des vitesses Doppler avec une période de 300 s (5 min) et une durée de vie de quelques périodes ; observations qui furent confirmées par Evans & Michard [38] la même année. Frazier (1968) [39] utilisa la transformée de Fourier (on passe ainsi des spectres (énergie) aux fréquences, nombres d'ondes (espace-temps)), indiquant alors une nature moins superficielle du phénomène. Ulrich, en 1970 [76], puis Leibacher & Stein, en 1971 [60], furent les premiers à essayer d'expliquer l'origine des oscillations en suggérant que les oscillations résultaient, d'ondes acoustiques (sonores), provenant de l'intérieur solaire. Wolf (1972) [78] et Ando & Osaki (1975) [2] développèrent les calculs dans cette voie et s'aperçurent que la gamme des fréquences (nombres d'ondes) était linéairement instable. Deubner (1975) [29] fut le premier à identifier des arrêtes dans le diagramme "nombre d'onde-fréquence". En 1975, à Cambridge, A. Hill annonce la détection d'oscillations dans le diamètre apparent solaire ; ce fut la première suggestion d'oscillations véritablement globales du soleil, ce qui a ainsi ouvert la possibilité d'employer les données dans l'étude de l'intérieur solaire. En 1979, Claverie [12] identifia la structure modale des oscillations de 5 min dans les observations des vitesses Doppler intégrées sous le disque solaire (observation sensible seulement aux oscillations bas degré (les plus pénétrantes)) ; ce fut la première identification des modes dit globaux. Entre 1979 et 1980, pendant l'été austral, Grec et son équipe [49], du laboratoire de Nice et afin d'éviter l'alternance jour-nuit, placèrent leur instruments au pôle sud et recueillirent des fréquences solaires d'une grande résolution que Tassoul [74] analysa en 1980 et résolut ainsi les multiplets individuels dans le spectre des oscillations à bas degrés. Ceci permit de comparer les données de fréquences (incluant le petit saut dans celles-ci) aux prévisions des modèles solaires. En 1982 Gough [47] précisa que le petit saut de fréquences était lié à la courbure des ondes sonores près du noyau solaire. Duvall & Harvey, en 1983 [31], observèrent en détail des degrés de modes intermédiaires, entre les bas observés par Claverie 1979 et les hauts degrés observés par Deubner en 1975, et démontrèrent ainsi que Christensen-Dalsgaard et Gough [11], qui suggérèrent la même origine aux deux modes, avaient raison. Cette observation permit une identification de l'ordre des modes et une meilleure connaissance des propriétés internes de notre soleil, telles que la rotation et la vitesse du son.

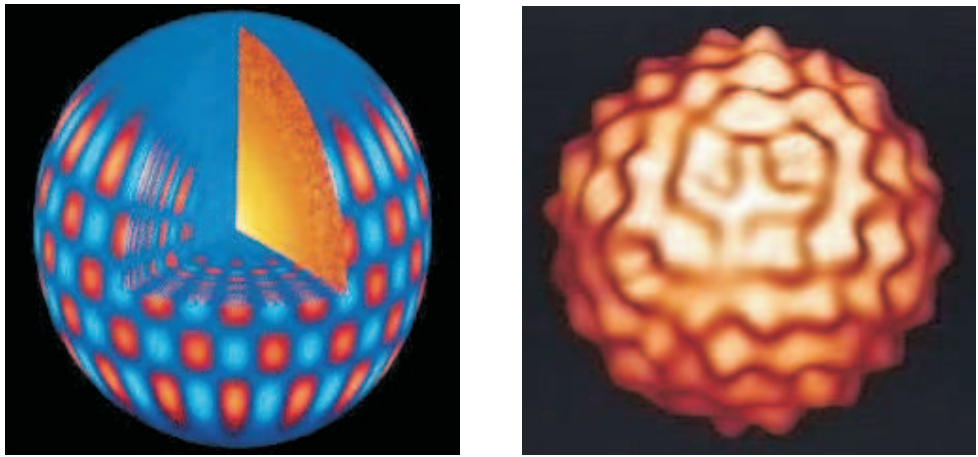

FIG. 2.6 – Le Soleil oscille : A gauche ; une représentation 3D des modes d'oscillations solaires. A droite ; une simulation théorique 3D des oscillations à la surface du soleil. [40].

### 2.6.1 Modes gravito-acoustiques

Le soleil est régi par deux forces majeures (la force de pression qui tend à le faire exploser et la force de gravité qui tend à le faire imploser). En conséquence il est le siège d'oscillations, interprétées comme étant la superposition d'ondes se propageant dans l'intérieur de l'étoile, conduisant à la formation par interférence d'ondes stationnaires, identifiables par le mouvement cohérent de la surface : appelées modes propres de vibration ; chacune est définie par une fréquence caractéristique. Chaque onde stationnaire (mode propre) est caractérisée essentiellement par 3 nombres entiers (les nombres harmoniques sphériques) décrivant la position des points et lignes de nœuds dans notre étoile :

1. **'n'** : L'ordre radial du mode, nombre de nœuds le long du rayon de l'étoile.

2. '$\ell$' : L'ordre du mode, nombre total de lignes de nœuds.

3. **'m'** : L'ordre azimutal du mode, nombre de lignes de nœuds qui passent par les pôles de vibrations.

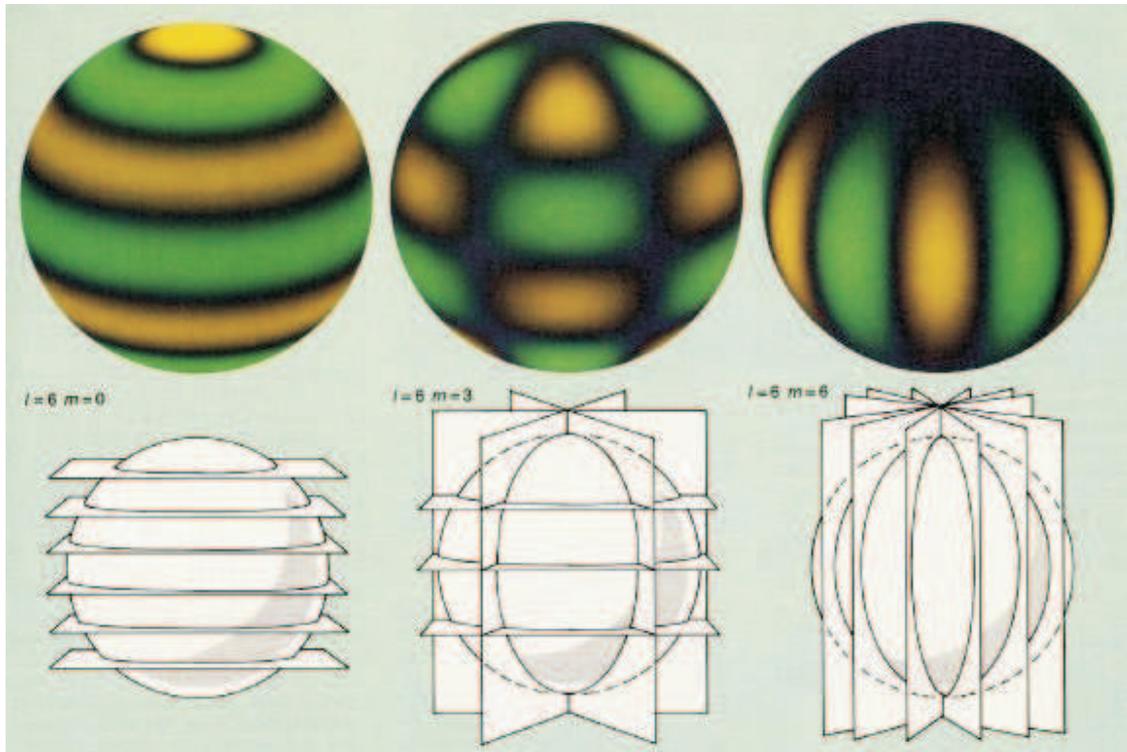

FIG. 2.7 – Trois exemples sur les nombres d'harmoniques sphériques : $(m,\ell) = (6,0), (6,3), (6,6)$ à partir de la gauche[40].

Comme pour une corde vibrante, qui compte des points fixes (nœuds), une surface vibrante (un tambour par exemple) compte aussi des lignes fixes appelées lignes de nœuds, '$\ell$' et 'm' traitent la partie tangentielle de la vibration 3D, tandis que 'n' traite la partie radiale.

On compte en général 2 types de modes :

1. **Les modes g :** modes de gravité, de fréquences relativement basses et d'ordre radial négatif ($n < 0$), ils sont créés par la force d'Archimède. Prédits théoriquement, mais jamais observés avec certitude, car essentiellement générés dans les zones les plus denses du soleil (le cœur plus une partie de la zone radiative), ils sont noyés et annihilés à la surface solaire par les modes p, de fréquences et d'intensité plus élevées.

2. **Les modes p :** modes de pression, de fréquences plus élevées et d'ordre radial positif ($n > 0$), ils sont générés près de la surface, plus précisément dans la zone de convection.

Un autre mode de fréquences intermédiaires et d'ordre radial nul ($n = 0$) existe. Appelé **mode f** (mode fondamental ou également connu sous le nom de mode de gravité de surface), il peut être classifié comme étant un troisième type de mode.

C'est en déterminant la relation de dispersion $\omega_{\ell,n}(k)$ du mode dans un milieu qu'on peut préciser les propriétés de ce dernier.

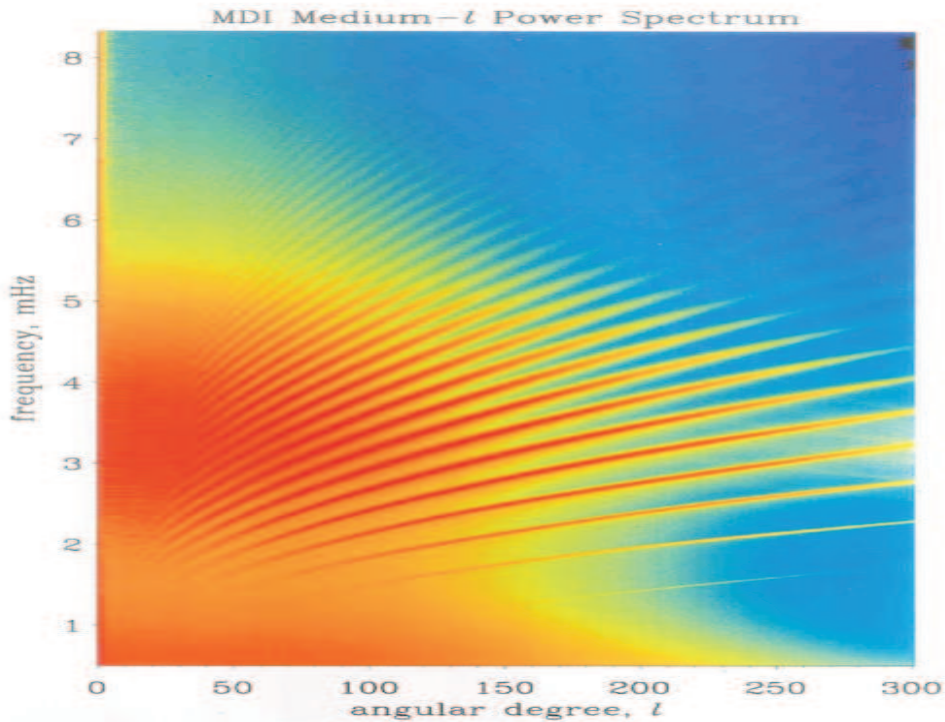

FIG. 2.8 – La fréquence ω en fonction du degré angulaire $L = \sqrt{\ell(\ell+1)}$ [44].

## 2.7 Héliosismologie locale

Les techniques heliosismiques sont employées pour étudier les caractéristiques globales de l'intérieur solaire après observation des oscillations acoustiques apparentes sur toute la photosphère. En observant de plus petites régions du soleil et en analysant les ondes qui se propagent, dans et en dehors de celles-ci, les chercheurs peuvent mieux déterminer les propriétés locales du Soleil. Par conséquent, de telles techniques d'analyse se nomment Héliosismologie locale. A la différence de l'Héliosimologie globale, qui étudie les phénomènes solaires globaux, en considérant le Soleil en tant que cavité résonnante, l'héliosismologie locale mesure les vitesses horizontales du fluide en bloc, comme ramenées à une moyenne au-dessus de la région d'intérêt, car les fréquences des ondes voyageant dans un milieu mobile seront décalées (décalage Doppler) par une quantité proportionnelle à la vitesse d'écoulement. Ce principe est employé dans les 3 principales techniques que compte l'Héliosismologie locale : Time-Distance, l'Holographie acoustique et Ring-Diagram.

### 2.7.1 Héliosismologie Ring-Diagram

L'idée de l'analyse héliosismique "Ring-diagram" a été proposée pour la première fois par Gough & Toomre en 1983 [45] puis reprise par la suite par Hill en 1988 [52]. Cette technique est basée sur l'analyse des spectres de puissance des "modes propres" des oscillations, calculés de la série chronologique des images de vitesses. Après le choix d'une région localisée dans la photosphère solaire, ces séries chronologiques sont décomposées grâce à la transformée de Fourier, transformant de ce fait les données de vitesse $(x,y,t)$ en données fréquentielles $(k_x, k_y, \omega)$. Les structures qui apparaissent comme des maxima dans l'espace de fréquence indiquent les fréquences propres d'oscillations. Les projections de ces structures dans le plan $(k_x, \omega)$, transforment celles-ci en le diagramme $(L, \nu)$, mais les projections dans le plan $(k_x, k_y)$ rapporte une série d'anneaux concentriques (d'où l'appellation de la méthode "Ring-Diagram" (voir Figure(2.9))). Le spectre de puissance peut être déterminé par la relation suivante [54] :

$$P = \frac{A}{(\omega - \omega_0 + k_x U_x + k_y U_y)^2 + W^2} + \frac{b_0}{k^3} \qquad (2.8)$$

Où $k_x U_x$ et $k_y U_y$ sont les 2 décalages Doppler, $b_0$ la puissance à l'état fondamental ; $\omega_0$ la fréquence centrale, alors que, $W$ la largeur et $A$ l'amplitude, sont des paramètres à approximer par inversion. Ainsi on tire les écoulements horizontaux, qui renferment de bien précieuses informations sur l'interieur solaire.

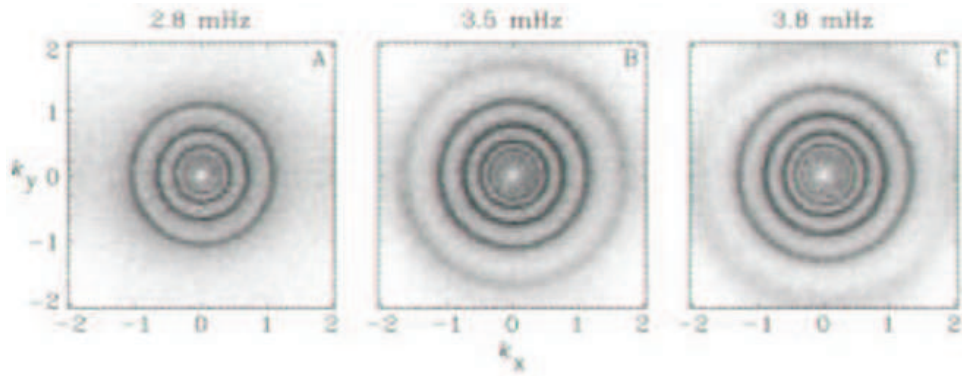

FIG. 2.9 – Section du spectre de puissance 3D à trois différentes fréquences. Ce graphe est tiré de la thèse de Junwei Zhao [54].

Les écoulements horizontaux dans la région localisée, près de la surface et en profondeur, causent des déplacements dans la position des anneaux. Ces derniers sont utilisés pour étudier la magnitude et la direction des flux.

### 2.7.2 Holographie acoustique

L'Holographie acoustique est une technique qui permet la reconstruction de la puissance du champ acoustique tri-dimensionnel de l'intérieur solaire, particulièrement sous les régions actives. Cette technique a été developée par Lindsey & Braun en 1997 [62] et, en parallèle par Chang [9] durant la même année. On doit à ce dernier l'introduction de "l'imagerie acoustique" en héliosismologie.

L'holographie acoustique est basée sur le calcul de :

$$H_{\pm}(\vec{r}, z, \nu) = \int_p' G_{\pm}(\vec{r}, \vec{r}', z, \nu) \psi(\vec{r}', \nu) d^2 r' \qquad (2.9)$$

Où $H_+$ et $H_-$ sont l'egression et l'ingression monochromatiques, $\psi$ est la perturbation acoustique locale à la surface localisée par $\vec{r}'$ et ayant la fréquence $\nu$. $G_+$ et $G_-$ sont les fonctions de Green, qui expriment comment le point monochromatique est perturbé à la position $\vec{r}'$ dans la surface de propagation en arrière et en avant dans le temps du foyer à la position $\vec{r}$ et à la profondeur $z$.

En calculant les puissances d'egression et d'ingression, "les fossés acoustiques" et "les brillances acoustiques" se trouvent directement associés avec les régions actives solaires. Cette technique est éventuellement utilisée pour détecter avec succès de larges régions actives, et ce en profondeur (voir Fig.(2.10)).

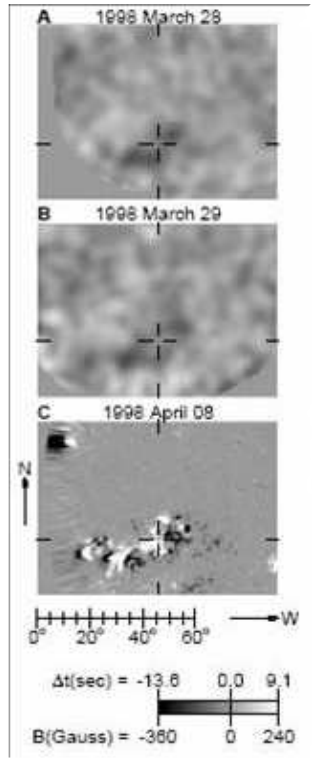

FIG. 2.10 – Ces images acoustiques profondes ont étés construites au moyen de dopplerogrammes du 28 et 29 mars 1998, et les magnétogrammes du 8 avril 1998 ; les anomalies acoustiques vues le 28 et 29 mars ont les mêmes longitudes de Carrigton que la région active vue par les magnétogrammes le 8 avril. Ces figures sont tirées de la thèse de Junwei Zhao [54].

L'holographie acoustique sensible à la phase (phase-sensitive acoustic holography) a été développée plus tard de la dérivée des différences de phase en corrélant les signaux egression et ingression.

$$C(\overrightarrow{r},z,t) = \int H_-(z,\overrightarrow{r},t')H_+(z,\overrightarrow{r},t'+\tau)dt' \qquad (2.10)$$

Les différences de phase peuvent apporter des informations sur la dynamique, qui alors peut être employée pour dériver le champ des écoulements subsurfaciques. Les champs des écoulements des super grannules sortant des taches solaires ont été déduits par une telle analyse, par Braun & Lindsey en 2003. De nouveaux modèles numériques pour une meilleure interprétation de l'holographie acoustique sont toujours en cours.

### 2.7.3 Héliosismologie Time-Distance

L'Heliosismologie time-distance a été développée pour la première fois par Duvall en 1993 [32], puis reprise par D'silva en 1996 [34]. Cette technique emploie une série chronologique d'images de vitesse, pour mesurer les temps de parcours des modes acoustiques se propageant le long des chemins d'ondes près de la surface. La méthode requiert le calcul de la corrélation entre les données en deux points séparés en changeant les distances et les périodes. Les endroits dans l'espace-temps où la corrélation est élevée indiquent les points tournants supérieurs des modes acoustiques après qu'ils aient traversé leur chemins d'ondes près de la surface. Dans l'approximation acoustique géométrique, le temps de parcours le long du chemin d'onde $\Gamma_i$ peut être écrit comme suit :

$$\tau_i(t) = \int_{\Gamma_i} \frac{ds}{c_s(\overrightarrow{r},t) + \overrightarrow{U}(\overrightarrow{r},t).\overrightarrow{n}} \quad (2.11)$$

Où $s$ est la distance le long du chemin de l'onde, $\overrightarrow{r}$ la coordonnée spatiale, $c_s$ la vitesse locale du son, $\overrightarrow{U}$ la vitesse locale de l'écoulement, et $\overrightarrow{n}$ le vecteur unitaire le long du chemin d'onde.

La différence des temps de parcours entre deux ondes traversant le même chemin d'onde dans des directions opposées indique la présence d'un écoulement, puisque les ondes voyageant à contre courant prennent plus de temps pour traverser la même distance qu'une onde voyageant avec l'écoulement, ces différences de temps de parcours peuvent être employées pour construire une carte des vitesses du fluide non seulement sur la surface mais également sur une gamme plus vaste de profondeurs. Cette technique a été employée pour mesurer des écoulements dans des régions à haute activité et sous les taches solaires (Duvall et al. 1996 [34]), aussi bien que pour caractériser des écoulements méridionaux près de la surface et en profondeur (Giles et al. 1997 [42] ; Giles 1999 [41]).

Les vitesses d'écoulement et du son sont obtenues par inversion des temps de parcours (méthode que nous développerons en détail dans le chapitre 6).

# Chapitre 3

# Rappel théorique et équations de base en héliosismologie

Après une brève introduction sur le Soleil, l'Héliosismologie globale et locale, et avant d'expliciter avec de plus amples détails la méthode héliosismique utilisée dans ce mémoire (l'héliosismologie Temps-Distance), nous nous devons, tout d'abord, d'avoir une idée sur la théorie et les équations de base utilisées en Héliosismologie ; celles-ci font l'objet du présent chapitre.

## 3.1 Équations hydrodynamiques de base

Pour caractériser les oscillations présentes dans l'intérieur solaire et en déduire ainsi la relation de dispersion des ondes $\omega_{\ell,n}(k)$, nous devons d'abord linéariser les équations hydrodynamiques ainsi que la l'équation de Poisson dans le cas des étoiles de type solaire.

Un système hydrodynamique est caractérisé en indiquant les quantités physiques comme fonctions de la position $r$ et du temps $t$. Ses propriétés incluent, par exemple, la densité locale $\rho(r,t)$, la pression locale $p(r,t)$ et la vitesse locale $\overrightarrow{v}(r,t)$. Pour l'Héliosismologie, les aspects les plus importants d'un système concernent surtout ses propriétés mécaniques. Les équations hydrodynamiques traduisent la conservation de la masse m, de l'impulsion $\overrightarrow{p}$ et de l'énergie $E$, **en négligeant le champ magnétique, la viscosité, la rotation et la convection** (très importantes hypothèses de base).

Ainsi **la conservation de la masse** est exprimée par **l'équation de continuité** [25] :

$$\frac{\partial \rho}{\partial t} + div(\rho \overrightarrow{v}) = 0 \tag{3.1}$$

D'après les hypothèses de base, les seules forces appropriées sont, dans la plupart des cas, la pression et la pesanteur. **L'équation du mouvement** (également connue sous le nom d'**équation d'Euler**) s'écrit alors comme suit :

$$\rho\left(\frac{\partial \overrightarrow{v}}{\partial t} + \overrightarrow{v}.\nabla \overrightarrow{v}\right) = -\overrightarrow{\nabla}p + \rho \overrightarrow{g} \tag{3.2}$$

Où la quantité entre parenthèses est la dérivée par rapport au temps de la vitesse d'un fluide en mouvement. Le premier terme du membre de droite est la force extérieure, donnée par la pression $p$, alors que le deuxième terme est donné par l'accélération de la gravité $\overrightarrow{g}$, obtenue à partir du gradient du potentiel de la gravité $\phi$, où $\overrightarrow{g} = -\overrightarrow{\nabla}\phi$ satisfait **l'équation de Poisson** :

$$\nabla^2 \phi = 4\pi G \rho \tag{3.3}$$

A partir de l'**équation de conservation de l'énergie** l'équa.(2.4), et sachant que notre soleil dispose d'une source d'energie propre (nucléaire), on trouve [7] :

$$\rho T \left(\frac{\partial}{\partial t} + \overrightarrow{v}.\overrightarrow{\nabla}\right)s = \rho(\varepsilon_N + \varepsilon_V) - \overrightarrow{\nabla}.\overrightarrow{F} \tag{3.4}$$

$s$ étant l'entropie spécifique, $\varepsilon_N$ le taux de production d'énergie nucléaire, $\varepsilon_V$ la chaleur créée par la viscosité et $\|\overrightarrow{F}\| = L/4\pi r^2$ le flux dénergie, avec :

$$\overrightarrow{F} = \overrightarrow{F}_R + \overrightarrow{F}_{Con} \tag{3.5}$$

Où $\overrightarrow{F}_R = -K.\overrightarrow{\nabla}T$ est le flux d'énergie radiatif, avec $K = 4ac_sT^3/3\kappa\rho$ la conductivité radiative, et $\overrightarrow{F}_{Con}$ le flux d'énergie convectif. Mais d'après nos hypothèses de base, on a : $\varepsilon_V = 0$ et $\overrightarrow{F}_{Con} = 0$, on peut donc écrire :

$$\rho T \left(\frac{\partial}{\partial t} + \overrightarrow{v}.\overrightarrow{\nabla}\right)s = \rho\varepsilon_N - \overrightarrow{\nabla}.\overrightarrow{F}_R \tag{3.6}$$

Cette dernière est la forme la plus simplifiée de l'équation (3.4)

## 3.2 Équations des oscillations

### 3.2.1 Expressions générales

On considère maintenant de petites perturbations autour d'un modèle stationnaire à l'équilibre, qui satisfait les hypothèses des équations de structure (2.1)-(2.4) à l'équilibre :

$$\frac{d m_0}{dr} = 4\pi r_0^2 \rho_0, \tag{3.7}$$

$$\frac{dp_0}{dr} = -\rho_0 g_0, \tag{3.8}$$

$$\frac{dT_0}{dr} = -\frac{3\kappa_0 \rho_0}{4ac_s T_0^3}\frac{L_{r,0}}{4\pi r_0^2}, \tag{3.9}$$

$$\frac{dL_{r,0}}{dr} = 4\pi r_0^2 \rho_0 \varepsilon_{N,0}. \tag{3.10}$$

Les quantités à l'équilibre sont caractérisées par l'indice inférieur '0', et $g_0 = Gm_0/r_0^2$ est l'accélération gravitationnelle.

A partir des équations hydrodynamiques de base précédemment citées, nous pouvons modifier l'état d'équilibre de notre étoile à l'aide de petites perturbations, en ne conservant que les termes du premier ordre. On peut approximer ces perturbations sous deux formes : eulérienne à une postion donnée ou lagrangienne en suivant l'élément fluide perturbé. On peut alors définir n'importe quelle quantité physique $f$ par [7] :

$$\begin{array}{c} f(\overrightarrow{r},t) = f_0(\overrightarrow{r}) + f'(\overrightarrow{r},t) \\ ou \\ f(\overrightarrow{r},t) = f_0(\overrightarrow{r_0}) + \delta f(\overrightarrow{r_0},t) \end{array} \tag{3.11}$$

Où la perturbation eulérienne est reliée à la perturbation lagrangienne par :

$$\delta f(\vec{r_0},t) = f'(\vec{r},t) + \vec{\xi}.\vec{\nabla} f_0(\vec{r}) \tag{3.12}$$

Au premier ordre du déplacement $\vec{\xi}$, avec :

$$\vec{\xi} = \delta \vec{r} = \vec{r} - \vec{r_0} \tag{3.13}$$

Avec $\vec{r}$ la variable de position lagrangienne d'un élément de fluide donné, $\vec{r} = \vec{r_0}$ à l'état d'équilibre. Les variations temporelles correspondantes s'écrivent $d\delta f(\vec{r},t)/dt$ et $\partial f'(\vec{r},t)/\partial t$, où :

$$\frac{d}{dt} = \frac{\partial}{\partial t} + \vec{v}.\vec{\nabla} \tag{3.14}$$

Dans le cas général, quand l'état non perturbé possède un courant stationnaire non nul $\vec{v} = \frac{d\vec{r}}{dt} \neq \vec{0}$, comme c'est le cas pour le Soleil en rotation par exemple, on relie les peruturbations eulérienne et lagrangienne du vecteur vitesse par :

$$\vec{v}' = \delta\vec{v} - (\vec{\xi}.\vec{\nabla})\vec{v_0} = \frac{\partial \vec{\xi}}{\partial t} + (\vec{v_0}.\vec{\nabla})\vec{\xi} - (\vec{\xi}.\vec{\nabla})\vec{v_0} \tag{3.15}$$

Dans notre cas simplifié ($\vec{v_0} = 0$), on a :

$$\vec{v}' = \delta\vec{v} = \frac{\partial \vec{\xi}}{\partial t} = \frac{d\vec{\xi}}{dt} \tag{3.16}$$

Pour plus de simplification, nous adopterons $\vec{v}'$ comme étant la perturbation du vecteur vitesse, dans tout ce qui suit dans ce chapitre.

Partant de ces définitions, et en tenant compte de l'état d'équilibre, le système d'équations linéarisées devient, sous forme eulérienne, comme suit :

$$\frac{\partial \rho'}{\partial t} + \vec{\nabla}.(\rho_0 \vec{v}) = 0, \tag{3.17}$$

$$\rho \frac{\partial \vec{v}}{\partial t} + \vec{\nabla} p' + \rho_0 \vec{\nabla} \Phi' + \rho' \vec{\nabla} \Phi_0 = 0, \tag{3.18}$$

$$\rho_0 T_0 \frac{\partial}{\partial t}(s' + \vec{\xi}.\vec{\nabla} s_0) = (\rho \varepsilon_N)' - \vec{\nabla}.\vec{F}, \tag{3.19}$$

$$\vec{\nabla}^2 \Phi' = 4\pi G \rho', \tag{3.20}$$

et :

$$\vec{F'} = -K_0 \vec{\nabla} T' - K' \vec{\nabla} T_0. \tag{3.21}$$

Ces équations sont linéaires, homogènes, aux dérivées partielles par rapport au temps $t$, et à la coordonnée spatiale $\vec{r}$ pour $\vec{v}$ et les variables perturbées notées prime. L'indice inférieur 0 caractérise les quantités à l'équilibre qui ne dépendent que de la coordonnée radiale $r$. En supposant que la dépendance temporelle pour toutes les variables physiques est de la forme $\exp(i\omega t)$, on peut remplacer $\frac{\partial}{\partial t}$ par $i\omega$ et écrire $\vec{v} = i\omega \vec{\xi}$. Puis en omettant l'indice 0 et en séparant l'équation d'Euler en ses deux composantes radiale et horizontale, où $\vec{\xi} = (\xi_r, \xi_\theta, \xi_\phi)$ est le déplacement et $\vec{\xi}_\perp = (0, \xi_\theta, \xi_\phi)$ sa composante horizontale, les équations (3.17), (3.18) et (3.20) deviennent :

$$\rho' + \vec{\nabla}.(\rho \vec{\xi}) = 0 \quad ou \quad \delta\rho/\rho + \vec{\nabla}.(\vec{\xi}_r + \vec{\xi}_\perp) = 0 \tag{3.22}$$

$$-\omega^2 \xi_r + \frac{1}{\rho}\frac{\partial p'}{\partial r} + \frac{\partial \Phi'}{\partial r} + \frac{\rho'}{\rho}\frac{d\Phi}{dr} = 0, \quad (3.23)$$

$$-\omega^2 \vec{\xi}_\perp + \vec{\nabla}_\perp\left(\frac{p'}{\rho} + \Phi'\right) = 0, \quad (3.24)$$

et :

$$\frac{1}{r^2}\left(r^2\frac{\partial \Phi'}{\partial r}\right) + \vec{\nabla}_\perp^2 \Phi' = 4\pi G \rho', \quad (3.25)$$

avec $\vec{\xi} = \vec{\xi}_r + \vec{\xi}_\perp$.

Sans oublier que $\vec{\nabla} = \frac{1}{r}(r\frac{\partial}{\partial r}, \frac{\partial}{\partial \theta}, \frac{1}{\sin\theta}\frac{\partial}{\partial \phi})$ est le gradient en coordonnées sphériques, et que $\vec{\nabla}_\perp = \frac{1}{r}(0, \frac{\partial}{\partial \theta}, \frac{1}{\sin\theta}\frac{\partial}{\partial \phi})$ est sa composante horizontale, avec $\vec{\nabla}_\perp^2 = \frac{1}{r^2 \sin^2\theta}[\sin\theta\frac{\partial}{\partial \theta}(\sin\theta\frac{\partial}{\partial \theta}) + \frac{\partial^2}{\partial \phi^2}]$.

En remplaçant $\xi_\perp$ de la deuxième expression de l'équation (3.22) par la divergence horizontale de l'équation (3.24), on obtient :

$$\frac{\delta\rho}{\rho} + \frac{1}{r^2}\frac{\partial}{\partial r}(r^2 \xi_r) + \frac{1}{\omega^2}\vec{\nabla}_\perp^2\left(\frac{p'}{\rho} + \Phi'\right) = 0 \quad (3.26)$$

Pour que notre description soit complète, nous avons besoin d'une relation de fermeture, et pour ce faire, nous devons rapporter un lien entre $p$ et $\rho$. En général, ceci exige la considération de l'énergétique du système, comme décrit par la première loi de la thermodynamique. Pour ce, utilisant $\rho$ et $T$ comme couple de variables indépendantes pour exprimer les quantités thermodynamiques, via la relation [7] :

$$\frac{\delta\rho}{\rho} = \frac{1}{\Gamma_1}\frac{\delta p}{p} - \nabla_{ad}\frac{\rho T}{p}\delta s \quad (3.27)$$

avec $\nabla_{ad} = (\frac{\partial \ln T}{\partial \ln p})_s$ étant le gradient adiabatique.

En utilisant la relation reliant la perturbation lagrangienne à la perturbation eulérienne (3.12), la dernière équation devient :

$$\frac{\rho'}{\rho} = \frac{1}{\Gamma_1}\frac{p'}{p} - A\xi_r - \nabla_{ad}\frac{\rho T}{p}\delta s \quad (3.28)$$

Où $A$ est appelé le critère de Schwarzchild, relié à la fréquence $N$, dite fréquence de Brunt-Väisälä, par [7] :

$$A = -N^2/g = \frac{d\ln\rho}{dr} - \frac{1}{\Gamma_1}\frac{d\ln p}{dr} \quad (3.29)$$

En usant des équations (3.27), (3.28) et (3.29), notre système d'équations devient :

$$\frac{1}{\rho}\left(\frac{\partial}{\partial r} + \frac{\rho g}{\Gamma_1 p}\right)p' - (\omega^2 - N^2)\xi_r + \frac{\partial \Phi'}{\partial r} = g\nabla_{ad}\frac{\rho T}{p}\delta s, \quad (3.30)$$

$$\frac{1}{r^2}\frac{\partial}{\partial r}(r^2 \xi_r) + \frac{1}{\Gamma_1}\frac{d\ln p}{dr}\xi_r + \left(\frac{\rho}{\Gamma_1 p} + \frac{\vec{\nabla}_\perp^2}{\omega^2}\right)\frac{p'}{\rho} + \frac{1}{\omega^2}\vec{\nabla}_\perp^2 \Phi' = \nabla_{ad}\frac{\rho T}{p}\delta s, \quad (3.31)$$

$$\left(\frac{1}{r^2}\frac{\partial}{\partial r}(r^2\frac{\partial}{\partial r}) + \vec{\nabla}_\perp^2\right)\Phi' - 4\pi G\rho\left(\frac{p'}{\Gamma_1 p} + \frac{N^2}{g}\xi_r\right) = -4\pi G\nabla_{ad}\frac{\rho^2 T}{p}\delta s, \quad (3.32)$$

$$F_r' = -K\frac{\partial T'}{\partial r} - K'\frac{\partial T}{\partial r}, \quad (3.33)$$

et

$$i\omega T\delta s = \rho'\varepsilon_N - \frac{1}{r^2}\frac{\partial}{\partial r}(r^2 F_r') + \vec{\nabla}_\perp^2(KT'). \quad (3.34)$$

Où $F_r'$ est la composante radiale de la perturbation eulérienne du flux radiatif $\vec{F'}_R$.

Sachant que toute fonction propre du type $f(\theta, \phi)$ satisfait l'équation suivante :

$$\nabla_\perp^2 f(\theta,\phi) = -\frac{1}{r^2}\Lambda f(\theta,\phi) \tag{3.35}$$

Où $\Lambda = L^2 = \ell(\ell+1)$ est une constante, en associant pour chaque $\ell$ (degré ou ordre du mode) une harmonique sphérique $Y_\ell^m(\theta,\phi)$. Dans ce cas, une séparation de variables est possible, et on peut écrire nos variables sous la forme (par exemple la pression ($f = p'$)) :

$$p'(r,\theta,\phi,t) = P'(r) Y_\ell^m(\theta,\phi) \exp(i\omega t) \tag{3.36}$$

Le vecteur déplacement s'écrit alors :

$$\vec{\xi} = [\xi_r(r), \xi_h(r)\frac{\partial}{\partial \theta}, \xi_h(r)\frac{\partial}{\sin\theta \partial \phi}] Y_\ell^m(\theta,\phi) \exp(i\omega t) \tag{3.37}$$

Où $\xi_h = 1/(\omega^2 r)[p'/\rho + \Phi']$ (tiré de l'équa.(3.24)) est le déplacement horizontal. En considérant uniquement la partie radiale de nos variables (i.e, $p'(r)$, $T'(r)$,...) des 3 premières équations de notre système, on écrit :

$$\frac{1}{\rho}\left(\frac{d}{dr} + \frac{g}{c_s^2}\right)p' + (N^2 - \omega^2)\xi_r + \frac{d\Phi'}{dr} = g\nabla_{ad}\frac{\rho T}{p}\delta s, \tag{3.38}$$

$$\frac{1}{r^2}\frac{d}{dr}(r^2 \xi_r) + \frac{1}{\Gamma_1}\frac{d\ln p}{dr}\xi_r + \left(1 - \frac{S_\ell^2}{\omega^2}\right)\frac{p'}{\rho c_s^2} - \frac{L^2}{\omega^2 r^2}\Phi' = \nabla_{ad}\frac{\rho T}{p}\delta s, \tag{3.39}$$

$$\left(\frac{1}{r^2}\frac{d}{dr}(r^2\frac{d}{dr}) - \frac{L^2}{r^2}\right)\Phi' - 4\pi G\rho\left(\frac{p'}{\rho c_s^2} + \frac{N^2}{g}\xi_r\right) = -4\pi G\nabla_{ad}\frac{\rho^2 T}{p}\delta s. \tag{3.40}$$

Avec $c_s^2 = \Gamma_1 p/\rho$ le carré de la vitesse du son du milieu, sans oublier les fréquences caractéristiques $S_\ell^2$ et $N$, qui sont à la fois la fréquence tangentielle, dite de Lamb, et la fréquence de flottabilité, dite de Brunt-Väisälä, et qui sont définies comme suit :

$$S_\ell = \frac{\ell(\ell+1)c_s^2}{r^2} \simeq k_h^2 c_s^2 \tag{3.41}$$

et :

$$N^2 = \left(\frac{1}{\Gamma_1 p}\frac{dp}{dr} - \frac{1}{\rho}\frac{d\rho}{dr}\right) \tag{3.42}$$

Où $\vec{k}_h$ est la partie horizontale du vecteur d'onde $\vec{k} = \vec{k}_r + \vec{k}_h$, avec $\vec{k}_r$ la partie radiale.

Pour $N^2 > 0$, la convection est dite stable (il y a retour vers l'équilibre) et elle est dite instable pour $N^2 < 0$.

### 3.2.2 Approximation adiabatique

Si on compare le temps d'échange de chaleur (dit : temps caractéristique thermodynamique) avec la période des oscillations (par exemple les oscillations de 5 min pour les ondes acoustiques solaires), on constate que le premier est considérablement plus grand, et on peut alors négliger l'échange de chaleur durant une période sur pratiquement toute la structure solaire (à l'exception de la région très externe dite superadiabatique). En faisant cela, on suppose alors que les oscillations sont adiabatiques (l'entropie $s = cste \Rightarrow \delta s = 0$). Par conséquent, on obtient alors un système d'ordre 4 pour les 4 variables $p'$, $\xi_r$, $\Phi'$, et $d\Phi'/dr$ :

$$\frac{1}{\rho}\left(\frac{d}{dr} + \frac{g}{c_s^2}\right)p' + (N^2 - \omega^2)\xi_r + \frac{d\Phi'}{dr} = 0, \tag{3.43}$$

$$\frac{1}{r^2}\frac{d}{dr}(r^2\xi_r) + \frac{1}{\Gamma_1}\frac{d\ln p}{dr}\xi_r + \left(1 - \frac{S_\ell^2}{\omega^2}\right)\frac{p'}{\rho c_s^2} - \frac{L^2}{\omega^2 r^2}\Phi' = 0, \tag{3.44}$$

$$\left(\frac{1}{r^2}\frac{d}{dr}(r^2\frac{d}{dr}) - \frac{L^2}{r^2}\right)\Phi' - 4\pi G\rho\left(\frac{p'}{\rho c_s^2} + \frac{N^2}{g}\xi_r\right) = 0. \tag{3.45}$$

### 3.2.3 Conditions aux limites

Les équations doivent être combinées avec quatre conditions aux limites : deux de ces dernières assurent la régularité au centre, $r = 0$, qui est un point singulier et régulier des équations. Une condition impose la continuité de champ gravitationnel $\phi'$ et de son gradient sur la surface, $r = R_\odot$. Finalement, la perturbation extérieure de pression doit satisfaire une condition dynamique, sous sa forme la plus simple, elle impose une perturbation nulle de la pression à la surface perturbée, i.e.

$$\partial p = 0 \quad \text{à} \quad r = R_\odot \tag{3.46}$$

Ainsi notre système d'équations différentielles du quatrième ordre et les conditions limites définissent un problème aux valeurs propres qui a comme solutions des valeurs discrètes de $\omega$. Où chacune d'elles est représentée par le couple $(\ell, m)$ ; ainsi nous obtenons un ensemble de fréquences propres $\omega_{n\ell m}$, caractérisé par leur ordre radial $n$.

Il faut noter que dans le cas d'une étoile sphériquement symétrique, les fréquences sont dégénérées dans l'ordre azimutal, car la définition de $m$ est attachée à l'orientation du système, ce qui n'a pour une étoile sphériquement symétrique, aucune signification physique. En effet, l'analyse des effets de la structure sphériquement symétrique du soleil a revélé que les équations et les conditions limites ne dépendent pas de $m$, et que les fréquences sont seulement caractérisées par $\ell$ et $n$, tandis que la dégénérescence en $m$ nous renseigne sur la rotation.

### 3.2.4 Approximation de Cowling

Cette approximation consiste à réduire l'ordre du système d'équations, en négligeant la perturbation eulérienne du potentiel gravitationnel $\Phi'$. Elle a été appliquée pour la première fois par Cowling en 1941 [13], et porte son nom. Elle n'est valable que pour des $n$ ou $\ell$ élevés. Notons que les modes observés au soleil sont d'orde radial élevé aussi. Cette approximation est justifiée, du moins en partie, en notant que la position de ces modes varie rapidement dans des régions à densité supposée quasi-constante, ce qui implique, qu'on peut écrire que $\rho' \approx 0$, et par conséquent, que $\Phi' \approx 0$ (voir l'équation de Poisson).

Le système se ramène donc à un ordre 2 pour les variables $P'$ et $\xi_r$, et il n'y a plus que deux conditions aux limites (l'une pour le centre et l'autre à la surface). Le système devient alors :

$$\frac{1}{\rho}\left(\frac{d}{dr} + \frac{g}{c_s^2}\right)p' + (N^2 - \omega^2)\xi_r = 0 \tag{3.47}$$

$$\frac{1}{r^2}\frac{d}{dr}(r^2\xi_r) - \frac{g}{c_s^2}\xi_r + \left(1 - \frac{S_\ell^2}{\omega^2}\right)\frac{p'}{\rho c_s^2} = 0 \tag{3.48}$$

Avec $g/c_s^2 = H_p^{-1}/\Gamma_1$, où $H_p^{-1} = -d\ln p/dr$ est l'échelle de pression.

## 3.3 Modes propres

A partir du système d'équations d'orde 2 (3.47), nous obtenons la relation de dispersion $\omega_{n,\ell} = \omega_{n,\ell}(\vec{k},\vec{r})$ des ondes se propageant dans le Soleil.

### 3.3.1 Cavité résonnante et quantification du spectre

Notre système d'équations réécrit différemment, est :

$$\frac{dp'}{dr} = \rho(\omega^2 - N^2)\xi_r - \frac{g}{c_s^2}p' \tag{3.49}$$

$$\frac{d\xi_r}{dr} = -\left(\frac{2}{r} - \frac{g}{c_s^2}\right)\xi_r + \left(\frac{S_\ell^2}{\omega^2} - 1\right)\frac{p'}{\rho c_s^2} \tag{3.50}$$

En ramenant ce système d'équations à une simple équation différentielle du second ordre, on verra plus tard (section (3.4)) comment résoudre rigoureusement ce système, en prenant en compte le fait que les fonctions propres des oscillations d'ordre radial élevé varient plus rapidement que les variables d'équilibre. Les termes contenant le produit de $g$ et de la vitesse du son $c_s^2 = (\partial p/\partial \rho)_s$, sont alors négligables, puisqu'ils contiennent un produit des dérivées des variables d'équilibre ($g/c_s^2 = H_p^{-1}/\Gamma_1$). En les annulant tout simplement et en négligeant le terme $2/r$ ($r$ est relativement grand), on obtient :

$$\frac{dp'}{dr} = \rho(\omega^2 - N^2)\xi_r \tag{3.51}$$

$$\frac{d\xi_r}{dr} = \left(\frac{S_\ell^2}{\omega^2} - 1\right)\frac{p'}{\rho c_s^2} \tag{3.52}$$

En dérivant la dernière équation une nouvelle fois par rapport à $r$, et en combinant avec la première, tout en négligeant à nouveau les dérivées des variables d'équilibre, on obtient :

$$\frac{d^2\xi_r^2}{dr^2} = \frac{\omega^2}{c_s^2}\left(1 - \frac{N^2}{\omega^2}\right)\left(\frac{S_\ell^2}{\omega^2} - 1\right)\xi_r \tag{3.53}$$

Cette équation différentielle du second ordre admet comme solution une équation d'onde, et afin de connaître le comportement de cette équation d'onde, il faut étudier le signe de son membre de droite :

1. S'il est négatif, la solution est alors sinusoïdale et l'onde est oscillante, et cela se produit :
   - pour $|\omega| < |N|$ et $|\omega| < |S_\ell|$, ou $|\omega| > |N|$ et $|\omega| > |S_\ell|$.

2. S'il est positif, la solution est alors exponentielle et l'onde est évanescente, et celà se produit :
   - pour $|\omega| < |N|$ et $|\omega| > |S_\ell|$, ou $|\omega| < |S_\ell|$ et $|\omega| > |N|$.

A la vue de ces inégalités, on se rend facilement compte de l'impossibilité que les ondes se propagent hors de la cavité délimitée par les fréquences de Lamb $S_\ell$ et Brunt-Väisälä $N$, et on dit alors que les ondes sont **piégées**. L'éxistence d'une telle **cavité résonnante** implique une quantification du spectre des modes propres d'oscillations (spectre discret (voir les Figures (3.1)-(3.2))).

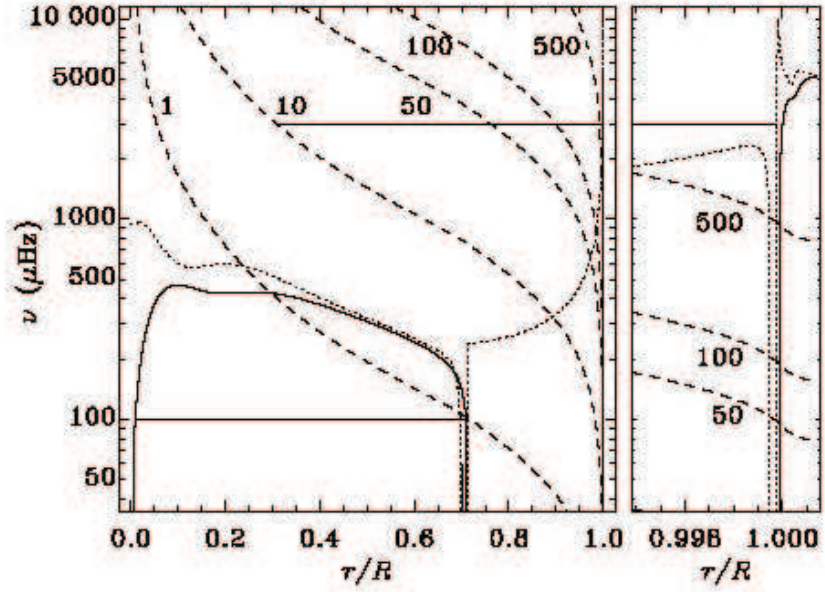

FIG. 3.1 – Les fréquences caractéristiques $N/2\pi$ (ligne continue), $\omega_c/2\pi$ (ligne en pointillés), $S_\ell/2\pi$ (ligne discontinue, identifiée par $\ell$) pour $\ell = 1, 10, 50, 100$. Les fréquences ont été calculées par le model S de Christensen-Dalsgaard et al. (1996). Les larges lignes horizontales délimitent les régions de piégeage des modes g à la fréquence $100\mu Hz$ et des modes $p$ à la fréquence $3000\mu Hz$ et $\ell = 10$ [25].

**Quantification du spectre des modes propres d'oscillations**

Du fait de la température qui augmente avec la profondeur, les ondes acoustiques se propageant près de la surface se réfractent vers la surface, puis du fait de la décroissance soudaine de la densité, celles-ci se réfléchissent, tout en sachant que la profondeur de pénétration ne dépend essentiellement que de la vitesse horizontale de phase de l'onde.

Ainsi le Soleil se trouve être une cavité résonnante pour les modes acoustiques, et ses oscillations sont généralement observées en mesurant leurs fluctuations d'intensité, ou bien leurs décalages Doppler à la surface solaire (fréquences, vitesses... etc.). Afin de représenter nos oscillations, on peut les considérer comme étant la somme d'ondes stationnaires ou de modes propres où le signal observé $f$, au point $(r, \theta, \phi)$ et à un temps $t$, est donné par [41] :

$$f(r, \theta, \phi) = \sum_{n\ell m} a_{n\ell m} \xi_{n\ell m}(r, \theta, \phi) \exp(i[\omega_{n\ell m} t + \alpha_{n\ell m}]) \tag{3.54}$$

Avec $n$ l'ordre radial, $\ell$ le degré angulaire et $m$ l'ordre azimutal, identifient chaque mode, $a_{n\ell m}$ l'amplitude du mode, $\omega_{n\ell m}$ est la fréquence propre (bien la fréquence caractéristique du mode), et $\alpha_{n\ell m}$ la phase. La fonction propre spatiale pour chaque mode est désignée par $\xi_{n\ell m}$, et comme on a émis au départ l'hypothèse d'un Soleil à symétrie sphérique, on peut alors séparer la partie radiale de la partie angulaire de notre fonction propre et on écrit :

$$\xi_{n\ell m}(r, \theta, \phi) = \xi_{n\ell}(r) Y_\ell^m(\theta, \phi) \tag{3.55}$$

Où $Y_\ell^m$ est l'harmonique sphérique, et $\xi_{n\ell}(r)$ la fonction propre radiale du mode.

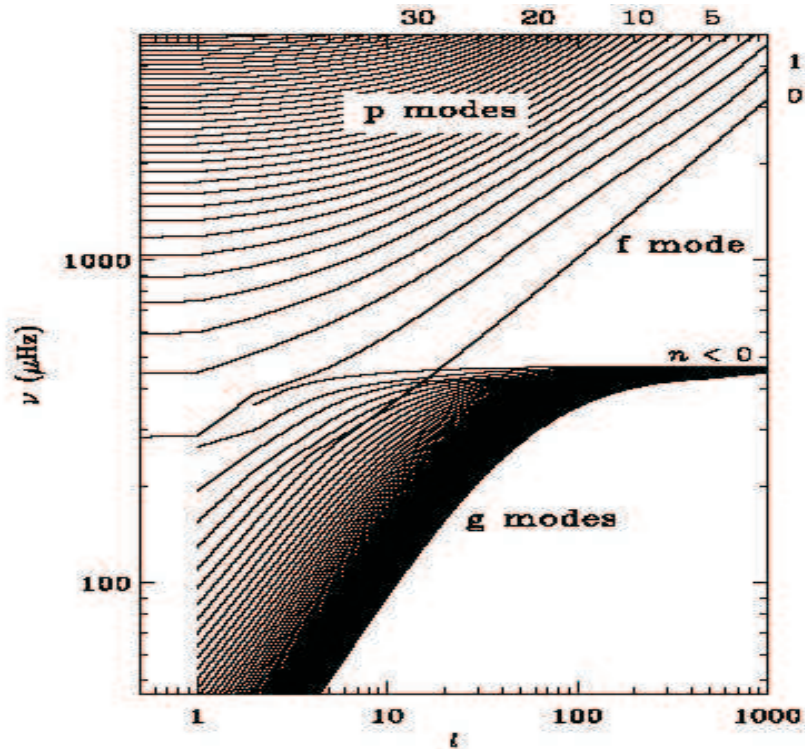

FIG. 3.2 – ν en fonction de $\ell$ pour tous les types de mode [25].

Nous allons à présent nous intéresser aux deux domaines de fréquences où les ondes sont oscillantes ; i.e. au domaine de hautes fréquences ($\omega > N, S_\ell$) et à celui des basses fréquences ($\omega < N, S_\ell$). Il faut noter que cette première analyse est simpliste mais qu'elle permet de comprendre la nature physique des deux types d'ondes qui parcourent le Soleil.

### 3.3.2  Modes de pression et loi de Duvall

Les oscillations de hautes fréquences ou de petites longueurs d'ondes (i.e. $\omega > N, S_\ell$) piégées dans une étoile (dans notre cas le Soleil) sont généralement appelées, les **modes de pression** ou modes "p". Modes purement acoustiques, la force de rappel de ces ondes est le gradient de pression (la compressibilité). Elles se propagent depuis la surface vers le centre, à l'intérieur d'une sorte de coquille sphérique, dont la profondeur dépend du degré $\ell$ du mode considéré. Parmi toutes les ondes excitées, seules les ondes retombant sur elles-mêmes (en phase) après un tour complet du soleil créent des ondes stationnaires (modes de vibration) caractéristiques de la coquille dans laquelle elles se propagent, nous renseignant ainsi sur les propriétés de l'intérieur solaire. Actuellement, dans le Soleil, on observe plus de 3000 de ces modes (20000 en considérant les multiplets de $m$). Les modes $\ell = 0$ sont des modes purement radiaux, ils correspondent à la déformation la plus simple de la sphère (gonflement) et ils sont les seuls à atteindre le centre du Soleil. Tous les autres modes tendent à dévier le Soleil de sa forme sphérique. Plus leur degré est élevé, moins les ondes pénètrent profondément dans le Soleil.

En supposant que $\omega \gg N$, on obtient pour l'équation (3.53) :

$$\frac{d^2\xi_r^2}{dr^2} \simeq \frac{1}{c_s^2}(S_\ell^2 - \omega^2)\xi_r \tag{3.56}$$

Où :

$$-k_r^2 = \frac{1}{c_s^2}(S_\ell^2 - \omega^2) \tag{3.57}$$

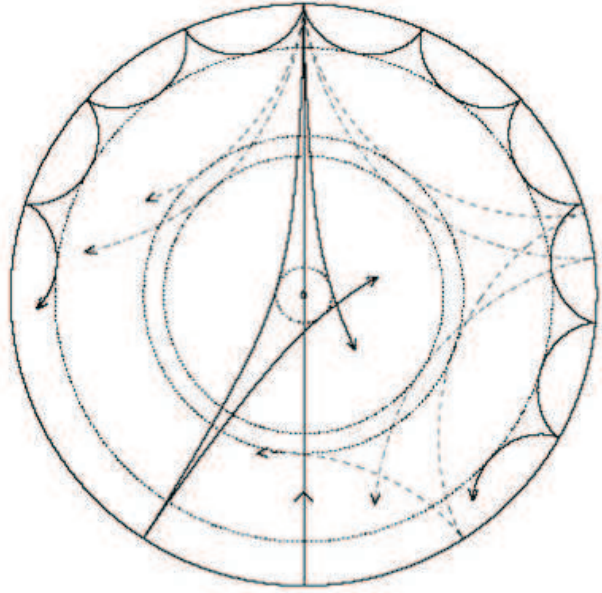

FIG. 3.3 – Représentation schématique des ondes acoustiques (modes "p") dans les couches internes solaires, et nous remarquons bien que pour $\ell = 0$ l'onde traverse le soleil de part en part [54].

En effet, on peut approximer les modes localement par des ondes planes sonores, satisfaisant la relation de dispersion suivante :

$$\omega^2 = c_s^2 |\vec{k}|^2 \tag{3.58}$$

Où $\vec{k} = \vec{k_r} + \vec{k_h}$ est le vecteur d'onde. Ainsi les propriétés des modes sont entièrement commandées par la variation de la vitesse du son adiabatique $c_s(r)$ et par le biais des équations (3.41) et (3.58) nous remontons aisément à l'équation (3.57).

A partir de l'équation (3.57) et en posant $\omega = S_\ell(r_t)$, on obtient une expression permettant de déterminer les points tournants internes (l'endroit où l'onde est réfléchie ($k_r = 0$)) :

$$\frac{c_s^2(r_t)}{r_t^2} = \frac{\omega^2}{\ell(\ell+1)} \tag{3.59}$$

La raison physique de cette réflexion interne est reliée à l'augmentation de la vitesse du son vers le centre solaire. Au fur et à mesure que l'onde s'enfonce, elle s'infléchit (car le vecteur d'onde radial $k_r$ diminue quand $c_s$ augmente), jusqu'à se réfléchir. Nous verrons dans la section (3.4) la raison du piégeage externe des ondes acoustiques, car celle-ci n'est pas incluse dans cette relation de dispersion présumée simpliste.

**Loi de Duvall**

L'équation (3.57) peut être employée pour justifier une approximation extrêmement utile, en une expression de fréquences d'oscillation acoustique, car la condition d'une onde se propageant dans la direction radiale implique que l'intégrale de $k_r$ au-dessus de la région de propagation, entre $r = r_t$ et $r = R_\odot$, doit être une valeur multiple de $\pi$, indépendamment des effets possibles de changement de phase aux limites de l'intervalle :

$$(n+\alpha)\pi \simeq \int_{r_t}^{R} k_r dr \simeq \int_{r_t}^{R} \frac{\omega}{c}\left(1 - \frac{s_\ell^2}{\omega^2}\right)^{1/2} dr, \quad (3.60)$$

où $\alpha$ est le déphasage aux points tournants. Ceci peut être également écrit comme suit :

$$\frac{\pi(n+\alpha)}{\omega} \simeq F\left(\frac{\omega}{L}\right), \quad (3.61)$$

où

$$F(\frac{\omega}{L}) = \int_{r_t}^{R} \left(1 - \frac{c^2}{\omega^2 r^2}\right)^{1/2} \frac{dr}{c} \quad (3.62)$$

Les fréquences observées de l'oscillation solaire satisfont la relation simple indiquée par l'équa.(3.61), généralement appelée loi de Duvall. Elle a été trouvée la première fois par Duvall en 1982 [31], et porte son nom.

### 3.3.3 Modes de gravité

Les modes de basses fréquences ou de grandes longueurs d'ondes (i.e. $\omega < N, S_\ell$) piégées dans le Soleil, sont généralement appelés : **modes de gravité** ou modes "g". Leur force de rappel est la poussée d'Archimède. Ils sont principalement confinés dans la zone centrale du soleil (la zone radiative) et la traversée de la zone convective atténue leur amplitude, les rendant de ce fait, plus difficilement observables que les modes acoustiques. Notons qu'à ce jour, les modes "g" n'ont pas été clairement détéctés, et sont fortement recherchés dans les données des expériences héliosismiques pour leur grande sensibilité au coeur du soleil.

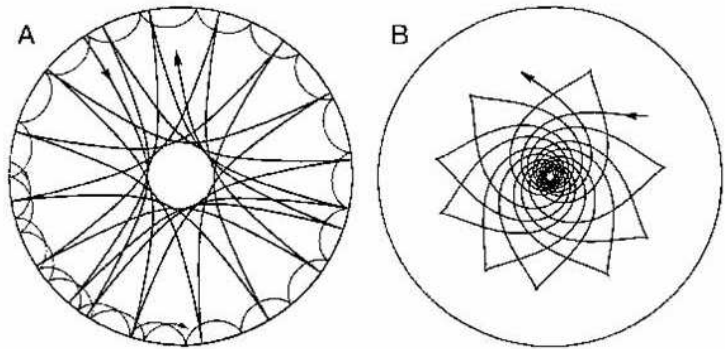

FIG. 3.4 – Différence entre les modes *p* et *g* : A gauche, représentation des modes *p*. A droite, représentation des modes g [26].

Comme pour les modes "p", nous allons simplifier l'équation (3.53), en supposant que $\omega \ll S_\ell$, ainsi on a :

$$\frac{d^2\xi_r}{dr^2} \simeq \frac{\ell(\ell+1)}{\omega^2 r^2}(\omega^2 - N^2)\xi_r \qquad ou \qquad -k_r^2 = \frac{\ell(\ell+1)}{\omega^2 r^2}(\omega^2 - N^2) \tag{3.63}$$

Les points tournants de ces modes se situent là où $\omega = N$ ($k_r = 0$). On remarque aussi que contrairement aux modes "p", leurs fréquences diminuent quand $\ell$ augmente, et qu'il n'existe pas le modes "g" de degré $\ell = 0$.

## 3.4 Relation de dispersion

Bien qu'instructives, les équations précédentes sont à peine satisfaisantes, dans le sens mathématique et physique, car elles négligent les effets de la variation de la structure solaire avec la position. En outre, les effets près de la surface menant à la réflexion sont purement postulés. Une description plus satisfaisante peut être faite sans approximation des équations d'oscillations.

De nombreux travaux basés sur des méthodes asymptotiques se sont attachés à trouver des solutions au système d'équations caractérisant les oscillations non radiales, avec ou sans l'approximation de Cowling. Nous nous contenterons ici de traiter rigoureusement le système d'équations afin de tirer la relation de dispersion, en suivant Christensen-Dalsgaard (1998) qui lui même s'est inspiré de Deubner et Gough (1984). En dérivant une deuxième fois par rapport à $r$ l'équation (3.48), en y injectant l'équation (3.47), tout en négligeant les dérivées de $r$ et de $g$ en plus du terme $2/r$, considérant ainsi que les oscillations et les variables thermodynamiques varient beaucoup plus rapidement (ce qui revient à étudier les oscillations d'une couche plane parallèle sous une gravité constante), et en posant le changement de variable :

$$\chi = div \vec{\delta r} = -\frac{1}{\Gamma_1 p}(p' - \rho g \xi_r), \tag{3.64}$$

on obtient, et après un certain nombre de manipulations algébriques, une équation différentielle du deuxième ordre pour $\chi$ :

$$\frac{d^2\chi}{dr^2} + \left(\frac{2}{c_s^2}\frac{dc_s^2}{dr} + \frac{1}{\rho}\frac{d\rho}{dr}\right)\frac{d\chi}{dr} + \left[\frac{1}{\Gamma_1}\frac{d^2\Gamma_1}{dr^2} - \frac{2}{\Gamma_1}\frac{d\Gamma_1}{dr}\frac{g\rho}{p} + k_h^2\left(\frac{N^2}{\omega^2} - 1\right) - \frac{1}{\rho}\frac{d\rho}{dr}\frac{1}{\Gamma_1}\frac{d\Gamma_1}{dr} + \frac{\rho\omega^2}{\Gamma_1 p}\right]\chi = 0 \tag{3.65}$$

Afin de faciliter l'analyse de cette équation et d'éliminer le terme $d\chi/dr$, on introduit la nouvelle variable $X = c_s^2 \rho^{1/2} \chi$, et on obtient :

$$\frac{d^2X}{dr^2} + \frac{1}{c_s^2}\left[S_\ell^2\left(\frac{N^2}{\omega^2} - 1\right) + \omega^2 - \omega_c^2\right]X = 0, \tag{3.66}$$

où :

$$k_r^2 = \frac{1}{c_s^2}(\omega^2 - \omega_c^2) - k_h^2\left(1 - \frac{N^2}{\omega^2}\right) \tag{3.67}$$

Avec $k_h^2 = S_\ell^2/c_s^2 = L^2/r^2$, où $\omega_c^2 = \frac{c_s^2}{4H_\rho^2}(1 - 2\frac{dH_\rho}{dr})$ est la fréquence de coupure acoustique, et $H_\rho^{-1} = -d\ln\rho/dr$ l'échelle de densité.

C'est à $\omega = \omega_c$ que s'effectue la réflexion de l'onde acoustique près de la surface.

**Important :**

L'équation (3.67) est la relation de dispersion la plus rigoureuse au soleil (en l'absence de champ magnétique). Elle est très importante, voire capitale, car elle renferme de très importantes informations sur les paramètres internes solaires ; on sera amené à l'utiliser souvent dans les prochains chapitres, surtout dans le cas des modes à hautes fréquences (i.e. $\omega > N$) qui nous intéressent particulièrement, afin de sonder l'intérieur solaire.

# Chapitre 4

# Héliosismologie Temps-Distance

Après une brève introduction sur le Soleil et sur l'Héliosismologie établie, et après une étude globale et détaillée des équations de base qui régissent l'Héliosismologie, nous nous consacrons dans ce chapitre à l'explication de la méthode héliosismique locale Temps-Distance et à l'étude approfondie de sa théorie.

## 4.1 Principe

L'Héliosismologie Temps-Distance (Duvall et al, 1993 [32]) vise à produire des cartes tridimensionnelles des écoulements sub-photosphèriques des inhomogénéités de la température et vraisemblablement du champ magnétique. Cette technique propose de mesurer le temps que prend un paquet d'onde pour voyager entre deux points quelconques sur la surface solaire, dans l'une ou l'autre direction. Un paquet d'onde acoustique atteindra des couches plus profondes à mesure que la séparation horizontale entre les deux points est grande. Des ondes de gravité de surface, qui se propagent horizontalement, peuvent être employées pour sonder près de la surface. Bien que l'onde observée, en un point donné, soit due à une superposition aléatoire d'ondes produites par des évènements éloignés de la source, le temps de parcours, entre deux endroits, peut être déterminé à partir de la fonction temporelle de corrélation du signal d'oscillation.

La Figure (4.1) montre une corrélation théorique croisée en fonction de la distance entre deux points et le temps de corrélation (calculé pour un modèle solaire sphériquement symétrique). La première arête correspond à une onde acoustique se propageant entre deux points à la surface solaire sans réflexion additionnelle. L'arête suivante correspond aux ondes qui arrivent après une réflexion à la surface, et les arêtes à plus grands temps sont le résultat d'ondes arrivant après des rebonds multiples. La branche en arrière liée à la deuxième arête correspond aux ondes réfléchies. Dans la plupart des applications, seuls les temps directs de parcours du premier rebond sont mesurés.

Les inhomogénéités locales au soleil affecteront les temps de parcours différemment selon le type de perturbation. Par exemple, les perturbations de la température et les perturbations d'écoulement ont des signatures très différentes. Pour deux points donnés sur la surface solaire, 1 et 2, la perturbation de temps de parcours due à une perturbation de la température est, en général, indépendante de la direction de propagation entre 1 et 2. Cependant un écoulement avec une composante dirigée le long de la direction $1 \rightarrow 2$ brisera la symétrie dans le temps de parcours pour les ondes se propageant dans des directions opposées : les ondes se déplacent plus rapidement le long de l'écoulement que contre lui. Les champs magnétiques présentent une anisotropie de vitesse d'onde et auront encore une autre signature du temps de parcours (celui-ci n'a pas encore été détecté ).

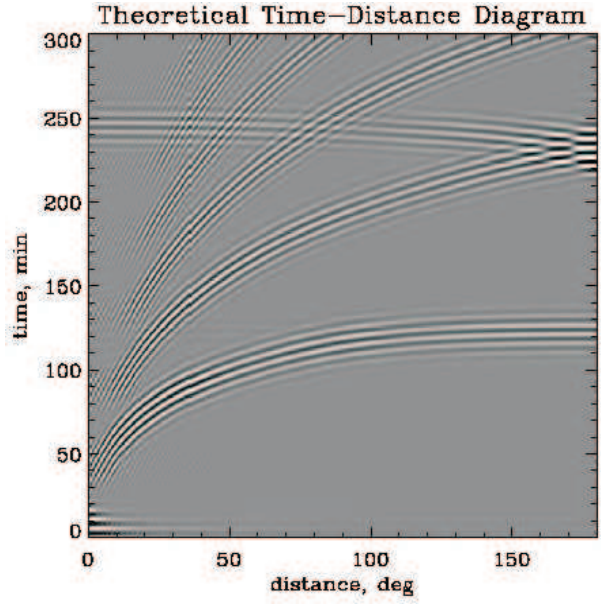

FIG. 4.1 – Diagramme théorique Time-Distance [44].

## 4.2 Théories de base de l'héliosismologie temps-distance

### 4.2.1 Temps de parcours d'onde et théorie de l'héliosismologie temps-distance

La notion de temps de parcours d'onde est un concept clef dans l'héliosismologie Temps-Distance. C'est un concept familier en Géophysique, où les ondes sont généralement générées et excitées par des sources spécifiques localisées dans l'espace et dans le temps. Dans le Soleil, l'excitation des ondes acoustiques est stochastique et il n'est pas possible d'isoler les sources dans l'espace ou dans le temps. Néanmoins Duvall, en 1993 [32], postula la possibilité de mesurer les temps de parcours d'onde en calculant la corrélation temporelle d'un signal à un point donné avec un autre signal à un autre point.

La fonction de corrélation des signaux d'oscillation $f$ en deux points aux coordonnées $\vec{r_1}(r_1;\theta_1,\phi_1)$ et $\vec{r_2}(r_2,\theta_2,\phi_2)$ et un déphasage temporel $\tau$ à la surface solaire (voir annexe (A.3)), s'écrit comme suit [17] :

$$\Gamma(\tau,\Delta) = \int_0^T f(\vec{r_2},t+\tau)f^*(\vec{r_1},t)dt \qquad (4.1)$$

Où $\Delta$ est la distance angulaire entre les deux points et $T$ la durée d'observation, sans oublier que notre signal $f$, et tel que vu dans le chapitre précédent (l'équa.(3.51)), peut être réécrit sous une forme plus générale :

$$f(\vec{r},t) = f(r,\theta,\phi,t) = \sum_{n\ell m} a_{n\ell m}\xi_{n\ell}(r)Y_\ell^m(\theta,\phi)e^{(i[\omega_{n\ell m}t+\alpha_{n\ell m}])} \qquad (4.2)$$

Avec $\xi_{n\ell}(r)$ la fonction radiale de notre signal, et $\alpha_{n\ell m} \propto \vec{k}.\vec{r}$ la phase de celui-ci.

Ainsi, chaque composante d'oscillation avec la fréquence ω et la longueur d'onde spatiale $k(\omega)$ peut être représentée par une onde plane : $A\cos(k.s - \omega t)$. Où $A(\omega)$ est l'amplitude, $\vec{k}$ le vecteur d'onde, $\vec{s}$ le vecteur de la tangente le long d'un certain chemin, $t$ le temps, et $\vec{k}.\vec{s}$ désigne l'intégrale linéaire $\int_s \vec{k}\, d\vec{s}$ le long du chemin $s$. Le signal d'oscillation est une somme de différentes composantes d'oscillations que l'on peut écrire comme suit :

$$V^{(r)}(\vec{r},t) = \sum_{i=0}^{\infty} A(\omega_i)\cos(\vec{k}_i.\vec{s} - \omega_i t) \tag{4.3}$$

Où $\vec{r}$ est le vecteur position indiquant la position le long du chemin à chaque instant, $\vec{s} = \vec{s}(\vec{r})$, et $k = k(\omega,\vec{r})$ ; l'indice $(r)$ indique que $V$ est réel. Il est plus facile de travailler dans l'espace complexe qu'avec les cosinus. Et pour ce faire, en prenant la transformation d'Hilbert du signal et en écrivant le signal analytique comme la somme du signal et de ces transformations d'Hilbert, la transformation d'Hilbert du $\cos(x)$ étant $(-\sin(x))$, le signal analytique s'écrit aussi :

$$V(\vec{r},t) = \sum_{i=0}^{\infty} A(\omega_i)\cos(\vec{k}_i.\vec{s} - \omega_i t) - \sum_{i=0}^{\infty} A(\omega_i)\sin(\vec{k}_i.\vec{s} - \omega_i t) \tag{4.4}$$

Dans le cas continu :

$$V(\vec{r},t) = \frac{1}{2\pi}\int_{-\infty}^{\infty} A(\omega)\exp[i(\omega t - \vec{k}.\vec{s})]d\omega \tag{4.5}$$

$V(\vec{r},t)$ peut être réarrangé comme une intégrale de Fourier, donnant ainsi :

$$V(\vec{r},t) = \frac{1}{2\pi}\int_{-\infty}^{\infty} v(\omega,\vec{r})e^{i\omega t}d\omega \tag{4.6}$$

Où $v(\omega,\vec{r}) = A(\omega)\exp[-i\vec{k}(\omega,\vec{r}).\vec{s}(\vec{r})]$, $A(\omega)$ est la distribution des amplitudes sur la fréquence, et $k(\omega,r)$ est la relation de dispersion du milieu dans lequel l'onde se propage. Si le milieu est homogène, $k$ est indépendant de $r$. Le signal étant l'amplitude du paquet d'onde en un point dans l'espace-temps indiqué par le vecteur position $\vec{r}$ et le temps $t$, le paquet d'onde se propage dans le milieu à un autre point $\vec{r} + \Delta\vec{r}$ après un temps $\Delta t$, et le signal en ce point s'écrit comme suit :

$$V(\vec{r}+\Delta\vec{r},t+\Delta t) = \frac{1}{2\pi}\int_{-\infty}^{\infty} v(\omega,\vec{r}+\Delta\vec{r})e^{i\omega(t+\Delta t)}d\omega \tag{4.7}$$

là où $v(\omega,\vec{r}+\Delta\vec{r}) = A(\omega)\exp[-i\vec{k}(\omega,\vec{r}+\Delta\vec{r}).\vec{s}(\vec{r}+\Delta\vec{r})]$, ainsi, la fonction de corrélation des signaux à $\vec{r}$ et $\vec{r}+\Delta\vec{r}$ est :

$$\Gamma(\vec{r},\Delta\vec{r},\tau) = \int_{-\infty}^{\infty} V(\vec{r}+\Delta\vec{r},t+\Delta t)V^*(\vec{r},t)dt \tag{4.8}$$

Où $\tau = \Delta t$ et l'astérisque (*) indique qu'on prend le conjugué complexe. Si les vecteurs position initiaux et finals $\vec{r}$ et $\vec{r}+\Delta\vec{r}$ sont très proches, $\Delta\vec{r} \mapsto d\vec{r}$, alors les extrémités des vecteurs de position représentent les fins de la tangente du chemin pris par le paquet d'onde $\vec{s}$, où $\vec{s} = d\vec{r}$, pour n'importe quelle valeur arbitraire de $\Delta\vec{r}$,

$$\Delta\vec{r} = \int_{\vec{r}}^{\vec{r}+\Delta\vec{r}} d\vec{r} \tag{4.9}$$

Où l'intégrale est prise le long du chemin $\vec{s}$. Ce faisant, $\vec{s}$, $\vec{r}$ et $\vec{r}+\Delta\vec{r}$ sont remplacés par $\vec{s}$ et $\vec{s}+\Delta\vec{s}$. Par conséquent, $\Gamma$ peut être écrit en fonction de la distance de chemin parcouru le long du chemin $\vec{s}$, on a $\Gamma(\vec{r}, \Delta \vec{r}, \tau) \equiv \Gamma(\vec{r}, \delta \vec{s}, \tau)$. Si la position initiale est prise à $s = 0$ et la position finale comme $s$; $\Gamma(\vec{r}, \Delta \vec{s}, \tau) \equiv \Gamma(\vec{r}, \vec{s}, \tau)$ et si $s = 0$ à $t = 0$, puis $\tau \equiv t$; en d'autres termes, tous les $\Delta s$ peuvent être ignorés; les $\delta s$ seront maintenus pour le moment et seront ignorés quand la position et le temps initiaux seront pris à $(0,0)$.

La corrélation dans le domaine du temps est équivalente à la transformation inverse de Fourier dans le domaine des phases (le produit de la transformée de Fourier (voir le théorème de Weinner-Khintchine en annexe A.3), et la fonction de corrélation est :

$$\Gamma(\vec{r}, \Delta s, \tau) = \frac{1}{2\pi} \int_{-\infty}^{\infty} \nu(\omega, \vec{r} + \Delta \vec{r}) \nu^*(\omega, \vec{r}) e^{i\omega\tau} d\omega, \tag{4.10}$$

qui peut être écrite comme suit :

$$\Gamma(\vec{r}, \Delta s, \tau) = \frac{1}{2\pi} \int_{-\infty}^{\infty} A^2(\omega) e^{i\mu(\omega)t} d\omega, \tag{4.11}$$

où

$$\mu(\omega) = \omega - \frac{1}{\tau}[\vec{k}(\omega, \vec{r} + \Delta \vec{r}).\vec{s}(\vec{r} + \Delta \vec{r}) - \vec{k}(\omega, \vec{r}).\vec{s}(\vec{r})]. \tag{4.12}$$

Par définition $A$ est réel, et $A^*A = A^2$. La fonction $\mu$ peut être encore simplifiée et remplacée par une forme intégrale pour donner :

$$\mu(\omega) = \omega - \frac{1}{\tau}\Delta[\vec{k}(\omega, \vec{r}).\vec{s}(\vec{r})] = \omega - \frac{1}{\tau}\Delta \int_s \vec{k}(\omega, \vec{r}).d\vec{r} \tag{4.13}$$

où $\int_s$ est l'intégrale le long de n'importe quel chemin $\vec{s}$. Par conséquent la fonction de corrélation se propage le long de $\vec{s}$.

En général, l'intégrale dans l'équation (4.11) ne renferme pas des expressions analytiques compliquées et son integrabilité dépend de la forme de $\mu$ et de $A$. La forme de $\mu$ dépend de la relation de dispersion du milieu dans lequel les ondes se propagent. Dans les prochaines sections, les solutions de l'équation (4.11) seront étudiées pour différentes relations de dispersion.

**Milieu non-dispersif**

Un milieu est non-dispersif quand la vitesse de groupe est égale à la vitesse de phase,

$$\frac{\partial \omega}{\partial k}.\hat{k} = \frac{\omega}{k}.\hat{k} \tag{4.14}$$

Ce qui implique que $|\vec{k}| = a\omega$, où $a$ pourrait être une fonction de $\vec{r}$, auquel cas, le milieu est non homogène, parce que $\vec{k}$ est une fonction de la position $\vec{r}$, $\vec{k} = \vec{k}(\omega, \vec{r})$. En remplaçant dans l'équation (4.13), $\mu$ devient :

$$\mu(\omega) = \omega\left(1 - \frac{1}{\tau}\Delta \int_s a\hat{k}.d\vec{r}\right) = \beta\omega \tag{4.15}$$

où $\beta$ est une constante fonction de la position et du temps, et $\hat{k}$ est le vecteur unitaire le long du vecteur $\vec{k}$. Par conséquent (4.11) devient :

$$\Gamma(\vec{r}, \Delta s, \tau) = \frac{1}{2\pi} \int_{-\infty}^{\infty} A^2(\omega) e^{i\omega\eta} d\omega \tag{4.16}$$

où $\eta = \beta\tau$ et $\Gamma$ est simplement la transformée de Fourier inverse de $A^2$. Le maximum de l'intégrale de Fourier se produit à $\eta = 0$, ou $\tau = \Delta \int_s a\hat{k}.dr$, ce qui implique que l'intégrale est exécuté le long du chemin donné par $d\overrightarrow{r}/dt = \omega/k.\hat{k}$. Dans un milieu non-dispersif le paquet d'onde voyage le long de la direction indiquée avec la vitesse de groupe ou la vitesse de phase, parce qu'elles sont identiques.

Si $A(\omega)$ est une gaussienne par exemple, $A(\omega) = A_c \exp[-b(\omega - \omega_0)^2]$, où $\omega_0$ est la fréquence centrale et $b$ est une mesure de la largeur de la gaussienne, on obtient après quelques changement de variables : $\Gamma = (A_c^2(\pi)^{1/2}/2(2b)^{1/2}) \exp(-\eta^2/8b) \exp(i\omega_0\eta)$. La fonction de corrélation a une enveloppe, dont le maximum se produit à $\eta = 0$ ou $\tau = \int_s a\vec{k}.dr$, ou à la fois $t = \int_s a\vec{k}dr$ si la position et le temps initiaux arbitraires sont $(0,0)$. Il est pratique de représenter $a$ comme $1/c_s$, où $c_s$ a la dimension d'une vitesse ; alors le temps pris par le paquet d'onde d'un point arbitraire à un autre le long du chemin $s$ est $t = \int_s (\hat{k}.dr/c_s)$. On note que le maximum de l'amplitude $A_c^2(\pi)^{1/2}/2(2b)^{1/2}$ à $\eta = 0$ n'est pas une fonction de $\Delta s$ ou $\tau$. Cependant, ce n'est pas le cas dans un milieu dispersif, où $\Gamma$ peut s'affaiblir avec la distance parcourue même dans un système non dissipatif.

**Milieu dispersif**

En général, $A(\omega)$ est une fonction qui est appréciable seulement dans un petit intervalle de fréquence autour de $\omega_0$. Par conséquent $\mu(\omega)$ peut s'écrire sous la forme d'une série de Taylor autour de $\omega_0$ et l'équation (4.11) devient :

$$\Gamma(\overrightarrow{r},\Delta s,\tau) = \frac{1}{2\pi}\int_{-\infty}^{\infty} A^2(\omega) \exp\left\{i\tau \sum_{p=0}^{\infty} \frac{\mu_0^{(p)}(\omega-\omega_0)^p}{p!}\right\} d\omega \tag{4.17}$$

où

$$\mu_0 = \omega_0 - \frac{1}{\tau}\Delta\int_s \overrightarrow{k}_0.d\overrightarrow{r}, \quad \mu_0' = 1 - \frac{1}{\tau}\Delta\int_s \left.\frac{\partial \overrightarrow{k}}{\partial \omega}\right|_0 .d\overrightarrow{r}$$

$$,\mu_0'' = -\frac{1}{\tau}\Delta\int_s \left.\frac{\partial^2 \overrightarrow{k}}{\partial \omega^2}\right|_0 .d\overrightarrow{r}, \quad et \quad \mu_0^{(p)} = -\frac{1}{\tau}\Delta\int_s \left.\frac{\partial^p \overrightarrow{k}}{\partial \omega^p}\right|_0 .d\overrightarrow{r} \tag{4.18}$$

Où l'indice supérieur désigne la p-ième dérivée de la fréquence $(\partial^p/\partial\omega^p)$, et $Q_0 = Q(\omega_0)$, $Q$ étant une quantité quelconque.

Dans le cas non-dispersif, $\mu$ s'écrit sous la forme : $\mu = \beta\omega$, où $\beta$ est indépendant de $\omega$ ; par conséquent, $\mu^{(p)} = 0, \forall p \geq 2$. Ainsi $\Gamma$ est la transformée inverse de Fourier de $A^2$.

En général, et dans le cas de nos relations de dispersion solaire, les dérivées d'ordre élevé de $\mu$ disparaissent. Prenons par exemple un milieu où $\mu^{(p)} = 0, \forall p \geq 3$. Il est facile de voir que le milieu avec la relation de dispersion $k = a\omega^2$ a cette propriété. Ceci laisse seulement les deux premiers termes de Taylor dans l'expression de $\mu$, parce que :

$$\frac{\partial \overrightarrow{k}}{\partial \omega} = \frac{2\overrightarrow{k}}{\omega}, \quad \frac{\partial^2 \overrightarrow{k}}{\partial \omega^2} = \frac{2\overrightarrow{k}}{\omega^2}, \quad \frac{\partial^p \overrightarrow{k}}{\partial \omega^p} = 0, \quad \forall p \geq 3. \tag{4.19}$$

Et par conséquent dans l'équation (4.18) :

$$\mu_0 = \omega_0 - \frac{1}{\tau}\Delta\int_s \overrightarrow{k}_0.d\overrightarrow{r}, \quad \mu_0' = 1 - \frac{2}{\tau\omega_0}\Delta\int_s \overrightarrow{k}_0.d\overrightarrow{r},$$

$$\mu_0'' = -\frac{2}{\tau\omega_0^2}\Delta\int_s \overrightarrow{k}_0.d\overrightarrow{r}, \quad \mu_0^{(p)} = 0, \quad \forall p \geq 3. \tag{4.20}$$

L'intégrale de l'équation (4.17) peut être évaluée si $A(\omega)$ est une fonction simple. Par exemple, si $A(\omega)$ est une gaussienne : $A(\omega) = A_c \exp[-b(\omega - \omega_0)^2]$. Comme pour le cas précédent, et en usant de la loi de Moivre et des lois sur les nombres complexes, on obtient la solution :

$$\Gamma(\vec{r}, \Delta s, \tau) = \frac{A_c^2}{\sqrt{2\pi} D^{1/4}} \exp\left[-\frac{2b(\mu_0' \tau)^2}{D}\right] \exp\left\{i\tau\left[\mu_0 - \frac{(\mu_0' \tau)^2 \mu_0''}{2D}\right] + \frac{i}{2} \tan^{-1}\left(\frac{\mu_0'' \tau}{4b}\right)\right\} \quad (4.21)$$

où $D = (\mu_0'' \tau)^2 + 16b^2$, $b^2$ étant le carré de la distance parcourue le long du chemin $\vec{s}$ ($b = 1/\delta\omega$). La propagation de la fonction de corrélation et de l'enveloppe ainsi que la phase qui lui est associée se fait le long du chemin $\vec{s}$. Si $\tau \neq 0$ et $s \neq 0$, alors le maximum de l'enveloppe de $\Gamma$ se produit à $\mu_0' = 0$. Maintenant $\mu'$ détermine le chemin $\vec{s}$ pris par $\Gamma$, et le chemin est :

$$\mu_0' = 0 \Rightarrow \frac{d\vec{r}}{dt} = \left.\frac{\partial \omega}{\partial \vec{k}}\right|_0 \quad (4.22)$$

Les propagations de la fonction de corrélation le long du chemin $\vec{s}$ sont déterminées par la vitesse de groupe $v_g$ à la fréquence centrale. Le paquet d'onde et sa corrélation se propagent le long du même chemin d'onde. En d'autres termes, la crête de l'enveloppe de la fonction de corrélation est obtenue par corrélation croisée du signal en deux différents points sur le chemin d'onde, fournissant le temps de parcours pris par le paquet d'onde entre, eux le long du chemin. Le temps de parcours est déterminé par l'équation de vitesse de groupe (4.22) et est appelé : temps de parcours du groupe $\tau_g$, donné par l'équation (4.22) :

$$\tau_g = \int_s \left.\frac{\partial \omega}{\partial \vec{k}}\right|_0 . d\vec{r} \quad (4.23)$$

Les crêtes de phase de la fonction de corrélation croisée se produisent au temps $\tau$ donné par la résolution de[17] :

$$\tau\left[\mu_0 - \frac{(\mu_0' \tau)^2 \mu_0''}{2D}\right] + \frac{1}{2} \tan^{-1}\left(\frac{\mu_0'' \tau}{4b}\right) = 2n\pi \quad (4.24)$$

Sachant que : $\exp\left[i\tau\left[\mu_0 - \frac{(\mu_0' \tau)^2 \mu_0''}{2D}\right] + \frac{i}{2} \tan^{-1}\left(\frac{\mu_0'' \tau}{4b}\right)\right] = 1$.

Où $n$ est un nombre entier quelconque. Remplacer $n$ par $(2n+1)$ donne le temps correspondant aux crêtes négatives ; prendre $n = (n + 1/2)$ donne le temps correspondant aux croisements zéro. Pour que la crête de phase coïncide avec la crête de l'enveloppe, il faut que $\mu_0' \to 0$ et la contribution due au deuxième terme peut être ignorée comparée à $\mu_0 \tau$. Aux très grandes distances $\vec{s}$, $b \gg |\mu_0''|\tau$, le troisième terme tend vers $\pm\pi/4$ selon que $\mu_0'' > 0$ ou $\mu_0'' < 0$. Et la solution approximative de l'équation (4.24) sera :

$$\tau = \left(2n \pm \frac{1}{4}\right)\frac{\pi}{\omega_0} + \frac{1}{\omega_0} \Delta \int_s \vec{k}_0 . d\vec{r} \quad (4.25)$$

Et le temps correspondant, si la position et le temps initiaux sont $(0,0)$, est désigné sous le nom du temps de phase $\tau_p$, et est indiqué par :

$$\tau_p = \left(2n \pm \frac{1}{4}\right)\frac{\pi}{\omega_0} + \frac{1}{\omega_0} \Delta \int_s \vec{k}_0 . d\vec{r} \quad (4.26)$$

où le chemin d'onde $\vec{s}$ est donné par l'équation (4.22). Le chemin d'onde est uniquement indiqué par l'équa.(4.22) si le système est homogène. Dans un système non homogène, pour toute dimension plus grande que 1, l'ensemble des équations des composantes de l'équa.(4.22) n'est pas fermé [Lighthill 1978]. Des équations additionnelles, qui indiquent uniquement le chemin d'onde $\vec{s}$ et qui décrivent l'évolution des composantes de $\vec{k}$, peuvent être dérivées de la relation de dispersion qui fermera l'ensemble des composantes de l'équa.(4.22) (voir les sections suivantes pour l'ensemble complet des équations d'onde dans la géométrie sphérique).

Même le terme $1/D^{1/4}$ le long du chemin d'onde contribue à un affaiblissement lent de $\Gamma$, pendant que le paquet d'onde se propage, quoique le système soit dissipatif. C'est dû à l'effet dispersif du milieu. Cet effet est lié à la fréquence et dépend de la relation de dispersion du milieu.

Dans le cas général où $A$ représente une fonction quelconque, non nulle en $\omega_0$ et appréciable seulement dans sa proximité, mais faible partout ailleurs, des solutions asymptotiques peuvent être obtenues pour l'intégrale du membre de droite de l'équation (4.17) pour la relation de dispersion $k = a\omega^2$. Laissant seulement le premier terme de l'expression de Taylor pour $A(\omega)$ à $\omega_0$ pour obtenir :

$$\Gamma(\vec{r}, \Delta s, \tau) \approx \frac{A^2(\omega_0)}{2\pi} \int_{-\infty}^{\infty} \exp\left\{i\tau \sum_{p=0}^{2} \frac{\mu_0^{(p)}(\omega - \omega_0)^p}{p!}\right\} d\omega, \qquad (4.27)$$

et en développant, on obtient :

$$\Gamma(\vec{r}, \Delta s, \tau) \approx \frac{A^2(\omega_0)}{\sqrt{2\pi|\mu_0''|\tau}} \exp\left\{i\left[\tau\left(\mu_0 - \frac{\mu_0'^2}{2\mu_0''}\right) \pm \frac{\pi}{4}\right]\right\}, \qquad (4.28)$$

le signe positif doit être choisi si $\mu_0'' > 0$ et le signe négatif si $\mu_0'' < 0$.

La solution (4.21) est la solution asymptotique générale pour toutes les relations de dispersion $k(\omega)$ avec $\mu_0'' \neq 0$, quoique $\mu_0^p \neq 0$, pour $p \geq 3$, si $A(\omega)$ est pris comme gaussienne. C'est une conséquence du lemme de Riemann-Lebesgue, où la contribution due aux termes d'ordres supérieurs doit être négligée comparée aux trois premiers termes dans l'expression de Taylor de $\mu$. Ceci a pour effet d'aboutir à des solutions asymptotiques pour n'importe quelle forme de fonction générale de $A(\omega)$ qui a la propriété que $A(\omega) \neq 0$ et reste appréciable seulement dans un petit intervalle de fréquence autour de la fréquence centrale $\omega_0$. Le lemme de Riemann-Lebesgue garantit que l'intégrale dans le membre de droite de l'équation (4.11) tende vers 0 quand $\tau \to \infty$, si $\int_{-\infty}^{\infty} |A(\omega)|^2 d\omega$ existe et $\mu(\omega)$ est continuellement différentiable pour $-\infty < \tau < \infty$. Ceci ne s'applique pas aux intégrales de la forme $A(\omega) \exp(i\omega\tau)$, quoique $A(\omega)$ est intégrable (par exemple, il ne s'appliquerait pas au cas non dispersif).

Si $\mu' = 0$ pour un certain chemin, alors l'intégrale dans l'équation (4.11) s'annule quand $\tau \to \infty$, et il s'annule moins rapidement que $1/\tau$ parce que la fonction à intégrer oscille moins rapidement près du chemin $\mu' = 0$ que le long de n'importe quel autre chemin $\mu' \neq 0$. Ce chemin s'appelle le chemin stationnaire, ou le chemin d'onde. En d'autres termes, il y a moins d'annulation entre les sous-intervalles adjacents près du chemin stationnaire. Explicitement, en choisissant le chemin stationnaire comme le chemin de l'intégration, la méthode de phase stationnaire (Born & Wolf 1964) prouve que l'intégrale dans l'équation (4.17) conduit à des solutions asymptotiques pour $\tau \to \infty$, ou pendant les périodes $\tau \gg (2\pi/\omega_0)$ ; cette méthode fournit la principale contribution à l'intégrale qui vient de la petite collection de chemins entourant le chemin donné par $\mu' = 0$.

Les détails de la méthode de phase stationnaire peuvent être obtenus ailleurs (Born & Wolf 1964). Ici la méthode est employée pour obtenir les solutions asymptotiques à l'équation (4.11) (quand $\tau \to \infty$). Comme avant, $A(\omega)$ est non nul à $\omega_0$ et est appréciable seulement dans un petit intervalle autour de lui. En remplaçant $A(\omega)$ par $A(\omega_0)$ dans l'équation (4.11), et comme dans l'équation (4.17), en ignorant toutes les dérivées d'ordre $p \geq 3$, parce que le lemme de Riemann-Lebesgue s'assure que leur contribution à l'intégrale est négligeable comparée aux trois premiers termes, l'intégrale est approximée à :

$$\Gamma(\overrightarrow{r}, \Delta s, \tau) \approx \frac{A^2(\omega_0)}{2\pi} \int_{-\infty}^{\infty} \exp\left\{i\tau\left[\frac{\mu_0''}{2}(\omega - \omega_0)^2\right]\right\} d\omega \qquad (4.29)$$

Et la solution asymptotique est :

$$\Gamma(\overrightarrow{r}, \Delta s, \tau) \approx \frac{A^2(\omega_0)}{\sqrt{2\pi |\mu_0''|\tau}} \exp\left\{i\left(\mu_0 \tau \pm \frac{\pi}{4}\right)\right\} \qquad (4.30)$$

Le signe positif doit être choisi si $\mu_0'' > 0$, et le signe négatif si $\mu_0'' < 0$. Dans un système dispersif $A(\omega)$ peut être une fonction de coordonnées spatiales et la méthode de phase stationnaire peut fournir les solutions asymptotiques pour la fonction de corrélation croisée quand toutes les autres échouent.

**Systèmes dissipatifs**

En présence de la dissipation (perte d'énergie), on remplace $A(\omega)$ par $A(\omega)\exp\{-\alpha(\overrightarrow{r})s(\overrightarrow{r})\}$. Le système est dispersif si $\alpha > 0$. Si le système contient des sources qui produisent des oscillations qui s'ajoutent au signal pendant que le paquet d'onde se propage à travers le milieu $\alpha < 0$. Ainsi il est aisé de prouver que la fonction de corrélation devient :

$$\Gamma(\overrightarrow{r}, \Delta s, \tau) = \frac{1}{2\pi} \int_{-\infty}^{\infty} A^2(\omega) e^{-\kappa} e^{i\mu(\omega)\tau} d\omega \qquad (4.31)$$

Avec :

$$\kappa = 2\alpha(\overrightarrow{r})s(\overrightarrow{r}) + \Delta\alpha(\overrightarrow{r})s(\overrightarrow{r}) \qquad (4.32)$$

Si la position et le temps initiaux sont fixés à $(0,0)$, alors $\kappa$ devient :

$$\kappa = \alpha(\overrightarrow{r})s(\overrightarrow{r}) = \int_s \alpha(\overrightarrow{r}) dr \qquad (4.33)$$

Où l'intégrale est le long de $s$. La démarche à suivre est la même que dans la section du cas non dispersif ; $\Gamma$ est la transformée de Fourier inverse de $A^2(\omega)\exp(-\kappa)$ si $\alpha$ dépend de la fréquence, sinon $\Gamma$ est égale à $\exp(-\kappa)$ fois la transformée inverse de Fourier de $A^2(\omega)$ pour le cas où $A(\omega)$ est une gaussienne $A_c \exp[-b(\omega - \omega_0)^2]$. En procédant comme dans la section du cas dispersif, la solution est le produit du membre de droite de l'équation (4.21) avec le facteur de dissipation $\exp(-\kappa)$. Si $\alpha$ dépend de la fréquence, alors des solutions asymptotiques peuvent être obtenues en utilisant la méthode de phase stationnaire et la solution est un produit de $\exp(-\kappa_0)$ par le membre droit de l'équation (4.30).

Strictement, les termes dispersifs apparaîtraient dans la relation de dispersion. Cependant, quand les termes dispersifs dans l'équation sont petits comparés au reste des termes, ils peuvent être ignorés tout en obtenant la relation de dispersion, mais pourraient être introduits plus tard.

L'amplitude de la fonction de corrélation croisée fournit des détails sur les forces dispersives le long du chemin du paquet d'onde et des paramètres liés à la convection ou à la génération d'ondes pourraient être extraits.

**Remarque :**

Il est important de souligner que dans tout ce qui suit, il sera considéré : que le milieu est dispersif (voir relation de disspertion ($\frac{\partial \omega}{\partial k} \neq \frac{\omega}{k}$)), que l'amplitude est sous forme gaussienne (voir équa.(3.54) et Fig.(2.4)) et que l'aspect de la dispertion est uniquement inclus dans les différentes relations de dispersion, étudiées en détail dans le chapitre précédent, rapprochant le plus possible notre modèle solaire de la réalité.

### 4.2.2 Équations d'onde de base en présence d'un écoulement subsurfacique

**Équations d'onde de base en présence d'un écoulement subsurfacique quelconque**

Les oscillations solaires à petites longueurs d'ondes ; i.e. à hautes fréquences (les mode *p*), comparées à la taille de l'échelle locale sont favorables aux traitements des ondes et ont été responsables du progrès considérable de l'Héliosismologie (Gough 1984). Le traitement des ondes suppose que les ondes gravito-acoustiques sont localement planes et que toute quantité physique dans un paquet d'onde s'écrit sous la forme : $\exp[-i\alpha(x_1,x_2,x_3,t)]$ (tel que vu précédémment dans le chapitre 3), où $\alpha$ est ici la phase, définie localement telle que : quelque soit un point donné $(x_1,x_2,x_3)$ de l'espace, la phase croit de $2\pi$ en une période, la fréquence locale étant $\omega = \partial \alpha / \partial t$, et quelque soit $t$, notre phase $\alpha$ décroît de $k_i$ pour chaque direction spatiale $x_i$.

Ainsi la variation spatiale de la phase dans la géometrie sphérique, comme propagation du paquet d'ondes, est donnée par [14] :

$$\frac{\partial \alpha}{\partial r} = -k_r, \tag{4.34}$$

$$\frac{1}{r}\frac{\partial \alpha}{\partial \theta} = -k_\theta, \tag{4.35}$$

$$\frac{1}{r\sin\theta}\frac{\partial \alpha}{\partial \phi} = -k_\phi, \tag{4.36}$$

et

$$\frac{\partial \alpha}{\partial t} = \omega. \tag{4.37}$$

où $\overrightarrow{k} = (k_r, k_\theta, k_\phi)$ est le vecteur d'onde et $\omega$ la fréquence de l'onde. La vitesse de groupe du paquet d'onde est donnée par : $\overrightarrow{v_g} = (v_{gr} = \frac{dr}{dt}, v_{g\theta} = r\frac{d\theta}{dt}, v_{g\phi} = r\sin\theta\frac{d\phi}{dt}) = (\frac{\partial \omega}{\partial k_r}, \frac{\partial \omega}{\partial k_\theta}, \frac{\partial \omega}{\partial k_\phi})$, donc :

$$\frac{dr}{dt} = \frac{\partial \omega}{\partial k_r} \tag{4.38}$$

$$r\frac{d\theta}{dt} = \frac{\partial \omega}{\partial k_\theta} \tag{4.39}$$

$$r\sin\theta\frac{d\phi}{dt} = \frac{\partial \omega}{\partial k_\phi} \tag{4.40}$$

Des équations (4.34)-(4.37), on a :

$$\frac{dk_r}{dt} = \frac{d}{dt}\left(-\frac{\partial \alpha}{\partial r}\right) = -\frac{\partial}{\partial r}\left(\frac{d\alpha}{dt}\right) \tag{4.41}$$

Avec :
$\alpha = \alpha(r,\theta,\phi,t) \Rightarrow d\alpha = \frac{\partial \alpha}{\partial r}dr + \frac{\partial \alpha}{\partial \theta}d\theta + \frac{\partial \alpha}{\partial \phi}d\phi + \frac{\partial \alpha}{\partial t}dt$

Sans oublier que :
$\frac{\partial^2 \alpha}{\partial x_i^2} = 0$ pour $i = 1, 2, 3$.

La même opération est effectuée sur $k_\theta$ et $k_\phi$, on obtient :

$$\frac{dk_r}{dt} = -\frac{\partial \omega}{\partial r} + k_\theta \dot\theta + k_\phi \sin\theta \dot\phi \tag{4.42}$$

$$\frac{dk_\theta}{dt} = -\frac{1}{r}\frac{\partial \omega}{\partial \theta} - \frac{\dot r}{r} k_\theta + k_\phi \cos\theta \dot\phi \tag{4.43}$$

$$\frac{dk_\phi}{dt} = -\frac{1}{r\sin\theta}\frac{\partial \omega}{\partial \phi} - \frac{\dot r}{r} k_\phi - \frac{k_\phi \cos\theta}{\sin\theta}\dot\theta \tag{4.44}$$

Les équations (4.38-4.44) forment l'ensemble des équations d'onde en géométrie sphérique.

Sans se préoccuper de la nature des oscillations, il est simple de démontrer que la fréquence du paquet d'onde se propageant dans ce milieu-ci est un invariant par rapport au temps, sachant que $\omega = \omega(k_r, k_\theta, k_\phi, r, \theta, \phi)$, donc :

$$\frac{d\omega}{dt} = \frac{\partial \omega}{\partial k_r}\frac{dk_r}{dt} + \frac{\partial \omega}{\partial k_\theta}\frac{dk_\theta}{dt} + \frac{\partial \omega}{\partial k_\phi}\frac{dk_\phi}{dt} + \frac{\partial \omega}{\partial r}\frac{dr}{dt} + \frac{\partial \omega}{\partial \theta}\frac{d\theta}{dt} + \frac{\partial \omega}{\partial \phi}\frac{d\phi}{dt} \tag{4.45}$$

D'après les équations d'onde et en remplaçant dans l'équa.(2.15), on trouve effectivement :

$$\frac{d\omega}{dt} = 0 \tag{4.46}$$

A présent, et pour trouver les diverses dérivées de $\omega$ dans les équations d'onde, on a besoin de la relation de dispersion locale des oscillations $\omega = \omega(k_r, k_\theta, k_\phi, r, \theta, \phi)$, déjà vue précédemment dans le chapitre 3 (équa.(3.67)). Dans le cas non rotationnel, en absence de champ magnétique, et en présence d'un écoulement subsurfacique (et près de la surface), $\vec{U} = (\vec{U_r}(r, \theta, \phi), \vec{U_\theta}(r, \theta, \phi), \vec{U_\phi}(r, \theta, \phi))$, où $U << c_s$ ($c_s$ la vitesse du son) ; ainsi les dérivées spatiales de ces écoulements peuvent être négligées dans les équations de mouvement comparées aux fréquences d'oscillations, et par conséquent les relations de dispersion vues précédemment restent les mêmes, à un détail près, car les fréquences ou vitesses mesurées ne sont pas dues qu'aux ondes seulement, mais il faut aussi rajouter à cela l'effet de la perturbation induite par ces écoulements horizontaux, qui provoquent un décalage Doppler des fréquences observées $\omega$ par advection.

Alors on a : $\vec{v} = \vec{\bar v} + \vec{U}$, avec $\vec{v}$ la vitesse de phase des ondes observée + advection, $\vec{\bar v}$ la vitesse sans advection, et bien sur $\vec{U}$ la vitesse d'écoulement horizontale (à cause de l'advection (voir annexe A.2)).

Donc on peut écrire aussi :

$$\frac{\bar\omega}{\vec{k}} = \frac{\omega}{\vec{k}} - \vec{U}, \tag{4.47}$$

ou encore :

$$\bar\omega = \omega - \vec{k}.\vec{U} \quad \Rightarrow \quad \bar\omega = \omega + \delta\omega \tag{4.48}$$

Ainsi notre relation de dispersion, et d'après l'équa.(3.67), s'écrit comme suit :

$$k_r^2 = \frac{\bar\omega^2 - \omega_c^2}{c_s^2} - k_h^2\left(1 - \frac{N^2}{\bar\omega^2}\right) \tag{4.49}$$

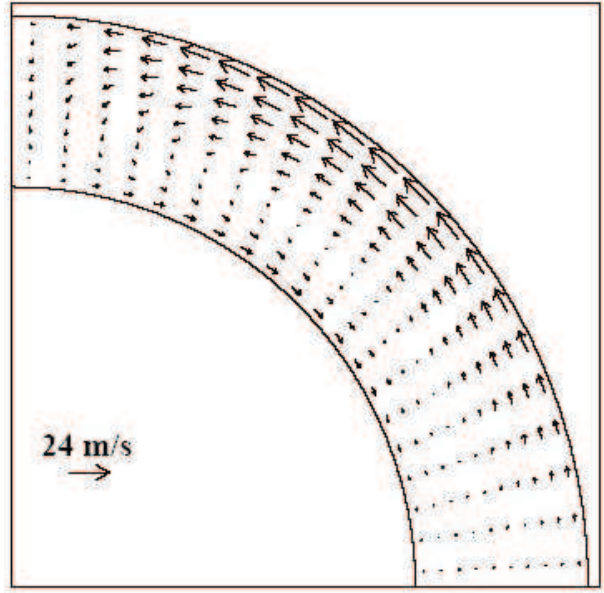

FIG. 4.2 – Coupe méridionale de la zone de convection solaire [41].

Avec $k_h^2 = k_\theta^2 + k_\phi^2$, et $\overrightarrow{U} = U_r \overrightarrow{r} + U_\theta \overrightarrow{\theta} + U_\phi \overrightarrow{\phi}$, ($\overrightarrow{r}$, $\overrightarrow{\theta}$ et $\overrightarrow{\phi}$ les vecteurs unitaires selon $r, \theta$ et $\phi$). On peut écrire alors :

$$\overline{\omega} = \omega - k_r U_r - k_\theta U_\theta - k_\phi U_\phi \tag{4.50}$$

De (4.47) on peut calculer les composantes de la vitesse de groupe $\partial \omega/\partial k_r, \partial \omega/\partial k_\theta, \partial \omega/\partial k_\phi$. En effet, on a :

$$k_r = \frac{1}{\overline{\omega} c_s}\sqrt{\overline{\omega}^4 - \overline{\omega}^2 \omega_c^2 - k_h^2 c_s^2 \overline{\omega}^2 + k_h^2 c_s^2 N^2} \quad \text{et} \quad \frac{\partial \omega}{\partial k_r} = \frac{\partial \overline{\omega}}{\partial k_r} + U_r \tag{4.51}$$

Et

$$k_h = \frac{1}{c_s}\sqrt{\frac{k_r^2 \overline{\omega}^2 c^2 - \overline{\omega}^4 - \overline{\omega}^2 \omega_c^2}{\overline{\omega}^2 - N^2}} \quad \text{et} \quad \frac{\partial \omega}{\partial k_h} = \frac{\partial \overline{\omega}}{\partial k_h} + U_h \tag{4.52}$$

En dérivant par rapport à $\overline{\omega}$ on obtient :

$$\begin{aligned}\frac{\partial \omega}{\partial k_r} &= U_r + \frac{\overline{\omega}^3 c_s^2 k_r}{\overline{\omega}^4 - k_h^2 c_s^2 N^2} \\ \frac{\partial \omega}{\partial k_h} &= U_h + k_h \overline{\omega} c_s^2 \left(\frac{\overline{\omega}^2 - N^2}{\overline{\omega}^4 - k_h^2 c_s^2 N^2}\right)\end{aligned} \tag{4.53}$$

Avec $\overrightarrow{U_h} = U_\theta \overrightarrow{\theta} + U_\phi \overrightarrow{\phi}$.

Finalement, en projetant le dernier terme sur $\overrightarrow{\theta}$ et $\overrightarrow{\phi}$ nous obtenons :

$$\frac{\partial \omega}{\partial k_r} = U_r + \frac{\overline{\omega}^3 c_s^2 k_r}{\overline{\omega}^4 - k_h^2 c_s^2 N^2} \tag{4.54}$$

$$\frac{\partial \omega}{\partial k_\theta} = U_\theta + k_\theta \overline{\omega} c_s^2 \left(\frac{\overline{\omega}^2 - N^2}{\overline{\omega}^4 - k_h^2 c_s^2 N^2}\right) \tag{4.55}$$

$$\frac{\partial \omega}{\partial k_\phi} = U_\phi + k_\phi \overline{\omega} c_s^2 \left(\frac{\overline{\omega}^2 - N^2}{\overline{\omega}^4 - k_h^2 c_s^2 N^2}\right) \tag{4.56}$$

A partir de l'éq.(4.48), on obtient les dérivées spatiales de $\omega$ :

$$\frac{\partial \omega}{\partial r} = k_r \frac{\partial U_r}{\partial r} + k_\theta \frac{\partial U_\theta}{\partial r} + k_\phi \frac{\partial U_\phi}{\partial r} + \frac{\overline{\omega}^3}{2(\overline{\omega}^4 - k_h^2 c_s^2 N^2)}\left[\frac{\partial \omega_c^2}{\partial r} + k^2 \frac{\partial c_s^2}{\partial r} - \frac{k_h^2}{\overline{\omega}^2}\left(c_s^2 \frac{\partial N^2}{\partial r} + N^2 \frac{\partial c_s^2}{\partial r}\right)\right] \tag{4.57}$$

$$\frac{1}{r}\frac{\partial \omega}{\partial \theta} = \frac{k_r}{r}\frac{\partial U_r}{\partial \theta} + \frac{k_\theta}{r}\frac{\partial U_\theta}{\partial \theta} + \frac{k_\phi}{r}\frac{\partial U_\phi}{\partial \theta} \tag{4.58}$$

$$\frac{1}{r\sin\theta}\frac{\partial \omega}{\partial \phi} = \frac{k_r}{r\sin\theta}\frac{\partial U_r}{\partial \phi} + \frac{k_\theta}{r\sin\theta}\frac{\partial U_\theta}{\partial \phi} + \frac{k_\phi}{r\sin\theta}\frac{\partial U_\phi}{\partial \phi} \tag{4.59}$$

**Équations d'onde de base en présence d'un écoulement horizontal subsurfacique**

L'écoulement horizontal est un cas particulier de l'étude du sous chapitre précédent, où $U_r = 0$, ainsi : $\vec{U} = U_\theta \vec{\theta} + U_\phi \vec{\phi}$. Alors :

$$\overline{\omega} = \omega - k_\theta U_\theta - k_\phi U_\phi \tag{4.60}$$

De la même manière on calcule les composantes de la vitesse de groupe $\partial\omega/\partial k_r, \partial\omega/\partial k_\theta, \partial\omega/\partial k_\phi$ :

$$k_r = \frac{1}{\overline{\omega}c_s}\sqrt{\overline{\omega}^4 - \overline{\omega}^2\omega_c^2 - k_h^2 c_s^2 \overline{\omega}^2 + k_h^2 c_s^2 N^2} \quad \text{et} \quad \frac{\partial\overline{\omega}}{\partial k_r} = \frac{\partial\omega}{\partial k_r} \tag{4.61}$$

$$k_h = \frac{1}{c_s}\sqrt{\frac{k_r^2 \overline{\omega}^2 c^2 - \overline{\omega}^4 - \overline{\omega}^2 \omega_c^2}{\overline{\omega}^2 - N^2}} \quad \text{et} \quad \frac{\partial\omega}{\partial k_h} = \frac{\partial\overline{\omega}}{\partial k_h} + U \tag{4.62}$$

Et en dérivant par rapport à $\overline{\omega}$ on obtient :

$$\begin{aligned} v_{gr} &= \frac{\partial\omega}{\partial k_r} = \frac{\overline{\omega}^3 c_s^2 k_r}{\overline{\omega}^4 - k_h^2 c_s^2 N^2} \\ v_{gh} &= \frac{\partial\omega}{\partial k_h} = U + k_h \overline{\omega} c_s^2 \left(\frac{\overline{\omega}^2 - N^2}{\overline{\omega}^4 - k_h^2 c_s^2 N^2}\right) \end{aligned} \tag{4.63}$$

En projetant encore le dernier terme sur $\vec{\theta}$ et $\vec{\phi}$ on obtient :

$$\frac{\partial\omega}{\partial k_r} = \frac{\overline{\omega}^3 c_s^2 k_r}{\overline{\omega}^4 - k_h^2 c_s^2 N^2} \tag{4.64}$$

$$\frac{\partial\omega}{\partial k_\theta} = U_\theta + k_\theta \overline{\omega} c_s^2 \left(\frac{\overline{\omega}^2 - N^2}{\overline{\omega}^4 - k_h^2 c_s^2 N^2}\right) \tag{4.65}$$

$$\frac{\partial\omega}{\partial k_\phi} = U_\phi + k_\phi \overline{\omega} c_s^2 \left(\frac{\overline{\omega}^2 - N^2}{\overline{\omega}^4 - k_h^2 c_s^2 N^2}\right) \tag{4.66}$$

Et les dérivées spatiales de $\omega$ sont aussi obtenues de la même manière que précédemment :

$$\frac{\partial\omega}{\partial r} = k_\theta \frac{\partial U_\theta}{\partial r} + k_\phi \frac{\partial U_\phi}{\partial r} + \frac{\overline{\omega}^3}{2(\overline{\omega}^4 - k_h^2 c_s^2 N^2)}\left[\frac{\partial\omega_c^2}{\partial r} + k^2 \frac{\partial c_s^2}{\partial r} - \frac{k_h^2}{\overline{\omega}^2}\left(c_s^2 \frac{\partial N^2}{\partial r} + N^2 \frac{\partial c_s^2}{\partial r}\right)\right] \tag{4.67}$$

$$\frac{1}{r}\frac{\partial\omega}{\partial\theta} = \frac{k_\theta}{r}\frac{\partial U_\theta}{\partial\theta} + \frac{k_\phi}{r}\frac{\partial U_\phi}{\partial\theta} \tag{4.68}$$

$$\frac{1}{r\sin\theta}\frac{\partial\omega}{\partial\phi} = \frac{k_\theta}{r\sin\theta}\frac{\partial U_\theta}{\partial\phi} + \frac{k_\phi}{r\sin\theta}\frac{\partial U_\phi}{\partial\phi} \tag{4.69}$$

**Équations d'onde de base dans le Soleil en rotation**

La rotation est considérée comme un cas spécial dans le traitement des vitesses des écoulements horizontaux. Ainsi dans le cas d'une rotation, la vitesse du milieu $\vec{U}$ est donnée par : $\vec{U} = \vec{\Omega} \times \vec{r}$, où la vitesse angulaire de rotation $\vec{\Omega} = (\Omega\cos\theta, \Omega\sin\theta, 0)$ (ici en coordonnées cartésiennes) et $\vec{\Omega} = \vec{\Omega}(r,\theta)$ est fonction de $r$ et $\theta$, alors, et d'après les équations vues dans le sous chapitre précédent, on remplace et on obtient [14] :

$$\vec{U} = \Omega r \sin\theta \, \vec{\phi} \tag{4.70}$$

$$\overline{\omega} = \omega - k_\phi \Omega r \sin\theta \tag{4.71}$$

De la même manière que pour les équa.(4.64-4.66), on obtient les différentes dérivées de ω par rapport aux composantes du vecteur d'onde :

$$\frac{\partial \omega}{\partial k_r} = \frac{\overline{\omega}^3 c_s^2 k_r}{\overline{\omega}^4 - k_h^2 c_s^2 N^2} \tag{4.72}$$

$$\frac{\partial \omega}{\partial k_\theta} = k_\theta \overline{\omega} c_s^2 \left( \frac{\overline{\omega}^2 - N^2}{\overline{\omega}^4 - k_h^2 c_s^2 N^2} \right) \tag{4.73}$$

$$\frac{\partial \omega}{\partial k_\phi} = \Omega r \sin\theta + k_\phi \overline{\omega} c_s^2 \left( \frac{\overline{\omega}^2 - N^2}{\overline{\omega}^4 - k_h^2 c_s^2 N^2} \right) \tag{4.74}$$

Et finalement, les dérivées spatiales de la fréquence s'écrivent (de la même manière que pour les équa.(4.67-4.69)) comme suit :

$$\frac{\partial \omega}{\partial r} = k_\phi \sin\theta \left( \Omega + r\frac{\partial \Omega}{\partial r} \right) + \frac{\overline{\omega}^3}{2(\overline{\omega}^4 - k_h^2 c_s^2 N^2)} \left[ \frac{\partial \omega_c^2}{\partial r} + k^2 \frac{\partial c_s^2}{\partial r} - \frac{k_h^2}{\overline{\omega}^2} \left( c_s^2 \frac{\partial N^2}{\partial r} + N^2 \frac{\partial c_s^2}{\partial r} \right) \right] \tag{4.75}$$

$$\frac{1}{r}\frac{\partial \omega}{\partial \theta} = k_\phi \left( \Omega \cos\theta + \frac{\partial \Omega}{\partial \theta} \sin\theta \right) \tag{4.76}$$

$$\frac{1}{r\sin\theta}\frac{\partial \omega}{\partial \phi} = 0 \tag{4.77}$$

### 4.2.3 Calcul des chemins d'onde

En supposant que l'onde entre deux sauts apparaît en deux points $\vec{r_1}(r_1, \theta_1, \phi_1)$ et $\vec{r_2}(r_2, \theta_2, \phi_2)$) à la surface de la sphère solaire, on peut décrire le carré de la distance droite séparant les deux points par :

$$(\vec{r_1}(r_1, \theta_1, \phi_1) - \vec{r_2}(r_2, \theta_2, \phi_2))^2 = \vec{r_1}^2 + \vec{r_2}^2 - 2\vec{r_1}\vec{r_2}\cos\Delta, \tag{4.78}$$

avec $\cos\Delta = \cos\theta_2\cos\theta_1 + \sin\theta_2\sin\theta_1\cos(\phi_2 - \phi_1)$

Ainsi, la distance réelle que parcourt une onde entre deux sauts en géométrie sphérique est déterminée par :

$$\Delta = \cos^{-1}[\cos\theta_2\cos\theta_1 + \sin\theta_2\sin\theta_1\cos(\phi_2 - \phi_1)] \tag{4.79}$$

Notons, que jusque là, on a travaillé qu'avec les coordonnées sphériques, mais comme les Dopllerogrammes sont des images 2D du disque solaire, travailler avec les coordonnées polaires $(r, \theta)$ s'avère plus approprié. Ainsi, et sans prendre en compte l'écoulement, les composantes de la vitesse de groupe sont :

$$v_{gr} = \frac{dr}{dt} = \frac{\partial \omega}{\partial k_r} = \frac{k_r \omega^3}{\omega^4 - k_h^2 c_s^2 N^2} \quad (4.80)$$

$$v_{gh} = r\frac{d\theta}{dt} = \frac{\partial \omega}{\partial k_\theta} = k_h \omega c_s^2 \left( \frac{\omega^2 - N^2}{\omega^4 - k_h^2 c_s^2 N^2} \right) \quad (4.81)$$

où $v_{gh}$ est la vitesse de groupe horizontale, avec $\vec{k} = k_r \vec{r} + k_\theta \vec{\theta}$, (ainsi $k_\theta = k_h = \frac{L}{r}$) et $\vec{r} = dr\vec{r} + rd\theta\vec{\theta}$.

Le rapport des deux dernières équations nous donne :

$$\frac{v_{gr}}{v_{gh}} = \frac{dr}{rd\theta} \quad \Rightarrow \quad d\theta = \frac{v_{gh}}{rv_{gr}} dr \quad (4.82)$$

Si on suppose que les coordonnées du point tournant supérieur sont $\vec{r_1}(r_1, \theta_1)$ et que celles du point tournant inférieur sont $\vec{r_2}(r_2, \theta_2)$, alors la distance de parcours d'onde $\Delta$, pour un saut qu'on définit par $\sigma = 1$, s'écrit comme suit :

$$\Delta = 2|\theta_2 - \theta_1| = 2\int_{\theta_1}^{\theta_2} d\theta = 2\int_{r_1}^{r_2} \frac{v_{gh}}{rv_{gr}} dr \quad (4.83)$$

Où $s$ dénote ici le chemin le long du chemin d'onde. Les temps de groupe $\tau_g$ et de phase $\tau_p$ peuvent être écrits ainsi :

$$\tau_g = \int_s \frac{d\vec{r}}{v_g} = \int_{r_1}^{r_2} \frac{dr}{v_{gr}} \quad (4.84)$$

et :

$$\tau_p = \int_\Gamma \frac{d\vec{r}}{v_p} = \int_\Gamma \frac{\vec{k} d\vec{r}}{\omega} \quad (4.85)$$

### 4.2.4 Équivalence entre la loi de Duvall et la courbure temps-distance

A partir de l'équation (4.26), qu'on peut réécrire sous la forme suivante [15] :

$$\tau_p + \beta(\omega) = \int_\Gamma \frac{d\vec{r}}{v_p} \quad (4.86)$$

Avec $\vec{v_p} = \omega/\vec{k}$ et le facteur d'ambiguïté $\beta(\omega) = -\left(2n \pm \frac{1}{4}\right)\frac{\pi}{\omega}$ ; le signe $(-)$ n'influence en rien ce dernier, on peut donc l'omettre.

Si $(r_1, \Delta/2)$ est le point tournant inférieur et $(R,0) \& (R,\Delta)$ les points tournants supérieurs, on peut réécrire notre temps de phase sous la forme :

$$\tau_p + \beta(\omega) = 2\int_0^{\Delta/2} \frac{rk_\theta}{\omega} d\theta + 2\int_{r_1}^R \frac{k_r}{\omega} dr \quad (4.87)$$

D'après l'équa.(4.83), l'équa.(4.87) devient :

$$\tau_p + \beta(\omega) = \frac{rk_\theta}{\omega}\Delta + 2\int_{r_1}^R \frac{k_r}{\omega} dr \quad (4.88)$$

En reliant la pente de courbure temps-distance à la vitesse tangentielle de phase : $w_p = \frac{\omega}{L} = \frac{\omega}{rk_\theta}$, on peut écrire aussi :

$$w_p = \frac{d\Delta}{d\tau_p} \quad \Rightarrow \quad \frac{d\tau_p}{d\Delta} = \frac{1}{w_p} = \frac{rk_\theta}{\omega} \quad (4.89)$$

4.2 Théories de base de l'héliosismologie temps-distance                                        49Et l'équa.(4.88) devient :

$$\tau_p + \beta(\omega) - \frac{d\tau_p}{d\Delta}\Delta = 2\int_{r_1}^{R} \frac{k_r}{\omega}dr \qquad (4.90)$$

En posant $E(\Delta) = \frac{1}{2}[\beta + \tau_p - \frac{d\tau_p}{d\Delta}\Delta]$ et connaissant l'expression de $\beta$, $\tau_p$ et $\frac{d\tau_p}{d\Delta}\Delta$, tout en sachant que $k_r = -\frac{\partial \alpha}{\partial r}$, avec $\alpha$ la phase, on obtient :

$$E(\Delta) = F(w_p) \simeq (n+\alpha)\frac{\pi}{\omega} \simeq \int_{r_1}^{R} \frac{k_r}{\omega}dr \qquad (4.91)$$

Ce qui correspond à :

$$(n+\alpha)\pi \simeq \int_{r_1}^{R} k_r dr \qquad (4.92)$$

Avec $\alpha = \alpha(R) - \alpha(r_1)$ le déphasage aux points tournants. Et là on retrouve bien, en effet, la loi de Duvall (1982) [31] (équa.(3.60)), à partir de la courbure temps-distance.

**Conclusion :**

Le fait de retrouver la loi de Duvall (une des lois fondamentales en Héliosismologie), simplement à partir des lois de courbures temps-distance, prouve que les hypothèses de base de celles-ci sont bien posées, et ceci conforte fortement notre modèle théorique précédemment établi (voir la Remarque de la section (4.2.1)).

### 4.2.5 Approche théorique du signal corrélé et fonction de Gabor

A partir des équations (3.54) du la section (3.3.1), et en supposant que $F$ soit la transformée de Fourier du signal $f$ [41][59] :

$$F(r,\theta,\phi,\omega) = \int_0^\infty f(r,\theta,\phi,t)e^{-i\omega t}dt = \sum_{n\ell m} a_{n\ell m}\xi_{n\ell}(r)Y_\ell^m(\theta,\phi)e^{i\alpha_{n\ell m}}\delta(\omega - \omega_{n\ell m}) \qquad (4.93)$$

Où $\delta(\omega - \omega_{n\ell m}) = \int_{-\infty}^{\infty} e^{-i(\omega-\omega_{n\ell m})t}dt$ est la fonction delta de Dirac. La fonction de corrélation, d'après le théorème de Weinner-Khintchine, peut être réécrite comme suit :

$$\Gamma(\tau,\Delta) = \int_0^T f(\vec{r_2},t+\tau)f^*(\vec{r_1},t)dt = \int_{-\infty}^{\infty} F(\vec{r_2},\omega)F^*(\vec{r_1},\omega)e^{i\omega\tau}d\omega \qquad (4.94)$$

En remplaçant par l'expression de $F$ dans la dernière équation, sans oublier l'usage d'un filtre afin de ne sélectionner que les fréquences du 'mode p' via la Gaussienne de transfert $G(\omega)$, avec :

$$G(\omega) = \exp\left[-\left(\frac{\omega-\omega_0}{\delta\omega}\right)^2\right] \qquad (4.95)$$

où $\omega$ est la fréquence cyclique, $\omega_0$ la fréquence centrale, et $\delta\omega$ la largeur de bande caractéristique de notre filtre (voir la section (5.6)),
sans oublier que les deux sommations $\int d\omega$ et $\sum_{n\ell m}$ sont éliminées par les deux fonctions de Dirac, qui résultent des expressions de $F$ et $F^*$, on obtient alors :

$$\Gamma(\tau,\Delta) = \sum_{n\ell} \sqrt{2\ell+1}\exp\left[-\left(\frac{\omega_{n\ell}-\omega_0}{\delta\omega}\right)^2 + i\omega_{n\ell}\tau\right]\sum_{m=-\ell}^{\ell}\sum_{m'=-\ell}^{\ell} Y_\ell^m(\theta_2,\phi_2)e^{i\alpha_{n\ell m}}Y_\ell^{m'*}(\theta_1,\phi_1)e^{-i\alpha_{n\ell m'}}$$

$$(4.96)$$

Avec, et pour des raisons de simplification, la supposition que les amplitudes ne dépendent que de $n$ et $\ell$ : $\sum_{n\ell m} a_{n\ell m} \xi_{n\ell}(R_\odot) \sum_{n'\ell'm'} a_{n'\ell'm'} \xi_{n'\ell'}(R_\odot) = \sum_{n\ell} \sqrt{2\ell+1}$, où $r = R_\odot$, car les mesures sont bien prises à partir de la surface solaire.

Les phases étant aléatoires, on suppose que le terme $e^{i(\alpha_{n\ell m} - \alpha_{n\ell m'})}$ en moyenne tend à s'annuler, excepté bien sur pour $m = m'$, et on réécrit notre dernière équation sous la forme :

$$\Gamma(\tau, \Delta) = \sum_{n\ell} \sqrt{2\ell+1} \exp\left[-\left(\frac{\omega_{n\ell} - \omega_0}{\delta\omega}\right)^2 + i\omega_{n\ell}\tau\right] \sum_{m=-\ell}^{\ell} Y_\ell^m(\theta_2, \phi_2) Y_\ell^{m*}(\theta_1, \phi_1) \quad (4.97)$$

D'après les équations des harmoniques sphériques [41], on a :

$$\sum_{m=-\ell}^{\ell} Y_\ell^m(\theta_2, \phi_2) Y_\ell^{m*}(\theta_1, \phi_1) = \left(\frac{2\ell+1}{4\pi}\right) P_\ell(\cos\Delta)$$
$$\&$$
$$P_\ell(\cos\Delta) \simeq J_0\left([2\ell+1]\sin\frac{\Delta}{2}\right) \simeq \sqrt{\frac{2}{\pi L\Delta}} \cos\left(L\Delta - \frac{\pi}{4}\right) \quad (4.98)$$

où $P_\ell$ est la fonction de Legendre d'ordre $\ell$, $J_0$ la fonction de Bessel au premier ordre [1], $\Delta = \cos^{-1}[\cos\theta_1\cos\theta_2 + \sin\theta_1\sin\theta_2\cos(\phi_1 - \phi_2)]$ est la distance entre les deux points $(\theta_1, \phi_1)$ et $(\theta_2, \phi_2)$, avec $L = \sqrt{\ell(\ell+1)} \simeq \ell + 1/2$ ($\ell \longrightarrow \infty$).

La dernière approximation, n'est valide que dans le cas où $\Delta$ est petit, mais $L\Delta$ grand ; ce qui est le cas pour notre étude, car nos distances angulaires sont petites par rapport à la taille de la surface solaire et notre ordre angulaire peut atteindre l'infini théoriquement et $10^4$ en pratique. Sur la base de ces dernières constatations et après diverses manipulations mathématiques, on obtient :

$$\Gamma(\tau, \Delta) = \frac{2}{\sqrt{\pi\Delta}} \sum_{n\ell} \exp\left(-\frac{\omega_{n\ell} - \omega_0}{\delta\omega}\right)^2 \cos\omega_{n\ell}\tau \cos L\Delta \quad (4.99)$$

La double sommation peut être réduite en une somme d'intégrales si on regroupe les modes de sorte que la somme externe soit sur le rapport $w_p = \omega/L$ (la vitesse de phase tangentielle : voir la section (4.2.4)) et la somme interne sur $\omega$. Sachant que la distance de parcours $\Delta$ est déterminée par $w_p$ ; $\Delta$ est autrement indépendante de $\omega$. D'après la nature de la bande limite de la fonction $G$, seules les valeurs de $L$ proches de $L_0 = \omega_0/w_p$ contribuent dans la sommation, et en développant $L$ près de la fréquence centrale via le développement de Taylor, on écrit :

$$L\Delta \simeq \Delta\left[L(\omega_0) + \frac{\partial L}{\partial \omega}(\omega - \omega_0)\right] = \Delta\left[\frac{\omega_0}{w_p} + \frac{\omega - \omega_0}{w_g}\right] \quad (4.100)$$

où $w_g = \partial\omega/\partial L$ est la vitesse de groupe tangentielle. En outre, en se rappelant que les sinus sont éliminés dans la sommation, on peut remplacer $\cos\omega\tau\cos L\Delta$ par $\cos(\omega\tau - L\Delta)$, et connaissant l'expression de $L\Delta$, on obtient :

$$\cos\left[\left(\tau - \frac{\Delta}{w_g}\right)\omega + \left(\frac{1}{w_g} - \frac{1}{w_p}\right)\Delta\omega_0\right] \quad (4.101)$$

Ainsi $\Gamma(\tau, \Delta)$ s'écrit comme suit :

$$\Gamma(\tau, \Delta) = \sum_{w_p} \frac{2}{\sqrt{\pi\Delta}} \sum_{\omega} \exp\left(-\frac{\omega - \omega_0}{\delta\omega}\right)^2 \cos\left[\left(\tau - \frac{\Delta}{w_g}\right)\omega + \left(\frac{1}{w_g} - \frac{1}{w_p}\right)\Delta\omega_0\right] \quad (4.102)$$

La somme interne peut être approximée par une intégrale sur $\omega$, car d'après Gradshteyn & Ryzhik [48] (page 480, 3.896/2), on a :

$$\int_{-\infty}^{\infty} e^{-q^2x^2} \cos[p(x+\lambda)]dx = \frac{\sqrt{\pi}}{q} e^{-\frac{p^2}{4q^2}} \cos p\lambda \quad (4.103)$$

Ainsi, et après avoir effectué le changement de variable $x = \omega - \omega_0$ on obtient :

$$\int_{-\infty}^{\infty} \exp\left(-\frac{\omega - \omega_0}{\delta\omega}\right)^2 \cos\left[\left(\tau - \frac{\Delta}{w_g}\right)\omega + \left(\frac{1}{w_g} - \frac{1}{w_p}\right)\Delta\omega_0\right] d\omega$$
$$= \sqrt{\pi\delta\omega^2} \exp\left[-\frac{\delta\omega^2}{4}\left(\tau - \frac{\Delta}{w_g}\right)^2\right] \cos\left[\omega_0\left(\tau - \frac{\Delta}{w_p}\right)\right] \quad (4.104)$$

Finalement, on peut résumer $\Gamma(\tau, \Delta)$ comme suit :

$$\Gamma(\tau,\Delta) \propto \sum_{w_p} \cos\left[\omega_0\left(\tau - \tau_p\right)\right] \exp\left[-\frac{\delta\omega^2}{4}\left(\tau - \tau_g\right)^2\right] \quad (4.105)$$

où $\tau_g = \Delta/w_g$ est le temps de groupe et $\tau_p = \Delta/w_p$ est le temp de phase, on peut aussi écrire :

$$\Gamma(\tau,\Delta) = A\cos\left[\omega_0\left(\tau - \tau_p\right)\right] \exp\left[-\frac{\delta\omega^2}{4}\left(\tau - \tau_g\right)^2\right] \quad (4.106)$$

*A* étant l'amplitude du signal corrélé. L'expression finale est plus communément connue sous le nom de : fonction de Gabor.

La démonstration, présentée dans cette section, nous sera trés utile dans le traitement des données, et nous permettra d'approximer notre signal corrélé par une fonction de Gabor et d'en tirer les temps de parcours qui, à leur tour une fois inversés, nous fournissent de précieuses informations sur l'intérieur solaire.

La relation, entre le temps de parcours de l'onde et les paramètres internes solaires, nous est permise via le principe de Fermat qui est bien explicité dans la section suivante :

### 4.2.6 Principe de Fermat et perturbation du temps de parcours

L'une des propriétés des chemins d'onde est qu'ils obéissent au principe de Fermat, qui stipule que le temps de parcours le long du rayon est stationnaire en ce qui concerne de petits changement de chemin et ceci implique, que si une petite perturbation est faite à l'état fondamental, le chemin reste inchangé et la perturbation du temps de parcours peut être écrite ainsi [41] :

$$\delta\tau = \tau - \tau_0 = \frac{1}{\omega} \int_\Lambda \delta k \, ds \quad (4.107)$$

Où $\delta k$ est la perturbation du vecteur d'onde dûe aux inhomogénéités dans l'état fondamental, et le principe de Fermat nous permet d'intégrer le long du chemin non perturbé $\Lambda_0$.

**Démonstration :**

Sachant que $v_p = \frac{ds}{dt} = \frac{\omega}{k} \Rightarrow \Delta\tau = \frac{k}{\omega}\Delta s$, et en posant le temps de phase à l'état fondamental $\tau_0 = \int_{\Lambda_0} \frac{k_0}{\omega} ds$, et à l'état perturbé $\tau = \int_\Lambda \frac{k}{\omega} ds'$, par le présent théorème on démontre que le temps de parcours est stationnaire $\Delta\tau = 0$ le long des chemins d'ondes pour de petits changement de chemin $\Delta s$ : $\Delta\tau = 0 \Rightarrow \Delta s = 0 \Rightarrow ds = ds'$ & $\Lambda = \Lambda_0$, ainsi la différence entre les deux états nous donne : $\tau - \tau_0 = \int_{\Lambda_0} \frac{k}{\omega} - \frac{k_0}{\omega} ds \Rightarrow \delta\tau = \frac{1}{\omega} \int_\Lambda \delta k \, ds$, ce qu'il fallait démontrer.

**Perturbation du vecteur d'onde**

D'après la relation de dispersion pour les modes "p" (voir l'équa.(3.67) avec ($\omega >> N$)), on a : $k_r^2 = \frac{1}{c_s^2}(\omega^2 - \omega_c^2) - k_h^2$ avec $k_h = \sqrt{k_r^2 + k_h^2}$, d'où on peut écrire alors :

$$k = \frac{1}{c_s}\sqrt{\omega^2 - \omega_c} \qquad (4.108)$$

Si on permet de petites perturbations (relatives à l'état fondamental) en $\omega$, $c_s$ et $\omega_c$, on peut écrire alors le développement en série de Taylor du vecteur d'onde perturbé $k$ :

$$\delta k = k - k_0 = \delta\omega \frac{\partial k}{\partial \omega} + \delta\omega_c \frac{\partial k}{\partial \omega_c} + \delta c_s \frac{\partial k}{\partial c_s} + \varepsilon \qquad (4.109)$$

En négligeant les termes au-delà du second ordre et en calculant les différentes dérivées de $k$, on trouve :

$$\delta k = \frac{\omega\delta\omega - \omega_c\delta\omega_c}{kc_s^2} - \frac{k\delta c_s}{c_s} \qquad (4.110)$$

Ainsi, on peut écrire :

$$\frac{\delta k}{\omega}ds = \left[\frac{\delta\omega}{c_s^2 k} - \left(\frac{\delta c_s}{c_s}\right)\frac{k}{\omega} - \left(\frac{\delta\omega_c}{\omega_c}\right)\left(\frac{\omega_c^2}{c_s^2\omega^2}\right)\frac{\omega}{k}\right]ds \qquad (4.111)$$

Avec, comme vu dans la section précédente (voir l'équa.(4.48)), on a : $\delta\omega = -k\vec{n}.\vec{U}$, où $\vec{n}$ est le vecteur unité tangent du chemin d'onde.

Ainisi, on peut écrire :

$$\delta\tau^{\pm} = -\int_{\Gamma_0}\left[\frac{\vec{U}.(\pm\vec{n})}{c_s^2} - \left(\frac{\delta c_s}{c_s}\right)\frac{k}{\omega} - \left(\frac{\delta\omega_c}{\omega_c}\right)\left(\frac{\omega_c^2}{c_s^2\omega^2}\right)\frac{\omega}{k}\right]ds \qquad (4.112)$$

On définit $\delta\tau^+$ comme étant le temps perturbé dans une direction le long du chemin d'onde où le vecteur unité est $+\vec{n}$ et $\delta\tau^-$ étant le temps perturbé dans la direction opposée ($-\vec{n}$).

A présent et afin de séparer les effets de la vitesse des flux (écoulements) de ceux de la vitesse du son et d'autres perturbations, on définit :

$$\delta\tau_{diff} = \tau^+ - \tau^- = -2\int_{\Gamma_0}\frac{\vec{U}.\vec{n}}{c_s^2}ds \qquad (4.113)$$

et :

$$\delta\tau_{mean} = \frac{\tau^+ + \tau^-}{2} = \int_{\Gamma_0}\left[\left(\frac{\delta c_s}{c_s}\right)\frac{k}{\omega} - \left(\frac{\delta\omega_c}{\omega_c}\right)\left(\frac{\omega_c^2}{c_s^2\omega^2}\right)\frac{\omega}{k}\right]ds \qquad (4.114)$$

On remarque bien ici que $\delta\tau_{diff}$ renferme des informations sur la vitesse des écoulements, alors que le $\delta\tau_{mean}$ contient des informations sur les vitesses du son et autres effets, qui seront explicités par la suite.

## 4.3 Perturbation en présence d'un champ magnétique

Les résultats précédents découlent d'une relation de dispersion qui a été calculée sans la prise en compte de l'effet du champ magnétique. Mais en considérant le champ magnétique, le résultat est tout autre, car la vitesse n'est plus seulement acoustique mais magnéto-acoustique.

En effet, et d'après la Magnéto-Hydro-Dynamique (MHD), un plasma animé d'une vitesse $\vec{v}$, à une pression $p$, de densité $\rho$, baignant dans un champ magnétique $\vec{B}$, soumis à la force de gravité $\rho.\vec{g}$ ainsi qu'à la force de Lorentz $\vec{j} \wedge \vec{B}$ (ce qui est le cas dans notre soleil), a pour équation de mouvement [24] :

$$\rho\left[\frac{\partial \vec{v}}{\partial t}+(\vec{v}.\vec{\nabla}).\vec{v}\right] = -\vec{\nabla}p + \rho\vec{g} + \vec{j} \wedge \vec{B} \qquad (4.115)$$

Cette dernière est plus connue sous le nom de l'équation de Navier-Stokes ; $\vec{j} = \frac{1}{\mu}\vec{\nabla} \wedge \vec{B}$ est la densité de courant, $\vec{B}$ étant l'induction magnétique et $\mu$ la perméabilité magnétique : $\mu = 4\pi\mu_0$[1].

Ainsi l'équ.(4.115) devient :

$$\rho\left[\frac{\partial \vec{v}}{\partial t}+(\vec{v}.\vec{\nabla}).\vec{v}\right] = -\vec{\nabla}p + \frac{1}{4\pi}(\vec{\nabla} \wedge \vec{B}) \wedge \vec{B} + \rho\vec{g} \qquad (4.116)$$

A l'équilibre on a : $\vec{v} = 0$, $\vec{B} = \vec{B_0}$, $p = p_0$, $\rho = \rho_0$, et en remplaçant dans l'équa.(4.116), on obtient :

$$\vec{\nabla}p = \frac{1}{4\pi}(\vec{\nabla} \wedge \vec{B}) \wedge \vec{B} + \rho\vec{g} \qquad (4.117)$$

Pour l'équation de continuité (voir démonstration en annexe), on écrit :

$$\frac{\partial \rho}{\partial t} + \vec{\nabla}(\vec{\rho}.\vec{v}) = 0 \qquad (4.118)$$

Et soit l'équation d'état $p = p(\rho,s)$, $s$ étant l'entropie :

$$dp = \left(\frac{\partial p}{\partial \rho}\right)_s d\rho + \left(\frac{\partial p}{\partial s}\right)_\rho ds \qquad (4.119)$$

Dans le cas adiabatique (à entropie constante) $(s = cst) \Rightarrow dp = \left(\frac{\partial p}{\partial \rho}\right)_s d\rho$, et sachant que $\frac{dp}{d\rho} = \left(\frac{\partial p}{\partial \rho}\right)_s = c_s^2$ ($c_s$ étant la vitesse du son), on peut écrire alors :

$$\frac{\partial p}{\partial t} + \vec{v}.\vec{\nabla}p = c_s^2\left(\frac{\partial \rho}{\partial t} + \vec{v}.\vec{\nabla}\rho\right) \qquad (4.120)$$

Dans le régime magnétique de convection, nous avons :

$$\frac{\partial \vec{B}}{\partial t} = \vec{\nabla} \wedge (\vec{v} \wedge \vec{B}) \qquad (4.121)$$

Et si on considère l'existence de petites perturbations eulériennes autour de la position d'équilibre :

$$\begin{aligned}
\vec{B} &= \vec{B}(\vec{r},t) + \vec{B_0}(\vec{r}) \\
\vec{v} &= \vec{v}(\vec{r},t) + 0 \\
p &= p(\vec{r},t) + p_0(\vec{r}) \\
\rho &= \rho(\vec{r},t) + \rho_0(\vec{r})
\end{aligned} \qquad (4.122)$$

En utilisant l'approximation MHD $\vec{\nabla} \wedge \vec{B} = \frac{\vec{B}}{D}$ (*D* étant la distance caractéristique ionique d'un plasma), le terme $(\vec{\nabla} \wedge \vec{B}) \wedge \vec{B}$ devient :

$$(\vec{\nabla} \wedge \vec{B}) \wedge \vec{B} = (\vec{\nabla} \wedge \vec{B_0}) \wedge \vec{B} + (\vec{\nabla} \wedge \vec{B}) \wedge \vec{B_0} \qquad (4.123)$$

---

[1] $\mu_0$ étant la perméabilité du vide, elle est prise égale à l'unité dans notre système de mesure.

En remplaçant nos grandeurs perturbées dans les équations (4.118) et (4.120), tout en négligeant les termes non linéaires et d'ordre 2, on obtient :

$$\frac{\partial \rho}{\partial t} + \vec{\nabla}(\rho_0 . \vec{v}) = 0 \quad (4.124)$$

$$\frac{\partial p}{\partial t} + \vec{v} . \vec{\nabla} p_0 = c_s^2 \left( \frac{\partial \rho}{\partial t} + \vec{v} . \vec{\nabla} \rho_0 \right) \quad (4.125)$$

De même dans l'équation de Navier-Stokes $\rho \left[ \frac{\partial \vec{v}}{\partial t} + (\vec{v} . \vec{\nabla}) . \vec{v} \right] = -\vec{\nabla} p + \frac{1}{4\pi} (\vec{\nabla} \wedge \vec{B}) \wedge \vec{B} + \rho \vec{g}$, sans oublier que $(\vec{\nabla} \wedge \vec{B}) \wedge \vec{B} = (\vec{\nabla} \wedge \vec{B_0}) \wedge \vec{B} + (\vec{\nabla} \wedge \vec{B}) \wedge \vec{B_0}$ et que $\vec{\nabla} p_0 = \frac{1}{4\pi} (\vec{\nabla} \wedge \vec{B_0}) \wedge \vec{B_0} + \rho_0 \vec{g}$, et en négligeant les termes d'ordre 2, on obtient :

$$\rho_0 \frac{\partial \vec{v}}{\partial t} = -\vec{\nabla} p + \rho \vec{g} + \frac{1}{4\pi} (\vec{\nabla} \wedge \vec{B}) \wedge \vec{B_0} + \frac{1}{4\pi} (\vec{\nabla} \wedge \vec{B_0}) \wedge \vec{B} \quad (4.126)$$

Et $\frac{\partial \vec{B}}{\partial t} = \vec{\nabla} \wedge (\vec{v} \wedge \vec{B})$ devient :

$$\frac{\partial \vec{B}}{\partial t} = \vec{\nabla} \wedge (\vec{v} \wedge \vec{B_0}) \quad (4.127)$$

En résumé, les équations MHD perturbées et linéairisées sont :

1- Equation de conservation du moment :

$$\rho_0 \frac{\partial \vec{v}}{\partial t} = -\vec{\nabla} p + \rho \vec{g} + \frac{1}{4\pi} (\vec{\nabla} \wedge \vec{B}) \wedge \vec{B_0} + \frac{1}{4\pi} (\vec{\nabla} \wedge \vec{B_0}) \wedge \vec{B} \quad (4.128)$$

2- Equation de conservation de l'énergie :

$$\frac{\partial p}{\partial t} + \vec{v} . \vec{\nabla} p_0 = c_s^2 \left( \frac{\partial \rho}{\partial t} + \vec{v} . \vec{\nabla} \rho_0 \right) \quad (4.129)$$

3- Equation de la magnéto-hydrostatique :

$$\vec{\nabla} p_0 = \frac{1}{4\pi} (\vec{\nabla} \wedge \vec{B_0}) \wedge \vec{B_0} + \rho_0 \vec{g} \quad (4.130)$$

4- Equation de continuité :

$$\frac{\partial \rho}{\partial t} + \vec{\nabla}(\rho_0 . \vec{v}) = 0 \quad (4.131)$$

5- Equation de l'induction :

$$\frac{\partial \vec{B}}{\partial t} = \vec{\nabla} \wedge (\vec{v} \wedge \vec{B_0}) \quad (4.132)$$

### 4.3.1 Ondes MAG : Magnéto-Acoustique-Gravité

La dérivée par rapport au temps de l'équation du mouvement nous donne :

$$\frac{\partial^2 \vec{v}}{\partial t^2} = -\frac{1}{\rho} \vec{\nabla} \left( \frac{\partial p}{\partial t} \right) + \frac{\vec{g}}{\rho_0} \frac{\partial \rho}{\partial t} + \frac{1}{4\pi\rho_0} \left[ \left( \vec{\nabla} \wedge \frac{\partial \vec{B}}{\partial t} \right) \wedge \vec{B_0} + (\vec{\nabla} \wedge \vec{B_0}) \wedge \frac{\partial \vec{B}}{\partial t} \right] \quad (4.133)$$

A partir des équation (4.129), (4.130) et (4.131) et en remplaçant dans (4.126), on obtient :

$$\frac{\partial^2 \vec{v}}{\partial t^2} = \frac{1}{\rho_0} \vec{\nabla}(c_s^2 \rho_0 \vec{\nabla} \vec{v}) + \frac{1}{\rho_0} \vec{\nabla} \left( \vec{v} \frac{1}{4\pi} (\vec{\nabla} \wedge \vec{B_0}) \wedge \vec{B_0} \right) + \frac{1}{\rho_0} \vec{\nabla}(\vec{v} \rho_0) \vec{g} -$$
$$- \frac{1}{\rho_0} \vec{g} \vec{\nabla}(\rho_0 \vec{v}) + \frac{1}{4\pi\rho_0} \left[ \left( \vec{\nabla} \wedge \frac{\partial \vec{B}}{\partial t} \right) \wedge \vec{B_0} + (\vec{\nabla} \wedge \vec{B_0}) \wedge \frac{\partial \vec{B}}{\partial t} \right] \quad (4.134)$$

Après développement, on a : $\frac{1}{\rho_0}\vec{\nabla}(\vec{v}\rho_0)\vec{g} - \frac{1}{\rho_0}\vec{g}\vec{\nabla}(\rho_0\vec{v}) = \vec{\nabla}(\vec{v}\vec{g}) - (\vec{\nabla}\vec{v})\vec{g}$, et $(\vec{\nabla}\wedge\vec{B_0})\wedge\vec{B_0} = -\frac{1}{2}\vec{\nabla}B_0^2 + (\vec{B_0}\vec{\nabla})\vec{B_0}$, avec $\frac{\partial\vec{B}}{\partial t} = \vec{\nabla}\wedge(\vec{v}\wedge\vec{B_0})$, et en remplaçant dans la dernière équation on obtient :

$$\frac{\partial^2\vec{v}}{\partial t^2} = \frac{1}{\rho_0}\vec{\nabla}(c_s^2\rho_0\vec{\nabla}\vec{v}) - (\vec{\nabla}\vec{v})\vec{g} + \vec{\nabla}(\vec{v}\vec{g}) - \frac{1}{\rho_0}\vec{\nabla}\left[\vec{v}\vec{\nabla}\left(\frac{B_0^2}{8\pi}\right) - \frac{1}{4\pi}\vec{B_0}.\vec{\nabla}\vec{B_0}\right] + $$
$$+ \frac{1}{4\pi\rho_0}[\vec{\nabla}\wedge(\vec{\nabla}\wedge(\vec{\nabla}\wedge\vec{B_0}))\wedge\vec{B_0} + (\vec{\nabla}\wedge\vec{B_0})\wedge(\vec{\nabla}\wedge(\vec{\nabla}\wedge\vec{B_0}))] \quad (4.135)$$

Dans le cas d'un champ magnétique uniforme $\vec{B_0}$, i.e. $\vec{\nabla}.\vec{B_0} = 0$, la dernière équation devient :

$$\frac{\partial^2\vec{v}}{\partial t^2} = \frac{1}{\rho_0}\vec{\nabla}(c_s^2\rho_0\vec{\nabla}\vec{v}) - (\vec{\nabla}\vec{v})\vec{g} + \vec{\nabla}(\vec{v}\vec{g}) + \frac{1}{4\pi\rho_0}[\vec{\nabla}\wedge(\vec{\nabla}\wedge(\vec{\nabla}\wedge\vec{B_0}))\wedge\vec{B_0}] \quad (4.136)$$

Sachant que $c_s^2 = \Gamma\frac{p_0}{\rho_0}$, et que pour $\vec{B_0}$ uniforme, l'équation magnéto-hydrostatique devient : $\vec{\nabla}p_0 = \rho_0\vec{g}$, on peut écrire : $\vec{\nabla}c_s^2 = \Gamma\vec{g} - \frac{c_s^2}{\rho_0}\vec{\nabla}\rho_0$, ($\Gamma$ étant le coefficient adiabatique).

Et en posant le vecteur unité du champ magnétique : $\vec{b_0} = \frac{\vec{B_0}}{\|\vec{B_0}\|}$ avec ($|\vec{b_0}| = 1$), et $c_a^2 = \frac{B_0^2}{4\pi\rho_0}$, $c_a$ étant la vitesse d'Alfven, notre dernière équation se réduit à :

$$\frac{\partial^2\vec{v}}{\partial t^2} = \Gamma\vec{g}\vec{\nabla}\vec{v} + c_s^2\vec{\nabla}(\vec{\nabla}\vec{v}) - (\vec{\nabla}\vec{v})\vec{g} + \vec{\nabla}(\vec{v}\vec{g}) + c_a^2[\vec{\nabla}\wedge(\vec{\nabla}\wedge(\vec{\nabla}\wedge\vec{b_0}))\wedge\vec{b_0}] \quad (4.137)$$

### 4.3.2 Ondes magnéto-acoustiques

On néglige, à présent l'effet de la gravité, en ne considérant de ce fait que les ondes magnéto-acoustiques. Et on suppose que notre milieu est homogène, $\rho_0$ est uniforme, ainsi que la vitesse du son et la vitesse d'Alfven (la stratification est négligée). Ainsi l'équa.(4.137) devient :

$$\frac{\partial^2\vec{v}}{\partial t^2} = c_s^2\vec{\nabla}(\vec{\nabla}\vec{v}) + c_a^2[\vec{\nabla}\wedge(\vec{\nabla}\wedge(\vec{\nabla}\wedge\vec{b_0}))\wedge\vec{b_0}] \quad (4.138)$$

A cause de l'uniformité, on peut privilégier l'axe des $z$, et écrire $\vec{b_0} = b_0\vec{e_z}$. Afin de simplifier l'écriture, on pose :

$$\Delta = \vec{\nabla}\vec{v},$$
$$\Lambda = \vec{e_z}.(\vec{\nabla}\wedge\vec{v}),$$
$$\text{et } \Omega = \frac{\partial v_z}{\partial z}.$$

Et en manipulant l'équa.(4.138), on obtient :

$$\frac{\partial^2\Delta}{\partial t^2} = (c_s^2 + c_a^2).\nabla^2\Delta - c_a^2.\nabla^2\Omega \quad (4.139)$$

$$\frac{\partial^2\Omega}{\partial t^2} = c_s^2\frac{\partial^2\Delta}{\partial z^2} \quad (4.140)$$

$$\left(\frac{\partial^2}{\partial t^2} - c_a^2\frac{\partial^2}{\partial z^2}\right)\Lambda = 0 \quad (4.141)$$

Les équations (4.139)-(4.141) sont appellées les équations de Lighthill (1960).

### 4.3.3 Relations de dispersion des ondes magnéto-acoustiques fast et slow

En dérivant la première relation de Lighthill par rapport au temps deux fois, et en usant de la deuxième, on obtient :

$$\frac{\partial^4 \Delta}{\partial t^4} - (c_s^2 + c_a^2)\frac{\partial^2}{\partial t^2}\nabla^2\Delta + c_a^2 c_s^2 \frac{\partial^2}{\partial z^2}\nabla^2\Delta = 0 \tag{4.142}$$

Comme nos vitesses sont assimilées à des ondes planes : $v = v_0 \exp[-i(\omega t - kr)]$, avec $k^2 = k_x^2 + k_y^2 + k_z^2$ (où $k_z = k\cos\theta$), on peut écrire alors : $\Delta = ikv_0 \exp[-i(\omega t - kr)]$, en remplaçant dans la dernière équation, et sachant que : $\vec{c_a}.\vec{k} = \|\vec{c_a}\|.\|\vec{k}\|\cos\theta$, on trouve :

$$\omega^4 - (c_s^2 + c_a^2)\omega^2 k^2 + c_s^2(\vec{c_a}.\vec{k})^2 k^2 = 0 \tag{4.143}$$

En posant $\chi = \omega^2$, et en calculant le discriminant puis les solutions de l'équation du $2^e$ degré de la variable $\chi$ qui en résulte, on obtient :

$$\chi_+ = \omega_+^2 = \frac{k^2}{2}\left[c_s^2 + c_a^2 + \sqrt{(c_s^2 + c_a^2)^2 - \frac{4c_s^2(\vec{c_a}.\vec{k})^2}{k^2}}\right] \tag{4.144}$$

qui représente la relation de dispersion magnéto-acoustique du mode "Fast".
Et :

$$\chi_- = \omega_-^2 = \frac{k^2}{2}\left[c_s^2 + c_a^2 - \sqrt{(c_s^2 + c_a^2)^2 - \frac{4c_s^2(\vec{c_a}.\vec{k})^2}{k^2}}\right] \tag{4.145}$$

qui représente la relation de dispersion magnéto-acoustiques du mode "Slow".
N'oublions pas qu'on cherche à déterminer les vitesses magnéto-acoustiques. Alors, des relations de dispersion précédentes, et sachant que la vitesse de phase $v_p = \omega/k$, on tire :

$$v_+^2 = \frac{1}{2}\left[c_s^2 + c_a^2 + \sqrt{(c_s^2 + c_a^2)^2 - \frac{4c_s^2(\vec{c_a}.\vec{k})^2}{k^2}}\right] \tag{4.146}$$

la vitesse de phase magnéto-acoustique du mode "Fast". Et

$$v_-^2 = \frac{1}{2}\left[c_s^2 + c_a^2 - \sqrt{(c_s^2 + c_a^2)^2 - \frac{4c_s^2(\vec{c_a}.\vec{k})^2}{k^2}}\right] \tag{4.147}$$

la vitesse de phase magnéto-acoustique du mode "Slow".

### 4.3.4 Étude des vitesses magnéto-acoustiques

Pour des rayons relativement petit $0 \leq r \leq R_\odot$ (où $R_\odot$ est le rayon du soleil), c'est le mode "Fast" qui prédomine sur le mode "Slow", mais pour $r > R_\odot$ (au delà de la photosphère (l'atmosphère solaire ; la chromosphère, la couronne,...etc.)) c'est plutôt le mode "Slow" qui prend le dessus sur le mode "Fast", et ceci est aisément vérifiable en étudiant de plus près les limites des expressions de $c_s^2, c_a^2$, et $v_\pm^2$ à des rayons relativement petits et grands. Ainsi et dans notre cas, on n'aura à travailler qu'avec la vitesse magnéto-acoustique du mode "Fast", et on pose $c_f = v_+$. En remplaçant alors la vitesse acoustique $c_s$ par notre vitesse magnéto-acoustique "Fast", $c_f$ dans la relation de dispersion du sous chapitre précédent $(\omega - \vec{k}.\vec{U})^2 = k^2 c_s^2 + \omega_c^2$, on obtient :

$$(\omega - \vec{k}.\vec{U})^2 = k^2 c_f^2 + \omega_c^2 \tag{4.148}$$

Avec $c_f^2 = \frac{1}{2}\left[c_s^2 + c_a^2 + \sqrt{(c_s^2 + c_a^2)^2 - \frac{4c_s^2(\vec{c_a}.\vec{k})^2}{k^2}}\right]$.

**Conclusion :**

En utilisant à présent le principe de Fermat et en perturbant notre vecteur d'onde $k$, comme dans la section (4.2.6), on obtient finalement :

$$\delta\tau_{diff} = -2\int_{\Gamma_0} \frac{\overrightarrow{U}.\overrightarrow{n}}{c_s^2} ds \tag{4.149}$$

et :

$$\delta\tau_{mean} = \int_{\Gamma_0} \Big[\Big(\frac{\delta c_s}{c_s}\Big)\frac{k}{\omega} - \Big(\frac{\delta\omega_c}{\omega_c}\Big)\Big(\frac{\omega_c^2}{c_s^2\omega^2}\Big)\frac{\omega}{k} + \frac{1}{2}\Big(\frac{c_a^2}{c_s^2} - \frac{(\overrightarrow{k}\,\overrightarrow{c_a})^2}{k^2 c_s^2}\Big)\Big] ds \tag{4.150}$$

On remarque bien qu'en l'absence de champ magnétique $\overrightarrow{B} = \overrightarrow{0} \Rightarrow c_a = 0$, notre relation de dispersion $\omega_\pm^2 = \frac{k^2}{2}\Big[c_s^2 + c_a^2 \pm \sqrt{(c_s^2 + c_a^2)^2 - \frac{4c_s^2(\overrightarrow{c_a}.\overrightarrow{k})^2}{k^2}}\Big]$ n'admet plus qu'une solution unique $v = c_s$ et on retrouve ainsi les mêmes équations que dans le sous chapitre précédent.

## 4.4 Écoulements radiaux et horizontaux

L'équation (4.113) ou (4.149) (qui ne change pas, pour le cas magnétique et pour le cas sans champ magnétique) prouve que la différence du temps $\delta\tau$ pour des ondes parcourant des chemins d'onde réciproques est sensible aux composants de l'écoulement le long du chemin d'onde. Si le flux radial est partout uniforme, alors la différence du temps due à la vitesse radiale sera nul car les ondes réciproques subiront le même écoulement. Cependant, la composante radiale de la vitesse peut ne pas être uniforme ; imaginons un chemin d'onde dans un plan méridional avec un champ de vitesse comme celui représenté sur la Fig.(4.2). Ce modèle de circulation est horizontal sur la surface et satisfait l'équation de continuité : $\partial\rho/\partial t + \nabla.(\rho.\overrightarrow{U}) = 0$. Puisque la densité dans la zone de convection solaire est inversement proportionnelle au rayon, une circulation méridionale qui satisfait la conservation de la masse doit avoir une petite composante radiale.

On peut décomposer $\delta\tau$ en [41] :

$$\delta\tau = \delta\tau_h + \delta\tau_r \tag{4.151}$$

où :

$$\delta\tau_r = 2\int_{r_1}^{r_2} \frac{U_r}{c^2} dr \tag{4.152}$$

$$\delta\tau_h = 2\int_{r_1}^{r_2} \frac{U_h v_{gh}}{v_{gr} c^2} dr \tag{4.153}$$

En utilisant ces équations, il est possible de calculer les contributions relatives des écoulements horizontaux et radiaux. Pour un choix des latitudes et des distances les résultats sont montrés dans la Fig.(4.3). Puisque la contribution des écoulements radiaux est toujours beaucoup plus petite que la contribution des écoulements horizontaux, ceux-ci peuvent être négligés, et le symbole $\overrightarrow{U}$ sera employé pour exprimer la vitesse horizontale d'écoulement. Et l'équa.(4.113) ou (4.149) devient alors :

$$\delta\tau_h = 2\int_{r_1}^{r_2} \frac{U v_{gh}}{v_{gr} c^2} dr \tag{4.154}$$

La méthode qui nous permettra l'obtention des $\overrightarrow{U}$ sera explicitée dans le chapitre 6.

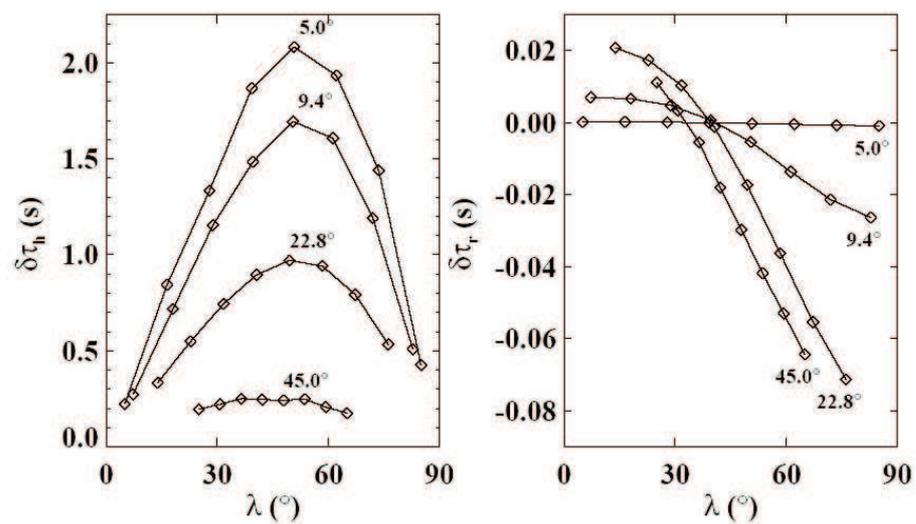

FIG. 4.3 – Comparaison entre la différence de temps de parcours radiale $\delta\tau_r$ et la différence de temps de parcours horizontale $\delta\tau_h$ en fonction de la latitude $\lambda$ [41].

# Chapitre 5

# Analyse des données et résultats

La théorie étant établie, nous allons maintenant aborder l'aspect application numérique, qui lui repose sur des données observationnelles. Dans ce chapitre, nous nous intéressons à la description des données, à leur provenance et surtout à leur traitement et analyse.

Nous insisterons plus particulièrement sur les résultats intermédiaires obtenus (temps de parcours des ondes traitées).

Les données traitées dans notre thèse, et sur lesquelles on tentera d'appliquer toute la théorie précédemment citée, sont des images de vitesse Doppler (Dopplerogrammes), obtenues par le réseau terrestre d'observation héliosismologique GONG.

## 5.1  Réseau GONG

Le Global Oscillation Network Group (GONG) entreprend une étude détaillée de la structure et de la dynamique internes solaires en utilisant l'héliosismologie et en développant un réseau de six stations d'acquisition d'images de vitesses solaires extrêmement sensibles et stables situées tout autour de la terre, pour obtenir des observations presque continues des oscillations ou des pulsations de "cinq-minutes" solaires. Le réseau GONG a également établi un système d'analyse, de réduction et de distribution des données afin de faciliter la coordination des recherches scientifiques sur ces mesures.

L'obstacle principal en utilisant ce nouvel outil est l'interruption des observations imposées par le cycle du jour et de la nuit à un simple observatoire terrestre, qui présente une incertitude dans la détermination de la période des ondes, en plus du bruit de fond créé par les fortes oscillations. Un aperçu, des conditions d'observation solaire actuelles, à de divers emplacements possibles, a été effectué pour permettre le choix des futurs endroits qui constitueront le réseau GONG. En employant des simulations, sur ordinateur, basées sur des statistiques climatologiques, en plus des évaluations des pannes d'équipements, les résultats révèlent que six emplacements au minimum, également espacés dans la longitude, étaient nécessaires (pour plus de détails sur le réseau GONG voir l'annexe B).

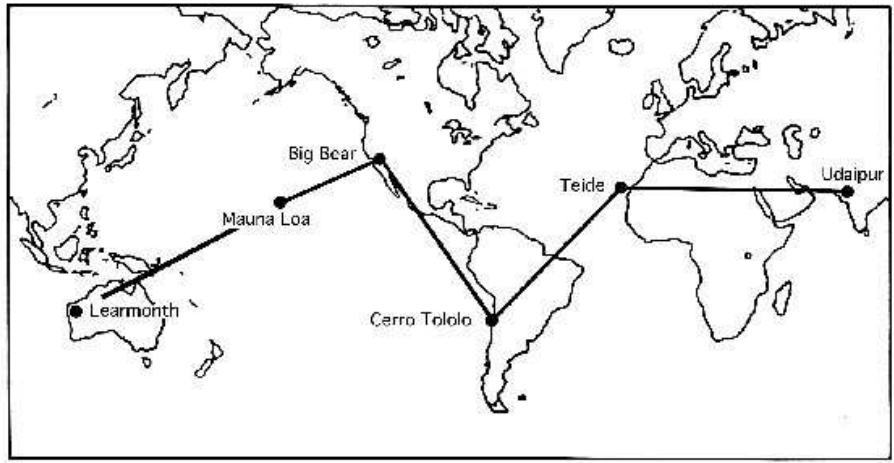

FIG. 5.1 – Le réseau GONG à travers le monde [40].

## 5.2  Description des instruments

Les différents modes solaires peuvent montrer des vitesses de moins de 0.2 mètres/seconde alors que la somme de tous les modes est seulement de quelque 100 mètres/seconde. Le but du réseau étant de faire des mesures sur les mouvements aléatoires de la surface du soleil, il a fallu développer six instruments stables, capables de faire des mesures reflétant des vitesses avec une précision de moins d'un métre/seconde. Un instrument basé sur un interféromètre de Michelson appelé "tachymètre de Fourier" (Beckers et Brown 1978), a été choisi et est soutenu par une installation fortement automatisée et portative ; sa philosophie de conception évoque celle d'un vaisseau spatial. L'instrument se compose de deux miroirs traquant le soleil et projetant la lumière horizontalement dans un container logeant le reste de l'équipement. Le système optique est fermé par une fenêtre filtrante ayant une ouverture efficace de 2.8cm. Le rayon lumineux passe ensuite par un dispositif optique qui peut être déplacé à l'intérieur et en dehors du faisceau lumineux. Tous ces mécanismes sont sous la gestion d'un ordinateur et fonctionnent automatiquement. Un filtre hybride de 1 Å passe bande isole la ligne de Ni à 6768 Å. Le coeur de l'instrument est un interféromètre de polarisation de Michelson (IPM) ayant une différence de chemin d'environ 30.000 ondes. Ceci est construit pour avoir un champ angulaire large et pour être thermiquement stable. Le modèle cosinus carré de transmission produit par l'interféromètre est balayé à travers le spectre filtré du soleil par un plat tournant d'onde ; ainsi, la modulation est produite par la présence de la ligne de Fraunhofer et la phase de la modulation est une bonne mesure d'effet Doppler. Le détecteur d'images est une caméra CCD qui a 2.5 Pixel d'arc seconde carré ; il donne une résolution de 5 arc seconde. Trois images sont obtenues pour chaque cycle de la modulation pour fournir la phase, l'amplitude et l'éclat de chaque Pixel. Une image est intégrée et le résultat enregistré sur bande numérique chaque soixante secondes.

Le container abrite également le support électronique, un ordinateur d'acquisition de données et les divers autres éléments électroniques de soutien pour rendre le dispositif capable d'être actionné par des sites éloignés. L'instrument est commandé par deux ordinateurs et une horloge précise ; il produit norma-lement un enregistrement chaque minute, jour et nuit. Pendant la nuit, seuls des paramètres instrumentaux et environnementaux sont enregistrés. Quand le programme détermine que le soleil s'est levé, l'embout avant de l'instrument est dégagé et dirigé vers le soleil de sorte qu'une sonde de guidage puisse fournir des informations de pointage précises. Si le ciel est nuageux, les évaluations de l'ordinateur désignent l'endroit où se trouve le soleil. Une fois que la lumière du soleil est suffisamment disponible, un ordre de calibrage est exécuté, et des observations sont faites jusqu'au coucher du soleil. Lorsque l'instrument s'arrime, il fonctionne automatiquement pour une semaine et le seul besoin d'intervention de l'utilisateur est de changer les bandes de données ou d'exécuter l'entretien.

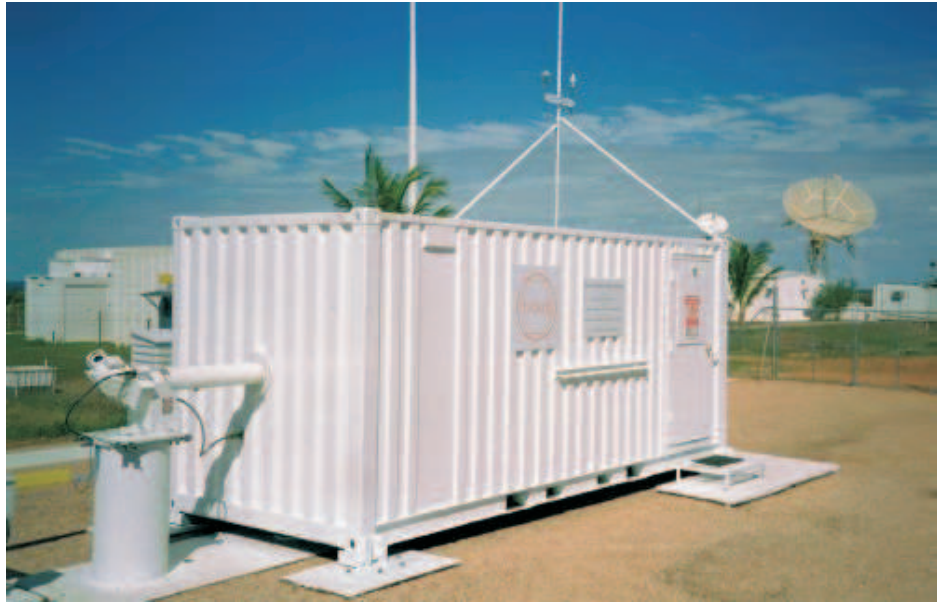

FIG. 5.2 – Une des six station GONG,(la station de BigBear) [40].

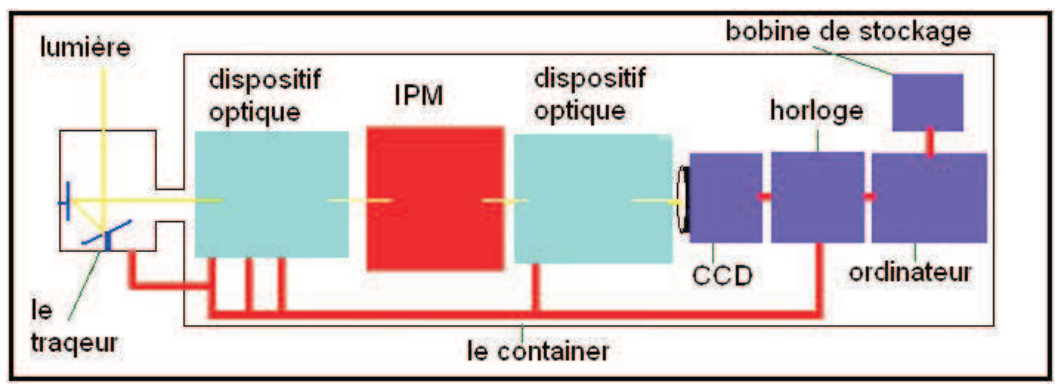

FIG. 5.3 – Schéma des instruments d'une station GONG.

Chaque station dans le réseau produit plus de 200 méga-octets de données chaque jour. Après trois ans d'observation continue, les données brutes excèdent un Terabyte (mille milliards d'octets).

Principalement, trois sortes d'images sont obtenues par ces stations : les images d'intensité, les images du champs magnétique (Magnétogrammes) et les images de vitesse Doppler (Dopplerogrammes).

Ces dernières sont justement les données traitées et utilisées dans notre travail.

## 5.3  Dopplerogrammes de GONG

Les images Doppler GONG sont des images de vitesse de ligne de vue (positives, pour les vitesses sortantes et négatives (entrantes)) de 860×860 pixels et d'une cadence d'échantillonnage d'une image par minute. Nos données sont sous la forme d'un cube de données "data cube" ($x \times y \times t$ voir Fig.(5.4)). Les images sont sous un format de fichier électronique FITS (Flixible Image Transport System) ; chaque pixel détient une information sur la vitesse. En plus de l'image, notre fichier FITS renferme une entête (FITS header) qui contient plusieurs informations sur l'image : allant de la taille en pixels, l'heure et date, jusqu'au lieu (station) de mesure,...etc. Les données choisies et fournies par GONG sont des "merged data" ; données assemblées par les six stations et couvrant deux journées entières et bien distinctes : la journée du 01 septembre 2001, qui correspond à une période de haute activité solaire et la journée du 10 janvier 2004 correspondant à une période de basse activité (voir Fig.(5.5) du cycle d'activité solaire ).

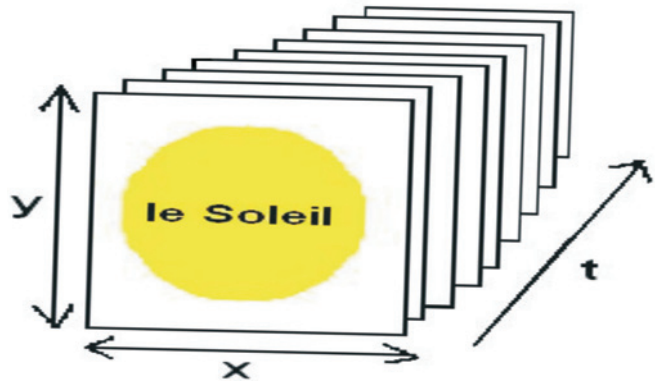

FIG. 5.4 – Représentation schématique de notre cube des données "data cube".

La cadence d'échantillonnage étant d'une minute ; le nombre de données par journée (24 heures) est donc de 1440 données, mais comme notre réseau est terrestre, le passage d'une station à une autre ou pour des causes climatologiques ou techniques, des trous dans notre échantillonnage peuvent surgir et sont même inévitables. Les minutes, voire les heures manquantes sont remplacées par des valeurs nulles (des données zéro). Ainsi, pour la journée de 01/09/2001 on a 321 données zéro et 1119 données exploitables sur les 1440 ; ce qui correspond à un rendement de 78%, alors que pour la journée du 10/01/2004, le rendement est de 95% pour 1370 données exploitables, et 70 données zéro sur les 1440.

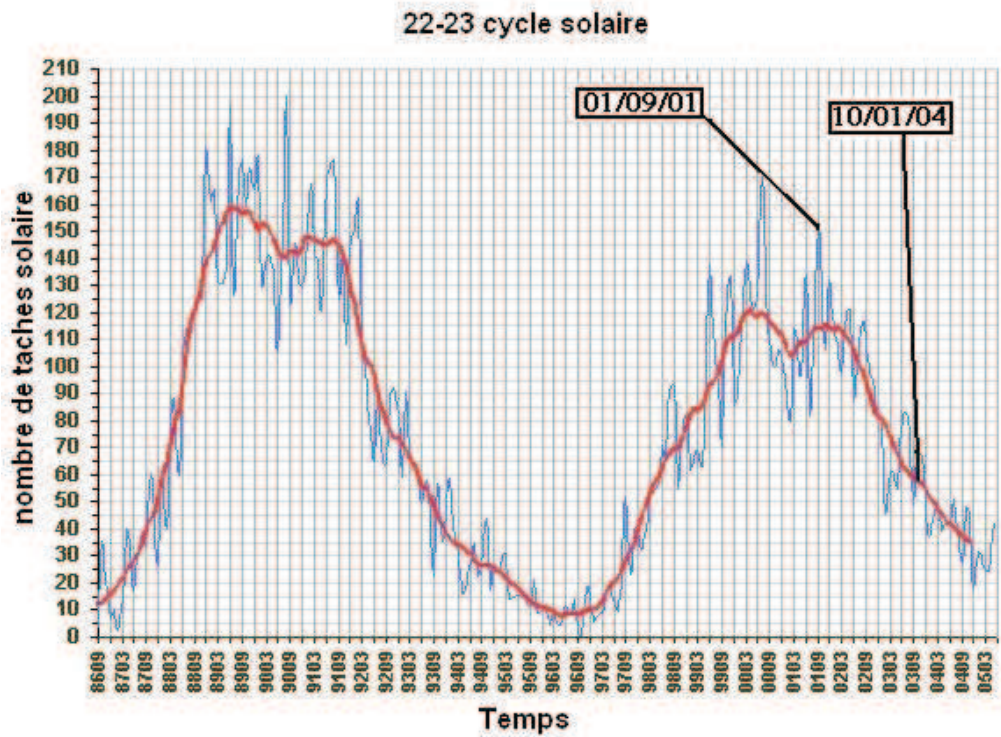

FIG. 5.5 – Figure de l'activité solaire au fil du temps [www.badastronomy.com].

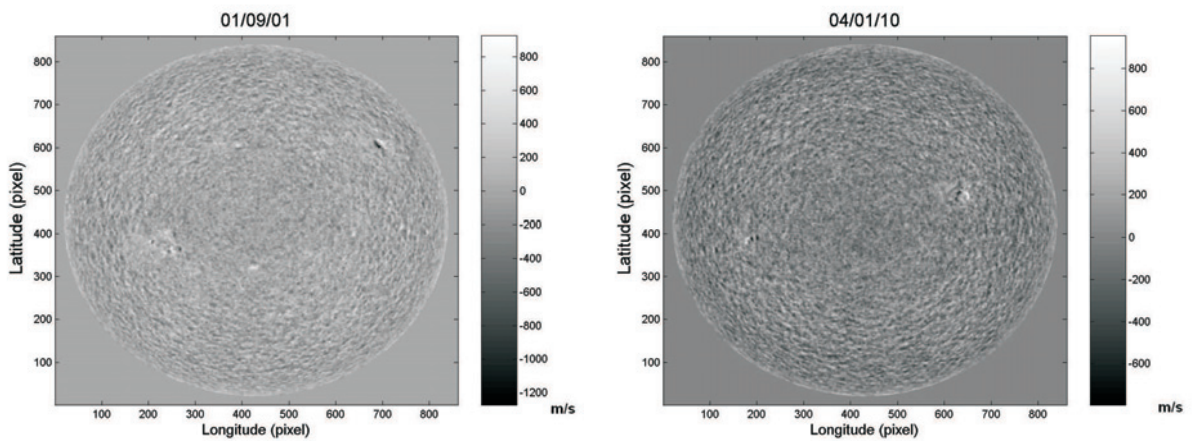

FIG. 5.6 – A gauche : merged velocity image de la journée 01/09/01 au temps de prise 0 min ("mrvzi010901t0000"), à droite : merged velocity image de la journée 10/01/04 au temps de prise 0 min ("mrvzi1040110t0000").

En suivant l'évolution des vitesses de n'importe quel point de la surface solaire sur les précédentes images Doppler, au cours du temps, et en prenant la transformée de Fourier du signal des vitesses du point dit, on retrouve facilement les oscillations trouvées par Robert Leighton en 1962 [61] (les fameuses oscillations de "5 min" (voir Fig.(5.7))).

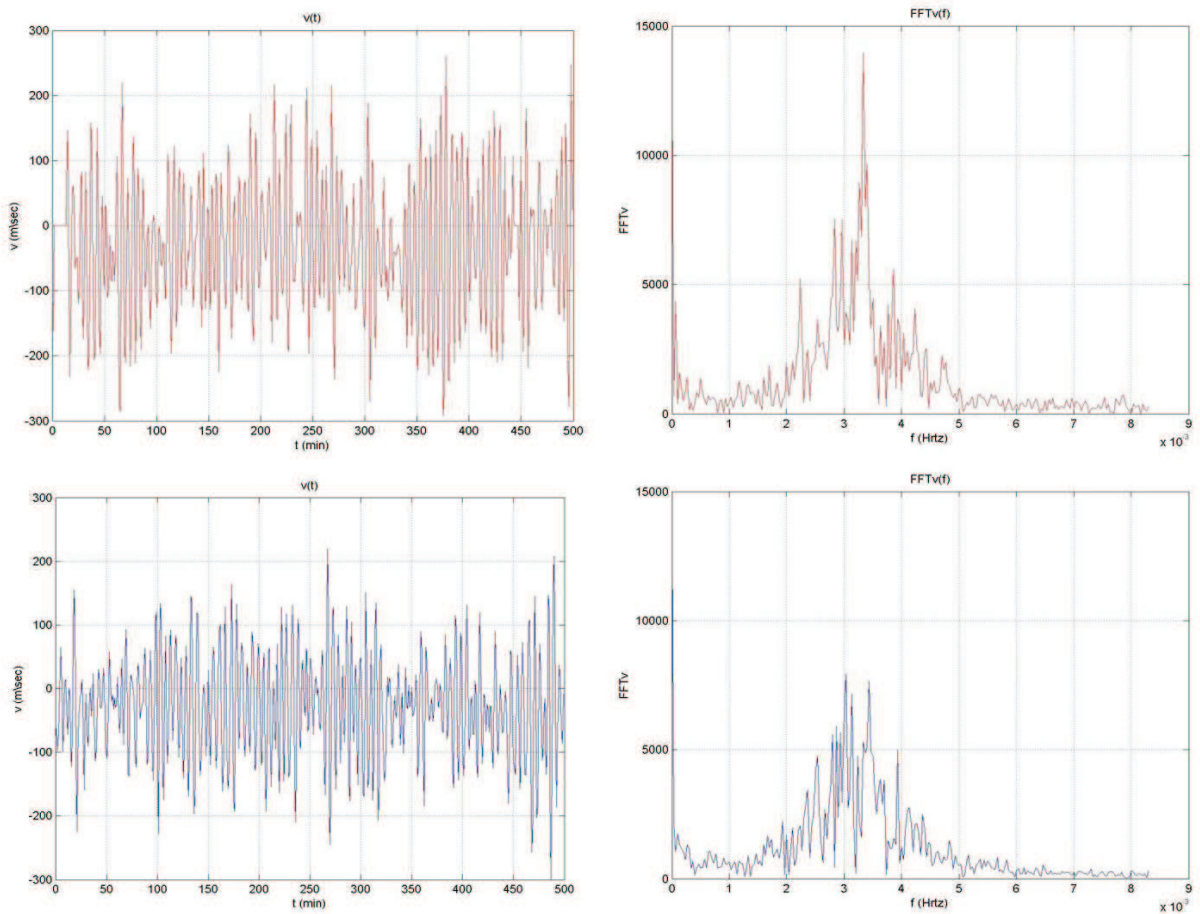

FIG. 5.7 – A gauche : les vitesses du point central ((430,430)pixel) au cours du temps, durant 500 min (8h20min), à droite : la transformé de Fourier du signal vitesse (là où on distingue clairement que les oscillations de 5min ($f \simeq 3mHz$) prédominentes), en rouge : les résultats de la journée à haute activité solaire (01/09/01), et en bleu : ceux de la journée à basse activité (10/01/04).

## 5.4 Remapping

Le soleil étant une sphère ; tout point à sa surface peut être localisé par des coordonnées sphériques. Mais, sur des régions spécifiques, il est beaucoup plus pratique de travailler avec les coordonnées cartésiennes, donc une transformation des coordonnées sphériques aux coordonnées cartésiennes s'impose ; c'est cette procédure qu'on nomme "Remapping".

Afin de mieux cerner cette méthode, il nous semble utile de faire une introduction sur les coordonnées héliographiques.

**Les Coordonnées Héliographiques**

La localisation d'un endroit ou d'un phénomène sur la surface solaire (par exemple, une tache solaire) est compliquée par le fait qu'il y a une inclinaison de 7.25° entre le plan de l'écliptique et le plan équatorial solaire, en plus de l'oscillation de l'axe de rotation solaire (le pôle nord solaire et le pôle nord céleste ne sont alignés que seulement deux fois par an). En conséquence, pour indiquer un endroit sur la surface solaire, trois coordonnées $(P, B, L)$ sont nécessaires pour définir une grille.

Les termes employés pour se rapporter aux coordonnées sont définis comme suit :

**L'angle $P$ :**

L'angle position entre le pôle nord géocentrique et le pôle nord de la rotation solaire, à mesurer vers l'est à partir du nord géocentrique (P variant entre $\pm 26.31°$).

**L'angle $B_0$ :**

La latitude Héliographique du point central du disque solaire ; également appelé l'angle B. La gamme de $B_0$ est de $\pm 7.23°$, corrigeant ainsi l'inclinaison de l'écliptique par rapport au plan équatorial solaire. Par exemple : Si $(P, B_0) = (-26.21, -6.54)$, la latitude héliographique du point central sur le disque solaire est de $-6.54°$ (le pôle nord de rotation n'est pas visible) et l'angle entre la projection sur le disque du pôle nord géocentrique et du pôle nord de la rotation solaire est de $26.21°$ vers l'ouest. Ce qui nous permet de corriger l'inclinaison des images.

**La longitude $L_0$ :**

La longitude héliographique du point central du disque solaire qui est déterminée sur un système de longitudes fixes tournant sur le soleil à un taux de $13.2°$/jour (la vitesse moyenne de rotation observée à partir du transit des taches solaires par rapport au méridien central). Le méridien standard du soleil est défini comme étant le méridien qui a traversé le noeud ascendant de l'équateur du soleil le 1 janvier 1854 à 1200 UTC et est calculé pour supposer aujourd'hui une période sidérale uniforme de rotation de $25.38°$ jours. (Le nœud ascendant ou the ascending node étant le point d'interaction entre le plan écliptique et l'équateur).

Une fois $P, B_0 \, et \, L_0$ connus, la latitude ($B$), la distance méridienne centrale et la longitude ($L$) d'un point solaire spécifique peuvent être déterminées comme suit :

**La Latitude ($L$ ou $\phi$) :**

Distance angulaire mesurée de l'équateur solaire vers le nord ou vers le sud le long d'un méridien.

**La Distance méridienne centrale (CMD) :**

Distance angulaire dans la longitude solaire, à mesurer à partir du méridien central. Cette position est relativement à la vue de la terre et changera pendant que le soleil tourne ; cette coordonnée ne devrait donc pas être confondue avec les positions héliographiques qui sont fixes sur la surface solaire.

**La Longitude ($\theta$) :**

La distance angulaire à partir du méridien standard (longitude héliographique de $0°$ ($L_0$)), mesurée de l'est à l'ouest (de $0°$ à $360°$) le long de l'équateur du soleil. Elle est calculée en combinant la CMD avec la longitude du méridien central à l'heure de l'observation, interpolant entre les valeurs d'éphémérides (pour 0000 UT) en employant le taux synodique de la rotation solaire (27.2753 jours, $13.2°$/jour).

Le point de vue synodique : un phénomène synodique est décrit dans un référentiel lié à la Terre, mais non géocentrique. Il se repère par rapport à un système d'axes dont l'un pointe en permanence vers le soleil.

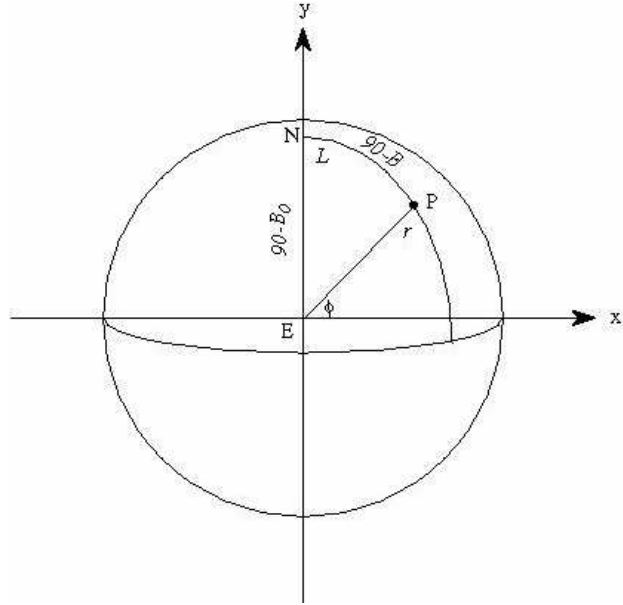

FIG. 5.8 – Système de coordonnées orthogonal, ayant une origine E confondue avec le centre du disque solaire, l'axe y étant pointé vers le nord, est parallèle à l'axe solaire de rotation et, et l'axe x est dirigé vers l'ouest ($\theta = 90° - B$, ($0 \leq B \leq 180$ et $90 \leq \theta \leq -90$ du nord au sud), et $\phi = L$) [40].

Le Remapping ici dans notre travail a été réalisé à partir de la méthode de la projection de Postel. Cette méthode reviendrait à appliquer sur le globe solaire une feuille de papier plat, obtenant ainsi une projection équidistante, dite de Guillaume Postel (1580). Ainsi des coordonnées $(\theta, \phi)$, on passe aux coordonnées cartésiennes $(x, y)$ et qui s'écrivent en fonction des coordonnées héliographiques comme suit (Williams 1994) :

$$x = x_c + R_\odot (X \cos P - Y \sin P) \tag{5.1}$$

$$y = y_c + R_\odot (Y \cos P - X \sin P) \tag{5.2}$$

Avec :

$$X = \frac{\sin\theta \sin L}{1 - 4.6524 \times 10^{-3}[\cos\theta \sin B_0 + \sin\theta \cos B_0 \cos L]} \tag{5.3}$$

$$Y = \frac{\cos\theta \cos B_0 - \sin\theta \sin B_0 \cos L}{1 - 4.6524 \times 10^{-3}[\cos\theta \sin B_0 + \sin\theta \cos B_0 \cos L]} \tag{5.4}$$

Où $x_c$, $y_c$ sont les coordonnées de la caméra au centre du disque et $R_\odot$ le rayon solaire.

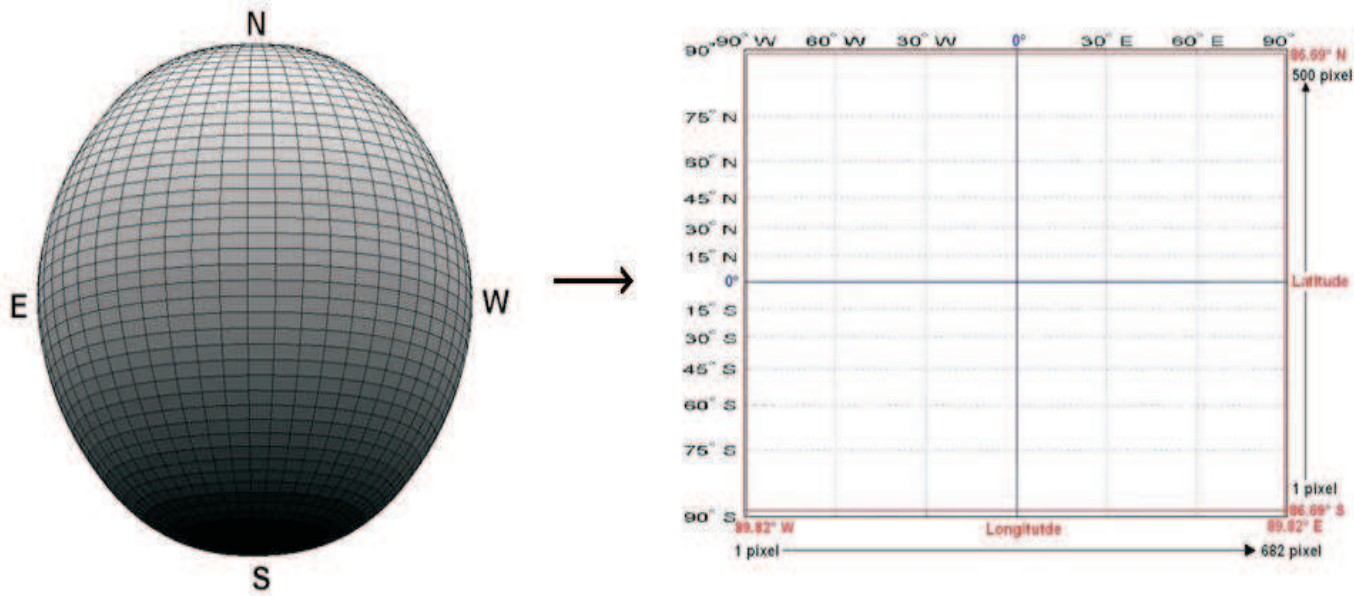

FIG. 5.9 – Figure schématique représentant la zone du remapping effectué (le cadre en rouge).

Le remapping a été effectué sur une région de la sphère solaire de dimension : 173.38° de latitude (de 86.69° Nord à 86.69° Sud) par 179.64° de longitude (de 89.82° East à 89.82° West), donnant ainsi une image de $500 \times 682$ pixels, ce qui nous fait une résolution de $0.2634° \times 0.3468°$ par pixels.

## 5.5 Tracking

Sachant que les vitesses mesurées sont la somme de plusieurs vitesses issues de différents modes, en plus de la vitesse due à la rotation différentielle, car notre soleil est en rotation sur lui même avec une vitesse de rotation équatoriale qui équivaut à $v_r = 2000$ m/s et comme celui-ci est une boule de gaz fluide, la vitesse de rotation de celle-ci nest pas la même à chaque latitude (la période de rotation du soleil à l'équateur est de 27 jours, alors qu'aux pôles, elle est de 36 jours.) ; c'est ce phénomène qui est appelé "rotation differentielle" et qui a été découvert pour la première fois par Carriongton en 1863 en observant le déplacement des taches solaires. Une correction des effets de ce phénomène (un "tracking") afin d'obtenir des vitesses purement modales s'impose.

Deux principales techniques sont employées afin de traiter la rotation différentielle ; l'une de ces méthodes et qu'on a utilisée ici dans cette thèse, est la vitesse angulaire de Snodgrass dépendante de la latitude (Snodgrass 1984), pour θ en gradient :

$$\Omega/2\pi(nHz) = 451 - 55\sin^2\theta - 80\sin^4\theta \qquad (5.5)$$

Où $\Omega = v_r d/R_\odot$ est la vitesse angulaire de rotation dépendante de la latitude θ, et $R_\odot = 6.961 \times 10^8 m$ le rayon solaire.

Ainsi, il ne reste plus qu'à parcourir notre disque solaire pour chaque image, de toute la série chronologique, et à soustraire, par rapport à une image référence au centre de l'axe temporel $T/2$, la vitesse de rotation appropriée de chaque vitesse, pixel par pixel.

$$v_m(\theta,\phi,t) = v_{m+rd}(\theta,\phi,t) - v_r d(\theta) \qquad (5.6)$$

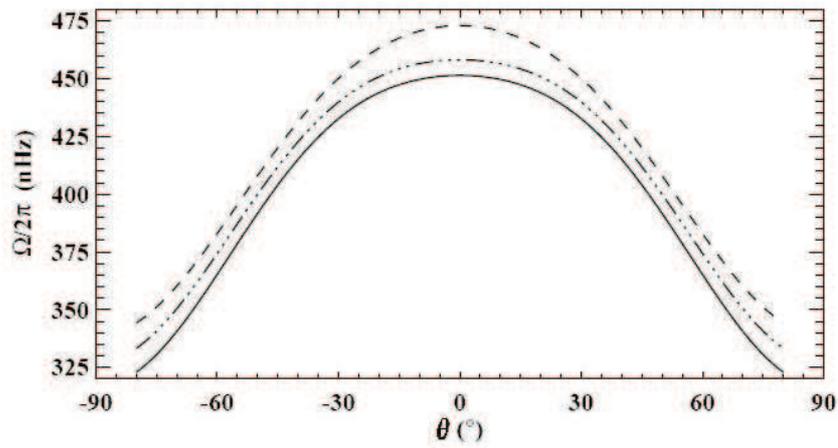

FIG. 5.10 – Quelques mesures de la rotation à la surface solaire. La ligne continue représente la vitesse angulaire de Snodgrass en fonction de la latitude ; la ligne discontinue est le résultat de mesures Doppler faites par Snodgrass et Ulrich (1990) et la ligne point-tiret, quant à elle, est le résultat de mesures magnétiques [41].

La Figure (5.10) démontre la proximité de la théorie avec les résultats expérimentaux, la concordance des différentes méthodes prouve la validité du modèle employé.

Le programme Fortran, sous Linux, qui traite simultanément le remapping et le tracking, qui nous a été généreusement fourni par Monsieur Kholikov Shukur (Chercheur au NSO (National Solar Observatory) GONG), nous permet d'avoir : soit des images 540x450 pixels ou 682x500pixels. Nous avons choisi les images 682x500pixels, car leur résolution est nettement meilleure. Et ainsi, les deux séries temporelles furent traitées (voir Fig.(5.11)).

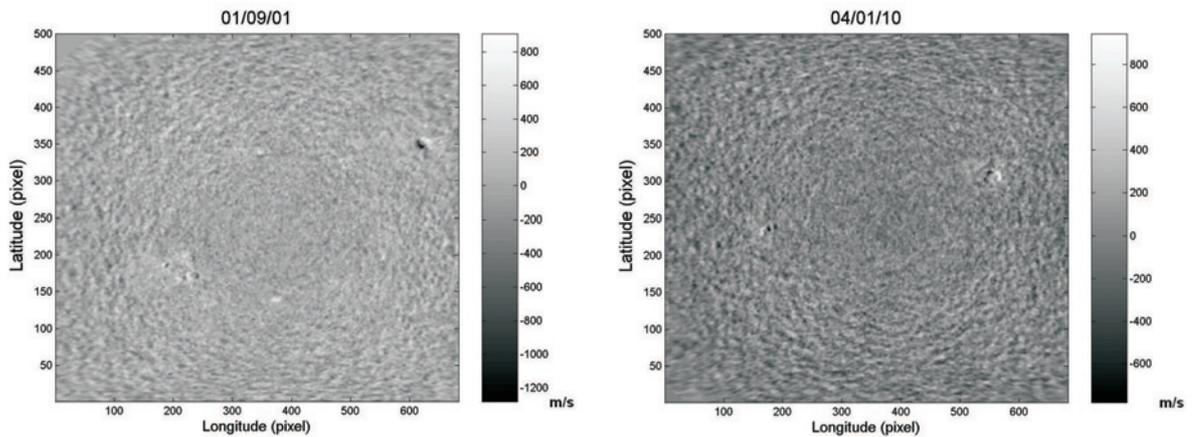

FIG. 5.11 – Les mêmes images de la Fig.(5.6), après remapping et tracking. A gauche l'image (682x500 pixels), à la période à haute activité solaire (01/09/01) et à droite l'image (682x500 pixels) à basse activité (10/01/04).

Concernant cette étape de traitement, le seul moyen de vérifier la validité des résultats sur toutes les données de la même série est de faire une sommation sur le carré de la valeur de chaque image, pixel par pixel, sur toute la série traitée. Ainsi, seules apparaissent les taches solaires et les régions actives (qui oscillent) dans une nouvelle images que l'on nomme : "carte de puissance" ("power map"). Il suffit ensuite d'une simple comparaison à l'œil nu, entre notre power map et une des images re-cartographiées (après remapping), pour s'assurer des résultats obtenus (voir Fig.(5.12)).

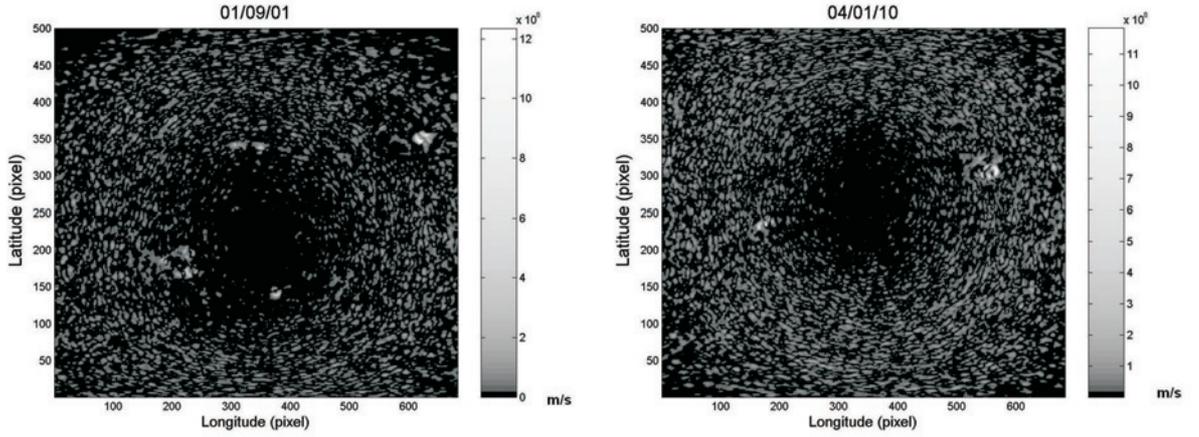

FIG. 5.12 – Les cartes de puissance (682x500 pixels). À gauche la power map correspondant à la période de haute activité solaire (01/09/01), et à droite celle à basse activité (10/01/04).

## 5.6 Filtrage

Après avoir traité les images et soustrait les effets dus à la rotation différentielle, il ne reste alors que les vitesses modales. Ainsi et comme vu précédemment (voir l'équation (3.54)), on peut réécrire et d'une manière plus générale le paramètre vitesse sous la forme suivante :

$$\mathrm{v}_{mod}(r,\theta,\phi,t) = \mathrm{v}(r,\theta,\phi,t) = \sum_{n\ell m} a_{n\ell m} \mathrm{v}_{n\ell}(r) Y_\ell^m(\theta,\phi) \exp(i[\omega_{n\ell m} t + \alpha_{n\ell m}]) \quad (5.7)$$

Ainsi, et en surface ($r = R_\odot$) :

$$\mathrm{v}(\theta,\phi,t) = \sum_{n\ell m} a_{n\ell m} Y_\ell^m(\theta,\phi) \exp(i[\omega_{n\ell m} t + \alpha_{n\ell m}]) \quad (5.8)$$

On voit que les vitesses observées ne sont en fait que la superposition de vitesses issues des oscillations simultanées des différents modes déja vus précédemment. Comme le mode fondamental qui est aussi bien appelé mode de gravité de surface est un mode intermédiaire entre les modes "g" et les modes "p" et que seuls les modes "p" et "f" arrivent en surface, seuls ces derniers peuvent être directement mesurés. Mais, comme notre étude est basée essentiellement sur les écoulements sub-surfaciques, on ne s'intéresse qu'aux modes à haut degré $\ell$ qui règnent dans la région de ces écoulements.

Un filtrage dans l'espace physique s'avère être fastidieux ; un passage donc par l'espace de Fourier via une transformation de Fourier s'impose. Pour celà, on utilise la **FFT à 3D** (voir annexe (A.4)) de notre cube de données "data cube" $\mathbf{v}(x,y,t)$, qui nous permet d'obtenir $\mathbf{V}(k_x,k_y,\omega)$ ; avec $V$ la transformée de Fourier de la vitesse v, $k_x$ et $k_y$ les vecteurs d'onde respectivement selon $x$ et $y$ et $\omega$ la pulsation.

De notre cube de données (data cube) $\mathbf{v}(x,y,t)$ et après avoir soustrait pour chaque image de la séquence temporelle, la moyenne de toutes les images, tout en faisant attention aux images nulles, ne gardant de ce fait que la perturbation de la vitesse. On fait subir à notre cube une FFT 3D passant ainsi de $\mathbf{v}(x,y,t)$ à $\mathbf{V}(k_x,k_y,\omega)$ dont on peut illustrer le module par la suivante figure :

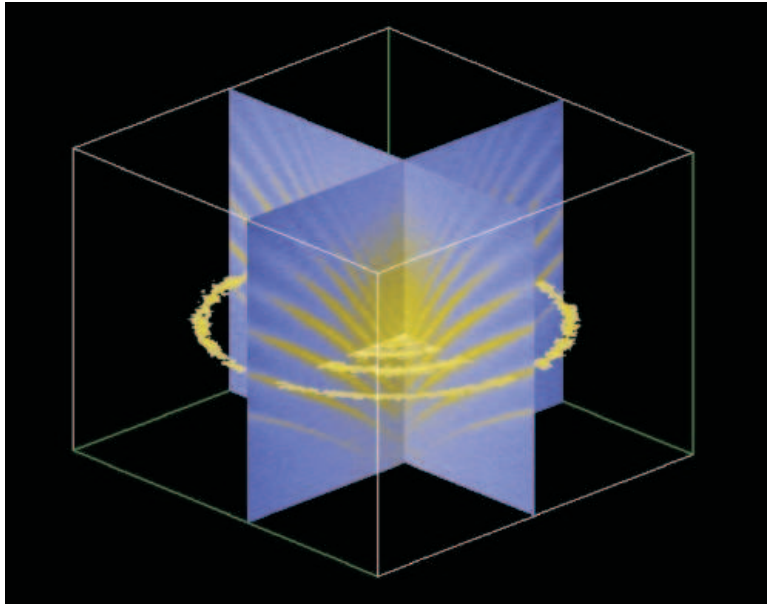

FIG. 5.13 – Représentation 3D du cube $\mathbf{V}(k_x,k_y,\omega)$ [Institut Keipenheuer de physique (Université de Freiburg)].

En appliquant la FFT 3D pour un cube $250 \times 250 \times 512$ ($x \times y \times t$), on obtient :

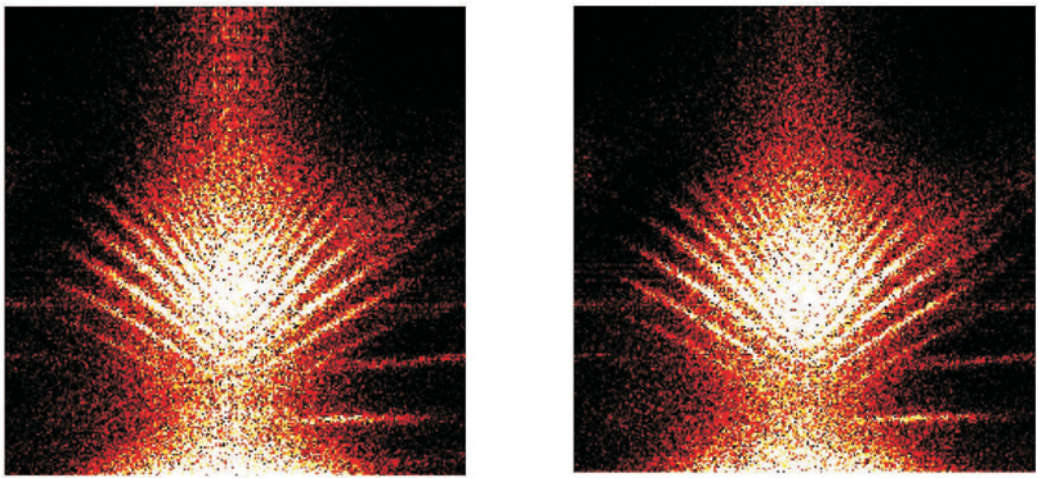

FIG. 5.14 – Voici une coupe longitudinale du module de la FFT 3D obtenu, à gauche pour la journée active et à droite pour la journée moins active.

La coupe transversale du dernier cube $(k_x, k_y, \omega)$ pour des fréquences bien données nous donne des schémas appelés "Ring diagram" :

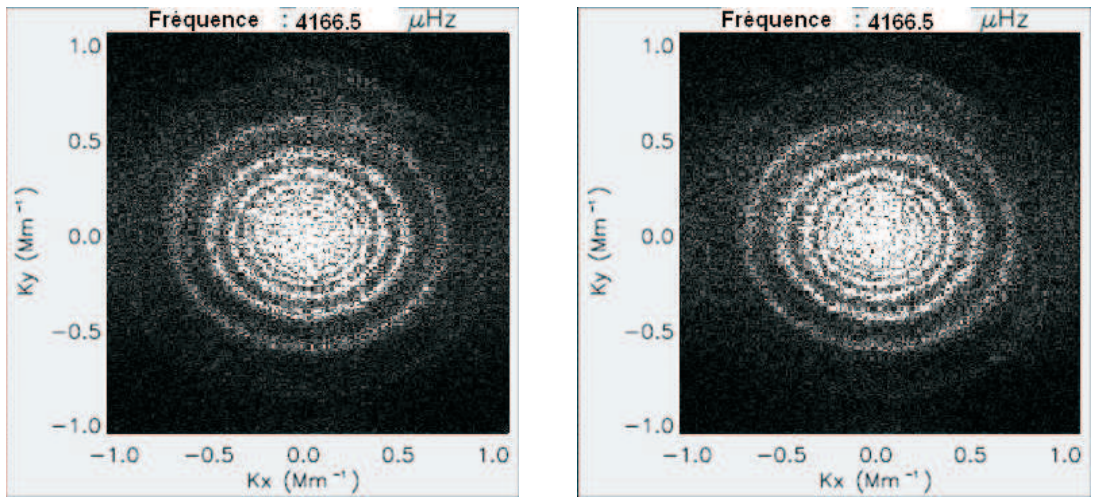

FIG. 5.15 – Le Ring diagram pour $\omega = 4166.5 \mu Hz$, à gauche, pour la journée active et à droite pour la journée moins active.

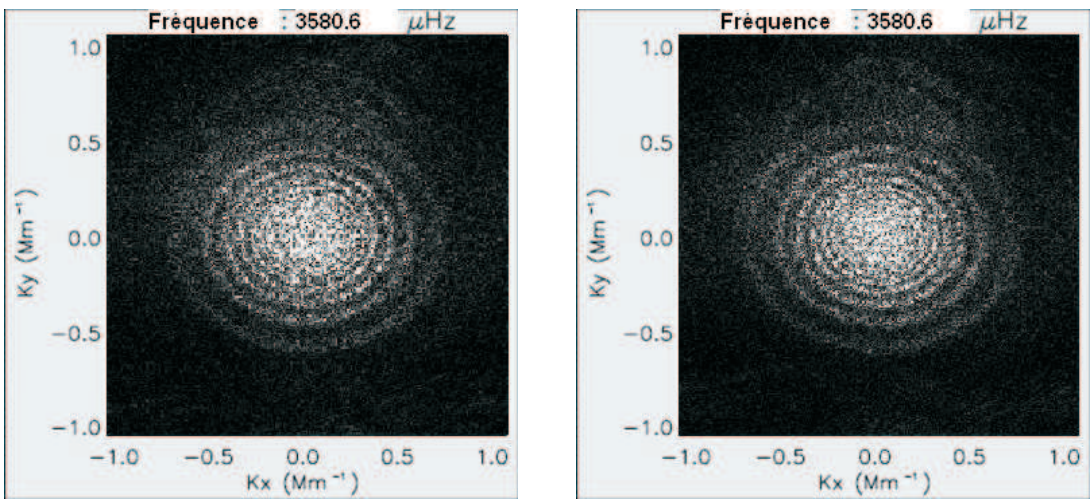

FIG. 5.16 – Le Ring diagram pour $\omega = 3580.6 \mu Hz$, à gauche, pour la journée active et à droite pour la journée moins active.

Nos derniers résultats sont similaires à ceux obtenus par les autres auteurs, par la même méthode (voir figure (2.9) de la section (2.7.1)).

La coupe longitudinale du même cube, pour un $k_x$ ou $k_y$ donné, nous révèle l'allure du diagramme $k - \nu$, c'est le diagramme $k_x - \nu$ :

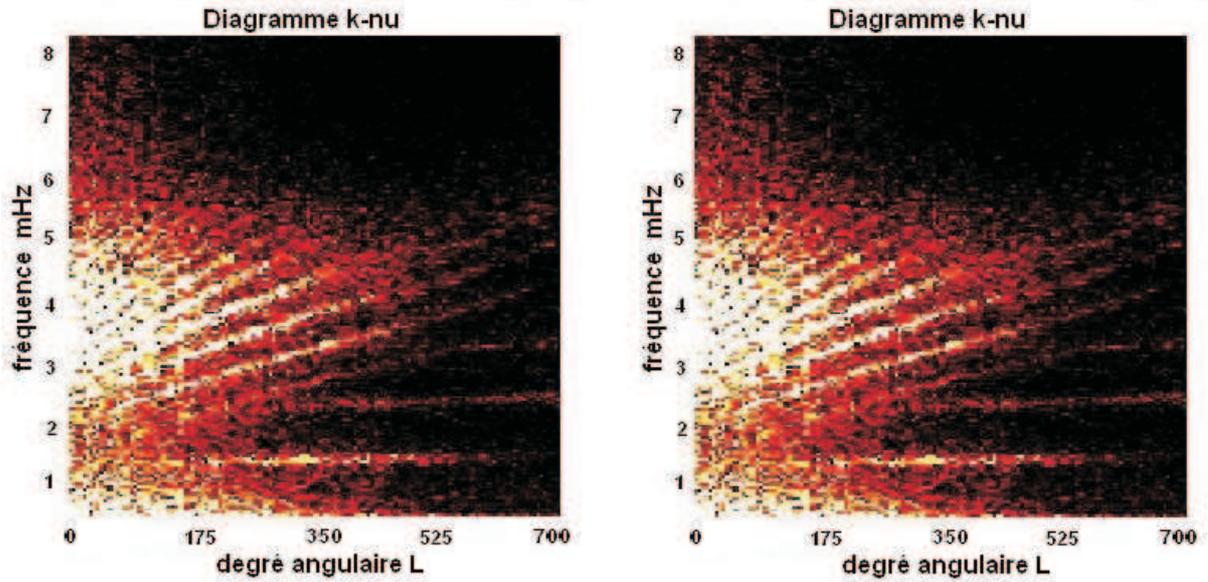

FIG. 5.17 – Le diagramme $k_x - \nu$, à gauche pour la journée active et à droite pour la journée moins active.

Le diagramme $k - \nu$ peut être aussi obtenu de manière plus rigoureuse en calculant la moyenne de chaque anneau du "Ring diagram" ($k = \sqrt{k_x^2 + k_y^2}$), plan par plan, pour chaque fréquence $\omega$ :

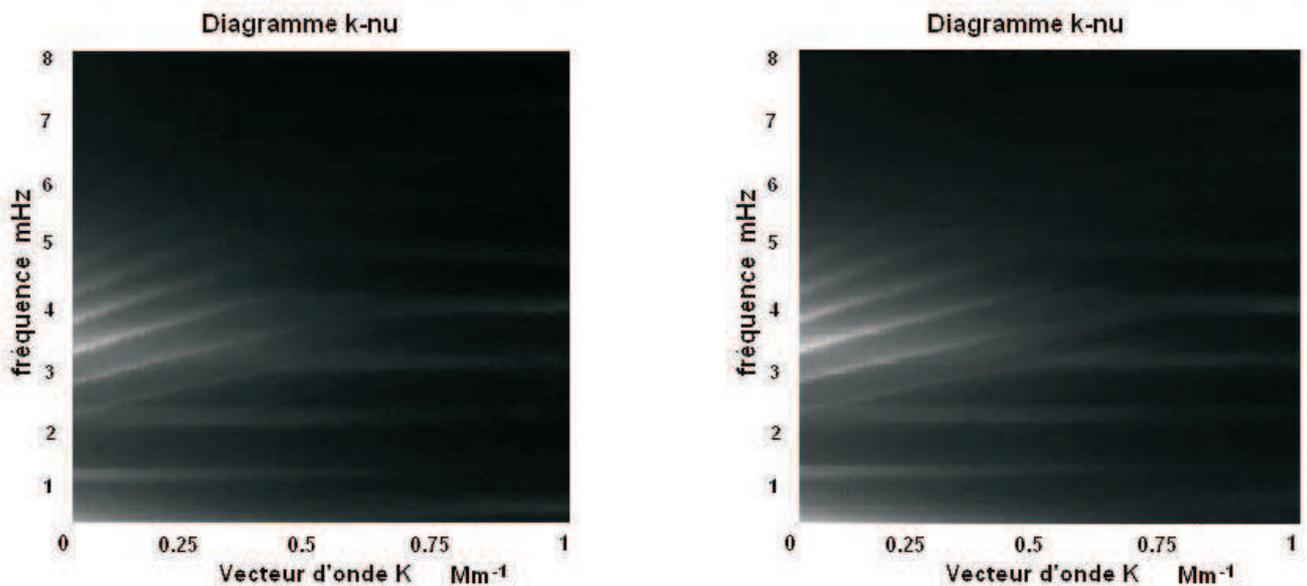

FIG. 5.18 – Le diagramme $k - \nu$, à gauche pour la journée active et à droite pour la journée moins active.

Avec : $K = L/R_\odot$, où $R_\odot = 6.961 \times 10^8 m$.

Le diagramme $k-\nu$ de la journée moins active est plus clair que celui de la journée active à cause du rendement de données exploitables de la première qui est plus élevé que celui de la dernière.

Sur les deux figures (5.18), on observe bien six arêtes, qui représentent les six mode observés : $f, p_1, p_2, \ldots, p_5$ de bas en haut.

Ce diagramme $k-\nu$ a été obtenu pour une résolution spatiale de $250 \times 250$ pixels sur une résolution temporelle de 512 minutes. Un meilleur résultat peut être obtenu avec l'augmentation de la résolution spatiale et temporelle. Comparons nos résultats avec le diagramme $k-\nu$ obtenu par le réseau GONG pour la journée du 10 mai 2000 (Fig.(5.19)) :

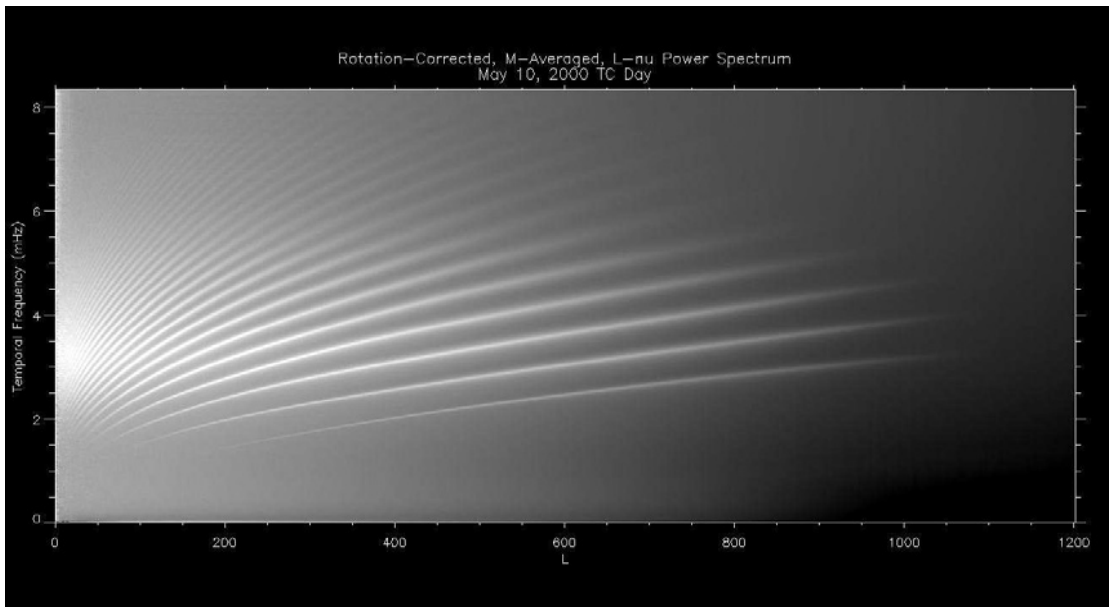

FIG. 5.19 – Le diagramme $\nu - L$ d'une journée d'observation obtenu par le reseau GONG, où, et d'après la loi de Shannon, la fréquence maximum d'observation $\nu_{Nyquist} = 1/2\delta t = 8.33 mHz$, avec $\delta t = 60 sec$ comme pas d'échantillonnage [40].

Avec $k_h = \sqrt{k_x^2 + k_y^2} = L/R_\odot$ : le vecteur d'onde horizontal et $\nu = \omega/2\pi$ la fréquence qui est mesurée en mhz, avec $L = \sqrt{\ell(\ell+1)}$.

C'est à partir de $\nu$ et $K_h$, qu'est obtenu le diagramme $(\nu - k_h)$ ou bien $(\omega - L)$, qu'on connaît plus communément sous le nom de diagramme de spectre de puissance. Ce dernier illustre bien les différents modes observés, comme on peut le constater sur la Figure (5.19) où les modes $f, p_1, p_2, p_3, p_4$ sont représentés par les cinq premières arêtes, à partir du bas. On remarque bien que les modes "p", sont, de loin, beaucoup plus présents à la surface que les modes "f".

En remarquant que les deux premières arêtes $f$ et $p_1$ représentent des modes à faible pénétration (i.e. se propageant près de la surface ($100 \leq L \leq 1000$)), de plus qu'elles sont les plus clairement distinguables sur un large intervalle de $\ell$, comparées aux autres. On décide alors de travailler sur ces deux modes et pour ce faire, les localisations du mode "f" et de la première arête $p_1$ de mode "p", dans le diagramme de puissance peuvent être approximées sous forme polynomiale par (Giles 1999) [41] :

$$\begin{aligned} l_0 &\approx R_\odot k_0 = 100\nu^2 \\ l_1 &\approx R_\odot k_1 = \sum_{i=0}^{4} c_i \nu^i \qquad c = \{17.4, -84.1, 95.6, -0.711, -0.41\} \end{aligned} \quad (5.9)$$

Le filtre est alors appliqué pour chaque fréquence, comme filtre passe bas dans le domaine spatial, avec, à chaque fois, le maximum de transmission à mi chemin entre $k_{0h}$ et $k_{1h}$ et 1% de transmission à $k_h = k_{0h}$ (voir Fig.(5.20)).

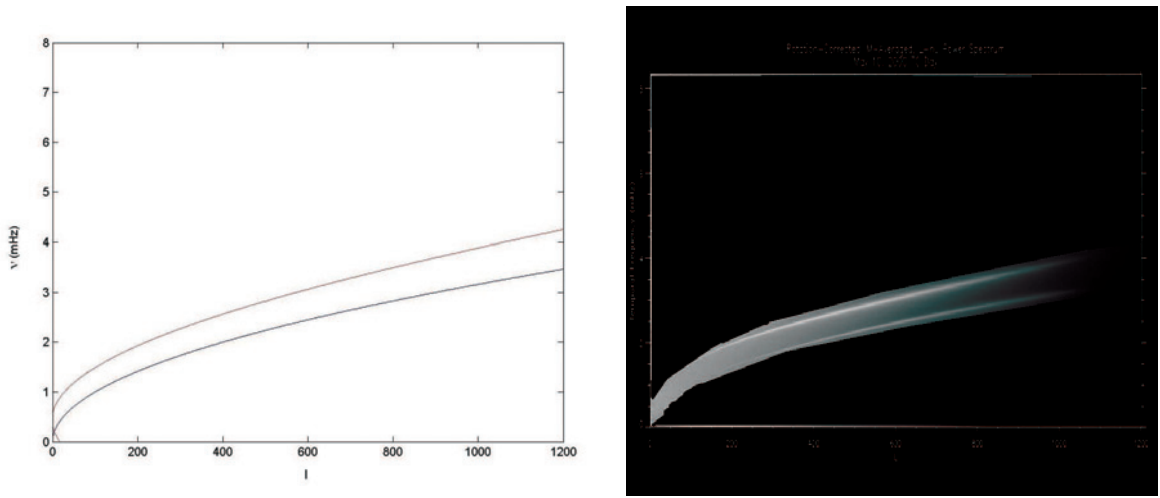

FIG. 5.20 – A gauche : représentation de l'approximation théorique de nos deux arêtes $f$(en noir), $p_1$ (en gris) en fonction du degré $\ell$. A droite le filtre de Giles appliqué à notre "$k-\nu$ diagram".

Une fois le filtrage dans l'espace des phases terminé, on n'a plus qu'à appliquer la **FFT 3D inverse** aux $\mathbf{V}(k_x, k_y, \omega)$ pour revenir à nos données vitesses $\mathbf{v}(x, y, t)$ filtrées.

**Remarques :**

1. Le filtrage utilisé ici est appelé " filtrage de fréquence temporelle " (temporary frequence filtre). Un second filtrage (optionnel) appelé filtrage de vitesse de phase et qui nous permet d'éliminer les ondes à longueur d'onde $\Delta$, relativement courtes ou bien longues, peut être utilisé (il est préférable de ne garder que les ondes comprises entre $3° \leq \Delta \leq 10°$ ; tout ce qui est en dehors de cet intervalle peut être considéré comme étant un signal parasite). La non utilisation de ce second filtre, dans notre travail, se traduirait par la présence d'un léger bruit dans nos mesures et risquerait d'affecter les résultats obtenus. Mais la corrélation des vitesses en deux points espacés par des distances angulaires $\Delta$ préalablement choisis, en plus d'un bon paramétrage résout parfaitement le problème (voir section (5.7)).

2. Une autre méthode, astucieuse et simple, consiste juste à appliquer une gaussienne, comme filtre

passe-bas autour de la fréquence $\nu = 3mHz$ pour un $\delta\nu = 0.5$ $mHz$, pour obtenir un résultat similaire au précédeant autour de $L = 500$ pour $\delta L = \pm 200$. Ainsi, en plus du filtrage des fréquences, on filtre aussi les longueurs d'onde, car sélectionner les $\ell$ reviendrait à sélectionner les $\Delta$, puisque le degré du mode $\ell$ détermine la profondeur de pénétrabilité de l'onde qui à son tour détermine la longueur de l'onde $\Delta$ à la surface du disque solaire ($300 \leq L \leq 700 \Leftrightarrow$ choisir $5° \leq \Delta \leq 10°$ ce qui à été fait dans la section (5.7)). Et c'est là que réside toute la simplicité et l'astuce de la méthode que nous avons utilisée dans notre travail (voir Fig.(5.21)).

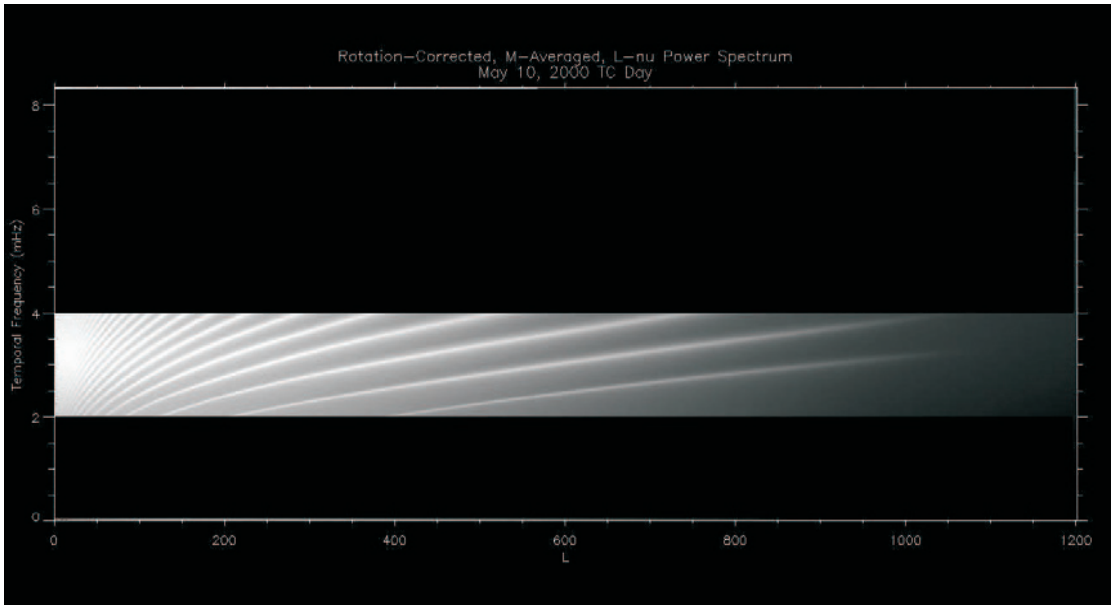

FIG. 5.21 – le résultat de notre filtre appliqué sur le "$k - \nu$ diagram".

Et les cartes de puissance (Fig.(5.12)) après filtrage deviennent :

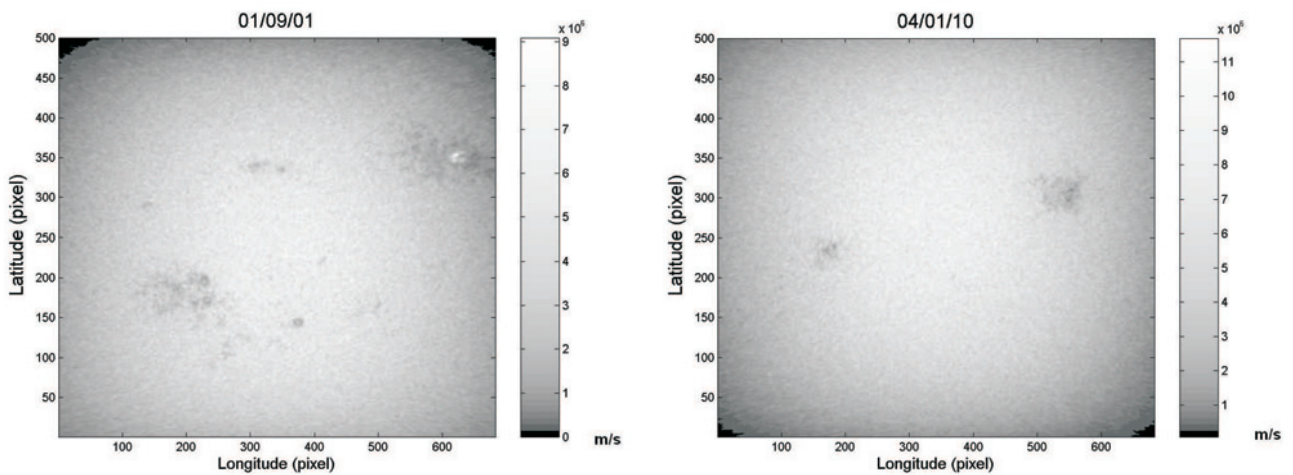

FIG. 5.22 – Les cartes de puissance (682x500 pixels) après filtrage. À gauche la power map correspondant à la période de haute activité solaire (01/09/01), et à droite celle à basse activité (10/01/04).

## 5.7  Calcul de la fonction de corrélation

Le calcul de la fonction de corrélation entre deux points différents du disque solaire, en deux temps différents, et comme vu précédemment, nous permet de voir les similitudes entre les deux points et ainsi suivre le parcours de l'onde en se référant juste aux maxima de la CCF.
La fonction de corrélation vue, dans le chapitre 4, n'est valable que dans le cas continu. Dans la réalité et pour des données régulièrement échantillonnées, on utilise le coefficient de corrélation $r_{xy}$ entre deux séries de signaux $x_i$ et $y_i$ ($i = 0...N-1$ (N étant le nombre de données)), avec :

$$r_{xy} = \frac{c_{xy}}{s_x s_y} \tag{5.10}$$

Où $s_x$ est l'écart type $x$. La variance de $s_x^2$ s'écrit alors :

$$s_x^2 = \frac{1}{N} \sum_{i=0}^{N-1} (x_i - \bar{x})^2 \tag{5.11}$$

Et $c_{xy}$ la covariance de $x$ et de $y$ :

$$c_{xy}(\tau) = \frac{1}{N} \sum_{i=0}^{N-1} (x_i - \bar{x})(y_i - \bar{y}) \tag{5.12}$$

Sachant qu'il existe un écart temporel $\tau$ et en l'injectant dans nos dernières relations, on peut écrire :

$$r_{xy}(\tau) = \begin{cases} \frac{\sum_{i=0}^{N-\tau-1}(x_i-\bar{x})(y_{i+\tau}-\bar{y})}{\sqrt{\sum_{i=0}^{N-\tau-1}(x_i-\bar{x})^2}\sqrt{\sum_{i=0}^{N-\tau-1}(y_{i+\tau}-\bar{y})^2}} & \text{pour } \tau \geq 0 \\ \frac{\sum_{i=0}^{N-\tau-1}(x_{i+\tau}-\bar{x})(y_i-\bar{y})}{\sqrt{\sum_{i=0}^{N-\tau-1}(x_i-\bar{x})^2}\sqrt{\sum_{i=0}^{N-\tau-1}(y_{i+\tau}-\bar{y})^2}} & \text{pour } \tau < 0 \end{cases}$$

Où un petit artifice de calcul, qui consiste à considérer notre écart temporel variant entre $-\tau$ et $+\tau$, nous permet de calculer la CCF pour l'onde entrante et sortante.

Quels que soient l'unité et les ordres de grandeur de $x$ et $y$, le coefficient de corrélation est un nombre sans unité, compris entre -1 et 1. Il traduit la plus ou moins grande dépendance linéaire de $x$ et $y$.

En utilisant un programme "C" sous linux fourni par GONG, en choisissant une région solaire de 459x390 pixels ($41 \leq x \leq 500$ et $79 \leq y \leq 469$ (région avec des taches solaires)) et en calculant la moyenne des CCF entre la moyenne des valeurs délimitant un cercle et la valeur au centre du cercle parcourant la latitude de la région choisie (voir Fig.(5.23)), pour chaque latitude, pour $5° \leq \Delta \leq 10°$ (avec un pas de $0.2°$), pour un temps total de 24h i.e. 1440 données, données sur les 1440 (la limite d'un ordinateur P IV possédant une RAM 1024 Mo), pour chaque journée et avec un écart temporel $\tau = 80min$ (voir Figures (5.24-5.25)), on obtient alors une série de 26 résultats ; un résultat pour chaque distance, sous format d'image FITS (161x390 pixels), où chaque ligne représente la CCF moyenne de chaque latitude variant entre $-\tau$ et $+\tau$ pour 161 valeurs.

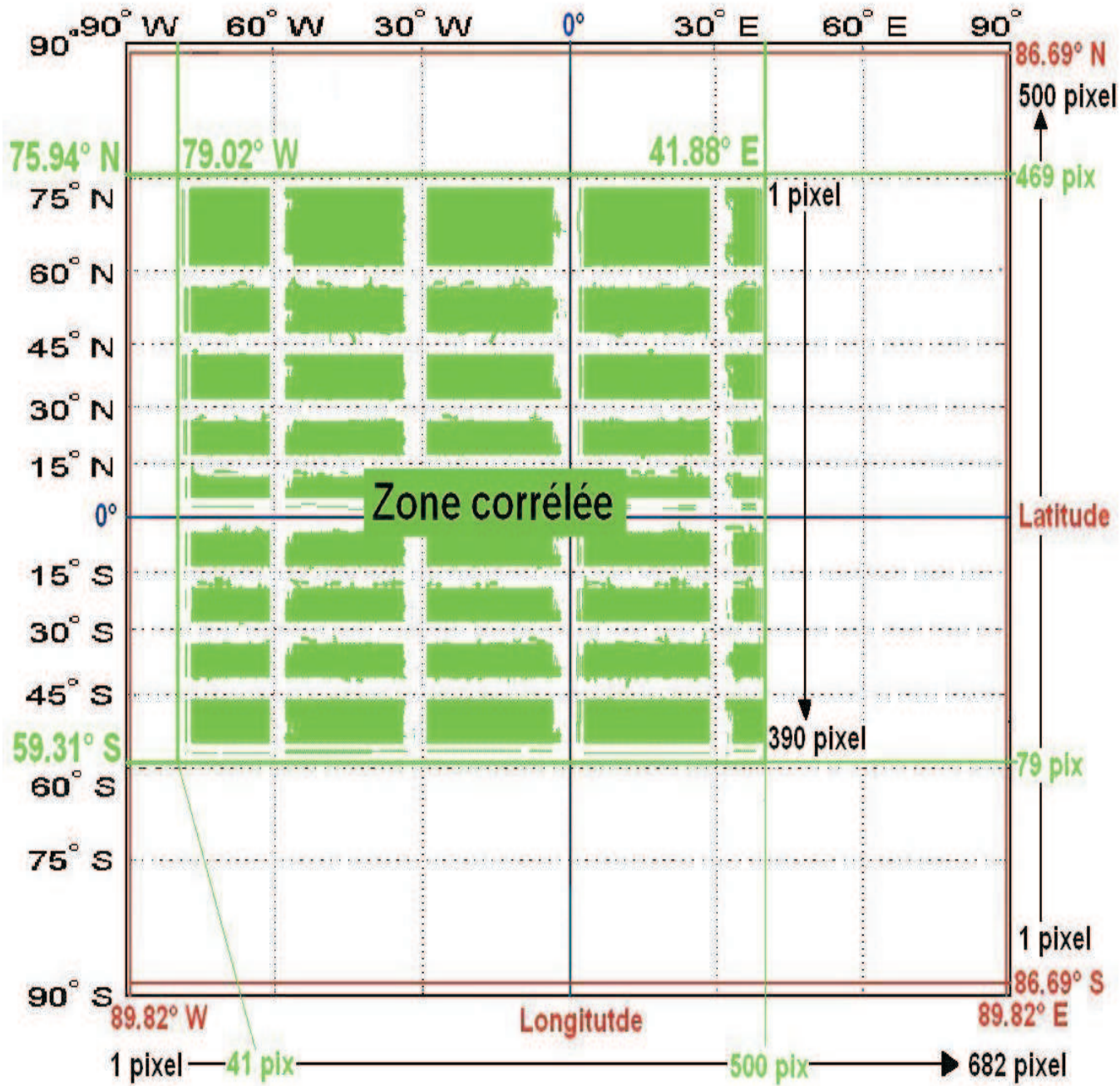

FIG. 5.23 – Figure schématique représentant la zone corrélée (la zone en vert), la zone en rouge représente la zone du remapping.

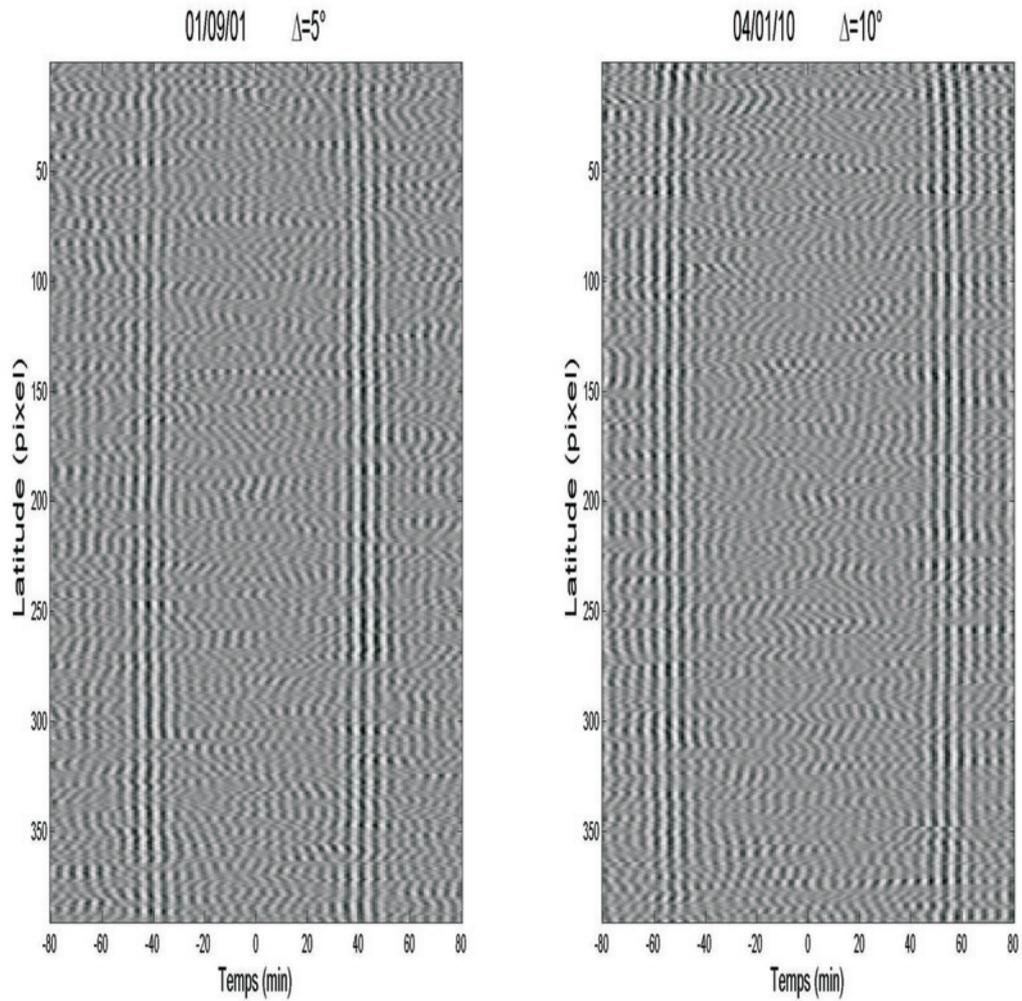

FIG. 5.24 – A gauche la carte de la CCF obtenue pour la journée active à $\Delta = 5°$ et à droite celle obtenue pour la journée moins active à $\Delta = 10°$ (les cartes sont toutes deux de dimension $390 \times 161$).

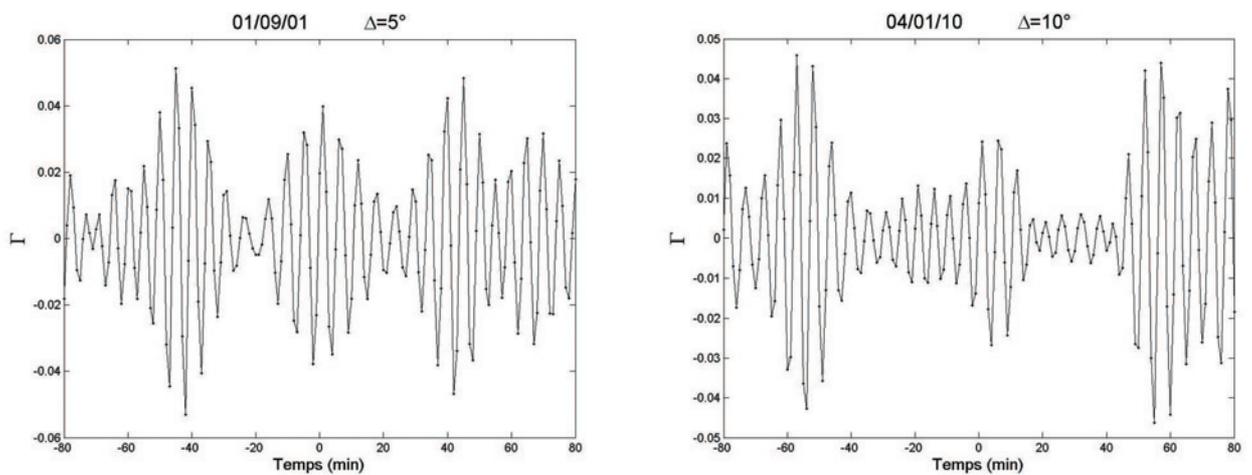

FIG. 5.25 – Représentation de la CCF correspondante à la ligne 195 des cartes précédentes (Fig.(5.15)).

## 5.7 Calcul de la fonction de corrélation 79

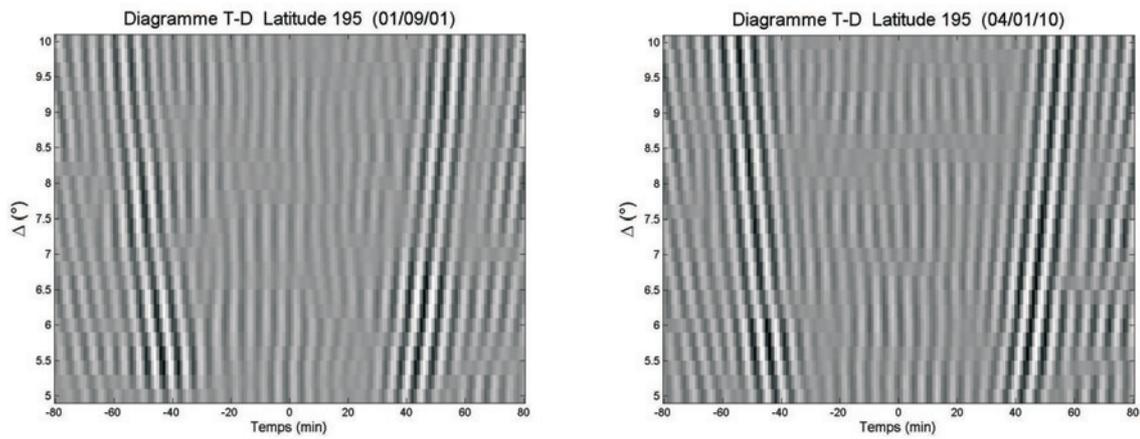

FIG. 5.26 – Nos diagrammes Temps-Distance. A gauche celui de la journée active et à droite celui de la journée moins actvie (résultats obtenus en utilisant la ligne 195 comme référence).

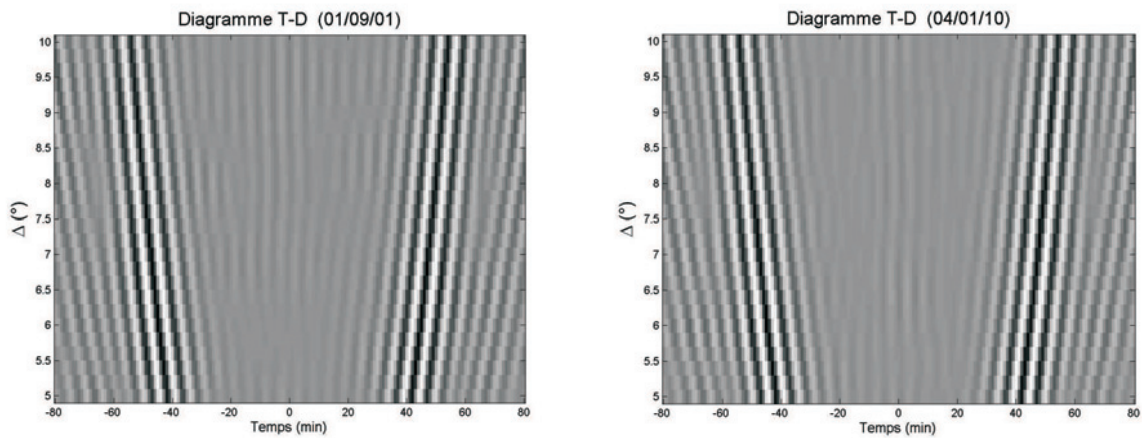

FIG. 5.27 – Nos diagrammes Temps-Distance. A gauche celui de la journée active et à droite celui de la journée moins active (résultats obtenus en moyennant toutes les latitudes).

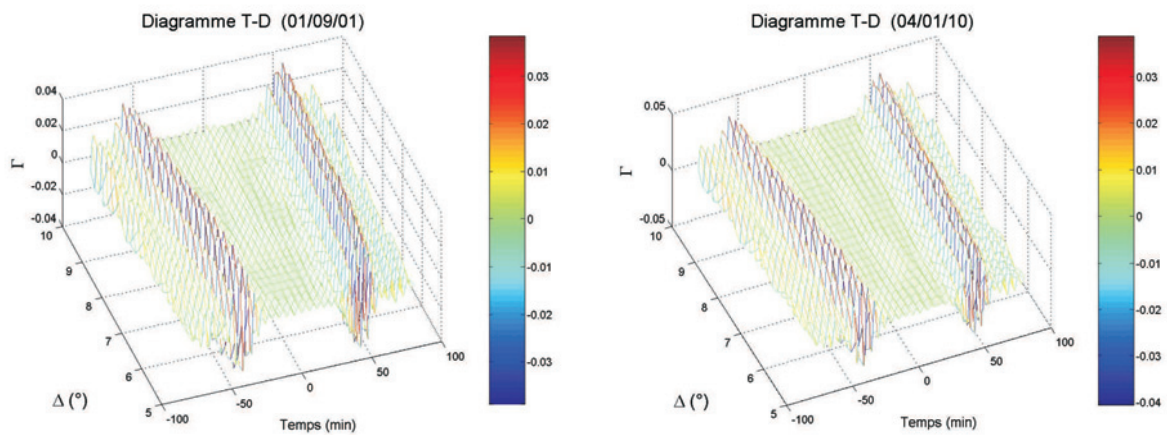

FIG. 5.28 – Nos diagrammes Temps-Distance 3D. A gauche celui de la journée active et à droite celui de la journée moins active (résultats obtenus en moyennant toutes les latitudes).

Les résultats trouvés sont semblalbes à ceux obtenus par les autres chercheurs (Fig.(5.29)) :

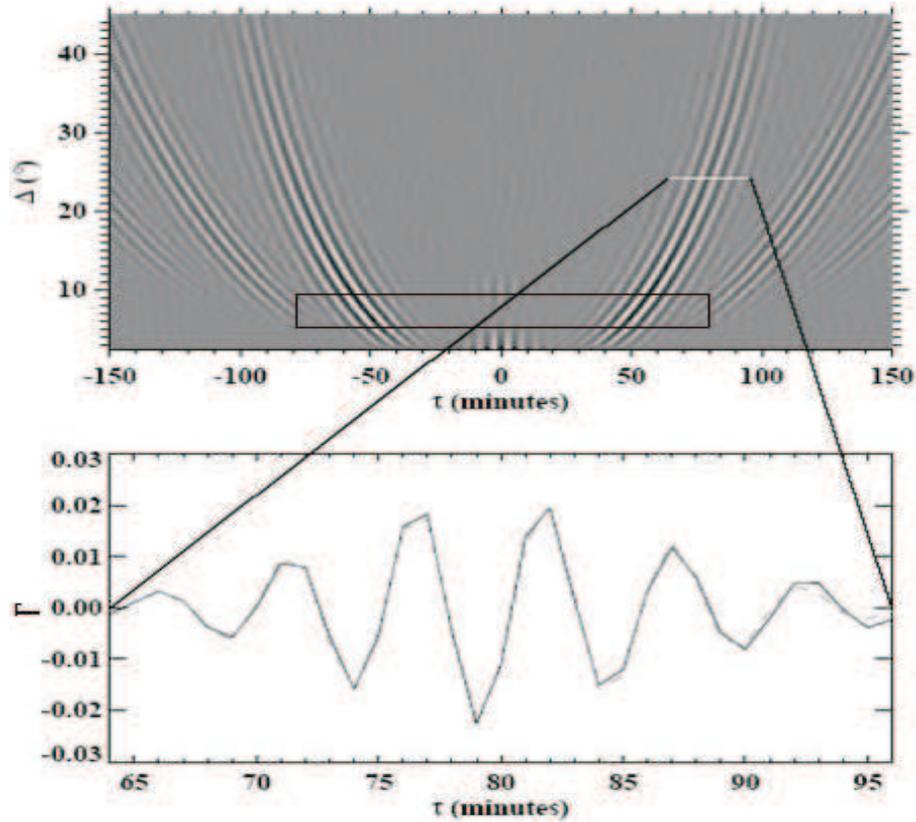

FIG. 5.29 – Diagramme Temps-Distance obtenu par Giles 1999 [41]. Le rectangle délimite nos résultats obtenus plus haut (de $-80$ à 80 min pour l'axe du temps et de $5°$ à $10°$ pour l'axe de la distance circulaire $\Delta$).

## 5.8   Temps de parcours à partir de la CCF via la fonction de Gabor

Les données corrélées, et comme vu dans le chapitre 4, peuvent être approximées par une fonction de Gabor. Pour ce faire, on utilise une approximation non linéaire des moindres carrés, et en prenant comme modèle théorique une fonction de Gabor, qu'on peut écrire comme suit :

$$G = f(x_i; a_1, a_2, a_3, a_4, a_5) = a_1 \cos[a_2(x_i - a_3)] \exp[-a_4(x_i - a_5)^2] \tag{5.13}$$

Où $a_1 = A$ est l'amplitude, $a_2 = 2\pi\nu$ est la pulsation, $a_3 = \tau_p$ le temps de phase, $a_4 = 1/2\sigma^2$ où $\sigma$ est la largeur de l'enveloppe a mi hauteur de la fonction de Gabor et $a_5 = \tau_g$ le temps de groupe.

Ayant des données empiriques $y_i$ (pour $i = 1...N$), on définit :

$$d\beta_i = y_i - f(x_i; a_1, a_2, a_3, a_4, a_5) \tag{5.14}$$

Qu'on peut écrire sous forme linéaire pour les différents $da_i$ qui doivent réduire l'écart $d\beta_i$ à zéro.

$$d\beta_i = \sum_{j=1}^{N} \frac{\partial f}{\partial a_j} da_j \bigg|_{x_i, a} \tag{5.15}$$

Avec une estimation de nos paramètres de départ $a = (a_1, a_2, a_3, a_4, a_5)$, on écrit alors :

$$d\beta_i = \alpha_{ij} da_i \tag{5.16}$$

Où $\alpha$ est une matrice $5 \times N$ :

$$\alpha_{ij} = \begin{pmatrix} \frac{\partial f}{\partial a_1}\Big|_{x_1,a} & \frac{\partial f}{\partial a_2}\Big|_{x_1,a} & \cdots & \frac{\partial f}{\partial a_5}\Big|_{x_1,a} \\ \frac{\partial f}{\partial a_1}\Big|_{x_2,a} & \frac{\partial f}{\partial a_2}\Big|_{x_2,a} & \cdots & \frac{\partial f}{\partial a_5}\Big|_{x_2,a} \\ \vdots & \vdots & \vdots & \vdots \\ \frac{\partial f}{\partial a_1}\Big|_{x_N,a} & \frac{\partial f}{\partial a_2}\Big|_{x_N,a} & \cdots & \frac{\partial f}{\partial a_5}\Big|_{x_N,a} \end{pmatrix}$$

Et sous forme matricielle :

$$d\beta_i = \alpha da \tag{5.17}$$

Ou $d\beta$ et $da$ sont des vecteurs à 5 dimensions. En multipliant des deux côtés par la matrice transposée de $\alpha$, on trouve :

$$Ada = B \tag{5.18}$$

Où $A = \alpha^T \alpha$ et $B = \alpha^T d\beta$. Suite à celà, il ne nous reste qu'à résoudre le système en usant des méthodes habituelles (Gauss, Gauss-Jordan, etc.), en itérant à chaque fois jusqu'à l'annulation de $d\beta$ (la méthode itérative utilisée ici celle "Levenberg-Marquardt" (sous Matlab)).

Les temps de groupe $\tau_g$ correspondent aux maxima des fonctions de corrélations. C'est d'ailleurs grâce à ces maxima là qu'on arrive à situer les autres paramètres de la fonction de Gabor tels que $\tau_p$ qui est généralement au voisinage de $\tau_g$ ($\tau_p > \tau_g$ d'une à trois minutes (voir Fig.(5.30))), $A$ qui se situe à proximité de l'ordonnée de $\tau_g$, $2\pi\nu = \omega$ est la fréquence du signal correspondant aux oscillations prédominantes (les oscillations de 5 min), et enfin la largeur de la gaussienne à mi-hauteur qui est proportionnelle au maxima par un certain facteur bien déterminé ($1/2\sigma^2 \propto \tau_g$).

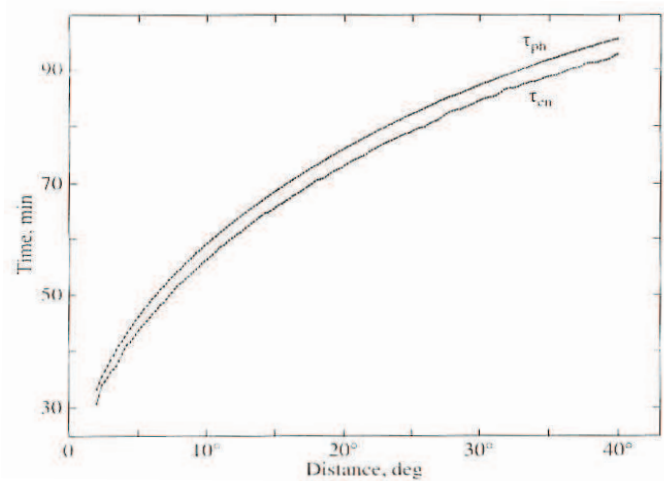

FIG. 5.30 – Courbes temps-distance obtenues par S.Kholikov et son équipe pour 12 jours de corrélation, pour $\Delta = 5°$ à $40°$. Le temps de phase $\tau_p$ est supposé être après le maximum de l'enveloppe du signal (le temps de groupe $\tau_g$), i.e. $\tau_g$ est toujours inferieur à $\tau_p$ [56].

Et ainsi nous obtenons les résultats suivants :

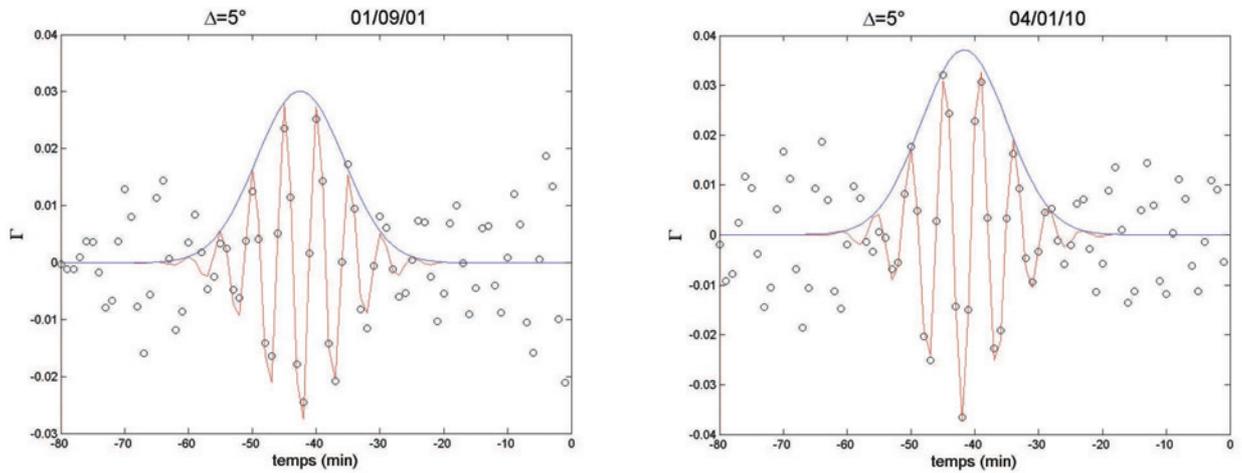

FIG. 5.31 – Pour $\Delta = 5°$ : à gauche pour la journée active, on obtient $\tau_p = 44,7645 min$ et à droite pour la moins active, on obtient $\tau_p = 44,6672 min$. En cercle le signal empirique, en rouge l'approximation théorique et en bleu l'enveloppe de l'approximation,

et

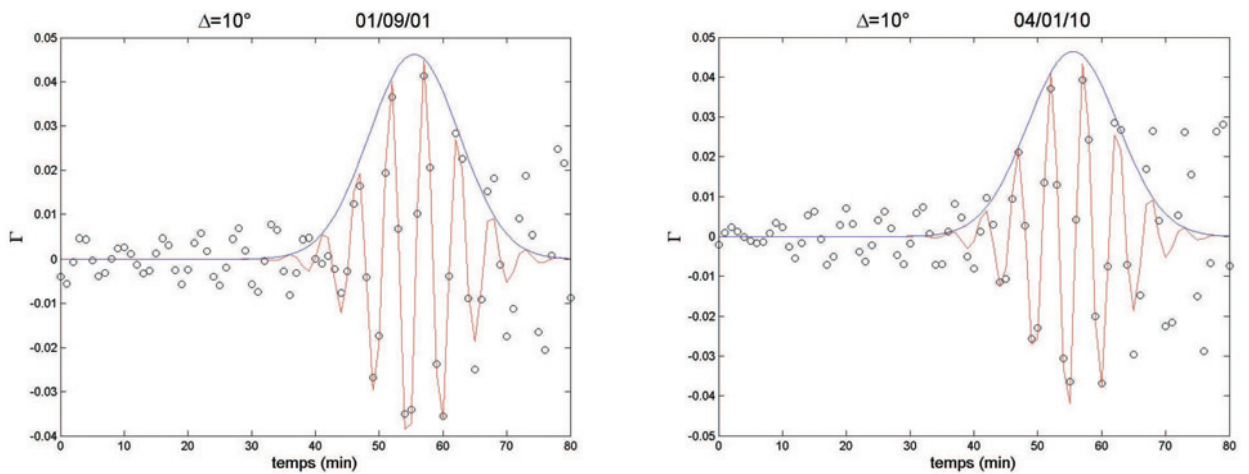

FIG. 5.32 – Pour $\Delta = 10°$ : à gauche pour la journée active, on obtient $\tau_p = 57,1124 min$ et à droite pour la moins active, on obtient $\tau_p = 57,2492 min$. En cercle le signal empirique, en rouge l'approximation théorique et en bleu l'enveloppe de l'approximation.

Avec :

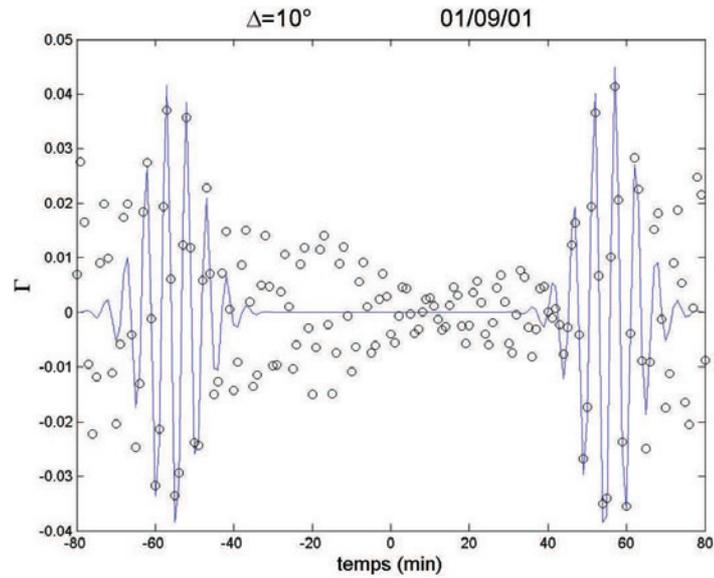

FIG. 5.33 – Exemple d'un signal corrélé entièrement ; signal de la ligne référence (195) de la carte CCF 2001 pour $\Delta = 10°$,

et

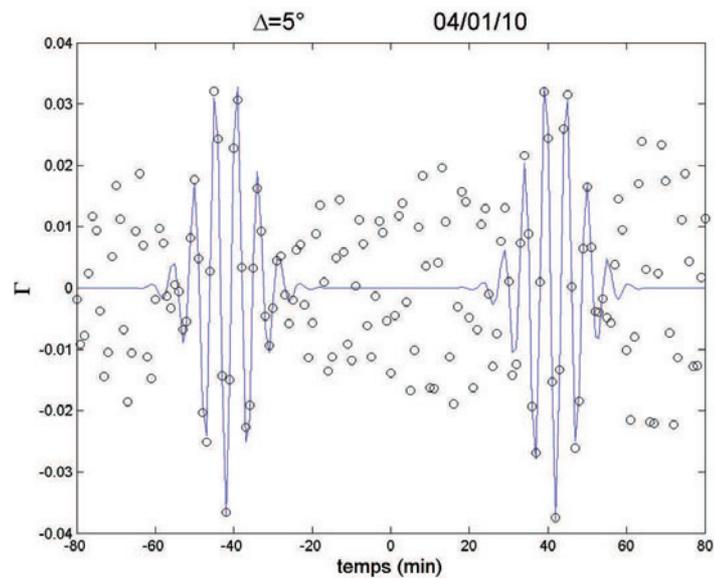

FIG. 5.34 – Exemple d'un signal corrélé entièrement ; signal de la ligne référence (195) de la carte CCF 2004 pour $\Delta = 5°$.

| Temps (min) | Journée active (01/09/01) | | Journée moins active (10/01/04) | |
|---|---|---|---|---|
| | $\Delta = 5°$ | $\Delta = 10°$ | $\Delta = 5°$ | $\Delta = 10°$ |
| $\tau_{p1}(min)$ | -44.8773 | -57.1735 | -44.7053 | -57.2956 |
| $\tau_{g1}(min)$ | -42.5000 | -55.5000 | -41.6827 | -55.5000 |
| $\tau_{p2}(min)$ | 44.7645 | 57.1124 | 44.6672 | 57.2492 |
| $\tau_{g2}(min)$ | 42.5000 | 55.5000 | 41.5000 | 55.5000 |
| $\delta\tau(sec)$ | 06.7682 | 03.6654 | 02.2870 | 02.7841 |
| $\tau_{mean}(min)$ | 44.8209 | 57.1429 | 44.6862 | 57.2724 |

TAB. 5.1 – Tableau représentant le temps de phase $\tau_{p1}$ pour l'onde sortante et $\tau_{p2}$ pour l'onde entrante, les temps de groupe $\tau_{g1}$ et $\tau_{g2}$ pour l'onde sortante et entrante respectivement ainsi que la différence des temps de phase $\delta\tau$ et la moyenne des temps $\tau_{mean}$.

Et si on trace $\tau_{mean}$ et $mean\{\tau_g\}$ en fonction de $\Delta$, pour la latitude 195 pixel, pour les deux journées, on obtient :

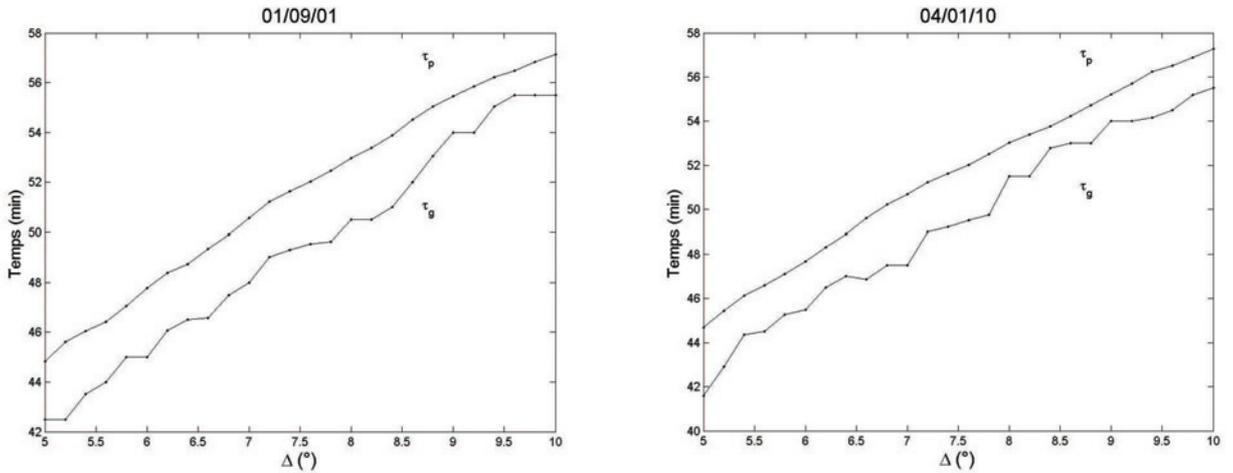

FIG. 5.35 – $\tau_p$ et $\tau_g$ en fonction de $\Delta = 5°$ à $10°$, à gauche pour la journée active et à droite pour la moins active.

D'après D.C. Braun [5], on peut approximer $\tau_{mean}$ en fonction de $\Delta$ par un polynôme logarithmique du 3 ième degré (voir Fig.(5.36)) :

$$\ln \tau_{mean} = \sum_{n=0}^{3} c_n (ln\Delta)^n \qquad (5.19)$$

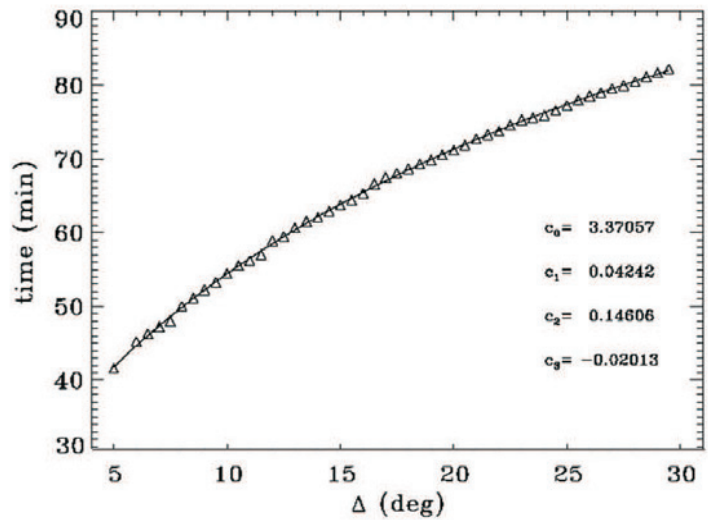

FIG. 5.36 – Cette figure représente la moyenne des temps de phase $\tau_p$ (en triangles) des ondes (sortantes et entrantes) qui sont déterminés par la fonction de corrélation. La ligne continue quant à elle représente l'approximation polynomiale des données (équa.(5.19)), avec $\Delta = 5°$ à $30°$. Les paramètres de l'approximation y sont affichés [5].

Et on obtient ainsi :

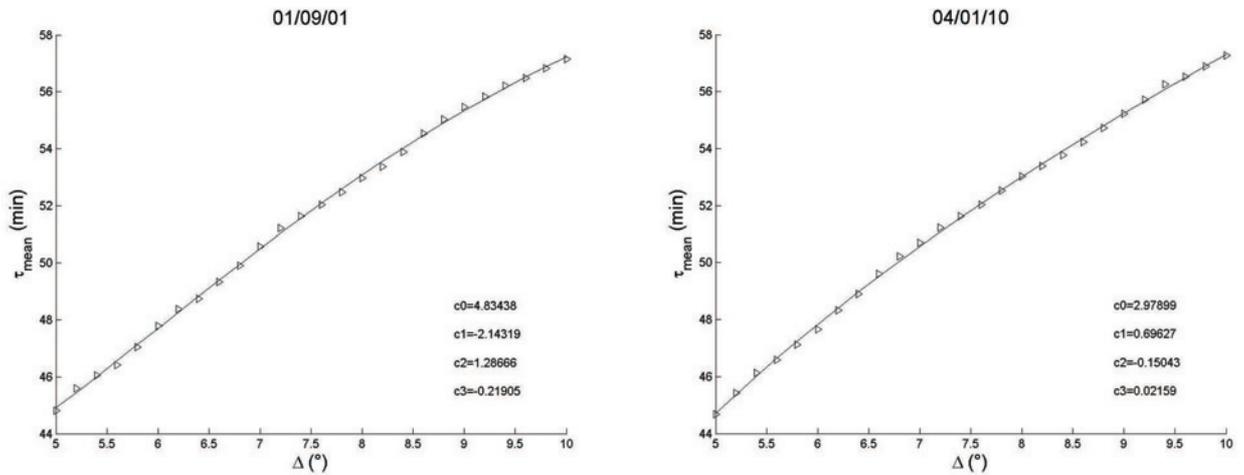

FIG. 5.37 – $\tau_p$ en fonction de $\Delta = 5°$ à $10°$. A gauche pour la journée active et à droite pour la moins active, pour la latitude 195.

D'après les figures (5.31-5.32) on constate que pour les deux cas $\Delta = 5°$ et $\Delta = 10°$ les temps de parcours de la journée active (à droite) sont plus petits que ceux de la journée moins active (à gauche). Selon les Fig.(5.33-5.34) le temps de parcours $\tau_{p2}$ de l'onde entrante à droite et beaucoup plus petit que celui de l'onde sortante à gauche ($\tau_{p1}$), ce qui est normal car celle-ci voyage à contre courant alors que la première voyage dans le même sens que l'écoulement. Dans notre cas, l'écoulement s'effectue de la gauche vers la droite.

Et d'après le tableau (5.1), on remarque que $\tau_{mean}(\Delta = 5°) < \tau_{mean}(\Delta = 10°)$, que $|\tau_{p1,g1}| > \tau_{p2,g2}$, et que $\tau_{p,g;1,2}(\Delta = 5°) < \tau_{p,g;1,2}(\Delta = 10°)$ ce qui est normal , car les ondes mettent moins de temps pour traverser une distance de $5°$ que celle de $10°$.

## 5.9 Constructions des cartes à partir des temps de parcours

Après calcul des temps de parcours entrants et sortants, pour une position bien spécifique (comme vu précédemment), on en tire le temps moyen de parcours et les différences de temps de parcours pour cette position :

$$\delta\tau_{mean}^{+-}(\overrightarrow{r},\Delta) = \frac{\tau^+ + \tau^-}{2}, \qquad \delta\tau_{diff}^{+-}(\overrightarrow{r},\Delta) = \tau^+ - \tau^-. \tag{5.20}$$

Où $\tau^+$ et $\tau^-$ indiquent les temps de parcours entrant et sortant, $\delta\tau_{mean}^{+-}$ et $\delta\tau_{diff}^{+-}$ sont les temps utilisés dans les inversions, pour obtenir les variations de la vitesse du son, ainsi que celle des écoulements des fluides à l'intérieur solaire. Si on change la position du point central ($p_c = \Delta/2$) et on répète la procédure, citée ci-dessus, alors $\delta\tau_{mean}$ et $\delta\tau_{diff}$ peuvent être mesurés par ce nouveau point. Ainsi, on peut sélectionner chaque pixel à l'intérieur de la région qui nous intéresse afin de calculer les temps de parcours correspondants, et obtenir ainsi une carte des temps de parcours.

Pour avoir plus d'informations (plus de mesures) à introduire dans les inversions, on prend une série de cercles autour du point central $p_c$, qu'on divise en 4 quarts de cercle (Est, Ouest, Nord, Sud) de 90° chacun. On calcule ensuite, pour chaque anneau, la fonction de corrélation entre le signal au point central et la moyenne des signaux des quarts de cercle (Est-Centre, Ouest-Centre, Nord-Centre, Sud-Centre) et on en tire les différences de temps de parcours $\delta\tau_{diff}$ en combinant les résultats (Est-Centre : $\tau^{ec}$, Ouest-Centre : $\tau^{oc}$) et (Nord-Centre : $\tau^{nc}$, Sud-Centre : $\tau^{sc}$), obtenant : $\delta\tau_{diff}^{ns} = \tau^{nc} - \tau^{sc}$ & $\delta\tau_{diff}^{eo} = \tau^{ec} - \tau^{oc}$.

Il est clair que $\delta\tau_{diff}^{ns}$ est plus sensible aux vitesses (Nord-Sud) et $\delta\tau_{diff}^{eo}$ aux vitesses (Est-Ouest).

Les cartes $\delta\tau_{diff}^{eo}$, $\delta\tau_{diff}^{ns}$ et $\delta\tau_{diff}^{+-}$ sont combinées puis inversées, afin d'obtenir des informations sur les écoulements des flux sub-surfaciques.

Alors que les $\delta\tau_{mean}^{oi}$ sont inversés pour obtenir des informations sur les variations des vitesses du son à l'intérieur du soleil.

Une bonne combinaison des cercles peut couvrir jusqu'à 20 voir 30 Mm du fond solaire, car plus le rayon du cercle est large, mieux on sonde le fond solaire (dans notre thèse et pour faire simple, on ne calcule que les $\delta\tau_{diff}^{oi}$, qu'on va tout simplement appeler $\delta\tau$, et $\delta\tau_{mean}$ juste $\tau_{mean}$).

Nous obtenons ainsi les résultats suivants :

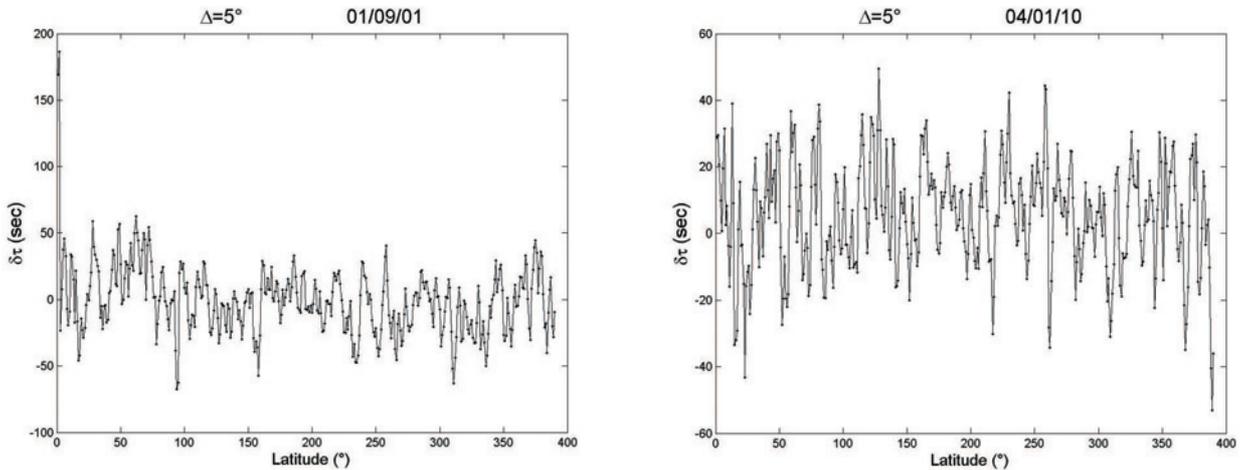

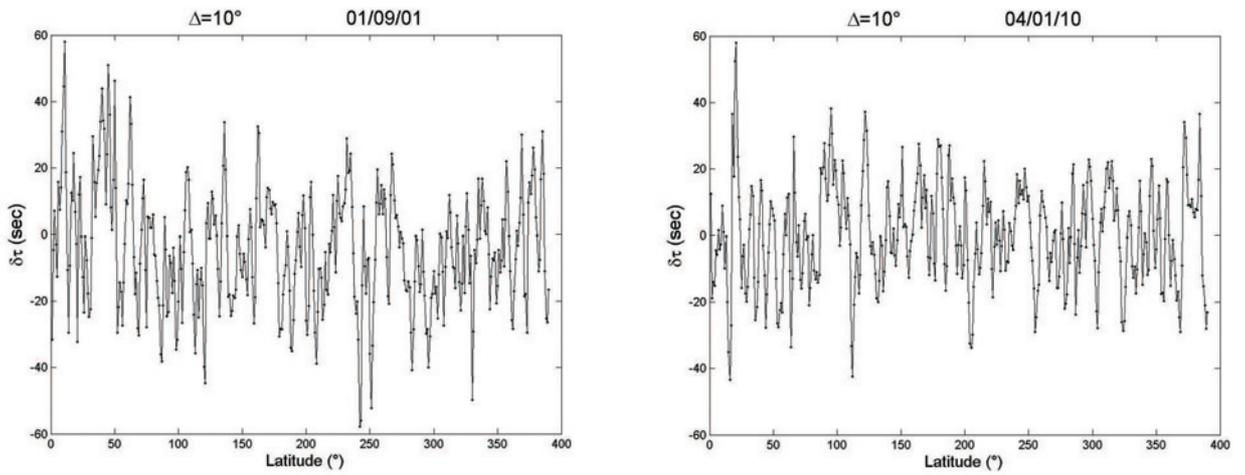

FIG. 5.38 – Figures représentant les différences de temps δτ pour toutes latitudes pour Δ = 5° en haut et 10° en bas, pour la journée active à droite et la journée moins active à gauche.

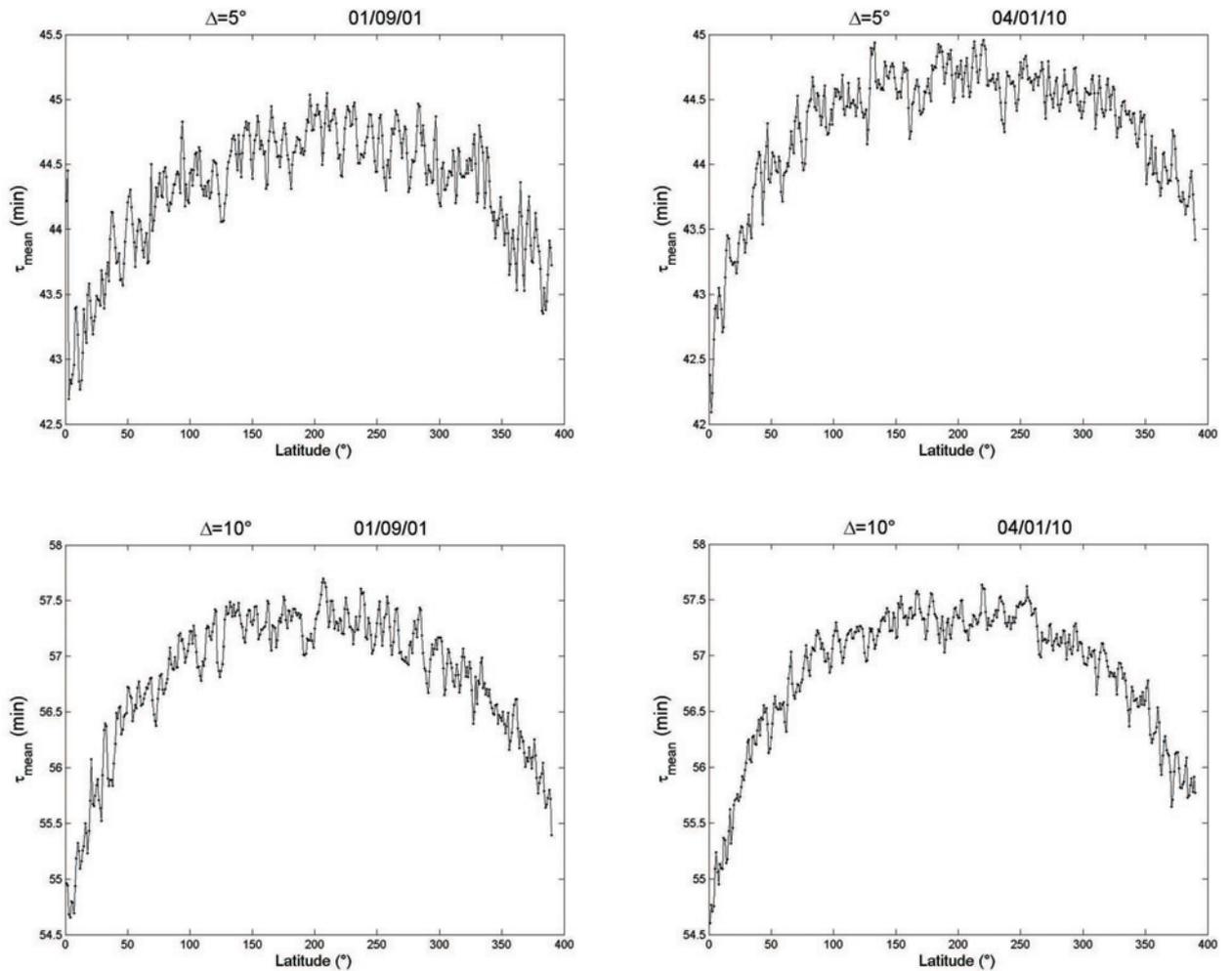

FIG. 5.39 – Figures représentant les temps moyens τ$_{mean}$ pour toutes les latitudes pour Δ = 5° en haut et 10° en bas, pour la journée active à droite et la journée moins active à gauche.

| Temps (min) | Journée active (01/09/01) | | Journée moins active (10/01/04) | |
|---|---|---|---|---|
| | $\Delta = 5°$ | $\Delta = 10°$ | $\Delta = 5°$ | $\Delta = 10°$ |
| $mean\{\delta\tau\}(sec)$ | -00.4118 | -04.2919 | 05.3017 | 00.3504 |
| $mean\{\tau_{mean}\}(min)$ | 44.3119 | 56.8108 | 44.3103 | 56.8414 |

TAB. 5.2 – Tableau représentant la moyenne des différences des temps de parcours $mean\{\delta\tau\}$ ainsi que la moyenne des temps moyens de parcours $mean\{\tau_{mean}\}$ des 390 latitudes étudiées.

Les pics des deux premières latitudes de la jourée active des figures $\delta\tau$ et $\tau_{mean}$ sont dus aux manque de données corrélées au bord (Au delà de 75.5° nord), tel que le montrent les figures suivantes :

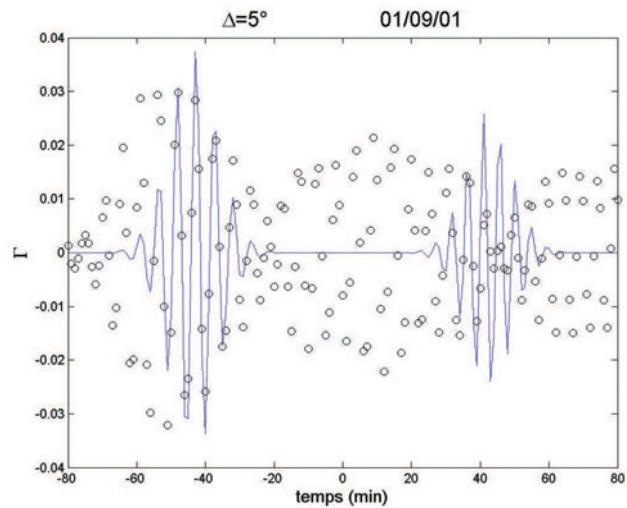

FIG. 5.40 – Signal corrélé entièrement ; de la latitude 1 (75.9° nord -voir Fig.(5.23)-) de la carte CCF 2001 pour $\Delta = 5°$.

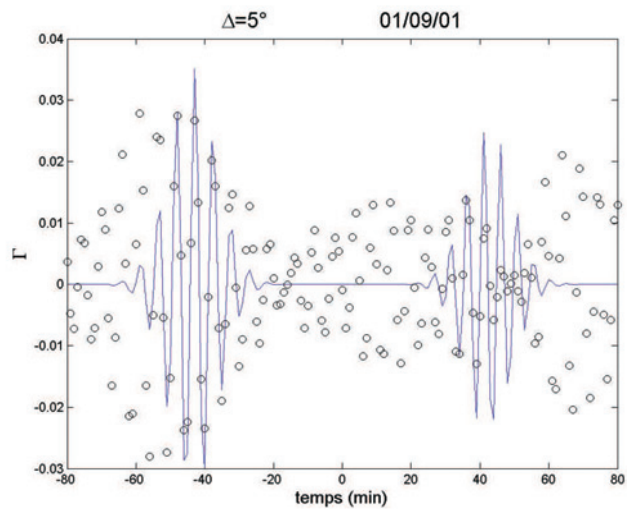

FIG. 5.41 – Signal corrélé entièrement ; de la latitude 2 (75.9° nord -voir Fig.(5.23)-) de la carte CCF 2001 pour $\Delta = 5°$.

Si on affiche maintenant les résultats de $\delta\tau$ et $\tau_{mean}$ de la journée active sans les deux premières latitudes contenant des pics (i.e. de 3 à 390 pixels), on obtient :

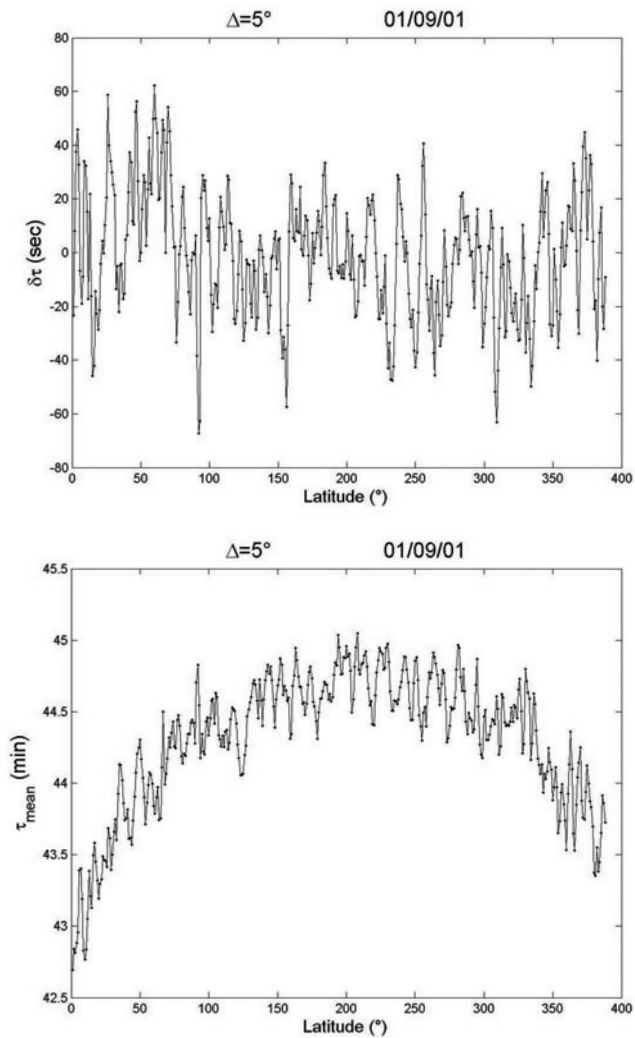

FIG. 5.42 – Figures représentant $\delta\tau$ à droite et $\tau_{mean}$ à gauche pour toutes les latitudes à part les deux premières pour $\Delta = 5°$ de la journée active, où ici $mean\{\delta\tau\} = -01.3298$ sec et $mean\{\tau_{mean}\} = 44.3118$ min.

Les figures de $\delta\tau$ -Fig.(5.38)- nous montrent des résultats qui oscillent en moyenne autour de $\pm 60$ sec, ces résultats contiennent de précieux renseignements sur la vitesse des écoulements (voir section (4.2.6)) qui circulent dans chaque latitude traitée, et qui ne peuvent être révélés qu'après inversion.

Les figures de $\tau_{mean}$ -Fig.(5.39)- nous montrent des résultats en forme de demi cercle ; demi cercle causé par la rotation du soleil, qui diffère d'une latitude à une autre (la période du soleil à l'équateur est de 27 jours, alors qu'aux pôles, elle est de 36 jours) , ce qui entraîne une influence différente sur la moyenne des temps de parcours des ondes pour chaque latitude. La région corrélée étant située entre 75.94° Nord (latitude 1 pix) et 59.31° Sud (latitude 390 pix) -voir Fig.(5.23)-, nos résultats -Fig.(5.39)- sont en parfaite concordance avec la zone de latitude traitée.

Les résultats de $\tau_{mean}$ et $\delta\tau$ pour toutes les latitudes et pour $\Delta = 5°$ à $10°$ nous donnent des cartes appelées : cartes temporelles de parcours "travel map" :

Pour la journée active :

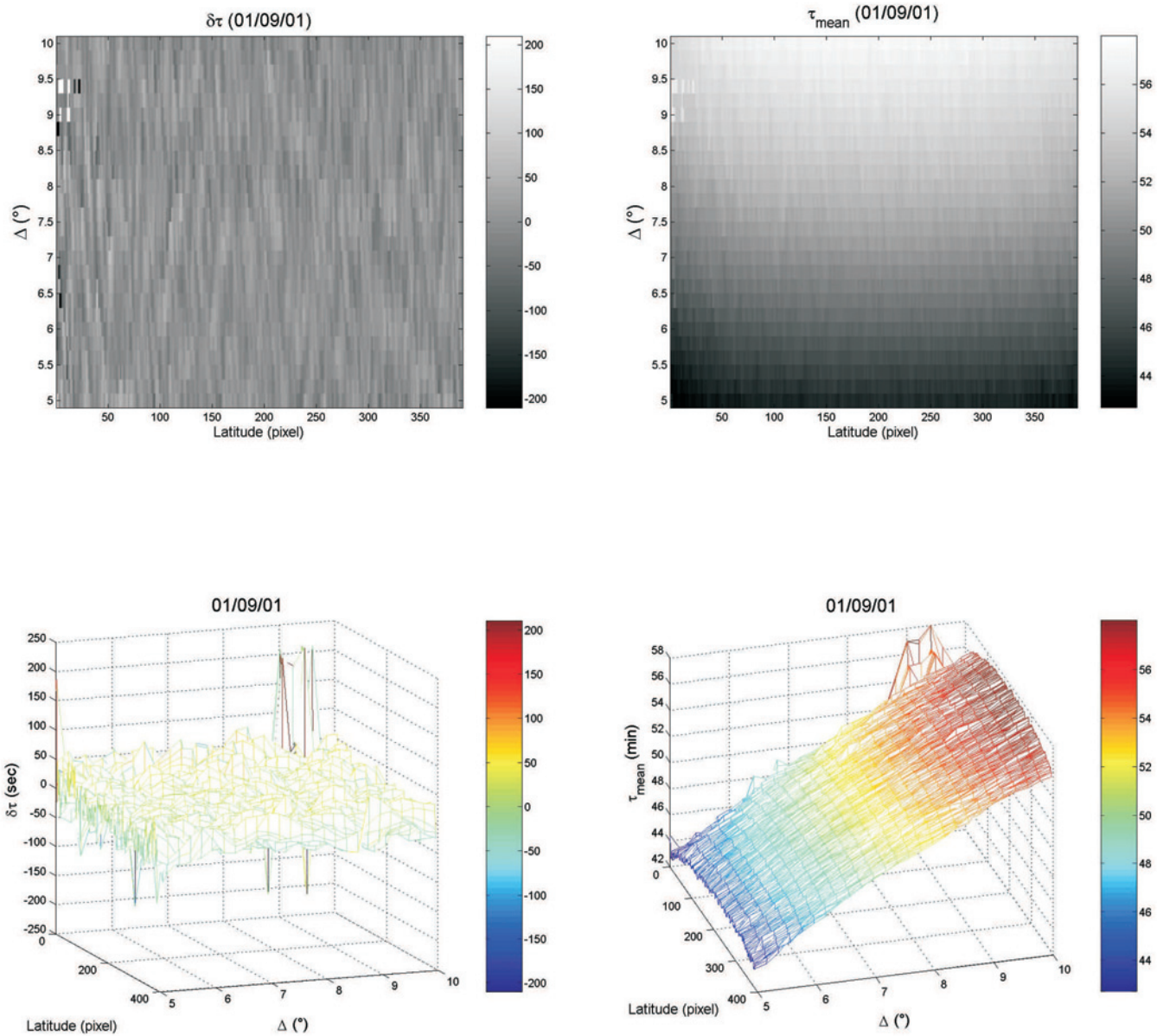

FIG. 5.43 – Figures représentant les cartes de parcours $\delta\tau$ (en seconde) à droite et $\tau_{mean}$ (en minute) à gauche pour toutes les latitudes pour $\Delta = 5°$ à $10°$ de la journée active, en haut, et leur représentation 3D en bas.

Et pour la journée moins active :

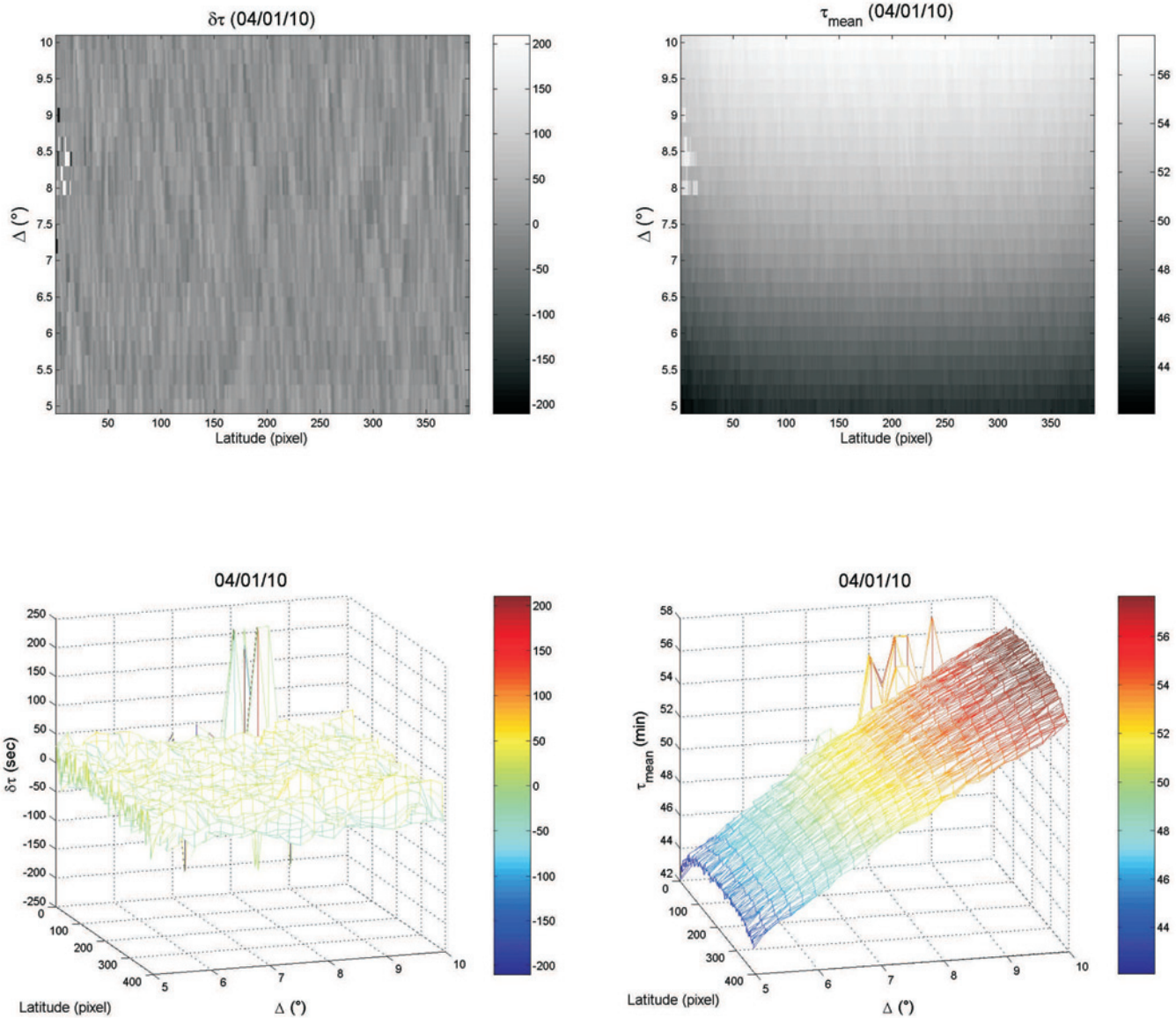

FIG. 5.44 – Figures représentant $\delta\tau$ (en seconde) à droite et $\tau_{mean}$ (en minute) à gauche pour toutes les latitudes de $\Delta = 5°$ à $10°$ pour la journée moins active, en haut, et leur représentation 3D en bas.

Résumons la moyenne de $\delta\tau$ et $\tau_{mean}$ des deux journées étudiées dans la tableau suivant :

| Temps (min) | Journée active (01/09/01) | Journée moins active (10/01/04) |
|---|---|---|
| $mean\{\delta\tau\}(sec)$ | 00.2710 | 00.6331 |
| $mean\{\tau_{mean}\}(min)$ | 51.0168 | 51.0382 |

TAB. 5.3 – Tableau représentant la moyenne des différences des temps de parcours $mean\{\delta\tau\}$ ainsi que la moyenne des temps moyens de parcours $mean\{\tau_{mean}\}$ des cartes de parcours précédentes.

On remarque sur les figures (5.44), qu'à part les pics qui apparaissent dans les 15 à 25 premières latitudes entre 8° et 9.5° des deux journées et pour des raisons précédemment citées, notre approximation est à 99.99% fiable ; pour 22 données mal approximées sur $390 \times 26 = 10140$ pour la journée active et 18 données mal approximées sur les 10140 pour la journée moins active. Nous remarquons bien qu'il y a moins d'erreurs pour la journée 04/01/10 que pour la journée 01/09/01, cela s'explique par le fait que le rendement des données exploitables de la première est bien supérieur à la dernière (voir section 5.3).

Voici à quoi ressemblent nos dernières cartes temporelles sans les premières latitudes comportant des pics, pour la journée active :

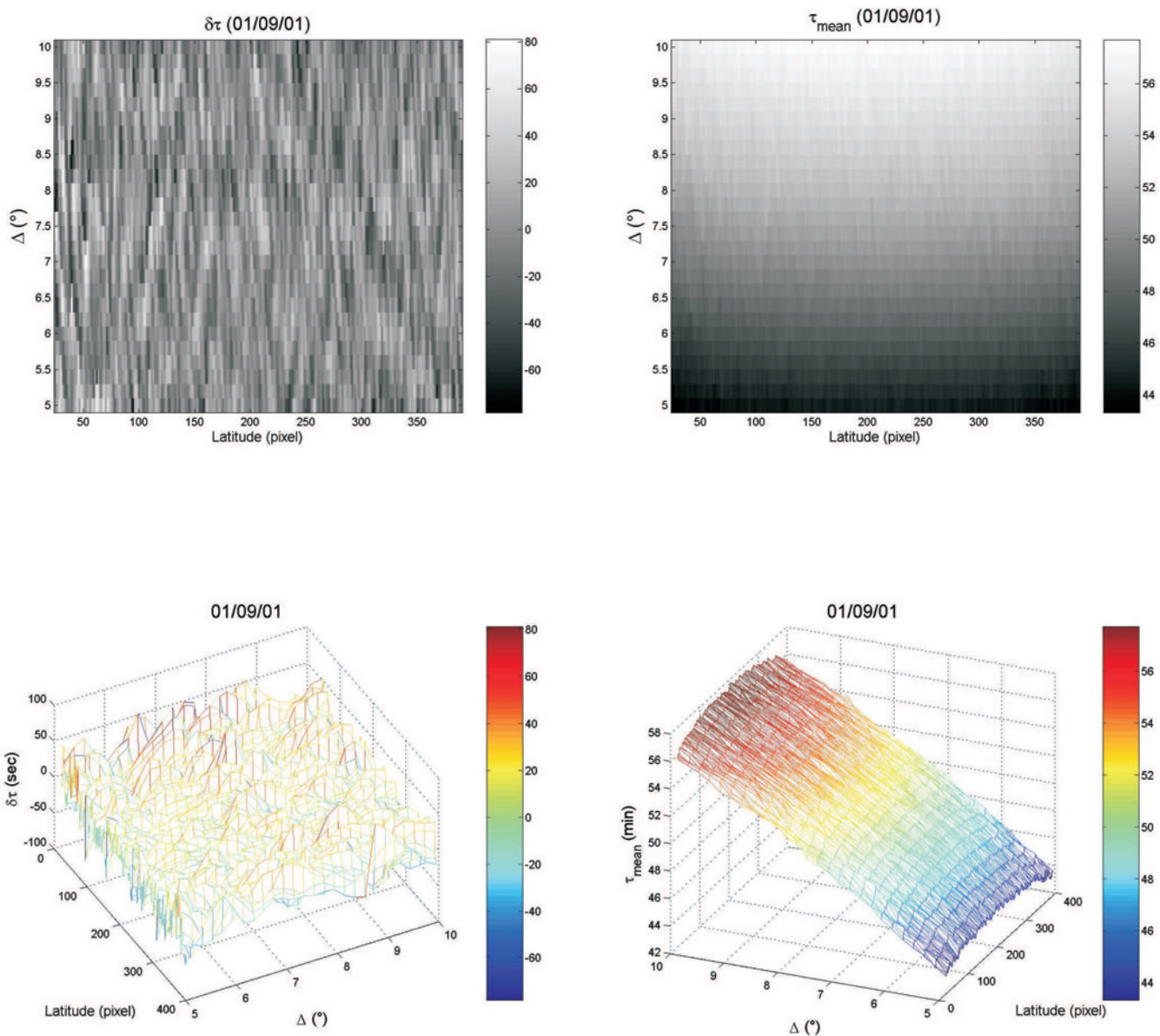

FIG. 5.45 – Figures représentant les cartes de parcours $\delta\tau$ à droite et $\tau_{mean}$ à gauche pour toutes les latitudes à part les 23 premières pour $\Delta = 5°$ à 10° de la journée active, en haut, et leur représentation 3D en bas.

Pour la journée moins active :

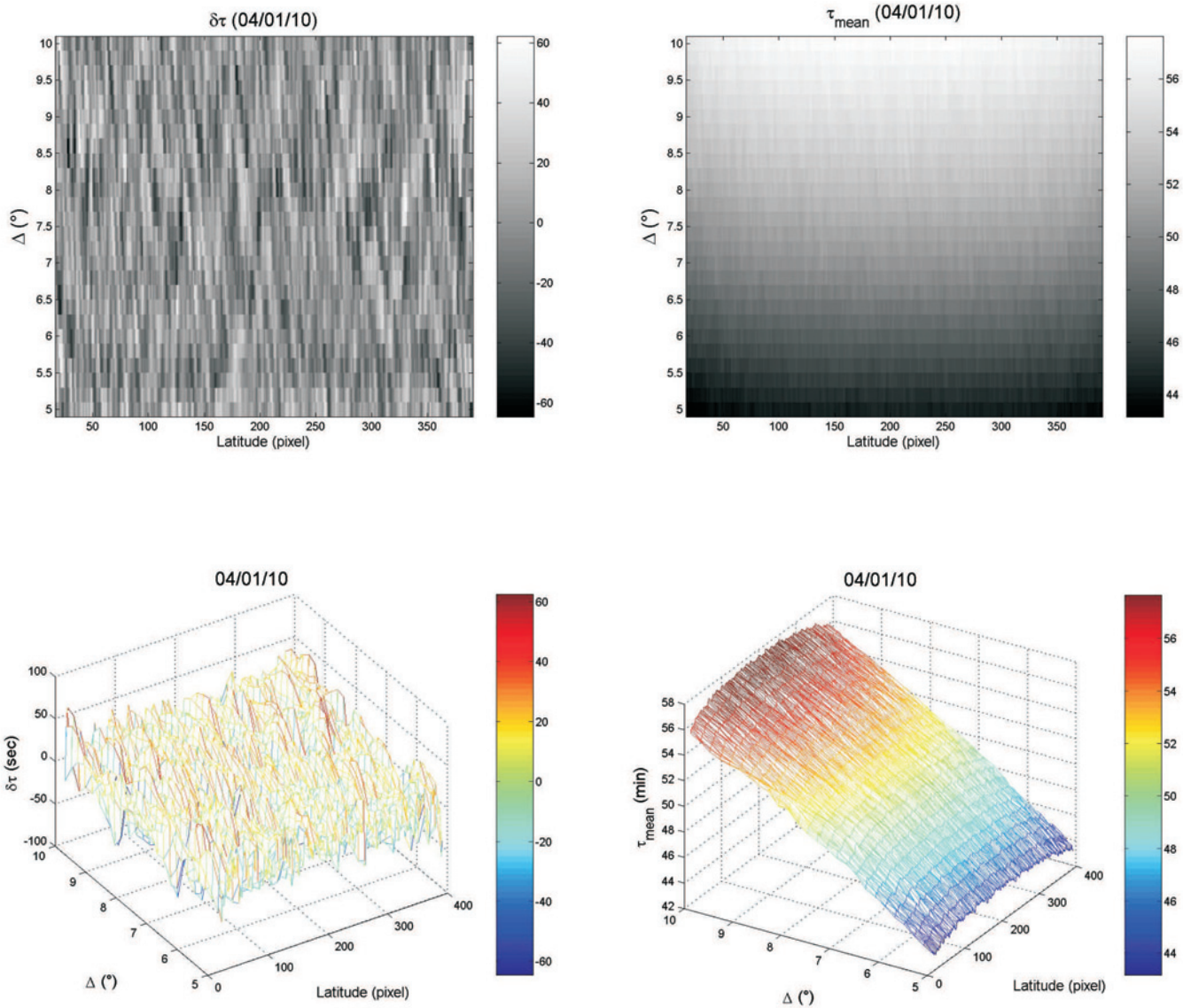

FIG. 5.46 – Figures représentant $\delta\tau$ à droite et $\tau_{mean}$ à gauche pour toutes les latitudes à part les 16 premières de $\Delta = 5°$ à $10°$ pour la journée moins active, en haut, et leur représentation 3D en bas.

Résumons là aussi la moyenne de $\delta\tau$ et $\tau_{mean}$ des deux journées étudiées dans la tableau suivant :

| Temps (min) | Journée active (01/09/01) | Journée moins active (10/01/04) |
| --- | --- | --- |
| $mean\{\delta\tau\}(sec)$ | 00.1476 | 00.5231 |
| $mean\{\tau_{mean}\}(min)$ | 51.0999 | 51.0929 |

TAB. 5.4 – Tableau représentant la moyenne des différences des temps de parcours $mean\{\delta\tau\}$ ainsi que la moyenne des temps moyens de parcours $mean\{\tau_{mean}\}$ des cartes de parcours précédentes.

Les tracés de la moyenne de toutes les latitudes (de 1 à 390 pixels) de $\tau_{mean}$ et de $mean\{\tau_g\}$ en fonction de $\Delta = 5°$ à $10°$, pour les deux journées, sont comme suit :

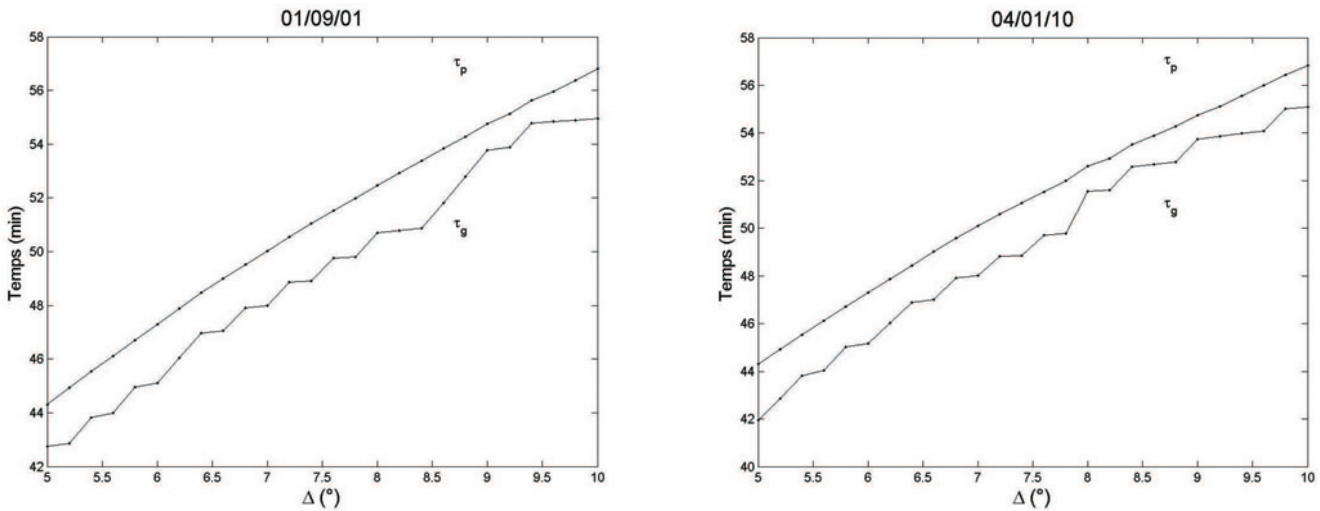

FIG. 5.47 – $\tau_p$ et $\tau_g$ en fonction de $\Delta = 5°$ à $10°$, à gauche pour la journée active et à droite pour la moins active.

Les résultats que nous avons obtenus dans la figure (5.47) sont comparables aux résultats obtenus par les autres chercheurs -voir Fig.(5.30)-.

L'approximation de la moyenne de toutes les latitudes (de 1 à 390 pixels) de $\tau_{mean}$ en fonction de $\Delta = 5°$ à $10°$ par un pôlynome du 3 ième degré, des deux journées est comme suit :

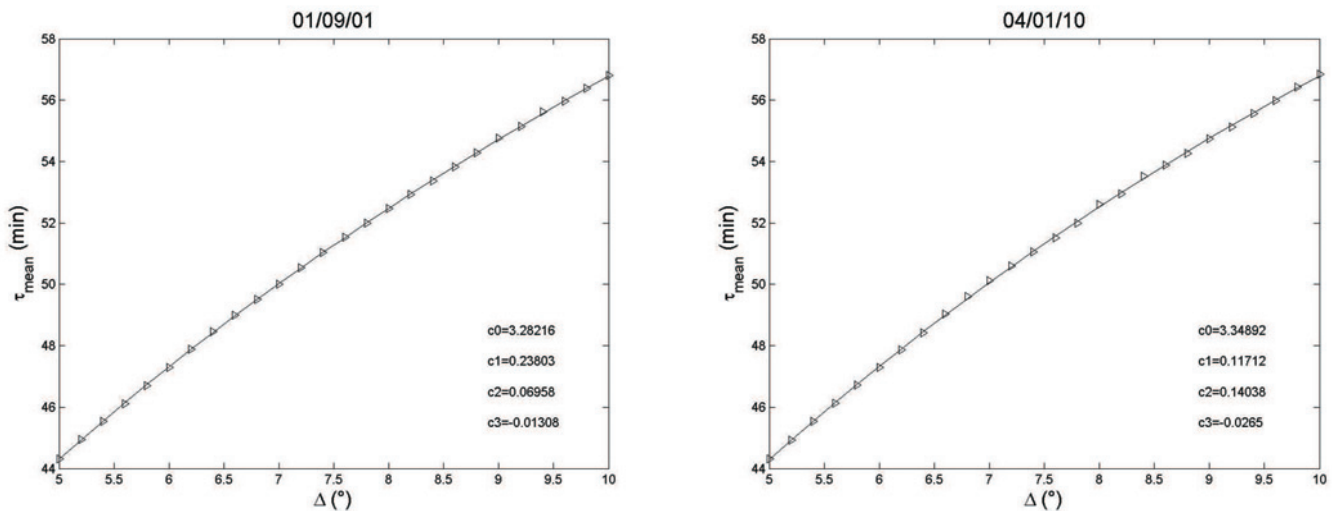

FIG. 5.48 – $\tau_p$ en fonction de $\Delta = 5°$ à $10°$, à gauche pour la journée active et à droite pour la moins active, pour la moyenne de toutes les latitudes.

Les résultats obtenus dans la figure (5.48) sont aussi comparables aux résultats obtenus par les autres chercheurs -voir Fig.(5.36)-.

## 5.10 Discussion des résultats

Les pics dus aux erreurs d'approximation des données corrélées, précédemment, peuvent être corrigés par le biais d'une corrélation plus vaste ; plus il y a de données à corréler, moins d'erreurs il y aura. La preuve en est que : la journée moins active, dont le rendement de données exploitables est de 95%, contient moins d'erreurs que la journée active qui, elle n'en contient que 78%. Pour un taux d'erreurs quasi similaire, la moyenne des temps moyens de parcours des ondes ($mean\{\tau_{mean}\}$) de la journée active, et du fait de l'augmentation de la température, serait toujours inférieure à celle de la journée moins active, comme on peut le constater sur les figures (5.43-5.44) ainsi que sur le tableau (5.3). Ce qui prouve que l'activité solaire a son influence sur les temps de parcours des ondes.

Les figures illustrant les cartes de parcours de $\delta\tau$ -Fig.(5.43-5.46)-, nous montrent des résultats qui varient de manière stochastique de $-60$ à $+60$ secondes en moyenne, mais dont la moyenne est quasiment nulle -voir le tableau (5.3-5.4)-. Le $mean\{\delta\tau\}$ qui tend vers zéro, n'obéit en fait, qu'à la loi de conservation de masse -voir équa.(3.1)-.

En Fig.(5.49) nous présentons quelques résultats de $\delta\tau$ obtenus par d'autres chercheurs par rapport à la longitude :

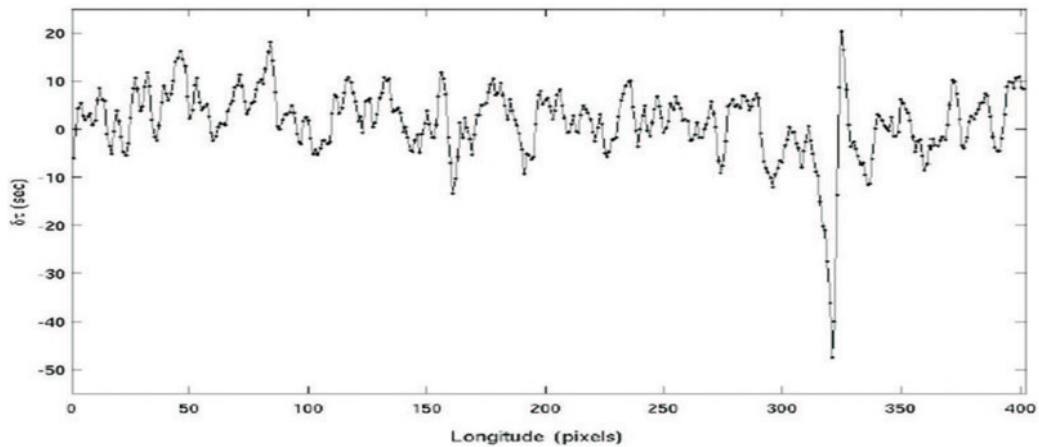

FIG. 5.49 – $\delta\tau$ en fonction de la longitude. Résultat obtenu par S.Kholikov : chercheur au réseau GONG [27].

On remarque bien que nos résultats -voir Fig.(5.38)- sont de même forme que les leurs, mais d'ordre moindre car le fait que la rotation du soleil soit transversale (de l'ouest vers l'est), les écoulements latitudinaux se trouvent plus affectés que les longitudinaux, et cela ne fait que se répercuter sur les $\delta\tau$.

Les figures illustrant les cartes de parcours de $\tau_{mean}$ -Fig.(5.43-5.46)-, nous montrent des résultats qui varient de 43 à 58 minutes selon $\Delta$. La projection de ces derniers, dans le plan "$\tau_{mean}$-Latitude" est sous forme de demi cercles -voir Fig(5.39)-. Quant à leur projection dans le plan "$\tau_{mean} - \Delta$", elle nous donne une courbe qui peut être approximée par un polynôme du 3 ième degré -voir Fig(5.48)-.

$\delta\tau$ et $\tau_{mean}$, et tel que vu dans la section (4.3.4), sont détenteurs de précieuses informations sur les vitesses d'écoulements et les vitesses du son respectivement. Notons que parmi tous les articles et thèses rencontrés au cours de notre étude, nous n'avons croisé aucun papier qui traite et illustre des figures de $\delta\tau$ et $\tau_{mean}$. Ces dernières sont très difficiles à interpréter sous cette forme, les papiers étudiés montrent tous les résultats aprés "inversion", qui eux par contre, sont très simples à interpréter physiquement.

# Chapitre 6

# Inversion

Le but ultime de la technique temps-distance étant d'obtenir certains paramètres internes du Soleil, et maintenant que nous avons pu -dans le précédent chapitre- retrouver les résultats intermédiaires (les temps de parcours), nous pouvons aisément via **"inversion"** de la formule du principe de Fermat (voir chapitre 4), remonter aux paramètres internes solaires.

Ce chapitre est résrvé à l'inversion et à tout le travail qui doit être effectué théoriquement afin d'y parvenir.

## 6.1 Solution du problème inverse

Une fois les différences de temps de parcours calculées en fonction de la distance de parcours $\Delta$ et de la latitude $\lambda$, il ne nous reste plus qu'à calculer la vitesse d'écoulement sub surfacique $U$. Ce processus est un peu plus compliqué qu'il en a l'air à première vue, et sa solution exige une bonne maîtrise des problèmes inverses, la terminologie et le formalisme de ce sujet sont brièvement décrits ici.

## 6.2 Équation intégrale

L'équation (4.154) relie la vitesse horizontale du plasma solaire $U$ à la différence mesurée particulière du temps $\delta\tau$ sous les hypothèses de l'approximation d'onde. L'objet est alors déterminé par la fonction $U(r,\lambda)$ qui est le meilleur ajustement des mesures. Une approche naïve serait de discrétiser le problème et puis de trouver le modèle de " la meilleure approximation" $U$ au sens des moindres carrés. Dans ce qui suit, nous décrivons le procédé avec plus de détail, où nous expliquons pourquoi celui-ci doit échouer, et puis nous décrivons l'utilisation de la régularisation afin d'obtenir une solution plus stable et plus raisonnable.

## 6.3 Méthode des moindres carrés régularisée

### 6.3.1 Discrétisation du problème

Pour convertir l'équation (4.154) en équation matricielle qui peut être résolue numériquement, il est d'abord nécessaire de modéliser (discrétiser) la région 3D de la propagation d'onde par une grille. La plus simple approche est de créer une grille équidistante dans le rayon et dans la latitude. Supposons qu'il y a $M_1$ points dans la grille de latitude et $M_2$ points dans la grille radiale où $M = M_1 M_2$. Ces $M$ vitesses modèles peuvent être considérées comme formant le vecteur $U$. Pour une mesure particulière $\delta\tau_i$ où l'indice $i$ dénote les coordonnées $(\lambda, \delta)$ qui déterminent le chemin d'onde. Le rapport entre le modèle et les données peut être exprimé comme suit [41] :

$$\delta\tau_i = \sum_{j=1}^{M} K_{ij} U_j \qquad (6.1)$$

Où le "Kernel" de sensibilité $K_{ij}$, est définie par (voir exemple de calcul des kernels (chapitre (6.5))) :

$$K_{ij} = 2 \int_i \frac{v_{gh}}{v_{gr} c_s^2} dr_j \qquad (6.2)$$

Où l'intégrale se fait le long de l'élément particulier du chemin d'onde déterminé par $i$ se trouvant en dessous de la cellule $j$ de la grille. En d'autres termes la vitesse $U$ est supposée être constante pour chaque cellule de la grille. Le choix de l'espacement (la discrétisation) de la grille est intimement lié à la question de la résolution de la solution qui est discutée dans la section (6.3.4). Cependant la coordonnée radiale mérite une attention spéciale. Le choix le plus simple est de faire une grille qui soit équidistante dans le rayon comme c'est le cas pour la latitude. Cependant des résultats d'inversion ont prouvé que ceci offrait une vue légèrement fallacieuse de la résolution des régions profondes de la zone de convection. Le meilleur choix est de créer une grille qui a un espacement égal dans "l'onde acoustique" défini comme suit :

$$r_\tau = \int_0^r \frac{dr'}{c_s} \qquad (6.3)$$

Dans ce cas-ci les points de grille sont équidistants dans le temps de parcours, supposant une propagation d'ondes radiales et ignorant la dispersion.

Puisque chaque mesure $\delta\tau_i$ est réellement la moyenne d'un grand nombre de corrélations, le kernel n'est pas construit réellement pour un chemin d'onde simple mais pour la moyenne pondérée de tous les chemins d'onde, ceux qui ont contribué à la moyenne. Le poids pour chaque chemin d'onde est proportionnel au nombre de paires de Pixel utilisées (pour celles qui ont exactement la même distance), latitude et direction. Une fois le problème discrétisé, l'équation (6.1) peut être ramenée à la forme matricielle :

$$KU = z \qquad (6.4)$$

Où les $N$ éléments du vecteur $z$ sont $z_i = \delta\tau_i$. Le vecteur $U$ a M composantes de vitesses et chaque rangée de la matrice $K(N \times M)$ est le paramètre de sensibilité avec des composantes données par l'équation (6.2). Si nous définissons deux nouvelles quantités $A$ et $B$ :

$$B_i = z_i/\sigma_i, \qquad i = (1,,N) \qquad (6.5)$$

$$A_{ij} = K_{ij}/\sigma_i, \qquad j = (1,,M) \qquad (6.6)$$

Nous devons trouver la solution $\hat{U}$ qui minimise :

$$\chi^2 = |AU - B|^2 \qquad (6.7)$$

Ceci est équivalent à résoudre un ensemble d'équations linéaires (Press et al (1992) [67]).

$$A^T A U = A^T B \tag{6.8}$$

Dans la pratique cependant ceci ne nous donnera pas une réponse utile de $\hat{u}$ sur toutes les réponses possibles à donner. Le problème ici; est que l'opérateur $K$ est un opérateur intégral, il fera en sorte de lisser la réponse des données au modèle. Ce "lissage externe" de l'information ne peut pas être récupéré en résolvant l'ensemble des équations linéaires (6.8). En outre puisque l'opérateur $K$ est un opérateur lissant, l'opérateur inverse (s'il existe) fonctionnant sur les mesures bruyantes rend la solution instable. Et finalement le problème peut être indéterminé. Nous pouvons essayer de résoudre le problème pour plus de paramètres modèles qu'il n'y a de mesures, mais ceci est strictement impossible mathématiquement parlant, mais cela peut être accompli si connaissance d'à priori il y a, qui peuvent être employées pour choisir la solution appropriée à partir de l'ensemble des solutions possibles données par les mesures.

### 6.3.2 Régularisation

Une approche simple, très utile à ce problème, consiste à ajouter une contrainte additionnelle qui rend la solution choisie "lisse" d'une certaine manière ; par exemple plutôt que de minimiser $\chi^2$ seulement, nous pouvons essayer de trouver la solution $\hat{U}$ qui minimise la quantité [41] :

$$|AU - B|^2 + \gamma U^T H U \tag{6.9}$$

Où le deuxième terme représente une sorte d'opération de lissage. Par exemple quatre formes sont employées pour la matrice $H$ :

$$U^T H_0 U \Rightarrow \int\int \|u\|^2 r dr d\theta \tag{6.10}$$

$$U^T H_1 U \Rightarrow \int\int \|\nabla u\|^2 r dr d\theta \tag{6.11}$$

$$U^T H_{U/r} U \Rightarrow \int\int \|\frac{\nabla u}{r}\|^2 r dr d\theta \tag{6.12}$$

$$U^T H_\Gamma U \Rightarrow \int\int \|\frac{\nabla u}{r\cos\lambda}\|^2 r dr d\theta \tag{6.13}$$

Où les opérations de la matrice du côté gauche sont des approximations discrètes des intégrales du côté droit. La matrice choisie H s'appelle l'opérateur de régularisation, et l'ajout de cet opérateur au problème contraint la solution à être lisse dans un certain sens. Minimiser la quantité (6.9) équivaut à résoudre l'ensemble des équations linéaires :

$$(A^T A + \gamma H)U = A^T B \tag{6.14}$$

La solution $U = \hat{U}$ dépend évidemment non seulement des données $z$, des erreurs $\sigma$ et du modèle $K$ mais également de l'arrangement de la régularisation choisi (choix de H). L'influence relatif de la régularisation est commandée par le paramètre libre $\gamma$ dans l'équation (6.14) qui s'appelle le paramètre de régularisation.

### 6.3.3 Paramètre de régularisation

Quand γ est grand l'influence de *H* est grande comparée à l'influence du modèle A et la solution obtenue, bien que très lisse, ne sera probablement pas un bon ajustement aux données. D'autre part, quand γ est petit, l'influence du terme de régularisation est relativement petit aussi. En général, les solutions modèles $\hat{U}$ sont habituellement calculées pour une gamme de différentes valeurs de γ et le "meilleur" modèle est alors choisi parmi l'ensemble des solutions. Aux deux extrémités, il est facile d'éliminer les solutions obtenues, mais la détermination de la "meilleure" solution est souvent une question de grand savoir faire.

L'un des moyens existant, nous permettant de voir l'effet du paramètre de régularisation afin de déterminer le meilleur γ, est de tracer les magnitudes des deux paramètres de l'équation (6.9) ($\chi^2$ et $U^THU$), voir la Fig.(6.1) qui est connue sous le nom de "courbe L". Le paramètre de régularisation croit de droite à gauche avec l'accroissement du lissage de la solution. Le point d'inflexion de la courbe L est la solution optimale pour ce choix particulier de régularisation ; c'est la solution la plus lisse possible avec une petite valeur de $\chi^2$. Si les erreurs σ dans les mesures sont fiables, alors on peut espérer que le point d'inflexion de la courbe L soit localisé à coté de la solution $\chi^2 = N$ le nombre de mesures.

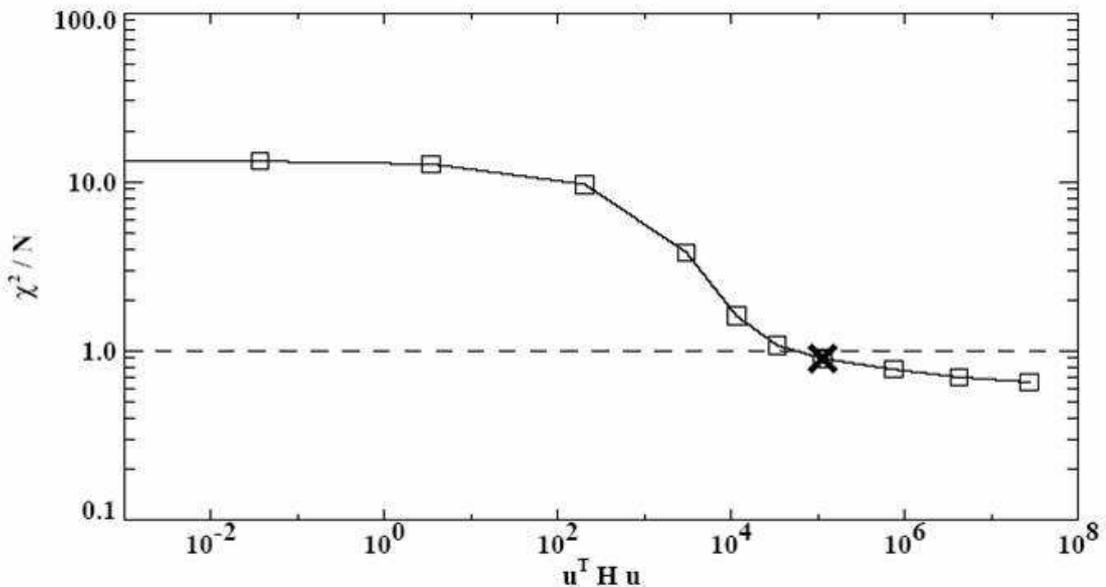

FIG. 6.1 – Cette courbe L montrant la relation entre la "douceur" et la "qualité de l'ajustement" des solutions d'une inversion pour des vitesses de rotation (courbe obtenue par Giles 1999 [29]). La ligne discontinue marque le point où $\chi^2$ (voir l'équa.(6.7)) est égal au nombre de mesures. L'axe horizontal représente la déviation de la douceur parfaite avec $U^THU$ étant mathématiquement décrit par l'équation (6.13). Le point identifié par × marque le point d'inflexion de la courbe ; la solution optimale qui désigne la meilleure valeur de γ à prendre pour l'inversion.

### 6.3.4 Moyenne spatiale des Kernels

Une autre manière d'étudier l'influence de γ est d'examiner l'effet entre la stabilité et la résolution de la solution d'inversion. Pour cela définissons la matrice *D* comme suit :

$$D = A^TA + \gamma H \qquad (6.15)$$

Nous pouvons exprimer la solution $\hat{U}$ de l'équation (6.14) comme :

$$\hat{U} = D^{-1}(A^T B) \tag{6.16}$$

Remplaçons $B$ par l'équation (6.8) :

$$\hat{U} = D^{-1} A^T A U \tag{6.17}$$

L'équation (6.17) prouve que chaque paramètre de la solution modèle $\hat{U}$ peut être exprimé comme combinaison des paramètres de la "vraie" vitesse $U$. Pour cette raison, chaque rangée $M \times M$ de la matrice $D^{-1} A^T A$ s'appelle un kernel moyenné de la solution. Pour le problème particulier, qu'on étudie en exemple ici, est que chaque élément du vecteur $U$ correspond à une latitude et un rayon particulier ; ainsi le kernel donne une représentation de la résolution spatiale de la solution.

### 6.3.5 Erreurs d'analyse

Comme dans le cas de toutes les techniques usuelles des moindres carrés les incertitudes dans les paramètres de modèle sont déterminées par les caractéristiques des kernels $K$. Dans les deux cas des formats standard et régularisé, l'information d'erreur est contenue dans la matrice de covariance de la solution. Cependant, dans le cas des moindres carrés régularisés, la formulation est légèrement compliquée. La variance dans la vitesse modèle $\hat{U}_j$ est définie par :

$$\sigma^2(\hat{U}_j) = \sum_{i=1}^{N} \sigma_i^2 \left( \frac{\partial \hat{U}_j}{\partial Z_i} \right)^2 \tag{6.18}$$

Où $\sigma_i^2$ est la variance de la mesure $Z_i$. En écrivant l'équation (6.16) sous forme indicielle on a :

$$\hat{U}_j = \sum_{k=1}^{M} [d_j k]^{-1} \sum_{i=1}^{N} \frac{Z_i K_{ki}}{\sigma_i^2} \tag{6.19}$$

De sorte que :

$$\frac{\partial \hat{U}_j}{\partial Z_l} = \sum_{k=1}^{M} [d_{jk}]^{-1} \frac{K_{kl}}{\sigma_l^2} \tag{6.20}$$

On a représenté les éléments $D^{-1}$ symboliquement comme $[d_{ij}]^{-1}$. L'insertion de ce résultat dans la définition (6.18) donne :

$$\sigma^2(U_j) = \sum_{k=1}^{M} [d_j k]^{-1} \sum_{k=1}^{M} [d_j k]^{-1} [A^T A]_{kl} \tag{6.21}$$

Ou sous forme matricielle :

$$\sigma^2(U_j) = [D^{-1} A^T A (D^{-1})]_{jj} \tag{6.22}$$

On note que dans le cas sans la régularisation ($\gamma = 0$), la variance se réduit à $[D^{-1}]_{jj}$ comme prévu. L'équation (6.12) prouve que la propagation d'erreur - la manière dont l'incertitude dans les mesures est transformée en incertitude dans les vitesses dépend non seulement du modèle mais également du choix du paramètre de régularisation $\gamma$ et de l'opérateur de régularisation $H$ de l'équation (6.15).

En fait, une autre manière de visualiser l'effet du paramètre de régularisation $\gamma$ est d'imaginer l'équilibre entre la propagation d'erreur et la résolution avec une petite régularisation. Les kernels moyennés sont bien localisés mais la solution est très sensible aux erreurs dans les mesures ; quand $\gamma$ est grand, la solution est stable en ce qui concerne les erreurs, mais la résolution est très pauvre.

## 6.4 Contraintes additionnelles

La section précédente décrit comment la régularisation est employée pour choisir la "bonne" solution à partir d'une infinité de modèles qui pourraient être compatibles avec un ensemble de mesures bruyantes. Cependant l'utilisation d'un tel mécanisme de choix n'est pas vraiment basée sur la physique mais sur quelques idées a priori au sujet de ce à quoi devrait ressembler le champ de vitesse. En plus d'un certain type de douceur il serait également souhaitable de choisir une solution modèle où le champ de vitesse $\vec{U}$ satisfait l'équation de continuité, et ceci ce traduit par :

$$\vec{\nabla}.(\rho(r)\vec{U})) = 0 \tag{6.23}$$

Malheureusement et comme vu dans la section (4.4) la composante radiale de la circulation méridionale est indétectable dans la technique de Temps-Distance de sorte qu'il ne soit pas possible de contraindre le champ d'écoulement en appliquant l'équation (6.23) à chaque point de grille. Cependant il n'y a qu'un cas important où la conservation de la masse peut être appliquée pour trouver une solution. Si on permet au modèle de se prolonger jusqu'au fond de la zone de convection, et si la circulation méridionale (voir Fig.(4.2)) est supposée non pénétrante à l'intérieur de la zone radiative. Puis l'énergie totale du fluide coulant du nord doit être égale à l'énergie totale du fluide montant du sud. En fait, il est possible d'énoncer la contrainte en termes de forces : où pour chaque latitude la quantité nette de croisement de masse doit être nulle. La méthode employée dans ce travail pour imposer cette condition est une "méthode de barrière". L'état de la conservation de masse est sous forme de matrice $C$ qui satisfait [41] :

$$U^T C U = \sum_{i=1}^{M_1} \Big( \sum_{j=1}^{M_2} \rho_j U_{ji} dr_j \Big)^2 = 0 \tag{6.24}$$

Ici le vecteur de vitesse $U$ est traité symboliquement comme matrice $\widetilde{U}$ ($M_1 \times M_2$) à l'intérieur de la somme ; les indices $i$ et $j$ indiquent respectivement le rayon et la latitude. Une fois que la matrice $C$ a été formée, la condition $U^T C U = 0$ est satisfaite dans un sens approximatif en résolvant une version modifiée de l'équation (6.14) :

$$(A^T A + \gamma H + \beta C) U = A^T B \tag{6.25}$$

Ainsi la condition que la masse soit conservée est traitée comme état supplémentaire de la régularisation ; l'advection de la continuité parfaite peut être rendue arbitrairement petite en rendant $\beta$ arbitrairement grand.

La solution des problèmes inverses est un art et une science en soi ; les méthodes décrites ici ont été employées comme un point de départ. Les méthodes d'inversion et les approximations utilisées dans le modèle seront considérablement améliorées au fur et à mesure que l'héliosismologie temps-distance se développera. Ainsi peuvent être obtenues les vitesses des écoulements en utilisant cette approche relativement simple.

### 6.4.1 Résultats qu'on peut obtenir

On peut bien sûr, et comme vu précédemment, obtenir les cartes des vitesses des écoulements $\vec{U}$ de la surface solaire étudiée (voir fig.(6.2)), comme on peut toujours obtenir celles des vitesses 3D du fond solaire (voir fig.(6.3)) en incluant la partie radiale des temps $\tau_{diff}$ (voir l'équation (4.151)).

On peut aussi, obtenir comme vu théoriquement (chapitre 4), les vitesses du son, en plus des vitesses d'Alfven $c_a$ en incluant le champ magnétique (voir l'équation(4.150)), à partir des temps $\tau_{mean}$.

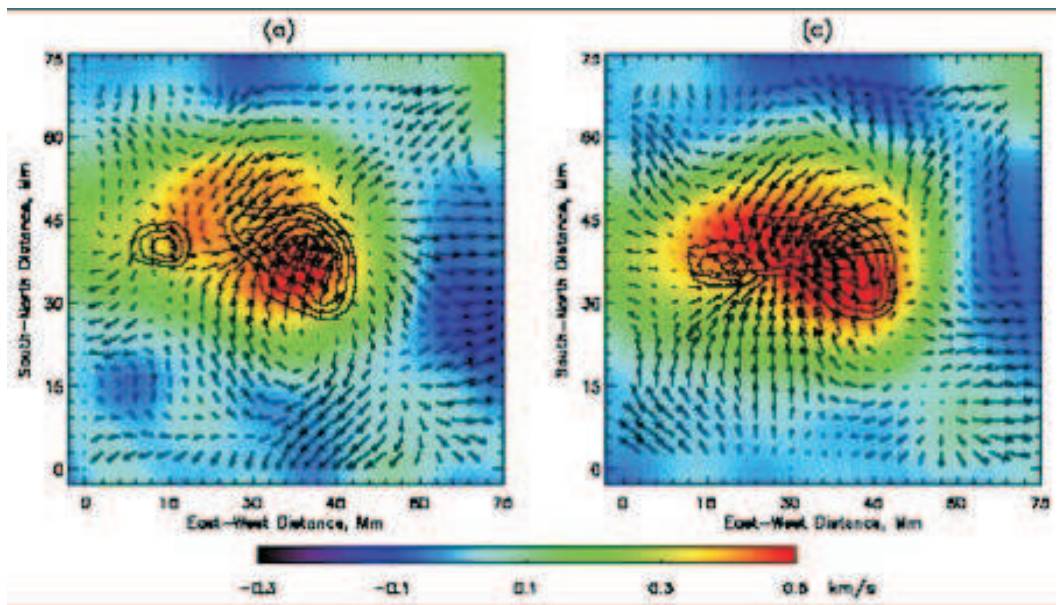

FIG. 6.2 – Figures montrant les variations de la vitesse d'écoulement 2D en surface (à droite) et à une surface un peu plus profonde (à gauche) [54].

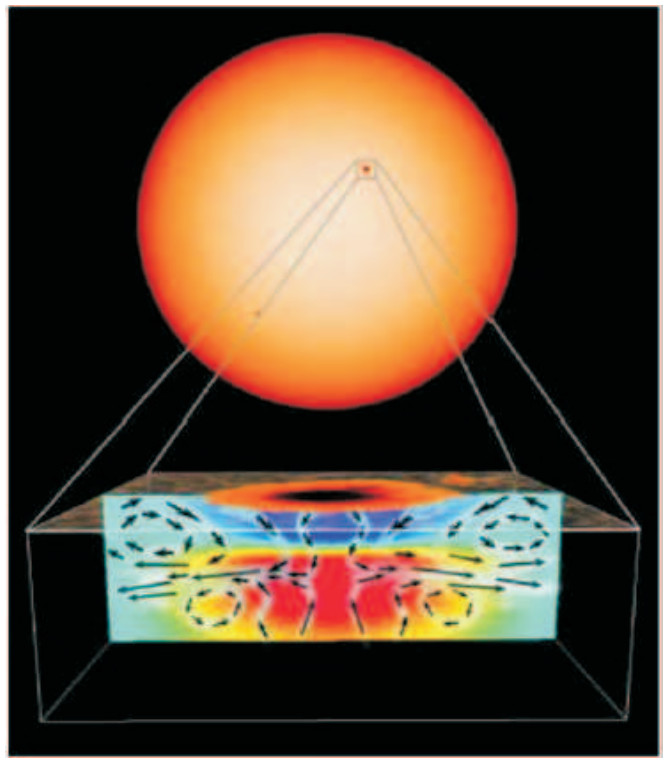

FIG. 6.3 – Figure montrant les variations de la vitesse d'écoulement sub surfacique autour d'une tache solaire (SOHO/MDI) [54].

## 6.5 Calcul des Kernels

L'interprétation des données Temps-Distance peut être divisée en un problème direct et un problème inverse. Le problème direct consiste à déterminer le rapport entre les données d'observation (perturbations des temps de parcours $\delta\tau$) et les propriétés internes solaires (notées ici $q_\alpha$). Généralement, cette relation est cherchée sous la forme d'une équation intégrale linéaire [4], [44] :

$$\delta\tau = \sum_\alpha \int_\odot d\vec{r}\, \delta q_\alpha(\vec{r}) K^\alpha(\vec{r}) \quad (6.26)$$

Où $\delta q_\alpha(\vec{r})$ représente la déviation dans les propriétés internes solaires d'un modèle théorique de référence. L'inidce $\alpha$ se rapporte à tous les types de perturbations indépendantes, tels que la vitesse du son, des écoulements, ou le champ magnétique. L'intégrale $\int_\odot dr$ désigne l'intégration spatiale sur le volume solaire. Les kernels des intégrales, $K^\alpha(\vec{r})$, donnent la sensibilité du parcours par rapport aux perturbations du modèle solaire. Le problème inverse, quant à lui, consiste a inverser l'équation ci-dessus, i.e. à estimer $\delta q_\alpha$, en fonction de la position $\vec{r}$, des $\delta t$ observés.

### 6.5.1 Modèle d'une seule source pour des sensibilités de temps de parcours

Dans cette section, on montrera le calcul du temps de parcours des kernels dans l'approximation d'une source simple. Dans cette approximation, le signal de vitesse d'une source simple est pris comme procuration pour la fonction de corrélation entre la "source" et le "récepteur". Le point de "source" (noté par le symbole $\vec{x_s}$) est l'endroit sur la surface solaire où la source est localisée et le point "récepteur" (noté par le symbole $\vec{x_r}$) est l'endroit d'où le signal source est observé. Dans cette section le signal est supposé être une vitesse radiale, qui est une bonne approximation près du centre du disque. Dans l'approximation de la source simple, la perturbation du temps de parcours est mesurée à partir du signal de vitesse. L'un des choix possibles pour la définition de la perturbation du temps de parcours est [4] :

$$\delta\tau(\vec{x_s}, \vec{x_r}) = argmax_\tau \left\{ \int_{t_0}^{t_1} dt \quad v_0(\vec{x_r}, t-\tau) v(\vec{x_r}, t) \right\} \quad (6.27)$$

La fonction $argmax_\tau\{f(\tau)\}$ renvoie la valeur de $\tau$ qui maximise la fonction $f(\tau)$. Ici $v_0(\vec{x_r},t)$ est la vitesse au récepteur dans le plus bas ordre du modèle solaire, et $v(\vec{x_r},t)$ est la vitesse perturbée à $\vec{x_r}$. La source d'onde est située à $\vec{x_s}$. Dans l'équation (6.27), $d\tau(\vec{x_s},\vec{x_r})$ dépend de $\vec{x_s}$ car $v(\vec{x_r},t)$ et $v_0(\vec{x_r},t)$ sont des signaux de vitesse dûs à la source à $\vec{x_s}$. L'intervalle du temps $[t_0,t_1]$ devrait être choisi afin d'isoler la partie particulière de la vitesse du signal d'intérêt, par exemple la partie du premier rebond. De l'équation (6.27), nous voyons que la perturbation du temps de parcours est le décalage de temps entre les signaux perturbés et non perturbés au récepteur, $\vec{x_s}$. Ces définitions du temps de parcours, de "source" et de "récepteur" sont standards dans la littérature de géophysique (par exemple Zhao et Jordan, 1998).

### 6.5.2 Dérivation des sensibilités des temps de parcours

Le but de cette section est de calculer comme exemple les kernels du temps de parcours $K(\vec{x_s}, \vec{x_r}; \vec{r})$ qui relient une perturbation locale, $\delta c_s(\vec{r})$, de la vitesse du son du modèle solaire, $c_s(\vec{r})$, à une perturbation du temps de parcours moyenné, notée $\delta\tau_{mean}(\vec{x_s}, \vec{x_r})$, entre les deux localisations $\vec{x_s}$ et $\vec{x_r}$ sur la surface solaire. Les kernels devraient satisfaire alors :

$$\delta\tau_{mean}(\vec{x_s},\vec{x_r}) = \int_{\odot} dr K(\vec{x_s},\vec{x_r};\vec{r})\frac{\delta c_s^2(\vec{r})}{c_s^2(r)} \qquad (6.28)$$

L'intégrale $\int_{\odot} dr$ signifie l'intégration sur le volume entier du soleil. La première étape dans la dérivation des kernels est de calculer les vitesses radiales $v_0$ et $v$ dans les modèles solaires perturbés et non perturbés, respectivement. Une relation linéaire entre la perturbation des temps de parcours et la perturbation au modèle solaire, sous la forme de L'équation (6.28), peut être obtenue par linéarisation de l'équation (6.27). Le reste de cette section accomplit cette tâche.

### 6.5.3 Équations gouvernantes

L'équation type du mouvement (par exemple Gough, 1987 [46]) dans l'approximation de Cowling, pour le déplacement du champ fluide $\xi$ dû à la source $S$, est (voir l'équation(3.2)) :

$$(\rho\delta_t^2 + L)\xi = S \qquad (6.29)$$

Où $\rho$ est la densité et le $L$ est la partie spatiale de l'opérateur d'équation d'ondes. Ici la fonction représente une source monopole impulsive ; ce que nous prenons comme :

$$S = A\vec{\nabla}\delta_D(\vec{x}-\vec{x_s})\delta_D(t-t_s) \qquad (6.30)$$

Où $\vec{x_s}$ est la localisation de la source, $t_s$ est la période de l'impulsion, $A$ l'amplitude de la source, $\delta_D$ est la fonction delta de Dirac, et $\vec{\nabla}$ est le gradient selon $\vec{x}$.

Afin d'appliquer l'approximation, nous devons diviser le problème en une partie non perturbée (indicée par 0) et une autre partie corrigée de premier ordre (précédée par un $\delta$). En général, nous acceptons que la perturbation au temps de parcours dépende des perturbations de la densité et de la vitesse de son, et comme dans le cas de l'héliosismologie globale, les deux kernels sont nécessaires afin d'effectuer des inversions. Dans ce travail, afin de montrer l'approche générale, nous calculons seulement la sensibilité des temps de parcours des perturbations de la vitesse du son à une densité fixe. Le problème peut être alors écrit ainsi :

$$(\rho\delta_t^2 + L_0 + \delta L)(\xi_0 + \delta\xi) = A\vec{\nabla}\delta_D(x-\vec{x_s})\delta_D(t-t_s) \qquad (6.31)$$

Les opérateurs non perturbés et perturbés $L^0$ et $\delta L$ sont :

$$L_0\xi = \vec{\nabla}\left[\delta c^2\rho\vec{\nabla}.\vec{\xi} - \xi_r\frac{dP_0}{dr}\right] - \vec{g_0}\vec{\nabla}.(\rho_0\vec{\xi}) \qquad (6.32)$$

$$\delta L\xi = \vec{\nabla}[\delta c^2\rho\vec{\nabla}.\vec{\xi}] \qquad (6.33)$$

Où $c_0$ est la vitesse du son adiabatique, $\vec{g_0}$ est l'accélération de la gravité (qui est une fonction de la profondeur), $P_0$ est la pression de gaz, et $\delta c$ la perturbation de la vitesse du son, qui est en général non sphérique. La composante radiale de $\vec{\xi}$ est notée par $\xi_r$.

Le plus bas ordre du problème est :

$$(\rho\delta_t^2 + L_0)\vec{\xi_0} = A\vec{\nabla}\delta_D(x-x_0)\delta_D(t-t_s) \qquad (6.34)$$

Et le problème au premier ordre est :

$$(\rho\delta_t^2 + L_0)\delta\vec{\xi} = -L\vec{\xi_0} \qquad (6.35)$$

### 6.5.4 Solution du problème au plus bas ordre

Pour le cas sphériquement symétrique, les fonctions propres et les valeurs propres de l'opérateur $(\rho\delta_t^2 + L_0)$ sont bien connues (par exemple Gough, 1987 [46]). Nous employons la convention standard en notant l'ordre radial par $n$, le degré angulaire par $\ell$ et l'ordre azimutal par $m$. Pour la normalisation des fonctions propres, nous choisissons :

$$\int_\odot d\vec{r}\, \rho_0(r)\vec{\xi}^{n\ell m}(\vec{r}) \cdot \vec{\xi}^{n\ell m}(\vec{r}) = 1 \tag{6.36}$$

Pour la convenance de la notation, suivant l'approche de Dahlen & Tromp (1998) [23], nous représentons des fonctions propres non perturbées en termes de vecteurs réels d'harmoniques sphériques $\vec{P}_{\ell m}$, et $\vec{B}_{\ell m}$

$$\vec{\xi}^{n\ell m}(r,\theta,\phi) = \vec{\xi}_r^{n\ell}(r)\vec{P}_{\ell m}(\theta,\phi) + \vec{\xi}_h^{n\ell}(r)\vec{B}_{\ell m}(\theta,\phi) \tag{6.37}$$

Où $\xi_r^{n\ell}$ et $\xi_h^{n\ell}$ sont les composantes radiales et horizontales du vecteur déplacement. Les vecteurs des harmoniques sphériques peuvent être écrits en termes d'harmoniques sphériques réels (scalaires), $Y_{\ell m}(\theta>,\phi)$ :

$$\vec{P}_{\ell m}(\theta,\phi) = \hat{r}Y_{\ell m}(\theta,\phi) \tag{6.38}$$

$$\vec{B}_{\ell m}(\theta,\phi) = -\frac{1}{\sqrt{\ell(\ell+1)}}\vec{\nabla}_h Y_{\ell m}(\theta,\phi) \tag{6.39}$$

Où $\nabla_h$ est le gradient horizontal sur la sphère d'unité. La solution du problème d'ordre zéro est alors obtenue comme la somme des fonctions propres. Pour la composante radiale de la vitesse d'oscillation, à $x_l$ au temps $t$ dûe à une source au temps $t_s = 0$ et à $x_0$, nous obtenons alors :

$$v_0(\Delta,t) = -A\sum_{n=0}^{\infty}\sum_{\ell=0}^{\infty}\frac{2\ell+1}{4\pi}D^{n\ell}(R_\odot)\xi_r^{n\ell}(R_\odot)P_\ell(\cos\Delta)Hea(t)\cos(\omega_{n\ell}t) \tag{6.40}$$

Où nous avons supposé que les points de source et d'observation $x_s$ et $x_r$ sont sur la surface solaire, que le déplacement du champ est nul à $t=0$ et que la surface solaire est libre. La distance circulaire entre $x_s$ et $x_r$ est notée par $\Delta$. Les $P_\ell$ sont les polynômes de Legendre, $Hea$ la fonction Heaviside et que :

$$D^{n\ell} = \frac{d\xi_r^{n\ell}}{dr} + \frac{1}{r}[2\xi_r^{n\ell} - \sqrt{\ell(\ell+1)}\xi_h^{n\ell}] \tag{6.41}$$

Qui découle de la divergence des fonctions propres. Notons que le signal de la vitesse d'ordre zéro dépend seulement de la grande distance circulaire entre $x_s$ et $x_r$ et du temps $t$.

### 6.5.5 Première approximation

Rappelons que le problème au premier ordre est :

$$(\rho\delta_t^2 + L)\delta\vec{\xi} = -\delta L\vec{\xi_0} \tag{6.42}$$

Avec $\vec{\xi_0}$ la solution du problème non perturbé. L'opérateur du côté gauche de l'équation est le même que pour le problème d'ordre zéro et nous pouvons résoudre ainsi le premier ordre du problème exactement de la même façon que nous avons fait pour problème non perturbé.

La solution au premier ordre pour la vitesse radiale est alors :

$$v(\vec{x_s}, \vec{x_r}, t) = A \int_\odot d\vec{r} \rho_0(r) \delta c^2(\vec{r}) Hea(t)$$
$$\times \sum_{n,n'=0}^{\infty} \sum_{n,n'=0}^{\infty} S_{n\ell}(\vec{x_s}, \vec{r}) R_{n'\ell'}(\vec{x_r}, \vec{r}) \frac{\cos(\omega_{n\ell}t) - \cos(\omega_{n'\ell'}t)}{\omega_{n'\ell'}^2 - \omega_{n\ell}^2} \qquad (6.43)$$

Où :

$$S_{n\ell}(\vec{x_s}, \vec{r}) = \frac{2\ell+1}{4\pi} D^{n\ell}(R_\odot) D^{n\ell}(r) P_\ell(\cos\Delta_1) \qquad (6.44)$$

$$R_{n\ell}(\vec{x_r}, \vec{r}) = \frac{2\ell+1}{4\pi} \xi_r^{n\ell}(R_\odot) D^{n\ell}(r) P_\ell(\cos\Delta_2) \qquad (6.45)$$

La grande distance circulaire entre $\vec{x_s}$ ($\vec{x_r}$) et $\vec{r}$ est désignée par $\Delta_1$ ($\Delta_2$).

### 6.5.6 Les Kernels

Maintenant que nous avons la solution du cas non perturbé, la solution de la première approximation que nous avons obtenue $v_0$ et $v_0 + \delta v$ respectivement, nous pouvons calculer le temps de retard entre eux. La linéarisation de la dépendance de $\delta\tau$ sur $\delta v$ dans l'équation (6.27), avec la supposition que la fenêtre $[t_0, t_l]$ de temps isole un rebond particulier dans la corrélation, donne :

$$\delta\tau = \frac{1}{N(\Delta)} \int_{t_0}^{t_1} dt v(t) \delta \dot{v}_0(t) \delta \qquad (6.46)$$

Avec :

$$N(\Delta) = \int_{t_0}^{t_1} dt \ddot{v}_0(\Delta, t) v_0(\Delta, t) \qquad (6.47)$$

En remplaçant l'équation(6.43) dans l'équation(6.46) nous obtenons :

$$\delta\tau = \frac{A}{N(\Delta)} \int_\odot d\vec{r} \rho_0(r) \delta c_s^2(\vec{r}) \frac{\delta c_s^2(\vec{r})}{c_s^2(r)}$$
$$\times \sum_{n,n'=0}^{\infty} \sum_{n,n'=0}^{\infty} S_{n\ell}(\vec{x_s}, \vec{r}) R_{n'\ell'}(\vec{x_r}, \vec{r}) \int_{t_0}^{t_1} dt \dot{v}_0(\Delta, t) \frac{\cos(\omega_{n\ell}t) - \cos(\omega_{n'\ell'}t)}{\omega_{n'\ell'}^2 - \omega_{n\ell}^2} \qquad (6.48)$$

Il est commode d'obtenir des kernels, pour la perturbation du carré de la vitesse du son, qui satisfont l'équation (6.28). De l'équation (6.48) nous pouvons voir celà :

$$\vec{K}(\vec{x_s}, \vec{x_r}, \vec{r}) = \frac{A}{N(\Delta)} \int_\odot d\vec{r} \rho_0(r) \delta c_s^2(\vec{r})$$
$$\times \sum_{n,n'=0}^{\infty} \sum_{n,n'=0}^{\infty} S_{n\ell}(\vec{x_s}, \vec{r}) R_{n'\ell'}(\vec{x_r}, \vec{r}) \int_{t_0}^{t_1} dt \dot{v}_0(\Delta, t) \frac{\cos(\omega_{n\ell}t) - \cos(\omega_{n'\ell'}t)}{\omega_{n'\ell'}^2 - \omega_{n\ell}^2} \qquad (6.49)$$

Cette équation peut être simplifiée en introduisant les indices $p = (n, \ell)$ et $q = (n', \ell')$ et en définissant la matrice :

$$G(\vec{x_s}, \vec{x_r}) = \frac{A}{N(\Delta)} \int_{t_0}^{t_1} dt \dot{v}_0(\Delta, t) \frac{\cos(\omega_p t) - \cos(\omega_q t)}{\omega_q^2 - \omega_p^2} \qquad (6.50)$$

Notons que $N \propto A^2$ et $v_0 \propto A$ de sorte que la matrice $G$ ne dépende pas de l'amplitude de la source $A$. Avec les définitions ci-dessus le kernel peut être alors écrit comme suit :

$$\vec{K}(\vec{x_s}, \vec{x_r}, \vec{r}) = -\rho_0(r) c_0^2(r) \vec{S}^T(\vec{x_s}, \vec{r}) \vec{G}(\vec{x_s}, \vec{x_r}) \vec{R}(\vec{x_r}, \vec{r}) \qquad (6.51)$$

Où l'indice supérieur $T$ désigne la transposée.

Notons que les kernels $K(\vec{x_s}, \vec{x_r}; \vec{r})$ ne sont cependant pas symétriques à l'échange de la source et du récepteur ; $\vec{x_s}$ et $\vec{x_r}$. L'asymétrie est un résultat du modèle d'une source simple, dans laquelle les modes ont une grande divergence à la source et sont pour la plupart excités et les modes qui ont les plus grandes vitesses radiales sur la surface contribuent plus fortement au signal observé. En réalité les temps de parcours sont déterminés à partir de la fonction de covariance. Le rapport exact entre le modèle d'une simple source et le modèle physique de la source distribuée n'est pas encore établi. Pour des discussions du problème voir (Rickett & Claerbout [68], 2000 ; Kosovichev et al, 2000 [58]). Par conséquent, afin d'obtenir la symétrie désirée, nous remplaçons simplement la divergence et le déplacement vertical sur la surface solaire qui apparaissent dans les vecteurs $\vec{s}$ et $\vec{r}$ par la racine carrée d'une certaine fonction $F^{n\ell}$. La racine carrée est employée de sorte que la vitesse à l'ordre zéro soit :

$$v_0(\Delta, t) = -A \sum_{n=0}^{\infty} \sum_{\ell=0}^{\infty} \frac{2\ell+1}{4\pi} F^{n\ell}(R_\odot) P_\ell(\cos\Delta) Hea(t) \cos(\omega_{n\ell} t) \tag{6.52}$$

La fonction $F^{n\ell}$ est l'amplitude avec laquelle chaque mode contribue au signal de vitesse à l'ordre zéro et est ainsi reliée au filtrage qui en cours d'application aux données de fabrication des mesures temps-distance (Kosovichev & Duvall, 1997). Dans de l'équation pour $v_0$, la fonction de Heaviside étant éliminée, qui est davantage une justification pour la substitution de $\sqrt{f^{n\ell}}$ pour la divergence de la vitesse radiale à la surface de soleil.

Nous remplaçons les vecteurs $\vec{S}$ et $\vec{R}$ avec un vecteur $\vec{H}$, qui est défini ainsi :

$$\vec{H}_{n\ell}(\vec{r'}, \vec{r}) = \frac{2\ell+1}{4\pi} \sqrt{F^{n\ell}} D^{n\ell}(r) P_\ell(\cos\Delta') \tag{6.53}$$

Où $\Delta'$ est la grande distance circulaire entre $\vec{r'}$ et $\vec{r}$. En termes de $\vec{H}$ les kernels de sensibilité du temps de parcours dans la première approximation sont :

$$\vec{K}(\vec{x_s}, \vec{x_r}, \vec{r}) = -\rho_0(r) c_0^2(r) \vec{H}^T(\vec{x_s}, \vec{r}) \vec{G}(\vec{x_s}, \vec{x_r}) \vec{H}(\vec{x_r}, \vec{r}) \tag{6.54}$$

Nous avons maintenant des kernels de sensibilité des temps de parcours, pour la vitesse du son, dans l'approximation d'une simple source. Ce résultat a été obtenu après une franche application de l'approximation et du à la linéarisation de la définition du temps de parcours. L'argument physique essentiel était que la vitesse d'une source simple se comporte comme la fonction de corrélation de temps-distance.

# Chapitre 7

# Discussion et perspectives

Dans cette thèse, nous avons étudié les temps de parcours d'ondes parcourant un milieu solaire pour deux différents cycles d'activité. Pour réaliser ce travail, nous avons tout d'abord étudié les principes et théories de la technique héliosismique locale "Temps-Distance", tout en incluant une description physique actualisée ; technique que nous avons essayée par la suite d'appliquer à des données héliosismiques traitées ; images Doppler "Dopplerograms" du réseau GONG, et ce dans la perspective de remonter via "inversion" aux propriétés internes du milieu traversé. Ces dernières nous apprennent beaucoup sur la dynamique et la structure du milieu ; nous permettant ainsi d'en apprendre plus sur les phénomènes locaux et leurs processus, tels que : les taches solaires, les écoulements sub-surfaciques à différentes latitudes et les zones d'éjections de masses, auxquels l'héliosismologie globale ne peut répondre de manière détaillée. Dans ce qui suit, nous récapitulons, sous forme de conclusions, les résultats majeurs obtenus au chapitre 5, tout en les comparant à la théorie étudiée précédemment (chapitres 3 et 4). Nous terminons enfin par énoncer les perspectives et présenter les autres applications possibles de la technique héliosismique étudiée.

## 7.1 Conclusions

De notre étude des temps de parcours des ondes héliosismiques, il ressort plusieurs constatations, notamment :

– En abordant l'aspect technique de la méthode héliosismique locale dans le chapitre 5, on a utilisé des images Doppler de deux journées différentes correspondant à des périodes différentes de l'activité solaire (voir Fig.(5.5) ). Les images étant sous forme de cube de données "data cube", elles nous ont permis :

1. De tirer et d'observer "les fameuses" oscillations de 5 min (voir Fig.(5.7)), où l'on remarque d'ailleurs que l'amplitude de la transformée de Fourier du signal de la journée active est plus importante de celle qui l'est moins.

2. Une fois traitées ("remapping", "tracking"), d'aboutir via une FFT 3D aux "ring diagrams" ainsi qu'aux diagrammes "$k - \nu$" (voir Fig.(5.14-5.18)), qui illustrent assez bien les modes d'oscillations observés du Soleil.

Ces résultats étant satisfaisants comparés à ceux obtenus par d'autres chercheurs, ils nous permettent d'attester la fiabilité des deux séries d'images utilisées.

– Une fois nos deux cubes de données traités et filtrés, nous avons appliqué sur ces derniers la théorie précédemment étudiée (chapitre 4), où nous avons corrélé différentes mesures de points à la surface solaire, pour différentes distances angulaires et pour différentes latitudes, dans un sens temporel puis dans l'autre, reliant ainsi les temps de parcours des ondes voyageant dans le sens du courant et à contre courant aux distances parcourues, ainsi qu'on peut le constater sur "les diagrammes temps-distance" (Fig.(5.26-5.28)). Ces derniers y ont la même allure i.e. pour les deux cycles d'activité, cela indique que quelleque soit la période d'activité solaire, le type de propagation des ondes reste le même. Sans oublier que, plus nous corrélerons de données, meilleurs seront les résultats obtenus.

– Les résultats corrélés obtenus, et comme souligné dans le chapitre 4, peuvent être approximés par fonction d'onde Gaussienne non linéaire à cinq paramètres (fonction Gabor). L'un des plus importants paramètres de cette dernière n'est autre que le temps de phase "temps réel de parcours des ondes", qui nous permet d'aboutir à la différence et moyenne des temps de parcours des ondes voyageant contre et dans le sens du courant $\delta\tau$ et $\tau_{mean}$. Ces derniers renferment de précieux renseignements sur l'intérieur du Soleil, où le premier ($\delta\tau$) nous renseigne sur la dynamique solaire via la vitesse des écoulements du milieu parcouru et le second ($\tau_{mean}$) nous informe sur la structure interne du Soleil via la vitesse du son dans le milieu traversé. En comparaison avec les résultats obtenus par d'autres chercheurs, nos résultats sont jugés satisfaisants, et ce pour les deux journées. La moyenne des $\delta\tau$ des latitudes traitées étant presque nulle démontre que l'hypothèse de la conservation de masse est respectée. Des $\tau_{mean}$ (voir Fig.(5.39)) nous pouvons avoir une idée sur la rotation du soleil à différentes latitudes. De plus, nous avons pu déduire que, pour un même nombre de données traitées, et pour un même taux de données exploitables, le $mean\{\tau_{mean}\}$ d'une période active doit toujours être inférieur à celui de la période moins active (voir section (5.10)).

Pour de plus fines conclusions sur les régions traitées, et afin de remonter aux propriétés dynamiques et physiques du milieu, on n'a d'autre choix, que d'inverser nos résultats de $\delta\tau$ et $\tau_{mean}$ obtenus. Le chapitre 6 illustre brièvement de quelle manière y parvenir dans le cas des propriétés dynamiques, via $\delta\tau$ par la méthode des moindres carrés et en se basant principalement sur des à priori. Cette étude a été faite théoriquement, car faute de temps, l'inversion des résultats à elle seule peut faire l'objet d'un sujet de doctorat.

## 7.2 Perspectives

La technique héliosisimique "Temps-Distance", comme on peut le constater, fait appel à plusieurs disciplines et spécialités telles que les mathématiques, l'informatique, l'astronomie et la physique. Ses applications à l'héliosismologie, ainsi que les outils nécessaires à leur réalisation sont, par conséquent, variés et multiples :

– Les principales théories de l'héliosismologie "temps-distance" traitées dans le chapitre 4 se basent surtout, et comme on a pu le constater, sur certaines approximations et hypothèses, sans les quelles on se retrouverait avec de longues et complexes équations. Néanmoins, une étude théorique plus poussée, incluant directement l'effet de l'advection et du champ magnétique et amenant à une relation de dispersion solaire plus complète, peut faire l'objet de futurs travaux.

– Afin de mieux maîtriser tous les tenants et aboutissants de notre étude, il serait souhaitable de réaliser nous même les programmes de traitement des données brutes, entre autres, "le remapping", "le tracking" et le filtrage ; un meilleur filtrage, par exemple, tel qu'il a été proposé par Giles en 1999 [41], peut être élaboré et appliqué. Les programmes de filtrage et de "tracking",à présent, sont relativement simples à réaliser, comparés à celui du "remapping".

Le programme nous permettant d'obtenir le diagramme "$k-\nu$" et le "ring diagram", quant à lui, peut faire l'objet de quelques améliorations. En effet, une augmentation de la résolution spatiale et temporelle nous aide à obtenir des résultats proches de ceux obtenus par le réseau GONG (voir Fig.(5.19)) par exemple.

Le programme mettant en œuvre la corrélation peut être aussi modifié ou élaboré pour des distances angulaires $\Delta$ plus larges, obtenant ainsi de larges diagrammes "temps-distance" (voir Fig.(5.29)). Comme on peut aussi adopter ce même programme afin d'étudier la sensibilité des temps de parcours à la longitude, ou tout bonnement, faire des corrélations circulaires autour d'un point central, ne privilégiant de ce fait aucune direction de propagation.

– De futurs travaux peuvent porter sur l'inversion des temps de parcours, en peaufinant la technique illustrée dans le chapitre 6 (technique des moindres carrés), ou en utilisant une autre ("the multi-channel deconvolution" [54]), ou bien encore en comparant les deux, ce qui a été fait tout récemment par Junwei Zhao en 2004 lors de sa thèse de doctorat [54]. L'étude des "Kernels" est tout aussi riche, elle peut être l'objet de futures investigations, obtenant de ce fait, notre propre tabulation des "kernels". L'étude seule de ces derniers a été l'objet des thèses de doctorat de A.C. Birch en 2002 [4] et de L. Gizon en 2003 [44]. L'inversion des données qui nous permet de retrouver les vitesses d'Alfven via le modèle qui tient en compte des effets du champ magnétique (voir section (4.3)) est très fastidieuse. Les chercheurs qui concentrent tous leurs efforts là dessus arrivent à certains résultats via certains à priori très spécifiques.

– Comme on peut le constater, la technique héliosismique locale "temps-distances" est très riche et peut être utilisée afin de dévoiler plusieurs aspects du soleil :

  1. De $\tau_{mean}$ ,par exemple et tel que vu précédemment, on peut remonter par inversion aux vitesses du son à différentes profondeurs au soleil, collectant ainsi plusieurs informations sur la physique solaire, telles que : la composition chimique, la densité et la température du milieu traversé. Comme on peut aussi, en considérant le modèle incluant les effets des champs magnétiques, tracer les cartes des flux magnétiques en surface et en profondeur.

  2. De $\delta\tau$ et en remontant aux vitesses d'écoulement en surface et en profondeur, on peut déduire : l'aspect des écoulements autour et dans les taches solaires, la rotation des différentes couches du Soleil à différentes profondeurs, la circulation des flux dans la zone convective, ainsi que l'aspect des super granules à la surface du Soleil, rien qu'en observant les écoulements qui y règnent.

– Comme on peut aussi utiliser cette technique pour observer les variations du diamètre solaire via la méthode "temps-distance" à multi rebonds. Cette technique est actuellement en cours d'étude et d'utilisation par A. Serebryanskiy, chercheur au réseau GONG.

# Annexe A

# Rappels

## A.1 Vitesse de groupe et vitesse de phase

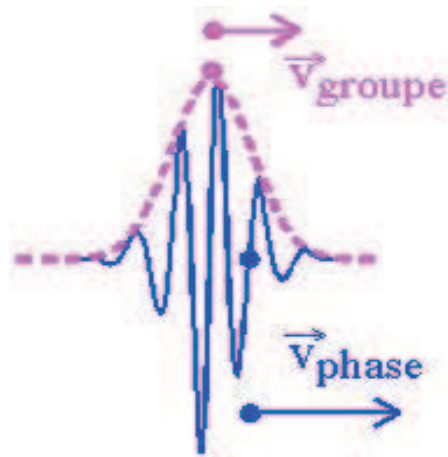

FIG. A.1 – ON PEUT DÉFINIR DEUX VITESSES pour une impulsion quelconque de longueur d'onde finie. **La vitesse de phase** est celle de la propagation des oscillations de l'onde proprement dite. **La vitesse de groupe** est celle de la propagation de l'enveloppe de l'impulsion, que l'on peut mesurer à partir du déplacement de son maximum.

## A.2 Advection

On appelle advection tout mouvement horizontal d'un fluide.

## A.3 La Corrélation

La fonction de corrélation ou bien la CCF (the cross correlation function) peut être définie comme étant la moyenne de tous les états à l'équilibre entre deux positions spatio-temporelles $A(\vec{r},t)$ et $A(\vec{r}',t')$ de la grandeur $A$, avec la position spatiale ici en coordonnées sphériques :

$$\widetilde{c}(\vec{r},t,\vec{r}',t') = \int_0^{+\infty} A(\vec{r}',t')A^*(\vec{r},t)dt \qquad (A.1)$$

Avec $A^*$ le complexe conjugué de $A$.

$\widetilde{c}(\overrightarrow{r}, t, \overrightarrow{r}', t')$ est appelé spectre de puissance ou bien densité spectrale de $A(\overrightarrow{r}, t)$.

Avec : $t' = t + \Delta t$, et $\overrightarrow{r'} = \overrightarrow{r} + \Delta \overrightarrow{r}$.

On définit aussi la fonction d'auto corrélation, (the auto-correlation function) comme la moyenne de tous les états à l'équilibre entre deux positions spatiales $A(\overrightarrow{r}, t)$ et $A(\overrightarrow{r}', t)$ de la grandeur $A$ :

$$c(\overrightarrow{r}, \overrightarrow{r'}, t) = \int_0^{+\infty} A(\overrightarrow{r'}, t) A^*(\overrightarrow{r}, t) dt \tag{A.2}$$

### A.3.1 Théorème de Weinner-Khintchine :

Ce théorème stipule que le spectre de puissance et la fonction d'auto-corrélation sont simplement reliés par la transformée de Fourier.

En effet, et en supposant que notre grandeur $A(\overrightarrow{r}, t)$ découle de la transformée de Fourier d'une autre grandeur $a(\overrightarrow{r}, \omega)$ :

$$A(\overrightarrow{r}, t) = \frac{1}{2\pi} \int_{-\infty}^{+\infty} a(\overrightarrow{r}, \omega) \exp^{i\omega t} d\omega \qquad \& \qquad a(\overrightarrow{r}, \omega) = \int_0^{+\infty} A(\overrightarrow{r}, t) \exp^{-i\omega t} dt \tag{A.3}$$

Le théorème dit que :

$$\widetilde{c}(\overrightarrow{r'}, t', \overrightarrow{r}, t) = \frac{1}{2\pi} \int_{-\infty}^{+\infty} a(\overrightarrow{r'}, \omega) a^*(\overrightarrow{r}, \omega) \exp^{i\omega\tau} d\omega \tag{A.4}$$

Et celà veut juste dire que la corrélation dans le domaine du temps est équivalente à la corrélation de la tarnsformation inverse de Fourier dans le domaine des phases :

$$\widetilde{c}(\overrightarrow{r'}, t', \overrightarrow{r}, t) = \int_0^{+\infty} A(\overrightarrow{r'}, t') A^*(\overrightarrow{r}, t) dt = \frac{1}{2\pi} \int_{-\infty}^{+\infty} a(\overrightarrow{r'}, \omega) a^*(\overrightarrow{r}, \omega) \exp^{i\omega\tau} d\omega \tag{A.5}$$

Et ce théorème nous sera d'une grande utilité dans cette thèse.

## A.4 Transformée de Fourier

### A.4.1 Définition mathématique

Pour une fonction temporelle à valeurs complexes $f(t)$ on peut associer par la transformée de Fourier une fonction $F(\nu)$ à valeurs complexes, que l'on appelle spectre de $f(t)$ et qui contient le poids de chacune des fréquences $\nu$ dans la fonction initiale.

$$F(\nu) = \int_{-\infty}^{\infty} f(t) e^{-2i\pi\nu t} dt \tag{A.6}$$

Bien que les deux fonctions soient toutes les deux de même nature, définies de **R** dans **C**, il est de coutume en physique de les noter avec des variables différentes pour noter la nature différente des quantités codées (ici temps et fréquences) qui sont appelées **réciproques** l'une de l'autre.

Sur cette transformée s'applique : **la linéarité, l'inversibilité, l'égalité de Perseval, la dérivée** ainsi que **le théorème de la convolution**.

### A.4.2 Transformée de Fourier numérique

Toutes les considérations du paragraphe précédant concernent des fonctions mathématiques continues de **R** dans **C**. Bien évidemment, la transformée de Fourier telle qu'elle est utilisée dans un ordinateur possède une définition numérique différente de celle de l'équation (A.6).

La transformée de Fourier numérique (Digital Fourier Transform (DFT) en anglais) possède la définition suivante : pour la série (de $n$ données) $X_j$, transformée de la série $x_k$ on a :

$$X_j = \frac{1}{n}\sum_{k=0}^{n-1} x_k \omega_n^{-jk} \tag{A.7}$$

Où $\omega_n$ est la racine complexe nième de l'unité : $\omega_n = \exp(\frac{2i\pi}{n})$.

La DFT possède des propriétés tout à fait équivalentes à la transformée de Fourier présentée plus haut. En particulier :
– C'est une opération linéaire, et à ce titre, peut se représenter sous forme matricielle, l'élément de la matrice de la DFT étant $\omega_n^{-jk}$.
– La matrice est carrée, les vecteurs $X_j$ et $x_k$ sont de même longeur.
– C'est une opération inversible, la matrice inverse a pour éléments $\omega_n^{jk}$.

### A.4.3 La FFT

L'opération de la DFT, telle qu'elle est présentée dans l'équation (A.7) demande $n$ opérations pour calculer la valeur d'un des points du spectre, le temps de calcul complet du spectre doit donc prendre l'ordre de $n^2$, c'est à dire un temps rapidement prohibitif (abusif) pour les $n$ grands. Par chance, il existe un algorithme extrêmement astucieux (l'algorithme du papillon) qui permet de réduire l'ordre du temps de calcul à $n\log(n)$, c'est à dire un temps beaucoup plus court. L'implémentation de cet algorithme ce fait dans la DFT rapide (Fast Fourier Transform : FFT en anglais). Cependant l'algorithme du papillon nécessite que le nombre de points de la série des données soit égal à une puissance de 2.

### A.4.4 La FT à N dimensions

Quand on fait une expérience de 2D ou de 3D, il faut bien sûr appliquer une transformée de Fourier à 2 ou 3dimensions.

Mathématiquement, la transformée de Fourier se généralise très naturellement aux fonctions de plusieurs variables de $R^N$ dans **C** (ici, un exemple en 3D) :

$$F(\nu_1,\nu_2,\nu_3) = \int\int\int_{R^3} f(t_1,t_2,t_3) e^{-2i\pi(\nu_1 t_1+\nu_2 t_2+\nu_3 t_3)} dt_1 dt_2 dt_3 \tag{A.8}$$

Et s'exprime tout naturellement numériquement :

$$X_{ijk} = \frac{1}{q}\frac{1}{p}\frac{1}{n}\sum_{l_1=0}^{n-1}\sum_{l_2=0}^{p-1}\sum_{l_3=0}^{q-1} x_{l_3 l_2 l_1} \omega_n^{-kl_1} \omega_p^{-jl_2} \omega_q^{-il_3} \tag{A.9}$$

## A.5 Remarques importantes

– La loi de Duvall (équa.(3.60)) et la formule du principe de Fermat (équa.(4.107)) découlent toutes deux du développement de l'équation (3.50), par la méthode WKB [25].
– Les programmes auxquels j'ai contribué dans cette thèse (sur matlab), dès l'obtention des données (dopplerogrammes) en février 2005, sont :

  1. le programme du diagramme $k - \nu$ et du "ring diagram" (Fig.(5.14-5.18)).
  2. Le programme des approximations des paramètres de la fonction de Gabor qui nous permet de retrouver le temps de parcours des ondes (Fig.(5.31-5.34) et Fig.(5.38-5.46)).
  3. Le diagramme temps-distance (voir Fig.(5.26-28)).
  4. Le programme des cartes de puissance "power maps" (voir Fig.(5.12) et Fig.(5.22)).
  5. Le programme qui nous permet de retrouver les oscillations de 5 min directement des images Doppler (voir Fig.(5.7)).
  6. Ainsi que quelques essais de programmations sur le tracking, et le calcul de la CCF, avant l'obtention des deux programmes ; remapping-tracking et filtrage-calcul de la CCF, qui ont été fournis par le réseau GONG en mars 2005.

– L'approximation des données de $\tau_{mean}$ de la latitude 195 pixel -voir Fig.(5.37)-, et de $mean\{\tau_{mean}\}$ de toutes les latitudes -voir Fig.(5.48)-, pour $\Delta = 5°$ à $10°$, par un polynôme d'ordre 3, a été faite via le logiciel " Origin ".
– Les données peuvent être aussi traitées par le logiciel IRAF/GRASP sous linux, ce qui a été fait, mais cela n'a hélas pas abouti à des résultats concluants.
– Cette thèse a été entièrement rédigée par Latex, logiciel qui peut être utilisé ou bien sous windows via winedit, ou bien directement sous linux.

# Annexe B

# Réseau GONG

Le but de conception du réseau étant l'acquisition de données presque ininterrompues, pendant trois ans au minimum, Quinze emplacements différents représentant les six ont été examinés avec de petits moniteurs automatiques solaires. Les emplacements qui constituent le réseau GONG sont :

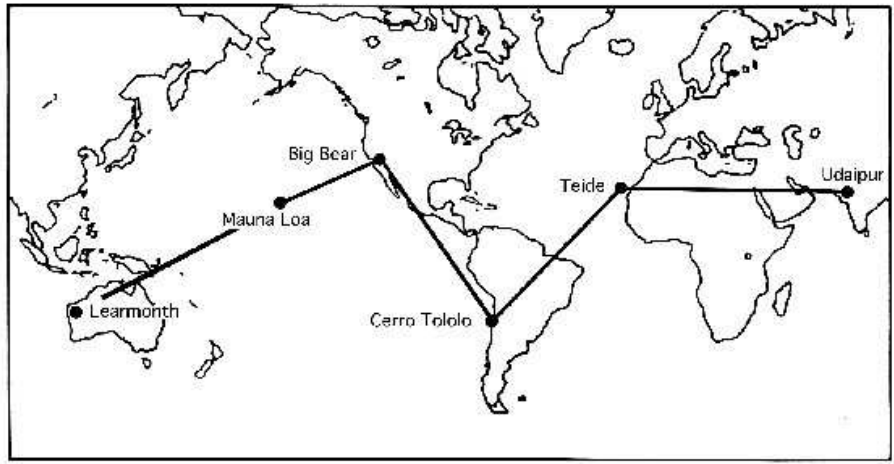

FIG. B.1 – Le réseau GONG à travers le monde [40].

1. **L'Observatoire solaire de Big Bear :** (BBSO) qui est situé sur une petite île près du rivage du nord du lac bigbear dans les montagnes de San Bernardino en Californie méridionale, à 120 kilomètres de Los Angeles Est-Nord-est.

2. **L'Observatoire solaire de Learmonth :** (LSO) qui est situé à une altitude de 10 m sur le rivage occidental du Golf d'Exmouth sur le cap occidental du nord de l'Australie.

3. **L'Observatoire solaire d'Udaipur :** (USO) qui est situé à une altitude de 750 m dans une région de montagne semi-aride de l'Inde occidentale, à mi-chemin entre Delhi et Bombay.

4. **L'Observatorio del Teide :** Cet emplacement d'Instituto Astrofisica de Canarias (IAC) est situé près d'Izaña à une altitude de 2398m sur l'île de Tenerife, dans les îles Canaries.

5. **L'Observatoire de Cerro Tololo Interamerican :** est situé à environ 500 kilomètres au nord de Santiago (Chili) à environ 70 kilomètres à l'est de La Serena, à une altitude de 2200 mètres.

6. **L'Observatoire de Mauna Loa :** (MLSO) qui est situé à une altitude de 3353m sur un gisement de lave sur le flanc nord-ouest du Mauna Loa sur l'île d'Hawaï.

Le rendement mesuré du réseau lors de l'enquête d'emplacement du réseau, était supérieur à 93%. Les données courantes et réelles sont autour de 87%. C'est moins que le rendement d'enquête d'emplacement, parce que des critères plus rigoureux on été employés pour un meilleur échantillonage des données, en plus des visites semi annuelles d'entretien préventif de chaque site qui réduisent le rendement des instruments pendant environ dix jours par visite.

Il y a actuellement 130 différents membres de GONG, de 67 établissements différents et 20 nations, qui sont organisés dans des équipes de recherche traitant des problèmes scientifiques spécifiques. Elles sont en activité dans la définition des possibilités du système aussi bien que dans le développement des outils pour l'analyse des résultats. Les scientifiques participant ont accès aux archives des données, aux logiciels et aux programmes de traitement et d'analyse. Le déploiement des instruments de champ a été accompli au début octobre 1995. L'acquisition de données a accompli les trois ans d'opérations et actuellement, le projet compte continuer d'actionner le réseau pour le reste du cycle solaire.

Le DMAC ( Data Management and Analysis Center) et le DSDS (Data Storage and Distribution System) du réseau GONG sont les deux organismes qui sont chargés de collecter les données, de les traiter et les distribuer.

Toutes les données du réseau GONG sont disponibles sur leur site officiel [40], sauf pour les "merged data" qui ne sont disponibles que depuis le 12/08/2005 sur le même site.

# Bibliographie